\documentclass[twocolumn,times,tighten]{aastex62}
\pdfoutput=1
\usepackage{graphicx}

\newcommand{\etal}{et al.}
\newcommand{\sinszc}{SINS/zC-SINF}

\newcommand{\niiha}{[\ion{N}{2}]/$\rm H\alpha$}
\newcommand{\uvrest}{$(U-V)_{\rm rest}$}
\newcommand{\psfave}{$\rm PSF_{2G,ave}$}
\newcommand{\psfgal}{$\rm PSF_{1G,gal}$}
\newcommand{\niigrad}{$\rm \Delta N2/\Delta r$}
\newcommand{\niigradpix}{$\rm {\Delta N2/\Delta r}_{pix}$}

\shorttitle{SINS/zC-SINF AO survey of 35 $z \sim 2$ galaxies}
\shortauthors{F\"orster Schreiber \etal}

\begin{document}

\title{THE SINS/zC-SINF SURVEY OF $z \sim 2$ GALAXY KINEMATICS:
        SINFONI ADAPTIVE OPTICS-ASSISTED DATA AND KILOPARSEC-SCALE
        EMISSION LINE PROPERTIES\,\footnote{
  Based on observations obtained at the Very Large Telescope
  of the European Southern Observatory, Paranal, Chile
 (ESO Programme IDs 
  075.A-0466, 076.A-0527, 079.A-0341, 080.A-0330, 080.A-0339, 080.A-0635,
  081.B-0568, 081.A-0672, 082.A-0396, 183.A-0781, 087.A-0081, 088.A-0202,
  088.A-0209, 091.A-0126).}
}

\author{N.M. F\"orster Schreiber}
\affiliation{Max-Planck-Institut f\"ur extraterrestrische Physik,
             Giessenbachstrasse, D-85748 Garching, Germany}
\author{A. Renzini}
\affiliation{INAF-Osservatorio Astronomico di Padova,
             Vicolo dell'Osservatorio 5, I-35122 Padova, Italy}
\author{C. Mancini}
\affiliation{Dipartimento di Fisica e Astronomia, Universit\`a di
          Padova, Vicolo dell'Osservatorio 2, I-35122 Padova, Italy}
\affiliation{INAF-Osservatorio Astronomico di Padova,
             Vicolo dell'Osservatorio 5, I-35122 Padova, Italy}
\author{R. Genzel}
\affiliation{Max-Planck-Institut f\"ur extraterrestrische Physik,
             Giessenbachstrasse, D-85748 Garching, Germany}
\affiliation{Department of Physics, Le Conte Hall,
             University of California, Berkeley, CA 94720}
\affiliation{Department of Astronomy, Hearst Field Annex,
             University of California, Berkeley, CA 94720}
\author{N. Bouch\'e}
\affiliation{Institut de Recherche en Astrophysique et Plan\'etologie
            (IRAP), Universit\'e de Toulouse, CNRS, UPS,
             31400 Toulouse, France}
\author{G. Cresci}
\affiliation{INAF-Osservatorio Astrofisico di Arcetri,
             Largo E. Fermi 5, I-50157 Firenze, Italy}
\author{E.K.S. Hicks}
\affiliation{Department of Physics and Astronomy,
             University of Alaska Anchorage, AK 99508-4664, USA}
\author{S.J. Lilly}
\affiliation{Institute of Astronomy, Department of Physiscs,
             Eidgen\"ossische Technische Hochschule,
             Z\"urich, CH-8093, Switzerland}
\author{Y. Peng}
\affiliation{Kavli Institute for Astronomy and Astrophysics,
             Peking University, 100871 Beijing, PR China}
\author{A. Burkert}
\affiliation{Universit\"ats-Sternwarte,
             Ludwig-Maximilians-Universit\"at M\"unchen,
             Scheinerstr. 1, D-81679 Munich, Germany}
\affiliation{Max-Planck-Institut f\"ur extraterrestrische Physik,
             Giessenbachstrasse, D-85748 Garching, Germany}
\author{C.M. Carollo}
\affiliation{Institute of Astronomy, Department of Physiscs,
             Eidgen\"ossische Technische Hochschule,
             Z\"urich, CH-8093, Switzerland}
\author{A. Cimatti}
\affiliation{Department of Physics and Astronomy (DIFA),
             Universit\`a di Bologna,
             Via Gobetti 93/2, I-40129 Bologna, Italy}
\affiliation{INAF-Osservatorio Astrofisico di Arcetri,
             Largo E. Fermi 5, I-50157 Firenze, Italy}
\author{E. Daddi}
\affiliation{Laboratoire AIM, CEA/DSM-CNRS-Universit\'e Paris Diderot,
             IRFU/Service d'Astrophysique, B\^at. 709, CEA Saclay,
             F-91191 Gif-sur-Yvette Cedex, France}
\author{R.I. Davies}
\affiliation{Max-Planck-Institut f\"ur extraterrestrische Physik,
             Giessenbachstrasse, D-85748 Garching, Germany}
\author{S. Genel}
\affiliation{Center for Computational Astrophysics,
             Flatiron Institute, 162 Fifth Avenue,
             New York, NY 10010, USA}
\affiliation{Columbia Astrophysics Laboratory,
             Columbia University, 550 West 120th Street,
             New York, NY 10027, USA}
\author{J.D. Kurk}
\affiliation{Max-Planck-Institut f\"ur extraterrestrische Physik,
             Giessenbachstrasse, D-85748 Garching, Germany}
\author{P. Lang}
\affiliation{Max Planck Institute for Astronomy,
             K\"onigstuhl 17, D-69117 Heidelberg, Germany}
\author{D. Lutz}
\affiliation{Max-Planck-Institut f\"ur extraterrestrische Physik,
             Giessenbachstrasse, D-85748 Garching, Germany}
\author{V. Mainieri}
\affiliation{European Southern Observatory,
             Karl-Schwarzschild-Strasse 2, 85748 Garching, Germany}
\author{H.J. McCracken}
\affiliation{Institut d'Astrophysique de Paris,
             98bis Boulevard Arago, F-75014 Paris, France}
\author{M. Mignoli}
\affiliation{INAF-Osservatorio di Astrofisica e Scienza dello Spazio di
             Bologna, via Piero Gobetti 93/3, I-40129 Bologna, Italy}
\author{T. Naab}
\affiliation{Max-Planck-Institut f\"ur Astrophysik,
             Karl-Schwarzschild-Strasse 1, D-85741 Garching, Germany}
\author{P. Oesch}
\affiliation{Observatoire de Gen\`eve, 51 Ch. des Maillettes,
             1290 Versoix, Switzerland}
\author{L. Pozzetti}
\affiliation{INAF-Osservatorio di Astrofisica e Scienza dello Spazio di
             Bologna, via Piero Gobetti 93/3, I-40129 Bologna, Italy}
\author{M. Scodeggio}
\affiliation{INAF-Istituto di Astrofisica Spaziale e Fisica
             Cosmica Milano, via Bassini 15, 20133 Milano, Italy}
\author{K. Shapiro Griffin}
\affiliation{Northrop Grumman Aerospace Systems,
             San Diego, CA 92150, USA}
\author{A.E. Shapley}
\affiliation{Department of Physics \& Astronomy,
             University of California, Los Angeles,
             430 Portola Plaza, Los Angeles, CA 90095, USA}
\author{A. Sternberg}
\affiliation{Raymond and Beverly Sackler School of
             Physics and Astronomy, Tel Aviv University,
             Ramat Aviv 69978, Israel}
\author{S. Tacchella}
\affiliation{Harvard-Smithsonian Center for Astrophysics,
             60 Garden Street, Cambridge, MA 02138, USA}
\author{L.J. Tacconi}
\affiliation{Max-Planck-Institut f\"ur extraterrestrische Physik,
             Giessenbachstrasse, D-85748 Garching, Germany}
\author{S. Wuyts}
\affiliation{Department of Physics, University of Bath,
             Claverton Down, Bath, BA2 7AY, UK}
\author{G. Zamorani}
\affiliation{INAF-Osservatorio di Astrofisica e Scienza dello Spazio di
             Bologna, via Piero Gobetti 93/3, I-40129 Bologna, Italy}

\begin{abstract}
We present the ``SINS/zC-SINF AO survey'' of 35 star-forming galaxies,
the largest sample with deep adaptive optics-assisted (AO) near-infrared
integral field spectroscopy at $z \sim 2$.  The observations, taken with
SINFONI at the Very Large Telescope, resolve the H$\alpha$ and [\ion{N}{2}]
line emission and kinematics on scales of $\sim$\,1.5 kpc.
In stellar mass, star formation rate, rest-optical colors and size,
the AO sample is representative of its parent seeing-limited sample and
probes the massive
($M_{\star} \sim 2 \times 10^{9} - 3 \times 10^{11}~{\rm M_{\odot}}$),
actively star-forming ($\rm SFR \sim 10 - 600~M_{\odot}\,yr^{-1}$)
part of the $z \sim 2$ galaxy population over a wide range in colors
($(U-V)_{\rm rest} \sim 0.15 - 1.5~{\rm mag}$) and half-light
radii ($R_{\rm e,H} \sim 1 - 8.5~{\rm kpc}$).
The sample overlaps largely with the ``main sequence'' of star-forming
galaxies in the same redshift range to a similar $K_{\rm AB} = 23$
magnitude limit; it has $\sim$\,0.3 dex higher median specific SFR,
$\sim$\,0.1 mag bluer median $(U-V)_{\rm rest}$ color, and $\sim$\,10\%
larger median rest-optical size. 
We describe the observations, data reduction, and extraction of basic
flux and kinematic properties.
With typically $3 - 4$ times higher resolution and $4 - 5$ times longer
integrations (up to 23\,hr) than the seeing-limited datasets of the same
objects, the AO data reveal much more detail in morphology and kinematics.
The now complete AO observations confirm the majority of
kinematically-classified disks and the typically elevated disk
velocity dispersions previously reported based on subsets of the data.
We derive typically flat or slightly negative radial \niiha\ gradients,
with no significant trend with global galaxy properties, kinematic nature,
or the presence of an AGN.  Azimuthal variations in \niiha\ are seen in
several sources and are associated with ionized gas outflows, and
possible more metal-poor star-forming clumps or small companions.
The reduced AO data sets are made publicly available.
\end{abstract}


\section{INTRODUCTION}
         \label{Sect-intro}

The advent of high-throughput near-infrared (near-IR) integral field
unit (IFU) spectrometers mounted on $\rm 8 - 10\,m$-class telescopes in
the past 15 years has made it possible to spatially resolve the kinematics
and distribution of the warm ionized gas in galaxies at redshift $z \ga 1$.
Near-IR IFU surveys have been instrumental in revealing that a significant
proportion ($\ga 50\%$) of massive $z \sim 1 - 3$ star-forming galaxies
(SFGs) are disks, characterized by high intrinsic local velocity dispersions
of $\rm \sigma_{0} \sim 25 - 100~km\,s^{-1}$ and typically irregular, often
clumpy emission-line morphologies
\citep[e.g.,][]
 {FS06,FS09,Gen06,Gen08,Sha08,Cre09,Law09,Wri09,Epi09,Epi12,Jon10b,Man11,
  Gne11a,Gne11b,Wis11,Wis12,Wis15,Swi12b,Swi12a,New13, Bui14,Sto14,Sto16,
  Lee16,Mol17}.
High-resolution rest-UV/optical and H$\alpha$ imaging of large
mass-selected samples out to $z \sim 2.5$ obtained with the
{\em Hubble Space Telescope\/} ({\em HST\/})
confirmed the prevalence of disk-like morphologies among massive SFGs,
with an increasingly important central stellar bulge-like component at
higher galaxy masses and clumpier appearances towards shorter wavelengths
\citep[e.g.,][see also \citealt{Elm07,Elm09,Law12a,Tacc15b,Tacc18}]
             {Wuy11b,Wuy12,Wuy13,Nel13,Nel16b,Lan14}.
Mapping and kinematics of the cold molecular gas in star-forming disks via
low-lying CO line emission also revealed clumpy distributions and elevated
intrinsic local velocity dispersions, showing that these characteristics
are not just a property of the ionized gas layer but of the entire ISM
\citep[e.g.,][]{Tac10,Tac13,Dad10,Swi11,Gen13,Ueb18}.
From surveys of CO line and cold dust continuum emission, it is now well
established that $z \sim 1 - 3$ SFGs have high gas mass fractions of
$\sim 30\% - 50\%$
\citep[see][and references therein]{Tac13,Tac18,Car13,Gen15,Sco16}.
 
A widely invoked theoretical framework to interpret these properties is
that of gas-rich, turbulent disks in which kpc-scale clumps form through
violent disk instabilities, and bulge formation takes place via efficient
secular processes and inward clump migration on timescales
$\rm \la 1~Gyr$
\citep[e.g.,][]
 {Nog99,Imm04a,Imm04b,Bour07,Bour08,Bour14,Gen08,Dek09,Jon10b,
  Genel12,Cev12,Cac12,Hop12,Dek14}.
Theory and numerical simulations indicate that telltale signatures of
physical processes at play appear on scales of $\rm \sim 1~kpc$ and
$\rm \sim 10 - 100~km\,s^{-1}$ or less.  Both high resolution and high
sensitivity are needed to separate different components from each other
in space and in velocity, such as bulges and clumps in disks, perturbations
induced by mergers, disk clumps or bar streaming, and narrow line profiles
tracing the potential well versus broader components associated with gas
outflows.  

Most of the near-IR IFU data at $z \sim 1 - 3$ published to date
comprise seeing-limited observations.  Good near-IR seeing conditions of
$\sim 0\farcs 5$ correspond to a spatial resolution of $\rm \sim 4~kpc$
at $z = 1 - 3$.  Observations aided by adaptive optics (AO) reach typically
$3 - 4$ times higher resolution or $\rm \sim 1~kpc$ (and even better in the
source-plane for rare, strongly lensed sources).  An important practical
limitation for AO-assisted near-IR IFU observations is that targets must
have both
(1) an accurate redshift to ensure the lines of interest fall within
atmospheric windows and away from the numerous bright night sky lines
in the near-IR, and
(2) a sufficiently bright nearby star providing a reference signal for
the AO correction.
Adding to the challenge, observing conditions must be favorable (good
seeing, long atmospheric turbulence coherence time) to achieve significant
image quality improvement with AO, and long integration times are required
for faint distant galaxies.  As a consequence, $z \sim 1 - 3$ near-IR IFU
AO samples are typically small, form a rather heterogeneous collection,
and few objects benefit from very sensitive data.

Using SINFONI at the ESO Very Large Telescope (VLT), we carried out a
substantial effort to collect a sizeable sample of 35 typical massive
$z \sim 2$ SFGs with deep AO observations.
The largest part of the data was obtained through an ESO Large Program
(LP) and a small pilot program, building on two previous major programs:
the ``SINS'' survey of $z \sim 1.5 - 3$ galaxies with SINFONI, and the
``zCOSMOS'' optical spectroscopic survey of $0 < z < 3$ galaxies.
The SINS survey observed a total of 80 galaxies mostly in seeing-limited
mode, and provided the first and largest near-IR IFU sample at $z \sim 2$
at the time \citep[][hereafter \citeauthor*{FS09}]{FS09}.
The galaxies were drawn from spectroscopically-confirmed subsets of various
samples selected by diverse photometric criteria.  While three-quarters of
the sources have a suitable AO reference star, routine AO operations with
SINFONI began roughly mid-way into the five years spanned by the SINS survey
observing campaigns, so that only a handful of targets were initially
followed-up with AO.
To expand and improve on the initial SINFONI$+$AO sample, we collected an
additional 30 $z \sim 2$ sources from the ``zCOSMOS-deep'' spectroscopic
survey, which spans $1.4 < z < 3$ and was conducted with VIMOS at the VLT
\citep{Lil07, Lil09}.  This ``zC-SINF'' sample was selected with uniform
criteria and such that {\em all objects} had an AO reference star and
accurate spectroscopic redshift \citep[][hereafter \citeauthor*{Man11}]{Man11},
capitalizing on the wide area (the central $\sim 1$ square degree of the COSMOS
field) and high sampling rate ($\rm \sim 50\%$ to $B_{\rm AB} = 25~{\rm mag}$)
of the zCOSMOS-Deep survey \citep{Die13}.

Taken together, the SINS and zC-SINF samples comprise a total of 110
massive $z \sim 1 - 3$ SFGs with seeing-limited SINFONI data, 35 of which
were followed-up with sensitive, high-resolution AO-assisted observations
targeting the H$\alpha$ and [\ion{N}{2}] emission lines.  After the initial
no-AO observations of the 30 new zC-SINF objects, the major part of our
SINFONI LP (and its pilot program) was devoted to the AO observations of
26 targets.  AO data of a further nine SINS targets were obtained as part
of complementary SINFONI$+$AO normal, open-time programs.
The underlying strategy for all these programs was to ultimately collect
high quality AO data for a reliable overview of kpc-scale kinematics and
emission line properties over a wide range in stellar mass ($M_{\star}$)
and star formation rate (SFR), and to obtain very deep AO data of a subset
most suitable to investigate particular dynamical/physical processes.

\looseness=-2
This strategy was motivated by a specific set of science goals:
(1) to quantify the fraction and structure of disks and mergers,
(2) to investigate what dynamical processes drive the early evolution
of galaxies,
(3) to determine the origin of the large turbulence in high-$z$ disks,
(4) to constrain the relative importance of gas accretion, and mass
and metals redistribution,
(5) to map the strength of galactic winds on galactic and sub-galactic
(clump) scales, and investigate the mechanisms and energetics of feedback
from star formation and active galactic nuclei (AGN),
(6) to unveil the nature of compact dispersion-dominated objects, and
(7) to shed light on the relative growth of bulges and supermassive
black holes.
The \sinszc\ project motivated three dedicated {\em HST\/} imaging
follow-up programs \citep{FS11a, Tacc15b, Tacc18} to map the stellar
light at a resolution comparable to that of the SINFONI$+$AO data,
enhancing the exploitation of the line emission and kinematics data.

Over the 12 years during which the SINFONI$+$AO data were taken,
key science results addressing the goals listed above based on a subset
of targets and/or of the observations were published in several papers,
which we summarize here. \\
\hphantom{}\hspace{0.8em}$\bullet$
Our first and deepest AO-assisted observations of a $z \sim 2$
galaxy to date revealed for the first time the prototypical properties of
massive high redshift SFGs: a large rotating disk with elevated intrinsic
gas velocity dispersion, several distinct star-forming clumps, and evidence
for a powerful gas outflow driven by AGN activity
\citep{Gen06, Gen08, Gen11, Gen14a, FS14}. \\
\hphantom{}\hspace{0.8em}$\bullet$
The disk dynamics, disk vs.~merger structure of the galaxies, and nature
of dispersion-dominated objects were further explored in several papers
(\citealt{Bou07, Sha08, Cre09}; \citeauthor*{FS09};
\citealt{New13, Gen14a, Gen17, Bur16}). \\
\hphantom{}\hspace{0.8em}$\bullet$
Bulge formation and incipient quenching of star formation in the most
massive galaxies of the sample rapidly became a scientific focus,
together with the nature and evolutionary role of the massive star-forming
clumps \citep{Gen08, Gen11, Gen14a, Tacc15a}. \\
\hphantom{}\hspace{0.8em}$\bullet$
The detection and characterization of the physical properties of the
ubiquitous star formation- and AGN-driven winds in typical $z \sim 2$
massive SFGs on galactic and sub-galactic scales -- as diagnosed by the
H$\alpha$$+$[\ion{N}{2}] ($+$[\ion{S}{2}]) line profiles --
was uniquely enabled by the sensitive SINFONI data
\citep{Sha09, Gen11, Gen14b, New12b, New12a, FS14}. \\
\hphantom{}\hspace{0.8em}$\bullet$
The SINFONI$+$AO data combined with {\em HST\/} imaging further
elucidated the structure of the galaxies through mapping of the
distribution of stellar mass, SFR, and dust extinction
\citep{FS11a, FS11b, Tacc15a, Tacc15b, Tacc18}. \\
\hphantom{}\hspace{0.8em}$\bullet$
Most recently, intriguing evidence was found for the largest galaxies
in the sample having falling rotation curves at large galactocentric radii,
at variance with the typically {\em flat\/} outer disk rotation curves that
are characteristic of local spiral galaxies \citep{Gen17, Lan17}.

We present here the complete sample of 35 \sinszc\ objects with SINFONI$+$AO
observations, the survey strategy, and the characteristics of the data sets.
The paper is organized as follows.
We describe the selection and global properties of the sample in
Section~\ref{Sect-sample}, the observations and data reduction in
Section~\ref{Sect-obsred}, and the extraction of maps and spectra in
Section~\ref{Sect-extract}.
We present the measurements of H$\alpha$ sizes and global surface
brightness distributions in Section~\ref{Sect-struct_meas}, and of
the kinematic properties in Section~\ref{Sect-kin_meas}.
In Section~\ref{Sect-class} we re-visit the nature of the galaxies
(i.e., disks vs. non-disks) based on their kinematics and morphologies,
and in Section~\ref{Sect-metal} we exploit the AO data to constrain
spatial variations in \niiha\ ratio.
The paper is summarized in Section~\ref{Sect-conclu}.
Several technical aspects of the AO observations and data analysis 
can be found in the appendices, including the presentation of the
full data set and a comparison of the results obtained from the
SINFONI AO and no-AO data.
Throughout, we assume a $\Lambda$-dominated cosmology
with $H_{0} = 70\,h_{70}~{\rm km\,s^{-1}\,Mpc^{-1}}$,
$\Omega_{\rm m} = 0.3$, and $\Omega_{\Lambda} = 0.7$.
For this cosmology, 1\arcsec\ corresponds to 8.3~kpc at $z = 2.2$.
Magnitudes are given in the AB photometric system unless otherwise
specified.

\section{\sinszc\ AO SAMPLE}
         \label{Sect-sample}

\looseness=-2
The SINFONI AO targets form a subset of the SINS~and zC-SINF H$\alpha$
samples at $z\sim 2$ initially observed in natural seeing.
Table~\ref{tab-sample} lists the galaxies of the AO sample,
their H$\alpha$ redshifts, $K$-band magnitudes, and several other global
properties.  An exhaustive description of the selection of the parent sample
and the derivation of stellar properties is given by \citeauthor*{FS09} and
\citeauthor*{Man11} (with updated results for six objects using new $H$-band
photometry presented by \citealt{FS11a}).  We refer the reader to these
papers for details and highlight the salient points in this Section.

The stellar mass, visual extinction ($A_{V}$), and SFR ($\rm SFR_{SED}$)
of the galaxies were derived from evolutionary synthesis modeling of
their optical to near-IR broad-band spectral energy distributions (SEDs)
supplemented with mid-IR $\rm 3 - 8~\mu m$ photometry when available.
Model assumptions dominate the uncertainties of the derived stellar
properties.  In general, however, for observed SEDs covering up to at
least near-IR wavelengths, and for similar model evolutionary tracks
and star formation histories, the relative ranking of galaxies in these
properties is fairly robust
(e.g., \citealt{FS04,Sha05,Wuy07,Wuy11a,Mar10}; \citeauthor*{Man11}).
For consistency, we adopted best-fit results for the SINS and zC-SINF
samples obtained with the same \citet{BC03} code, a \citet{Cha03} IMF,
solar metallicity, the \citet{Cal00} reddening law, and constant or
exponentially declining SFRs.\,\footnote{
 Declining star formation histories may not be appropriate for $z \sim 2$
 SFGs \citep[e.g.,][]{Ren09, Mar10}.  The objects modelled with an
 exponentially declining SFR of $e$-folding timescale $\rm \tau = 300~Myr$
 by \citeauthor*{FS09} have $\rm age/\tau \sim 1$ on average with a
 central 68\% distribution between 0.3 and 2, and in most cases cannot
 be statistically distinguished from fits with a constant SFR.
 The resulting typically young ages indicate that most stars formed in
 the recent past of these galaxies, meaning that exponentially declining
 models are actually trying to mimic a secularly increasing SFR.
}

\looseness=-2
Table~\ref{tab-sample} also lists the rest-frame $U-V$ color of the
galaxies and, for objects in fields with mid- or far-IR coverage with
the {\em Spitzer\/}/MIPS or {\em Herschel\/}/PACS instruments
(COSMOS, GOODS-South, and Deep3a)
and detected in at least one of the 24, 70, 100, or $\rm 160\,\mu m$ bands,
the SFR from the emergent rest-UV and IR emission ($\rm SFR_{UV+IR}$).
The rest-frame photometry was computed from interpolation of the observed
photometry using the code EAZY \citep{Bra08}.
The $\rm SFR_{UV+IR}$ estimates were derived following the prescriptions
of \citet{Wuy11a}: the UV contribution was calculated from the rest-frame
2800\,\AA\ luminosity and the IR contribution was derived from the longest
wavelength at which an object is detected.\,\footnote{
 IR photometry was taken from the PEP survey catalog in GOODS-South and
 COSMOS \citep{Lut11, Ber11} with updated MIPS data from the COSMOS2015
 catalog \citep{Lai16} where relevant.
}
Throughout the paper, we use the $\rm SFR_{UV+IR}$ when available and
the $\rm SFR_{SED}$ otherwise.

The \sinszc\ AO sample includes six galaxies with evidence for a Type 2 AGN,
noted in Table~\ref{tab-sample}.
Two of them were known to host an AGN based on the detection of rest-UV
spectral signatures and a 1.4 GHz radio emission excess \citep{Gen06,FS14}.
As discussed by \citet[see also \citealt{New14}]{FS14}, the high-resolution
SINFONI$+$AO H$\alpha + $[\ion{N}{2}] data, supplemented with seeing-limited
maps of [\ion{O}{3}] and H$\beta$ obtained for four objects, further indicate
the presence of an AGN through the line ratios and high-velocity gas outflow
signatures in the central few kpc of all of the AO targets more massive than
$M_{\star} \ga 10^{11}~{\rm M_{\odot}}$, including four cases previously
unidentified as hosting an AGN.

\tabletypesize{\scriptsize}
\begin{deluxetable*}{lllccccrrrrcc}[!ht]
\renewcommand\arraystretch{0.83}
\tablecaption{\sinszc\ AO Survey: Sample Galaxies
              \label{tab-sample}}
\tablecolumns{13}
\tablewidth{0pt}
\setlength{\tabcolsep}{4pt}
\tablehead{
   \multicolumn{13}{c}{} \\[-2.20ex]
   \colhead{Source} &
   \colhead{R.A.} &
   \colhead{Decl.} &
   \colhead{$K_{\rm AB}$\,\tablenotemark{a}} &
   \colhead{$z_{\rm H\alpha}$\,\tablenotemark{b}} &
   \colhead{$M_{\star}$} &
   \colhead{$A_{V}$} &
   \colhead{$\rm SFR_{SED}$} &
   \colhead{$\rm sSFR_{SED}$} &
   \colhead{$\rm SFR_{UV+IR}$\,\tablenotemark{c}} &
   \colhead{$\rm sSFR_{UV+IR}$} &
   \colhead{$(U-V)_{\rm rest}$\,\tablenotemark{d}} &
   \colhead{Notes\,\tablenotemark{e}} \\[-2.7ex]
   \colhead{} &
   \colhead{} &
   \colhead{} &
   \colhead{(mag)} &
   \colhead{} &
   \colhead{($\rm 10^{10}~M_{\odot}$)} &
   \colhead{(mag)} &
   \colhead{($\rm M_{\odot}\,yr^{-1}$)} &
   \colhead{($\rm Gyr^{-1}$)} &
   \colhead{($\rm M_{\odot}\,yr^{-1}$)} &
   \colhead{($\rm Gyr^{-1}$)} &
   \colhead{(mag)} &
   \colhead{}
}
\startdata
Q1623-BX455  & 16:25:51.7 & $+$26:46:55 & 23.41 & 2.4078 &  1.03   &  
  0.6  & 15     & 1.5  & \ldots   & \ldots   &  0.67  & \ldots \\
Q1623-BX502  & 16:25:54.4 & $+$26:44:09 & 23.89 & 2.1557 &  0.23   &
  0.4  & 14     & 6.0  & \ldots   & \ldots   &  0.14  & \ldots \\
Q1623-BX543  & 16:25:57.7 & $+$26:50:09 & 22.39 & 2.5209 &  0.94   &
  0.8  & 145    & 15.5 & \ldots   & \ldots   &  0.24  & \ldots \\
Q1623-BX599  & 16:26:02.6 & $+$26:45:32 & 21.78 & 2.3312 &  5.66   &
  0.4  & 34     & 0.6  & \ldots   & \ldots   &  0.93  & \ldots \\
Q2343-BX389  & 23:46:28.9 & $+$12:47:34 & 22.04 & 2.1724 &  4.12   &
  1.0  & 25     & 0.6  & \ldots   & \ldots   &  1.29  & \ldots \\
Q2343-BX513  & 23:46:11.1 & $+$12:48:32 & 21.95 & 2.1082 &  2.70   &
  0.2  & 10     & 0.4  & \ldots   & \ldots   &  0.93  & \ldots \\
Q2343-BX610  & 23:46:09.4 & $+$12:49:19 & 21.07 & 2.2107 &  10.0   &
  0.8  & 60     & 0.6  & \ldots   & \ldots   &  0.93  & 1      \\
Q2346-BX482  & 23:48:13.0 & $+$00:25:46 & (22.34)\,\tablenotemark{f} & 2.2571 &  1.84   &
  0.8  & 80     & 4.3  & \ldots   & \ldots   &  0.77  & \ldots \\
Deep3a-6004  & 11:25:03.8 & $-$21:45:33 & 20.79 & 2.3871 &  31.6   &
  1.8  & 214    & 0.7  & 355      & 1.1      &  1.52  & 1      \\
Deep3a-6397  & 11:25:10.5 & $-$21:45:06 & 19.96 & 1.5133 &  12.0   &
  2.2  & 563    & 4.7  & 214      & 1.8      &  1.28  & 1      \\
Deep3a-15504 & 11:24:15.6 & $-$21:39:31 & 21.03 & 2.3830 &  10.9   &
  1.0  & 150    & 1.4  & 146      & 1.3      &  0.71  & 1      \\
K20-ID6      & 03:32:29.1 & $-$27:45:21 & 22.13 & 2.2348 &  2.67   &
  1.0  &  45    & 1.7  &  47      & 1.8      &  0.91  & \ldots \\
K20-ID7      & 03:32:29.1 & $-$27:46:29 & 21.47 & 2.2240 &  3.95   &
  1.0  & 112    & 2.8  & 101      & 2.6      &  0.49  & \ldots \\
GMASS-2303   & 03:32:38.9 & $-$27:43:22 & 22.78 & 2.4507 &  0.72   &
  0.4  &  21    & 2.9  & IR-undet & IR-undet &  0.46  & \ldots \\
GMASS-2363   & 03:32:39.4 & $-$27:42:36 & 22.67 & 2.4520 &  2.16   &
  1.2  &  64    & 2.9  & 45       & 2.1      &  0.87  & \ldots \\
GMASS-2540   & 03:32:30.3 & $-$27:42:40 & 21.80 & 1.6146 &  1.89   &
  0.6  &  21    & 1.1  & 32       & 1.7      &  0.96  & \ldots \\
SA12-6339    & 12:05:32.7 & $-$07:23:38 & 22.00 & 2.2971 &  2.57   &
  2.0  & 620    & 24   & \ldots   & \ldots   &  0.61  & \ldots \\
ZC400528     & 09:59:47.6 & $+$01:44:19 & 21.08 & 2.3873 &  11.0   &
  0.9  & 148    & 1.3  & 556      & 5.1      &  0.84  & 1      \\
ZC400569     & 10:01:08.7 & $+$01:44:28 & 20.69 & 2.2405 &  16.1   &
  1.4  & 241    & 1.5  & 239      & 1.5      &  1.29  & 1,2    \\
ZC400569N    & 10:01:08.7 & $+$01:44:28 & \ldots & 2.2432 & 12.9 &
\ldots & 157    & 1.2  & 156      & 1.2      & \ldots & 1,2    \\
ZC401925     & 10:01:01.7	& $+$01:48:38 & 22.74 & 2.1413 &  0.58   &
  0.7  & 47     & 8.2  & IR-undet & IR-undet &  0.40  & \ldots \\
ZC403741     & 10:00:18.4       & $+$01:55:08 & 21.02 & 1.4457 &  4.45   &
  1.6  & 113    & 2.5  & IR-undet & IR-undet &  0.99  & \ldots \\
ZC404221     & 10:01:41.3	& $+$01:56:43 & 22.44 & 2.2199 &  1.57   &
  0.7  &  61    & 3.9  & IR-undet & IR-undet &  0.40  & \ldots \\
ZC405226     & 10:02:19.5	& $+$02:00:18 & 22.33 & 2.2870 &  0.93   &
  1.0  & 117    & 12.6 & IR-undet & IR-undet &  0.56  & \ldots \\
ZC405501     & 09:59:53.7	& $+$02:01:09 & 22.25 & 2.1539 &  0.84   &
  0.9  &  85    & 10.1 & IR-undet & IR-undet &  0.33  & \ldots \\
ZC406690     & 09:58:59.1       & $+$02:05:04 & 20.81 & 2.1950 &  4.14   &
  0.7  & 200    & 4.8  & 296      & 7.2      &  0.56  & \ldots \\
ZC407302     & 09:59:56.0	& $+$02:06:51 & 21.48 & 2.1819 &  2.44   &
  1.3  & 340    & 13.9 & 358      & 14.7     &  0.50  & \ldots \\
ZC407376     & 10:00:45.1	& $+$02:07:05 & 21.79 & 2.1729 &  2.53   &
  1.2  &  89    & 3.5  & 124      & 4.9      &  0.80  & 3      \\
ZC407376S    & 10:00:45.1       & $+$02:07:05 & \ldots & 2.1730 & 1.39   &
\ldots &  67    & 4.8  & 93       & 6.7      & \ldots & 3      \\
ZC407376N    & 10:00:45.2       & $+$02:07:06 & \ldots & 2.1728 & 1.14   &
\ldots &  22    & 1.9  & 31       & 2.7      & \ldots & 3      \\
ZC409985     & 09:59:14.2	& $+$02:15:47 & 22.30 & 2.4569 &  1.61   &
  0.6  &  51    & 3.2  & IR-undet & IR-undet &  0.64  & \ldots  \\
ZC410041     & 10:00:44.3	& $+$02:15:59 & 23.16 & 2.4541 &  0.46   &
  0.6  &  47    & 10.2 & IR-undet & IR-undet &  0.36  & \ldots  \\
ZC410123     & 10:02:06.5	& $+$02:16:16 & 22.80 & 2.1986 &  0.42   &
  0.8  &  59    & 13.9 & IR-undet & IR-undet &  0.31  & \ldots  \\
ZC411737     & 10:00:32.4	& $+$02:21:21 & 22.81 & 2.4442 &  0.34   &
  0.6  &  48    & 13.9 & IR-undet & IR-undet &  0.53  & \ldots  \\
ZC412369     & 10:01:46.9       & $+$02:23:25 & 21.39 & 2.0281 &  2.17   &
  1.0  &  94    & 4.3  & IR-undet & IR-undet &  0.88  & \ldots  \\
ZC413507     & 10:00:24.2       & $+$02:27:41 & 22.52 & 2.4800 &  0.88   &
  1.1  & 111    & 12.6 & IR-undet & IR-undet &  0.57  & \ldots  \\
ZC413597     & 09:59:36.4	& $+$02:27:59 & 22.58 & 2.4502 &  0.75   &
  1.0  &  84    & 11.3 & IR-undet & IR-undet &  0.51  & \ldots  \\
ZC415876     & 10:00:09.4	& $+$02:36:58 & 22.38 & 2.4354 &  0.92   &
  1.0  &  94    & 10.2 & IR-undet & IR-undet &  0.68  & \ldots  \\[0.65ex]
\enddata
\parskip=-2.2ex
\tablecomments
{
The stellar properties are taken from \citet{FS09, FS11a} and \citet{Man11},
and were derived using \citet{BC03} models with a \citet{Cha03} IMF, solar
metallicity, the \citet{Cal00} reddening law, and either constant or
exponentially declining SFRs (see text).
Stellar masses correspond to those in live stars and remnants.
Uncertainties for the stellar properties derived from SEDs are dominated
by systematics from model assumptions; for this sample and the analyses
throughout this paper, we adopt typical uncertainties of
$\rm 0.2~dex$ in $\log(M_{\star})$, $\rm 0.3~mag$ in $A_{V}$, and
$\rm 0.47~dex$ in $\log({\rm SFR_{SED}})$.
}
\tablenotetext{a}
{
Typical uncertainties on the $K$-band magnitudes range from 0.05~mag for the
brightest quartile (mean and median $K_{\rm AB} \approx 21.0~{\rm mag}$) up
to 0.15~mag for the faintest quartile (mean and median
$K_{\rm AB} \approx 22.9~{\rm mag}$).
}
\tablenotetext{b}
{
Spectroscopic redshift based on the source-integrated H$\alpha$ emission.
}
\tablenotetext{c}
{
For sources in fields observed in the mid- or far-IR with
{\em Spitzer\/}/MIPS and/or {\em Herschel\/}/PACS (GOODS-South, COSMOS,
Deep3a) and detected in at least one band, the SFR is also computed from the
emergent rest-frame 2800\,\AA\ and IR luminosities following \citet{Wuy11a}.
Sources undetected with MIPS and PACS are indicated explicitly with
``IR-undet'', to distinguish them from objects in fields without MIPS
and PACS observations.  For the analysis, we adopt typical uncertainties
of $\rm 0.47~dex$ in $\log({\rm SFR_{UV+IR}})$.
}
\tablenotetext{d}
{
Rest-frame $U-V$ colors derived following the procedure described by
\citeauthor{Wuy12}(2012; see also \citealt{Tay09,Bra11}).
Uncertainties are dominated by those of the observed photometry and
systematics from the set of templates used for interpolation, and are
estimated to be typically $\rm 0.1~mag$.
}
\tablenotetext{e}
{
1. These objects have evidence for an AGN (Section~\ref{Sub-sample_AO}).
\hspace{0.2em}
2. ZC400569 has a complex morphology characterized by a brighter northern
source and a southern clumpy extension; the stellar mass and SFR of the
dominant northern component are scaled from the total values according to
the fractions estimated from the stellar mass map (based on {\em HST\/}/WFC3
near-IR imaging) and the observed H$\alpha$ line map, respectively.
\hspace{0.2em}
3. ZC407376 is an interacting pair; the stellar mass and SFR of each
component are scaled from those of the total system according to the
fractions estimated from the {\em HST\/}/WFC3-based stellar mass map
and the observed H$\alpha$ line map, respectively.
}
\tablenotetext{f}
{
For Q2346-BX482, no $K$ band photometry is available; we give here
the $H$ band magnitude from {\em HST\/}/NICMOS imaging through
the F160W filter \citep{FS11a}.
}
\vspace{-4.5ex}
\end{deluxetable*}

\subsection{Selection of the Parent \sinszc\ Seeing-limited Sample}
            \label{Sub-sample_parent}

The SINS H$\alpha$ targets were drawn from large optical spectroscopic surveys
of $z \sim 1.5 - 2.5$ candidates selected by their $U_{\rm n}G{\mathcal R}$
optical colors (``BX/BM'' objects; \citealt{Ste04}), $K$-band magnitudes
(the ``K20'' survey at $K_{\rm s,Vega} < 20~{\rm mag}$, \citealt{Cim02};
the Gemini Deep Deep Survey or ``GDDS'' at $K_{\rm s,Vega} < 20.6~{\rm mag}$,
\citealt{Abr04}), $\rm 4.5\,\mu m$ magnitudes (the Galaxy Mass Assembly
ultra-deep Spectroscopic Survey or ``GMASS'' at
$m_{\rm 4.5\mu m,AB} < 23.0~{\rm mag}$, \citealt{Cim08, Kur13}),
or a combination of $K$-band and $BzK$ color criteria (from the survey
by \citealt{Kon06} of the ``Deep3a'' field of the ESO Imaging Survey,
\citealt{Ren97}).
The SINS BX/BM targets were more specifically taken from the near-IR
long-slit spectroscopic follow-up with NIRSPEC at the Keck~II telescope
of \citet{Erb06b} and the SINS K20 objects comprised all five $z > 2$
sources discussed by \citet{Dad04}.
In addition to a reliable optical redshift, the selection criteria common
to all SINS targets were that H$\alpha$ falls in wavelength intervals of
high atmospheric transmission and away from bright night sky lines, object
visibility during the observing runs, and an observed integrated line flux
of $\rm \ga 5 \times 10^{-17}~erg\,s^{-1}\,cm^{-2}$ in existing long-slit
spectroscopy (for the BX/BM objects) or expected based on the best-fit
$\rm SFR_{SED}$ and $A_{V}$ values.
Fainter sources were discarded because of prohibitively long integration
times for detection but we note that this criterion, applied last in the
selection, removed very few objects.
As argued by \citeauthor*{FS09}, the diversity of criteria employed for
the surveys from which the SINS targets were drawn makes the resulting
sample less biased than any of its constituent subsamples.

The zC-SINF sample was drawn from the spectroscopic zCOSMOS-Deep survey
carried out with VLT/VIMOS \citep{Lil07, Lil09}, covering the central
1~square degree of the 2~square degree COSMOS field \citep{Sco07}.
Reliable optical redshifts ($z_{\rm opt}$) were derived for $\sim 6000$
objects at $1.4 < z < 2.5$, pre-selected at $K_{\rm s} < 23.5~{\rm mag}$
and according to the $BzK$ or BX/BM color criteria.  The zC-SINF SINFONI
targets were then culled among those within $30^{\prime\prime}$ of a
$g_{\rm AB} < 17~{\rm mag}$ star enabling natural guide star AO (with one
exception), a sufficiently secure redshift, avoidance of spectral regions
with bright night sky lines and/or low atmospheric transmission for the
H$\alpha$ line, and a minimum SFR of $\rm \sim 10~M_{\odot}\,yr^{-1}$
(see \citeauthor*{Man11} for details).  This SFR roughly matches the
minimum H$\alpha$ flux criterion for the SINS galaxies at $z = 2$ assuming
$A_{V} = 1~{\rm mag}$ and, as for the SINS sample, this cut removes only a
few objects.  Of the 62 viable SINFONI targets thus culled, the best 30 in
terms of all criteria combined were observed.  Of this zC-SINF sample, 29
belong to the zCOSMOS-Deep subset pre-selected with the $BzK$ criteria
and one is from the ``BX'' subset.

The resulting SINS and zC-SINF H$\alpha$ seeing-limited samples range in
redshift from 1.35 to 2.58, with median $z = 2.22$; $76\%$ of the targets
are at $z > 2$.  The fraction of objects with H$\alpha$ detected in the
seeing-limited data is $84\%$.
Although non-AGN targets were preferentially selected, six of the galaxies
observed were known to host an AGN based on diagnostic features in their
rest-UV spectra, their strong mid-IR excess, or their brightness in X-ray or
radio emission depending on the multi-wavelength and spectroscopic coverage
of the sources in different fields.
As noted above, our SINFONI data added
four cases with evidence for an AGN from their rest-optical
emission line properties.  The different diagnostics available among the
fields prevent a reliable assessment of AGN fraction and biases with respect
to this population for the \sinszc\ sample.  The increase in AGN fraction
with higher galaxy mass is however fully consistent with the trends observed
from much larger multiwavelength and spectroscopic surveys at $z \sim 2$
\citep[e.g.,][]{Red05, Dad07, Bru09, Hai12, Bon12, Man15}.

\subsection{Selection of the \sinszc\ AO Sample}
            \label{Sub-sample_AO}

For the SINFONI$+$AO observations, 17 targets were taken from the
parent SINS seeing-limited survey and 18 from the zC-SINF sample.
They were mostly chosen to lie at $z > 2$ although three objects at
$z \sim 1.5$ were considered.
The redshifts range from 1.45 to 2.52, with a median $z = 2.24$,
approximately the same as for the parent seeing-limited sample.
The goal of the SINFONI LP was to collect AO data of $\sim 25$
sources covering the $M_{\star}-{\rm SFR}$ plane as widely as possible
and obtain a set of ``benchmark objects'' to relate resolved kinematics,
star formation, and physical conditions with global properties of the
massive star-forming galaxy population.  Some of the objects had been
previously observed with AO as part of other programs; for them, the
additional LP observations aimed at a substantial increase in the
sensitivity of the data sets.  In addition to the final 26 AO targets
of the LP, nine other galaxies were observed during the course of various
SINFONI$+$AO programs addressing more specific, though related, science
goals.  The total number of targets results from the trade-off between
sample size and S/N of the data within the constraints set by the
available observing time.

The SINS objects were taken among those with a suitable AO reference
star ($R_{\rm Vega} < 18~{\rm mag}$ and distance from the galaxy
$< 60^{\prime\prime}$) except for one without nearby bright star
that was observed in the so-called ``Seeing Enhancer'' mode (see
Section~\ref{Sub-obs}).  Some preference was given to brighter
SINS objects with a source-integrated H$\alpha$ flux
$\rm \ga 10^{-16}~erg\,s^{-1}\,cm^{-2}$ although three fainter sources
were included (GMASS-2303, GMASS-2363, and K20-ID6, with fluxes of
$\rm (3 - 5) \times 10^{-17}~erg\,s^{-1}\,cm^{-2}$).  In general, for these
SINS AO targets, emphasis was given to larger disk-like systems or to compact
objects with velocity dispersion-dominated kinematics in seeing-limited data.
One object identified as merger from seeing-limited kinematics (K20-ID7;
see \citealt{Sha08}) was also observed.  These choices were driven by the aim
of obtaining very deep kpc-scale resolution IFU data of objects in different
kinematic classes, of unveiling the nature of the most compact sources, and
of studying in detail the dynamical and star formation processes in early
disks \citep[see][]{Gen06,Gen08,Gen11,Gen14a,New12b,New13}.

The choice of zC-SINF objects for the AO observations was more objective
and dictated by the targets being detected in the $\rm 1 - 2\,h$ natural
seeing observations and the goal of ensuring wide $M_{\star} - {\rm SFR}$
coverage in combination with the SINS AO targets.  No explicit H$\alpha$
flux or S/N cut was applied for the zC-SINF AO targets;
their integrated H$\alpha$ fluxes are in the range
$\rm \sim 4 \times 10^{-17} - 6 \times 10^{-16}~erg\,s^{-1}\,cm^{-2}$.
H$\alpha$ morphology and kinematics were not considered, except to exclude
two candidate Type~1 AGN based on their bright and point-like morphologies
in H$\alpha$ and {\em HST\/} broad-band imaging in conjunction with their
X-ray emission and rest-UV spectral features.  For both SINS and zC-SINF
AO targets, no explicit requirement on the averaged H$\alpha$ surface
brightness was used; instead, the integration times were adjusted to
optimize the S/N per resolution element.

\begin{figure*}[!ht]
\begin{center}
\includegraphics[scale=0.71,clip=0,angle=0]{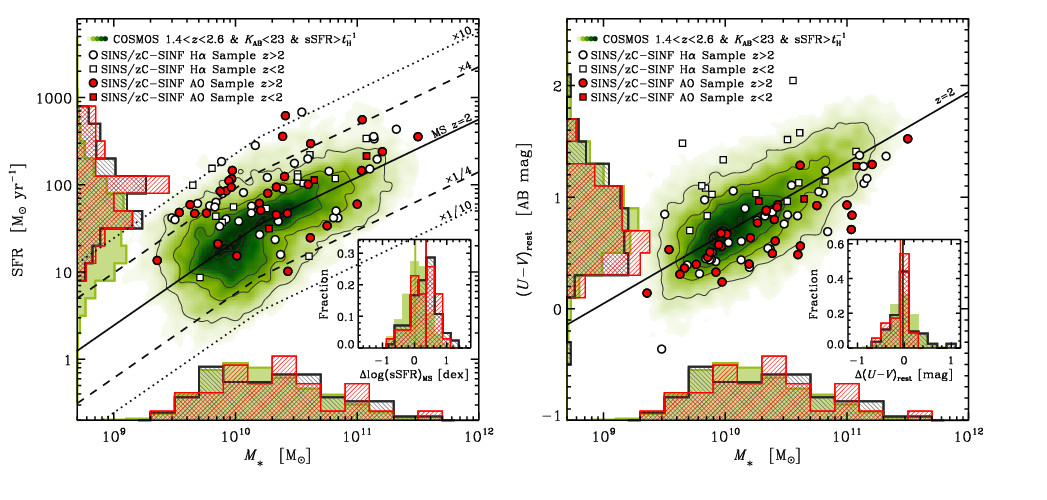}
\end{center}
\vspace{-0.9cm}
\renewcommand\baselinestretch{0.5}
\caption{
\small
Distributions in stellar and color properties of the \sinszc\ AO sample and
of the parent H$\alpha$ sample observed with SINFONI in seeing-limited mode.
The SINFONI samples are compared to $1.4 < z < 2.6$ SFGs in the COSMOS field
at $K_{\rm s,AB} < 23.0~{\rm mag}$ and with inverse specific SFR lower than
the Hubble time at the redshift of each object.
The \sinszc\ galaxies are plotted as large circles ($z > 2$) and squares
($z < 2$), with red symbols showing those observed with AO.
The density distribution of COSMOS SFGs is shown in green colors, with
contours corresponding to fractions of 0.1, 0.3, 0.5, 0.7, 0.9, and 0.98
of the maximum.
The green-filled, grey-hatched, and red-hatched histograms show the
fractional distributions projected onto each axis for the reference SFG
population and the SINFONI seeing-limited and AO samples, respectively
(with median $z$ of 1.73, 2.22, and 2.24).
{\em Left:\/} Stellar mass versus star formation rate.
The solid line indicates the ``main-sequence'' (MS) of SFGs at $z = 2$
from \citet{Whi14}; dashed and dotted lines correspond to offsets in SFR
relative to the MS as labeled in the plot.
The inset shows the distributions of the offsets in specific SFR
(in logarithmic units) of the reference SFG sample, and the \sinszc\
no-AO and AO targets, relative to the MS at the mass and redshift of
each individual source.
{\em Right:\/} Stellar mass versus rest-frame $U-V$ color.
The solid line indicates the mean \uvrest\ as a function of $M_{\star}$
of the reference SFG sample around $z = 2$ (see Section~\ref{Sub-sample_prop}).
The inset shows the distributions of the reference SFG, SINFONI no-AO,
and AO samples in color offset from the $M_{\star}$ vs \uvrest\ relation
accounting for its zero-point evolution ($\propto -0.24\times z$).
The \sinszc\ AO sample covers similar ranges in $M_{\star}$, SFR,
$\rm \Delta\log(sSFR)_{MS}$, and \uvrest\ as the parent no-AO sample,
and emphasizes somewhat the bluer targets at fixed stellar mass.
The \sinszc\ objects have a similar coverage in $M_{\star}$ as the
reference SFG sample but preferentially probe the more actively
star-forming and bluer part of this population.
\label{fig-sampleA}
}
\vspace{2.0ex}
\end{figure*}

\subsection{Global Stellar and Color Properties of the Samples}
            \label{Sub-sample_prop}

In terms of stellar mass, SFR, and color, the \sinszc\ AO sample is
fairly representative of its parent seeing-limited sample.  This is
shown in Figure~\ref{fig-sampleA}, which compares their distributions
in $M_{\star}$ versus SFR and \uvrest\ color diagrams.
To place the samples in a broader context, Figure~\ref{fig-sampleA} also
shows the distributions of the underlying galaxy population in the COSMOS
field at $1.4 < z < 2.6$ to $K_{\rm AB} < 23~{\rm mag}$ --- similar to the
ranges for the \sinszc\ objects.  Since we are here primarily interested
in massive SFGs, objects in the reference sample with a specific
${\rm SFR} < t_{\rm H}^{-1}$ are excluded, where $t_{\rm H}$ is the Hubble
time at the redshift of each source.  This cut removes a small fraction
($17\%$) of all $K_{\rm AB} < 23~{\rm mag}$ sources at $1.4 < z < 2.6$
(and only $12\%$ for the $2 < z < 2.6$ interval encompassing $\ge 80\%$
of the parent \sinszc\ sample and the AO subset).
The stellar properties and colors for the reference sample are taken
from \citet{Wuy11b}, where they were computed using the source catalog
of \citet{Ilb09} supplemented with {\em Spitzer\/}/MIPS and
{\em Herschel\/}/PACS mid- and far-IR photometry \citep{LeF09,Lut11,Ber11}
in a similar fashion as for the \sinszc\ objects.

The \sinszc\ AO and the parent seeing-limited samples cover
nearly identical ranges in stellar mass and star formation rate
($M_{\star} \sim 2 \times 10^{9} - 3 \times 10^{11}~{\rm M_{\odot}}$,
$\rm SFR \sim 10 - 650~M_{\odot}\,yr^{-1}$).
The median (mean) values in stellar mass and SFR are nearly the same,
with $M_{\star} \sim 2 \times 10^{10}~{\rm M_{\odot}}$
($\rm \sim 4 \times 10^{10}~M_{\odot}$) and
$\rm SFR \sim 80~M_{\odot}\,yr^{-1}$
($\rm \sim 125~M_{\odot}\,yr^{-1}$).
The \uvrest\ colors of the AO targets range from $\sim 0.15$ to
$\rm \sim 1.5~mag$, compared to $\sim -0.35$ to $\rm \sim 2.1~mag$
for the full seeing-limited sample.  The median and mean colors of the AO
sample are 0.67 and 0.71~mag, slightly bluer than for the parent no-AO
sample (0.83 and 0.82, respectively).
For comparison, the reference SFG sample has median values of
$M_{\star} \sim 1.5 \times 10^{10}~{\rm M_{\odot}}$,
$\rm SFR \sim 30~M_{\odot}\,yr^{-1}$, and
$(U-V)_{\rm rest} \sim 0.82~{\rm mag}$.
The ranges in properties for the \sinszc\ samples overlap largely with
those of the more general $z \sim 2$ SFG population.  It is however
apparent from Figure~\ref{fig-sampleA} that they probe preferentially
higher specific SFRs and bluer colors at fixed stellar mass, as
illustrated by the distributions in the inset of each panel and
obtained as follows.

The trend in specific SFR can be quantified through the offset relative
to the ``main sequence'' (MS) delineated by the distribution of SFGs in
the $M_{\star} - {\rm SFR}$ plane, $\rm \Delta \log(sSFR)_{MS}$.  For this
purpose, we used the MS parametrization of \citet{Whi14} and calculated
the offsets at the redshift and stellar mass of each individual object.
The \sinszc\ no-AO sample extends down to 0.7~dex below and up to 1.2~dex
above the MS, with a median (mean) $\rm \sim 0.35~dex$ ($\rm \sim 0.30~dex$)
above the MS.
The $\rm \Delta \log(sSFR)_{MS}$ distribution for the AO sample spans a
similar range, from 0.7~dex below up to 1.0~dex above the MS, with a median
(mean) offset of $\rm \sim 0.35~dex$ ($\rm \sim 0.25~dex$).
About $75\%$ of the \sinszc\ seeing-limited and AO targets lie above the MS;
several of them have a sSFR $> 4$ times larger than the MS, and would qualify
as ``starbursts'' according to \citet{Rod11}.
The bias towards higher sSFRs is more important at lower $M_{\star}$
(a result of the target selection introducing an effective lower SFR limit
of $\rm \sim 10~M_{\odot}\,yr^{-1}$).

The trend in colors of the \sinszc\ samples relative to the bulk of
$z \sim 2$ SFGs can be quantified in an analogous manner.  At fixed stellar
mass and redshift, the distribution of the reference SFG sample is well
approximated by a Gaussian.  The mean \uvrest\ color from Gaussian fits
in stellar mass bins within a redshift slice\,\footnote{
 We used a bin width in $\log(M_{\star})$ of 0.3~dex but the result is
 not very sensitive to this choice as long as each bin contains $\ga 1000$
 sources; our approach is inspired by that of \citet{Rod11} in deriving their
 MS relationship in $M_{\star} - {\rm SFR}$.  For the redshift bins, we used
 a width of 0.5.
}
increases as a function of $\log(M_{\star})$ with approximately constant
slope.  We thus defined the locus of SFGs in $M_{\star} - (U-V)_{\rm rest}$
space by fitting a line to these values, with the evolution over $z \sim 1-3$
being mostly in the zero-point.  The color offset relative to relationship
at the mass and redshift of an object is denoted $\Delta(U-V)_{\rm rest}$.
The $\Delta(U-V)_{\rm rest}$ values of the \sinszc\ seeing-limited sample
span from $\rm -0.6~mag$ to $\rm 0.9~mag$, with a small median and mean
offset of $\rm \sim -0.03~mag$.  The AO subset covers a narrower range
from $-0.55$ to $\rm 0.25~mag$, with median and mean offsets of
$\rm \approx -0.10~mag$.
About $65\%$ of the no-AO sample lies on the blue side of the SFG locus in
the $M_{\star} - (U-V)_{\rm rest}$ plane, and $70\%$ of the AO targets do.

\looseness=-2
The preferentially higher sSFRs and bluer rest-optical colors of the
\sinszc\ samples compared to the underlying population of massive SFGs
results from the combination of selection criteria, as extensively
discussed by \citeauthor*{FS09} and \citeauthor*{Man11}.  In particular,
even for a primary selection at near-/mid-IR wavelengths (as for the
majority of our targets), the mandatory $z_{\rm opt}$ means in practice
an optical magnitude cut to ensure sufficient S/N for a reliable redshift
determination.  This limit is typically $\rm \sim 25 - 26~mag$ for the
spectroscopic surveys from which the galaxies were drawn, and implies that
on average, the objects have bluer colors than an unbiased, purely $K$
(or mass) selected sample in the same redshift range.  In addition, the
requirement of minimum H$\alpha$ flux (or SFR) likely emphasizes younger,
less obscured, and more actively star-forming systems but we note again
that few objects were discarded as SINFONI targets by this criterion,
applied last in the selection process.  An examination of the optical
brightness and SFR distributions of photometrically-selected candidate
$z \sim 1.5-2.5$ galaxies and spectroscopically-confirmed subsets in
public datasets for popular deep fields (GOODS, COSMOS) confirms that
biases towards higher sSFRs and bluer colors result largely from the
optical magnitude limits of the spectroscopic surveys.

\subsection{Rest-Optical Size Distribution of the Samples}
            \label{Sub-sample_size}

The strategies for both SINS and zC-SINF relied on the detection of
H$\alpha$ emission in $\rm 1 - 2\,h$ on-source integrations in natural
seeing.  Therefore, while no surface brightness criterion was applied,
there is an implicit bias towards galaxies with at least some regions above
a minimum H$\alpha$ surface brightness.  This effect could plausibly set the
upper envelope in the H$\alpha$ size versus flux distribution of the parent
no-AO sample (\citeauthor*{FS09}; \citeauthor*{Man11}) and, consequently,
the AO sample may be missing the largest objects at any given integrated
line flux.  It is also possible that the most extended objects at a given
optical magnitude are underrepresented because a low surface brightness
may have prevented reliable redshift measurements.  
On the other hand, some of the more subjective choices made in particular
for the AO follow-up of the SINS targets may have favored overall larger
than average objects (see Section~\ref{Sub-sample_AO}).

\begin{figure}[!ht]
\begin{center}
\includegraphics[scale=0.70,clip=1,angle=0]{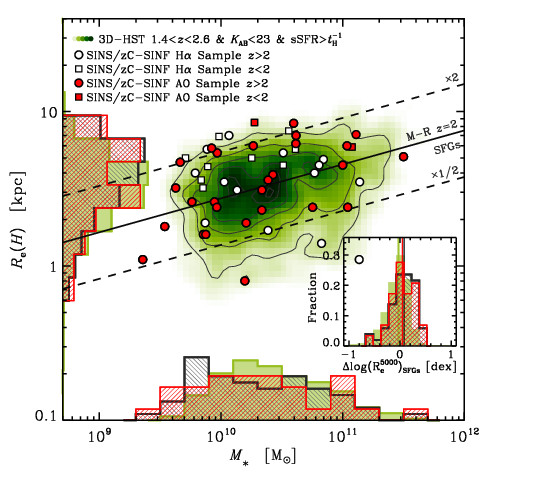}
\end{center}
\vspace{-1.0cm}
\renewcommand\baselinestretch{0.5}
\caption{
\small
Distribution in stellar mass versus size of the \sinszc\ AO sample and
of the parent H$\alpha$ sample observed in seeing-limited
mode.  The SINFONI samples are compared to $1.4 < z < 2.6$ SFGs in the 
CANDELS/3D-HST fields at $K_{\rm s,AB} < 23.0~{\rm mag}$ and with inverse
specific SFR lower than the Hubble time at the redshift of each object.
Symbols and colors are the same as in Figure~\ref{fig-sampleA}.
In the main plot, the sizes are the major axis effective radii derived
from {\em HST\/} $H$-band imaging, available for 51 targets in the full
\sinszc\ sample, and 29 targets from the AO subset.
The solid and dashed lines indicate the mass-size relation for SFGs at
$z = 2$ from \citet{vdW14b}, and the offsets by factors of two around it.
The inset shows the distributions of the offsets in $\log(R_{\rm e})$
from the mass-size relation at the mass and redshift of each individual
source, where the sizes are here corrected to a reference rest-frame
wavelength of 5000\,\AA\ to account for the (small) effects of color
gradients.
The \sinszc\ AO sample covers similar ranges in $R_{\rm e}(H)$ and
$\rm \Delta\log(R_{\rm e}^{5000})_{SFGs}$ as the parent no-AO sample
and the reference population of SFGs, with only a small shift towards
larger sizes at fixed mass (by $\sim 10\%$ on average).
\label{fig-sampleB}
}
\end{figure}

Figure~\ref{fig-sampleB} compares the distributions in $M_{\star}$
versus effective radius $R_{\rm e}$ of our SINFONI samples to that of
the underlying population of SFGs.  We considered rest-frame optical
continuum sizes obtained from $H$-band observations, less prone to the
effects of extinction and localized star formation than the rest-UV or
H$\alpha$. 
The reference SFG population is here taken from the CANDELS/3D-HST surveys
\citep{Gro11, Koe11, Ske14, Mom16}, which provide the largest sample with
sizes measured in the near-IR from {\em HST\/} imaging
\citep{vdW12, vdW14b}.
The same redshift, $K$ magnitude, and sSFR cuts are applied as for the
COSMOS reference sample in the previous subsection.
$H$-band sizes are available for 51 objects in our full \sinszc\ sample
and 29 of the AO targets; we use the measurements presented by \citet{FS11a}
and \citet{Tacc15b}, as well as by \citet{vdW12} for the other objects 
that fall within the CANDELS/3D-HST fields.

\looseness=-2
The size distribution of the \sinszc\ galaxies overlaps well with that of
the underlying SFG population.  The full and AO samples cover the same
range in $R_{\rm e}(H)$ from 0.8 to 8.5~kpc, with the same average of
4.1~kpc, and comparable median values of 4.0 and 3.6~kpc, respectively.
These average and median sizes are modestly larger than those of the
reference SFG sample, which are 3.4 and 3.2~kpc, respectively.

To further quantify possible size biases, we considered the offsets in
effective radius relative to the mass-size relation for SFGs at the redshift
and stellar mass of individual objects, as parametrized by \citet{vdW14b}.
For consistency with this relation, we accounted for the effects of average
color gradients following the prescriptions given by \citeauthor{vdW14b} to
derive effective radii at rest-frame 5000\,\AA.
The offsets, $\rm \Delta \log(R_{\rm e}^{5000})_{SFGs}$, range from $-0.6$
to $\rm 0.4~dex$ for the parent seeing-limited \sinszc\ sample and the AO
subset.  The mean and median offsets are $\rm \sim 0.07~dex$ for the full
sample and $\rm 0.05~dex$ for the AO targets (with a scatter of 0.24~dex,
identical to that of the reference SFG sample used here).
Based on the comparison presented here, there is no substantial size
bias for our \sinszc\ samples relative to the underlying SFG population,
at least for the objects (in majority) with {\em HST\/} $H$-band imaging.

\subsection{Comparison to Other AO Samples}
            \label{Sub-compAOsamples}

To further place our \sinszc\ AO sample in context, we compare it here
with several other published near-IR IFU AO samples at $0.8 < z < 3.5$.
We consider galaxies that were detected with AO observations obtained with
comparable pixel scales of $\rm 50 - 100~mas$, i.e. with data usable for
analysis and well-sampled angular resolution of $0\farcs 3$ or better.
The emission lines targeted were H$\alpha$ and
[\ion{N}{2}]\,$\lambda\lambda 6548,6584$ for $z \la 3$ objects, and H$\beta$
and [\ion{O}{3}]\,$\lambda\lambda 4959,5007$ for those at $z \ga 3$.  Because
such observations are time-consuming, the samples are still fairly modest in
size.

Using SINFONI, \citet[][see also \citealt{Gne11a, Gne11b, Tro14}]{Man09}
obtained deep AO data and detected nine LBGs at $2.9 < z < 3.4$ from the
``Lyman-break galaxies Stellar populations and Dynamics'' (LSD) project.
In the ``Mass Assembly Survey with SINFONI in VVDS'' (MASSIV) survey,
five of the eleven AO targets were observed at the 50~mas pixel scale
and detected, four at $1 < z < 1.6$ selected based on their
[\ion{O}{2}]\,$\lambda 3727$ emission line flux and equivalent width,
and one at $z = 2.24$ selected based on rest-UV properties
\citep[][see also \citealt{Epi12, Que12, Ver12}]{Con12}.
Twenty H$\alpha$-selected galaxies from the HiZELS narrow-band survey
were observed with AO as part of the ``sHiZELS'' project, with six
$z \approx 0.8$, eight $z \approx 1.47$, and six $z \approx 2.23$
galaxies \citep{Swi12b,Swi12a,Mol17}.
Using OSIRIS$+$AO at the Keck~II telescope, \citet{Law09,Law12a}
presented data of thirteen $2 < z < 2.5$ BX-selected objects and one
Lyman-break galaxy (LBG) at $z = 3.32$; three of the BX objects are in
common with our SINS AO subset.  At lower redshifts, \citet{Wri07,Wri09}
studied seven $1.5 < z < 1.7$ BX/BM-selected objects.  \citet{Mie16}
analyzed data of 16 galaxies at $0.8 < z < 1.0$ and one at $z = 1.4$ drawn
from optical spectroscopic surveys in well-studied extragalactic fields in
their ``Intermediate Redshift OSIRIS Chemo-Kinematic Survey'' (IROCKS).
Thirteen SFGs at $1.2 < z < 1.5$ selected from the WiggleZ Dark Energy
Survey based on their [\ion{O}{2}]\,$\lambda 3727$ strength and rest-UV
colors were detected with sufficient S/N for analysis \citep{Wis11, Wis12}.
In addition, AO data obtained with OSIRIS, SINFONI, and NIFS on Gemini North
of a collection of 35 strongly lensed objects at $1.0 < z < 3.7$ have been
published \citep{Sta08,Jon10b,Jon10a,Jon13,Yua11,Yua12,EWuy14b,Liv15,Lee16}.

\begin{figure}[!t]
\begin{center}
\includegraphics[scale=0.55,clip=1,angle=0]{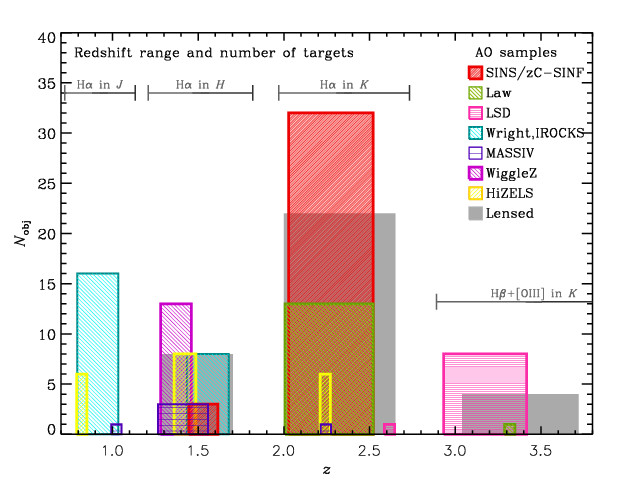}
\end{center}
\vspace{-0.5cm}
\renewcommand\baselinestretch{0.5}
\caption{
\small
Redshift distributions of $z \sim 1 - 3$ SFGs observed and detected with
AO-assisted near-IR IFUs.  The different samples are plotted with color
and filling schemes as labeled in the legend.  The samples are split in
redshift slices corresponding to the near-IR band in which the main lines
of interest were observed (H$\alpha$ for $0.8 < z < 2.8$,
H$\beta$+[\ion{O}{3}] for $2.8 < z < 3.5$), and the bin width spans the
redshift range of the sources within each band.
In addition to \sinszc, the samples shown include the galaxies observed
with SINFONI$+$AO from the LSD \citep{Man09,Tro14}, MASSIV \citep{Con12},
and sHiZELS \citep{Swi12b,Swi12a,Mol17} surveys, and with OSIRIS$+$AO
from the work of \citet{Law09, Law12a}, of \citet{Wri07, Wri09}, the
IROCKS survey \citep{Mie16}, and the WiggleZ sample \citep{Wis11, Wis12}.
A collection of strongly-lensed objects observed with AO using SINFONI,
OSIRIS, and Gemini/NIFS is also plotted
\citep{Sta08,Jon10b,Jon10a,Jon13,Yua11,Yua12,EWuy14b,Liv15,Lee16}.
With 32 of 35 SFGs at $2 < z < 2.7$, \sinszc\ is the largest near-IR
IFU AO sample in this redshift slice.
\label{fig-AOsurveys_z}
}
\end{figure}

\begin{figure}[!t]
\begin{center}
\includegraphics[scale=0.54,clip=1,angle=0]{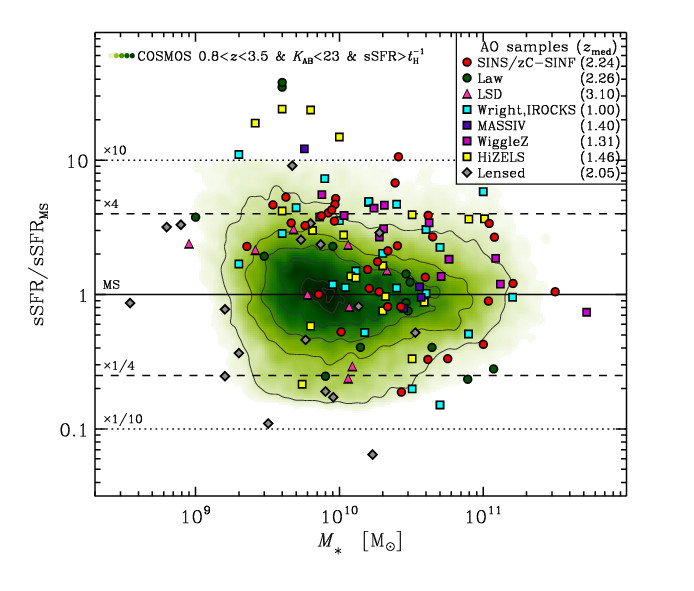}
\end{center}
\vspace{-0.5cm}
\renewcommand\baselinestretch{0.5}
\caption{
\small
Distributions in stellar mass and offset in specific SFR from the MS of
$z \sim 1 - 3$ SFGs obtained with AO-assisted near-IR IFUs.  The samples
are the same as plotted in Figure~\ref{fig-AOsurveys_z}, with symbols and
median redshifts as labeled in the legend (excluding the 17 lensed objects
without an $M_{\star}$ estimate).
The density distribution of COSMOS SFGs at $0.8 < z < 3.5$,
$K_{\rm s,AB} < 23.0~{\rm mag}$ and ${\rm sSFR} > t_{\rm H}^{-1}$
is shown in green colors, with contours marking fractions of 0.1, 0.3,
0.5,0.7,0.9, and 0.98 of the maximum value.
The MS population is fairly well probed by all AO surveys down to
$\sim 10^{10}~{\rm M_{\odot}}$.  Below this mass, non-lensed samples
are biased towards higher specific SFRs while lensed samples reach
more easily the lower-mass population also below the MS.
\label{fig-AOsurveys_MdlSFR}
}
\end{figure}

Including our \sinszc\ survey, these samples provide near-IR AO-assisted
IFU data at kpc-scale resolution, or better for lensed sources, for 152 SFGs
at $0.8 < z < 3.7$ (not counting twice the objects in common between SINS and
\citealt{Law09}). 
Figure~\ref{fig-AOsurveys_z} illustrates the redshift coverage of the AO
samples.  Figure~\ref{fig-AOsurveys_MdlSFR} shows the distribution in
specific SFR relative to the MS as a function of $M_{\star}$, excluding 17
lensed objects without $M_{\star}$ estimate.  The stellar properties are
adjusted to our adopted \citet{Cha03} IMF where relevant.  SFRs from SED
modeling are plotted whenever available for the published samples, otherwise
the extinction-corrected H$\alpha$ or H$\beta$-based SFRs are used.
The reference SFG population in the COSMOS field, defined as in
Section~\ref{Sub-sample_prop} but over $0.8 < z < 3.5$, is also plotted.

With 32 of 35 targets at $2 < z < 2.7$, the \sinszc\ AO survey constitutes
the largest near-IR AO-assisted IFU sample in this redshift slice.  In
the same redshift interval, lensed galaxies form the next largest sample,
comprising 22 objects.  At $M_{\star} \la 10^{10}~{\rm M_{\odot}}$, the
unlensed samples preferentially probe the galaxy population above the MS.
While there is a large overlap in $M_{\star}$ and sSFR ranges, the lensed
samples unsurprisingly extend to the lowest masses and levels of star
formation.  Comparable efforts as those made at $z \sim 2$ would be
desirable at lower and higher redshift to better constrain the evolution
of the kinematic, star formation, and physical properties from near-IR IFU
data with $\rm 1 - 2~kpc$ resolution (or better) across the broad peak epoch
of cosmic star formation activity.
Future progress will benefit from improved statistics in combination with
target selections that will ensure a more complete and uniform coverage
in galaxy parameters.

\section{SINFONI OBSERVATIONS AND DATA REDUCTION}
         \label{Sect-obsred}

\subsection{Observations}
            \label{Sub-obs}

The observations of the \sinszc\ AO sample were carried out with SINFONI
\citep{Eis03, Bon04} mounted at the Cassegrain focus of the VLT UT4
telescope.  The AO correction by the MACAO module \citep{Bon03} was
performed in Natural Guide Star (NGS) or in Laser Guide Star (LGS) mode.
The choice of mode depended on the brightness of the reference star and
its distance to the science target; 18 targets were observed in NGS mode,
and 15 in LGS mode.  For the two other galaxies, we used the LGS
``Seeing Enhancer'' mode (LGS-SE); one of these targets has no suitable
nearby AO reference star and the other was at too high airmass for stable
tip-tilt correction during the observations.  In LGS-SE mode, the higher-order
wavefront distortions are corrected for on the laser spot at the position
of the science target but not the tip-tilt motions that require a reference
star.

For all objects, we selected the intermediate $\rm 50~mas~pixel^{-1}$
scale of SINFONI, with nominal pixel size of $\rm 50 \times 100~mas$
and total field of view $\rm FOV = 3\farcs 2 \times 3\farcs 2$.  This
pixel scale offers the best trade-off between high angular resolution
and surface brightness sensitivity for our faint distant galaxies.
Depending on the redshift of the sources, we used the $K$-band ($z > 2$)
or $H$-band ($z < 2$) grating to map the main emission lines of interest
(H$\alpha$ and the [\ion{N}{2}]\,$\lambda\lambda 6548,6584$ doublet).
The nominal FWHM spectral resolution for the adopted pixel scale is
$R \sim 5090$ in $K$ and $\sim 2730$ in $H$.

The data were collected between 2005 April and 2016 August, as part of
our ESO LP and of other normal open time programs and MPE guaranteed
time observations.
Observing runs were scheduled in both Visitor and Service mode.
The observing conditions were generally good to excellent, with clear
to photometric sky transparency, median seeing of $0\farcs 87$ in the
optical ($0\farcs 55$ in the near-IR at the airmass of the observations),
and median atmospheric coherence time $\rm \tau_{0}$ of 3.5~ms.
Table~\ref{tab-obs} summarizes the observations for each target, with
the band/grating, AO mode, instrument's position angle (PA) on the sky,
total on-source integration time, observing strategy and angular resolution
of the data (see below), and runs during which the data were taken.
Table~\ref{tab-obslog} lists the observing dates for every object.

\tabletypesize{\footnotesize}
\begin{deluxetable*}{llclrcrcl}[!ht]
\renewcommand\arraystretch{0.8}
\tablecaption{Summary of the SINFONI AO Observations
              \label{tab-obs}}
\tablecolumns{9}
\tablewidth{0pt}
\setlength{\tabcolsep}{7pt}
\tablehead{
   \multicolumn{9}{c}{} \\[-2.20ex]
   \colhead{Source} &
   \colhead{$z_{\rm H\alpha}$} &
   \colhead{Band} &
   \colhead{AO mode\,\tablenotemark{a}} &
   \colhead{PA\,\tablenotemark{b}} &
   \colhead{Dithering\,\tablenotemark{c}} &
   \colhead{$t_{\rm int}$\,\tablenotemark{d}} &
   \colhead{PSF FWHM\,\tablenotemark{e}} &
   \colhead{Run ID\,\tablenotemark{f}} \\[-2.5ex]
   \colhead{} &
   \colhead{} &
   \colhead{} &
   \colhead{} &
   \colhead{(degrees)} &
   \colhead{} &
   \colhead{(s)} &
   \colhead{} &
   \colhead{}
}
\startdata
Q1623-BX455  & 2.4078 & $K$  & LGS    & $+90$ & OO & 12600 & $0\farcs 11$ & 087.A-0081 \\
Q1623-BX502  & 2.1557 & $K$  & NGS    &   $0$ & OO & 24000 & $0\farcs 15$ & 075.A-0466, 080.A-0330, 081.B-0568 \\
Q1623-BX543  & 2.5209 & $K$  & NGS    &   $0$ & OO & 25200 & $0\farcs 30$ & 087.A-0081 \\
Q1623-BX599  & 2.3312 & $K$  & LGS    &   $0$ & OO & 37200 & $0\farcs 25$ & 183.A-0781 \\
Q2343-BX389  & 2.1724 & $K$  & LGS-SE & $-45$ & OO & 25200 & $0\farcs 20$ & 183.A-0781 \\
Q2343-BX513  & 2.1082 & $K$  & LGS    &   $0$ & OO & 10800 & $0\farcs 15$ & 087.A-0081 \\
Q2343-BX610  & 2.2107 & $K$  & LGS-SE & $+20$ & OS & 30000 & $0\farcs 24$ & 183.A-0781, 088.A-0202 \\
Q2346-BX482  & 2.2571 & $K$  & LGS    &   $0$ & OS & 44400 & $0\farcs 18$ & 080.A-0330, 080.A-0339, 183.A-0781 \\
Deep3a-6004  & 2.3871 & $K$  & LGS    &   $0$ & OS & 19200 & $0\farcs 16$ & 183.A-0781, 091.A-0126 \\
Deep3a-6397  & 1.5133 & $H$  & LGS    &   $0$ & OS & 30600 & $0\farcs 19$ & 082.A-0396 \\
Deep3a-15504 & 2.3830 & $K$  & NGS    &   $0$ & OO & 82800 & $0\farcs 16$ & 076.A-0527, 183.A-0781 \\
K20-ID6      & 2.2348 & $K$  & LGS    & $+45$ & OO & 13200 & $0\farcs 20$ & 080.A-0339 \\
K20-ID7      & 2.2240 & $K$  & LGS    & $+90$ & OS & 25800 & $0\farcs 15$ & 183.A-0781 \\
GMASS-2303   & 2.4507 & $K$  & LGS    & $+90$ & OO & 15600 & $0\farcs 17$ & 080.A-0635 \\
GMASS-2363   & 2.4520 & $K$  & NGS    &   $0$ & OO & 49200 & $0\farcs 17$ & 080.A-0635 \\
GMASS-2540   & 1.6146 & $H$  & LGS    & $+90$ & OS & 36000 & $0\farcs 17$ & 183.A-0781 \\
SA12-6339    & 2.2971 & $K$  & LGS    & $-20$ & OO & 28200 & $0\farcs 14$ & 087.A-0081 \\
ZC400528     & 2.3873 & $K$  & NGS    &   $0$ & OO & 14400 & $0\farcs 15$ & 183.A-0781 \\
ZC400569     & 2.2405 & $K$  & NGS    & $+30$ & OS & 81000 & $0\farcs 15$ & 183.A-0781, 091.A-0126 \\
ZC401925     & 2.1413 & $K$  & NGS    &   $0$ & OO & 21000 & $0\farcs 25$ & 183.A-0781 \\
ZC403741     & 1.4457 & $H$  & NGS    &   $0$ & OO & 14400 & $0\farcs 16$ & 183.A-0781 \\
ZC404221     & 2.2199 & $K$  & NGS    &   $0$ & OO & 14400 & $0\farcs 20$ & 183.A-0781 \\
ZC405226     & 2.2870 & $K$  & NGS    &   $0$ & OO & 69000 & $0\farcs 24$ & 081.A-0672, 183.A-0781 \\
ZC405501     & 2.1539 & $K$  & NGS    &   $0$ & OO & 20400 & $0\farcs 18$ & 183.A-0781 \\
ZC406690     & 2.1950 & $K$  & NGS    &   $0$ & OS & 36000 & $0\farcs 17$ & 183.A-0781 \\
ZC407302     & 2.1819 & $K$  & LGS    &   $0$ & OO & 68400 & $0\farcs 16$ & 079.A-0341, 183.A-0781, 088.A-0209 \\
ZC407376     & 2.1729 & $K$  & NGS    &   $0$ & OO & 21600 & $0\farcs 22$ & 183.A-0781 \\
ZC409985     & 2.4569 & $K$  & NGS    &   $0$ & OO & 18000 & $0\farcs 13$ & 081.A-0672 \\
ZC410041     & 2.4541 & $K$  & NGS    &   $0$ & OO & 21600 & $0\farcs 16$ & 183.A-0781 \\
ZC410123     & 2.1986 & $K$  & LGS    &   $0$ & OO &  7200 & $0\farcs 18$ & 081.A-0672 \\
ZC411737     & 2.4442 & $K$  & LGS    &   $0$ & OO & 15000 & $0\farcs 19$ & 183.A-0781 \\
ZC412369     & 2.0281 & $K$  & LGS    &   $0$ & OO & 14400 & $0\farcs 15$ & 183.A-0781 \\
ZC413507     & 2.4800 & $K$  & NGS    &   $0$ & OO & 29400 & $0\farcs 14$ & 183.A-0781 \\
ZC413597     & 2.4502 & $K$  & NGS    & $+90$ & OO & 21000 & $0\farcs 17$ & 183.A-0781 \\
ZC415876     & 2.4354 & $K$  & NGS    & $+90$ & OO & 21000 & $0\farcs 14$ & 183.A-0781 \\[0.65ex]
\enddata
\parskip=-1.8ex
\tablecomments
{
The SINFONI AO data for all sources were taken with the
intermediate $\rm 50~mas~pixel^{-1}$ scale and nominal
FOV of $3\farcs 2 \times 3\farcs 2$.
}
\tablenotetext{a}
{
AO mode used, either with a Natural Guide Star (NGS) or Laser Guide
Star (LGS).  For two objects, the LGS observations were carried out
in ``Seeing Enhancer'' (SE) mode, with higher-order corrections
performed on the laser spot but without correction for tip-tilt
motions on a reference star.
}
\tablenotetext{b}
{
Position angle of SINFONI for the AO observations, in degrees East
of North.
}
\tablenotetext{c}
{
Observing strategy followed for adequate sampling of the background.
``OO'' refers to on-source dithering in which the object is present
in all individual exposures and  ``OS'' refers to the scheme applied
for the largest galaxies, in which the background emission is sampled
away from the target in 1/3 of the exposures (see Section~\ref{Sub-obs}).
}
\tablenotetext{d}
{
Total on-source integration time of the combined data sets used for
the analysis, excluding low-quality exposures for some sources (e.g.,
taken under poorer observing conditions leading to poorer AO correction).
}
\tablenotetext{e}
{
The PSF FWHM corresponds to the effective resolution of all
observations for a given object.  It is estimated from the combined
data of the acquisition star taken regularly during the observations
of a science target, fitting a circularly symmetric 2D Gaussian profile
(i.e., it is the \psfgal\ defined in Section~\ref{Sub-PSF}).
}
\tablenotetext{f}
{
ESO program number under which the data were taken (see Table~\ref{tab-obslog}
for more details).
}
\parskip=0ex
\vspace{-4.5ex}
\end{deluxetable*}

\tabletypesize{\footnotesize}
\begin{deluxetable*}{lll}[!ht]
\renewcommand\arraystretch{0.8}
\tablecaption{Log of the Observations
              \label{tab-obslog}}
\tablecolumns{3}
\tablewidth{0pt}
\tablehead{
   \multicolumn{3}{c}{} \\[-2.20ex]
   \colhead{Source} &
   \colhead{Run ID\,\tablenotemark{a}} &
   \colhead{Observing dates\,\tablenotemark{b}}
}
\startdata
Q1623-BX455  & 087.A-0081(B) & 2011 Jul 24 \\
Q1623-BX502  & 075.A-0466(A) & 2005 Apr 07 \\
\ldots             & 080.A-0330(B) & 2008 Mar 25,26 \\
\ldots             & 081.B-0568(A) & 2008 Apr 04 \\
Q1623-BX543  & 087.A-0081(A) & 2011 Apr 28,29,30 \\
Q1623-BX599  & 183.A-0781(E) & 2010 Apr 11,12; 2013 May 03; 2013 Aug 15; 2014 Jul 13; 2015 Jun 06,15; 2016 Jun 22; 2016 Jul 27,29,31; 2016 Aug 01\\
Q2343-BX389  & 183.A-0781(E) & 2012 Sep 10,11,12; 2012 Oct 09; 2012 Dec 12 \\
Q2343-BX513  & 087.A-0081(B) & 2011 Jul 24 \\
Q2343-BX610  & 183.A-0781(D) & 2011 Sep 27; 2012 Aug 09,10,13 \\
\ldots             & 088.A-0202(A) & 2011 Oct 24,26 \\
Q2346-BX482  & 080.A-0330(A) & 2007 Oct 28,29 \\
\ldots             & 080.A-0339(A) & 2007 Oct 31; 2008 Jul 27,28,30 \\
\ldots             & 183.A-0781(D) & 2009 Nov 10,11,12,17; 2010 Sep 07,08; 2010 Nov 04; 2011 Nov 25 \\
Deep3a-6004  & 183.A-0781(B) & 2010 Jan 09,13,14; 2010 Mar 15; 2011 Jan 01; 2011 Mar 26,27,30; 2012 Mar 19 \\
\ldots             & 091.A-0126(A) & 2013 Apr 04 \\
Deep3a-6397  & 082.A-0396(B) & 2008 Dec 23; 2009 Feb 21; 2009 Mar 21,24; 2009 Jun 18; 2010 Jan 08,10; 2010 Feb 09; 2010 Mar 09 \\
Deep3a-15504 & 076.A-0527(B) & 2006 Mar 19,20 \\
\ldots             & 183.A-0781(B) & 2009 Apr 29,30; 2009 May 15; 2009 Jun 15; 2010 Apr 01; 2011 Mar 26,31; 2011 Apr 12,19 \\
\ldots             & 183.A-0781(G) & 2010 Feb 10,11,12; 2010 Mar 04,08 \\
K20-ID6      & 080.A-0339(A) & 2008 Jan 01; 2008 Dec 11,20,22 \\
K20-ID7      & 183.A-0781(F) & 2012 Sep 11; 2013 Oct 27,30; 2013 Nov 05,12,21,22 \\
GMASS-2303   & 080.A-0635(A) & 2007 Nov 13,14 \\
GMASS-2363   & 080.A-0635(B) & 2007 Dec 17; 2008 Jan 19; 2008 Feb 03,08,09,12,13,14,21,22 \\
GMASS-2540   & 183.A-0781(F) & 2010 Nov 01; 2010 Dec 05; 2011 Oct 27; 2011 Nov 27,30; 2011 Dec 01,19,21 \\
SA12-6339    & 087.A-0081(A) & 2011 Apr 29,30 \\
ZC400528     & 183.A-0781(B) & 2010 Jan 15,24,25 \\
ZC400569     & 183.A-0781(B) & 2010 Feb 12; 2010 Mar 09,14; 2012 Jan 18; 2012 Feb 15,25,26 \\
\ldots             & 183.A-0781(I) & 2012 Jan 31; 2012 Feb 16,22,23 \\
\ldots             & 091.A-0126(A) & 2013 Apr 04,05,06,07,08,09 \\
ZC401925     & 183.A-0781(I) & 2011 Jan 25; 2011 Apr 04,17; 2011 Dec 18,28 \\
ZC403741     & 183.A-0781(B) & 2009 May 15,24,25,27 \\
ZC404221     & 183.A-0781(G) & 2011 Feb 14 \\
\ldots             & 183.A-0781(H) & 2012 Dec 14; 2013 Jan 07 \\
ZC405226     & 081.A-0672(B) & 2008 Jun 01,04,05; 2009 Jan 08,20 \\
\ldots             & 183.A-0781(G) & 2011 Feb 12; 2012 Jan 06,07,16,17 \\
\ldots             & 183.A-0781(H) & 2013 Jan 12,15,17,24 \\
ZC405501     & 183.A-0781(G) & 2011 Jan 06,17,18 \\
ZC406690     & 183.A-0781(G) & 2010 Apr 17; 2010 May 24; 2010 Nov 30 \\
\ldots             & 183.A-0781(H) & 2010 Dec 07,10,29,30; 2011 Jan 02,03,17 \\
ZC407302     & 079.A-0341(A) & 2007 Apr 17,22 \\
\ldots             & 183.A-0781(B) & 2009 Apr 17; 2010 Jan 08,12; 2010 Feb 09 \\
\ldots             & 088.A-0209(A) & 2012 Mar 14,15,16,17 \\
ZC407376     & 183.A-0781(H) & 2012 Jan 17,23,27; 2012 Feb 20,22 \\
ZC409985     & 081.A-0672(B) & 2008 May 09,24,26,31 \\
ZC410041     & 183.A-0781(H) & 2011 Jan 16; 2011 Feb 10,23; 2011 May 25 \\
ZC410123     & 081.A-0672(B) & 2009 Mar 22 \\
ZC411737     & 183.A-0781(E) & 2011 Feb 26; 2011 Mar 03; 2011 Dec 19 \\
ZC412369     & 183.A-0781(E) & 2010 Dec 03,06; 2011 Jan 04,27 \\
ZC413507     & 183.A-0781(I) & 2011 Mar 28; 2011 Dec 29; 2012 Mar 03; 2012 May 19; 2013 Jan 08,17 \\
ZC413597     & 183.A-0781(I) & 2011 Jan 04,27; 2011 Mar 04,05 \\
ZC415876     & 183.A-0781(I) & 2011 Jan 06,19,29 \\[0.65ex]
\enddata
\parskip=-1.8ex
\tablenotetext{a}
{
ESO observing run under which the data were taken.
}
\tablenotetext{b}
{
The date corresponds to that when the observing night started.
}
\vspace{-3.0ex}
\end{deluxetable*}

The observations were carried out in series of ``observing blocks''
(OBs) consisting typically of six exposures.  For the majority of the 
targets, we adopted an efficient ``on-source dithering'' strategy with
typical nod throws between successive exposures of about half the SINFONI
FOV so as to image the source in all frames, and jitter box widths of
about one-tenth the FOV to minimize the number of redundant positions on
the detector array.  Integer$+$fractional pixel offsets ensured adequate
sampling of the AO PSF, which has a FWHM typically $1 - 2$ times the
largest size of the nominal rectangular pixels.
For eight of the largest sources, with H$\alpha$ emission extending over
$\ga 1\farcs 5$, we followed an ``offsets-to-sky'' strategy where the
exposures for background subtraction were taken at positions generally
$20^{\prime\prime}$ away from the target.  In this scheme, the telescope
pointing was alternated between the object (``O'') and adjacent sky regions
(``S'') empty of sources in an ``O-S-O-O-S-O'' pattern for each OB.  The
pointing on the object and sky positions was also varied by about one-tenth
of the FOV, ensuring an adequate sampling of the sky signal subtracted from
the two object frames sharing the same sky frame.  The deepest area covered
by all dithered exposures of a galaxy is hereafter referred to as the
``effective FOV.''

The individual exposure times were of 600\,s, optimizing the quality
of the background subtraction while remaining in the background-limited
regime in the wavelength regions around the emission lines of interest.
The total on-source integration times ranged from 2\,hr to 23\,hr, with
an average of 8.1\,hr and median of 6.0\,hr.
Our science requirements and the H$\alpha$ properties of the sources
(based on the initial seeing-limited data) dictated on-source integration
times of typically $\rm \geq 4\,hr$ to reach an average $\rm S/N \sim 5$
per resolution element.
Only four targets were observed for shorter times.
Long integration times of $\rm 10 - 23\,hr$ (typically around 15\,hr)
for nine targets were driven by specific science goals requiring higher S/N
to reliably distinguish low contrast features (in flux, in velocity)
indicative of gas outflows, fainter clumps, or perturbations in the
kinematics induced by non-axisymmetric structures
\citep[see, e.g.,][]{Gen06,Gen11,Gen17,New12b,FS14}.

Exposures of the acquisition stars used for the AO correction and for blind
offsetting to the galaxies were taken to monitor the angular resolution and
positional accuracy throughout the observations.  For flux calibration and
atmospheric transmission correction, B type stars and early-G dwarfs with
near-IR magnitudes in the range $\rm \sim 7 - 10~mag$ were observed.
The telluric standards data were taken every night, as close in time and
airmass as possible to each target observed during the night.  Acquisition
stars and telluric standards were always observed with AO and the same
instrument setup as for the science objects.

\bigskip
\subsection{SINFONI Data Reduction}
            \label{Sub-datared}

We reduced the data using the software package {\em SPRED} developed
for SINFONI \citep{Sch04, Abu06}, complemented with additional custom
routines to optimize the reduction for faint high-redshift targets.
We followed the same steps as described by \citeauthor*{FS09}, to which
we refer for details.  Key features of the procedure include the method
developed by \citet{Dav07} for accurate wavelength registration and
background subtraction, which helps to reduce residuals from the night
sky emission lines.  Each individual exposure was background-subtracted,
flat-fielded, wavelength-calibrated, distortion-corrected, and
reconstructed into a three-dimensional cube with spatial sampling
of $\rm 50 \times 50~mas\,pixel^{-1}$.  These pre-processed cubes
were then corrected for atmospheric transmission, flux-calibrated,
spatially aligned, and co-averaged (with a $2.5\,\sigma$ clipping
algorithm) to produce the final reduced cube of a given science target.
A ``noise cube'' was also generated, containing the rms deviation over
all combined cubes of each pixel corresponding to the same spatial and
spectral coordinate (see Section 4.2 and Appendix C of \citeauthor*{FS09}).

The data of the telluric standard stars and the acquisition (i.e. PSF
calibration) stars were reduced in a similar way as the science data.
Flux calibration of the galaxies' data was performed on a night-by-night
basis using the broad-band magnitudes of the telluric standards.  The
integrated spectrum of the telluric standard stars was used to correct
the science data for atmospheric transmission.
The reduced cubes of the acquisition star's data associated with all OBs
of a target were co-averaged (with $\sigma$-clipping) into a final PSF cube.
Broad-band images were made by averaging together all wavelength channels
of the reduced cubes to create the final PSF image.

As described in detail in Appendix~C of \citeauthor*{FS09}, the noise
behavior of our SINFONI data is consistent with being Gaussian for a given
aperture size and spectral channel, but scales with aperture size faster
than for pure Gaussian noise due to correlations present in the data.
The noise scaling was derived for each galaxy individually from the
analysis of the fluctuations of the fluxes in apertures placed randomly
over regions empty of source emission within the effective FOV, and with
a range of sizes.  This analysis was carried out for each spectral channel
separately, and represents the average behaviour over the FOV at each
wavelength.  For all measurements in apertures larger than a pixel,
the noise spectrum was calculated from the average pixel-to-pixel rms
multiplied by the aperture scaling factor derived from the noise
analysis.

\subsection{Effective Angular Resolution}
            \label{Sub-PSF}

\begin{figure}[!t]
\begin{center}
\includegraphics[scale=0.50,clip=1,angle=0]{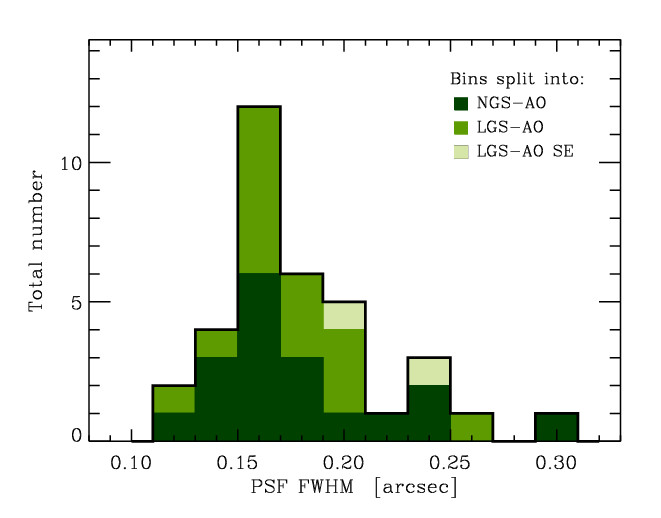}
\end{center}
\vspace{-0.6cm}
\renewcommand\baselinestretch{0.5}
\caption{
\small
Distribution of the PSF FWHMs for the \sinszc\ AO sample.
Each histogram bin shows in different colors the number of galaxies observed
in NGS, LGS, and LGS-SE AO modes added vertically on top of each other (dark,
middle, and light green, respectively).
The FWHMs correspond to the single-component 2D circular
Gaussian fit to the PSF image associated with each galaxy.
\label{fig-PSFfwhm}
}
\end{figure}

We characterized the effective angular resolution of our AO-assisted data
by fitting a two-dimensional (2D) Gaussian profile to the final PSF image
associated with the combined OBs of each galaxy.  As reported in
Table~\ref{tab-obs}, the FWHMs of the best-fit circular Gaussians range
from $0\farcs 11$ to $0\farcs 30$, with a mean of $0\farcs 18$ and a
median of $0\farcs 17$.
Figure~\ref{fig-PSFfwhm} shows the distribution of the PSF FWHMs.  There is
overall no significant difference in FWHMs between the PSFs observed in NGS
or LGS mode (the mean and median are identical within $0\farcs 01$).
Perhaps more remarkably, the resolution for the two LGS-SE data sets,
taken under similar typical (median) seeing, coherence time, and airmass
conditions as the NGS and LGS data sets, is very comparable with PSF FWHMs
of $0\farcs 20$ and $0\farcs 24$.

Due to a combination of the instrument's optics, the nominal rectangular
pixel shape, and anisoplanatism \citep[e.g.][]{Cre05, Dav12}, the PSF
deviates slightly from circularity.
To quantify these deviations, we also fitted 2D elliptical Gaussians.
The major axis FWHM, minor-to-major axis ratio, and PA of these fits
are given in Appendix~\ref{App-AOperf} (Table~\ref{tab-AOperf}).
The major axis FWHMs are in the range $0\farcs 13 - 0\farcs 33$,
with mean and median of $0\farcs 19$ and $0\farcs 18$, respectively.
On average, the axis ratios are 0.88 (median of 0.89), implying that
a circular Gaussian provides a good approximation of the effective
PSF shape.

Since the PSF calibrations were not obtained simultaneously with the science
data, and the AO correction for the objects is not fully on-axis (except in
LGS-SE mode), the PSF characteristics represent approximately the effective
angular resolution of the data sets.  Inspection of the individual exposures
of the stars indicate typical OB-to-OB variations of $\sim 30\%$ in PSF
FWHM and $\sim 10\%$ in axis ratio for our AO data sets.
Examination of the light profile of the most compact sources and of the
smallest substructures seen in our galaxies' data (bright clumps, nuclei)
suggest the degradation in FWHM resolution due to tilt anisoplanatism is
modest and within the typical OB-to-OB variations.

In Appendix~\ref{App-AOperf}, we analyze the PSF properties in more detail
based on higher S/N co-averaged images of the individual PSFs.  The high
S/N profiles better reveal the broad halo from the uncorrected seeing
underneath the AO-corrected narrow core component.
The peak amplitude of the narrow core is five times higher
than that of the broad halo, its width is three times narrower, and
it contains about $40\%$ of the total flux.  The best-fit parameters of
the PSF core component are very close to those derived from a single 2D
Gaussian fit to the PSFs associated with individual galaxies.
Although significant in terms of photometry, the broad halo only
dominates the total enclosed flux at radii larger than $0\farcs 3$.
Despite the limitations on AO performance stemming from practicalities when
observing faint high-$z$ galaxies (i.e., the choice of pixel scale driven by
the trade-off between resolution and surface brightness sensitivity, and the
brightness of AO reference stars around the scientifically selected targets),
the gain from the PSF core/halo contrast for our data is important in terms
of revealing substructure on physical scales as small as $\rm 1 - 2~kpc$.

We further examined from the PSF images the impact of the reference star
brightness, seeing, atmospheric coherence time $\tau_{0}$, and airmass on
the AO performance.  The analysis is presented in Appendix~\ref{App-AOperf},
where the star properties and average conditions during the PSF observations
associated with each target are also given.  The AO star brightness influenced
most importantly the angular resolution (and Strehl ratio).  The airmass
(affecting the actual seeing at a target's elevation) and the coherence
time played a noticeable but lesser role.  The AO performance for our
observations is weakly, if at all, coupled to the seeing as measured in
the optical at zenith from the DIMM telescope at the VLT.

While the PSFs associated with individual galaxies more closely track
variations in angular resolution among the objects and are appropriate
to characterize substructure within them, the scatter in best-fit FWHMs
is $\sim 25\%$ of the average, somewhat smaller than the typical
OB-to-OB variations of $30\%$.  For the quantitative analyses presented
in this paper, we adopted the results derived with the double-Gaussian
model of the average PSF to account for the extended and photometrically
important halo, and compared with results obtained with the individual
single-Gaussian PSFs where relevant.  For conciseness, the notation
\psfave\ and \psfgal, respectively, indicates the choice of PSF.

\subsection{Effective Spectral Resolution}
            \label{Sub-LSF}

We determined the resulting spectral resolution of the reduced data
based on night sky line measurements, described in Appendix~\ref{App-LSF}.
The empirical line spread function (LSF) is well approximated by a Gaussian
profile and fits give an effective spectral resolution corresponding to a
velocity FWHM of about $\rm 85~km\,s^{-1}$ across the $K$ band and
$\rm 120~km\,s^{-1}$ across the $H$ band for the SINFONI
$\rm 50~mas\,pixel^{-1}$ scale.
The LSFs show a small excess in low-amplitude wings, which however contain
only about 12\% and 5\% of the total flux in $K$ and $H$, respectively.

\section{EXTRACTION OF EMISSION LINE AND KINEMATIC MAPS AND PROFILES}
         \label{Sect-extract}

In this section, we describe the methodology followed to extract the
H$\alpha$ and [\ion{N}{2}] emission line flux and kinematic maps,
position-velocity diagrams, axis and radial profiles, and integrated
spectra.  The main products of the extraction are presented in
Figures~\ref{fig-SBHaMSFR}, \ref{fig-VFsMSFR}, and in the series of
figures described in more detail in Appendices~\ref{App-FullDataSets},
\ref{App-AOnoAO}, and \ref{App-metal}.  The source-integrated
spectral properties are reported in Table~\ref{tab-Hameas}.

\subsection{Fitting Methodology}
            \label{Sub-meth}

We extracted the emission line properties and kinematics following procedures
similar to those described by \citeauthor*{FS09} and \citeauthor*{Man11}.
We employed the code {\em LINEFIT} developed for SINFONI applications
\citep{Dav11}.  The core of the algorithm fits a Gaussian line profile
to an input spectrum within a specified wavelength interval after
continuum subtraction and $2\sigma$-clipping rejection of outliers.
The continuum corresponds to the best-fit first-order polynomial through
adjacent spectral intervals free from possible line emission from the
sources.  The line fits are weighted based on the input noise spectrum
(we used Gaussian weighting, appropriate for our data) to account for the 
variations with wavelength of the noise due to the near-IR night sky line
emission.  The instrumental spectral resolution is implicitely taken into
account by convolving the template profile derived from sky lines with
the assumed intrinsic Gaussian prior to the fitting.

Formal fitting uncertainties were computed from 100 Monte Carlo simulations,
where the points of the input spectrum are perturbed according to a Gaussian
distribution of dispersion given by the associated noise spectrum (see
Section~\ref{Sub-datared}).  For the objects with undetected [\ion{N}{2}] line
emission, upper limits were computed based on the noise spectrum for the
appropriate aperture size and wavelength interval.  Throughout this paper,
we quote formal measurement uncertainties and limits thus derived.
The uncertainties from the absolute flux calibration are estimated to be
$\sim 10\%$ and those from the wavelength calibration, $\la 5\%$.  Other
sources of uncertainties include the continuum placement and the wavelength
intervals used for line and continuum fits, which were gauged by varying
the continuum and line intervals and by inspecting the curve-of-growth
behaviour at large radii (see Section~\ref{Sub-spectra}).  These tests suggest
that the associated uncertainties amount to $20\%$ typically, and up to
$\sim 50\%$ in some data sets with lowest S/N or potentially missing faint
extended emission falling close to the edges or outside of the effective
FOV of the AO data.

With the assumption of a single Gaussian profile, the fits are primarily
sensitive to the narrower line emission component dominated by star
formation but the fluxes and line widths may be overestimated due to
the presence of an underlying broader, lower-amplitude component tracing
for instance outflowing gas
\citep[e.g.,][]{Sha09, Gen11, Gen14b, New12b, New12a, FS14}.
This component is generally not noticeable in the lower S/N spectra of
individual pixels or small apertures.  In Appendix~\ref{App-broad}, we
quantify the possible impact of such a component on our measurements,
and conclude that it is unlikely to significantly affect the overall
results.  We thus neglected these effects throughout this paper but
discuss potential biases where relevant.

\begin{figure*}[!ht]
\begin{center}
\includegraphics[scale=0.78,clip=1,angle=0]{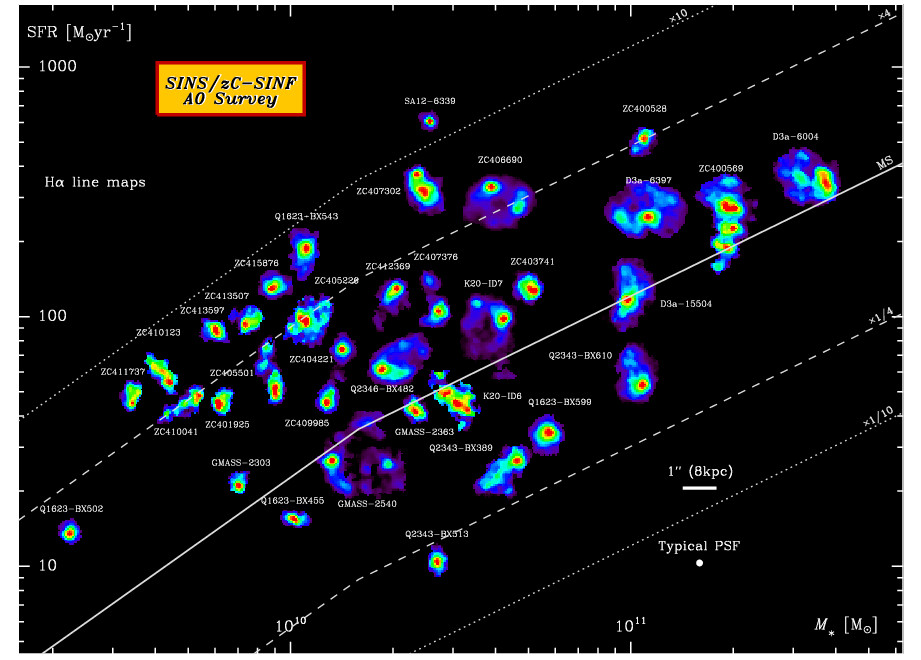}
\end{center}
\vspace{-0.30cm}
\renewcommand\baselinestretch{0.5}
\caption{
\small
H$\alpha$ surface brightness distributions for all 35 galaxies of the
\sinszc\ AO sample.
The galaxies are plotted in the stellar mass versus star formation rate
plane.  For clarity, objects in crowded parts of the diagram are slightly
shifted, by $\rm < 0.1~dex$ in $\log(M_{\star})$ and $\rm < 0.15~dex$ in
$\rm \log(SFR)$.
Dark blue to red colors correspond to linearly increasing line fluxes,
scaled up to the maximum value of each galaxy individually.
For reference, the solid line indicates the main sequence of SFGs at $z = 2$
from \citet{Whi14}; dashed and dotted lines correspond to offsets by factors
of 4 and 10 in SFR from the MS.  
All sources are shown on the same angular scale, as indicated by the
white bar of length $1^{\prime\prime}$, or about $\rm 8~kpc$ at $z = 2$;
North is up and East is to the left for all objects.
The mean and median FWHM resolution of the maps is $0\farcs 20$, or
$\rm 1.65~kpc$, shown by the white-filled circle.
\label{fig-SBHaMSFR}
}
\vspace{2.5ex}
\end{figure*}

\begin{figure*}[!ht]
\begin{center}
\includegraphics[scale=0.78,clip=1,angle=0]{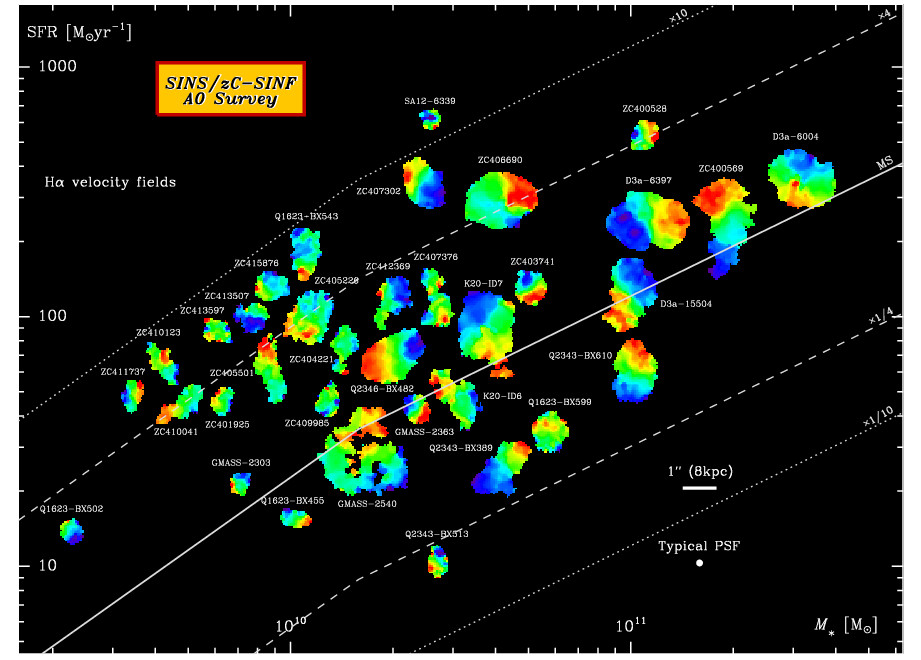}
\end{center}
\vspace{-0.30cm}
\renewcommand\baselinestretch{0.5}
\caption{
\small
H$\alpha$ velocity fields for all 35 galaxies of the \sinszc\ AO sample.
The galaxies are displayed in the stellar mass versus star formation rate
plane as in Figure~\ref{fig-SBHaMSFR}.  The $z = 2$ MS of \citet{Whi14}
and offsets thereof in SFR by factors of 4 and 10 are overplotted with
the solid, dashed, and dotted lines for reference.
The color-coding is such that blue to red colors correspond to the
blueshifted to redshifted line emission with respect to the systemic
velocity, scaled from the minimum to maximum value of each galaxy.
All sources are shown on the same angular scale, with North up and
East to the left.  The mean and median FWHM resolution of the maps
is $0\farcs 20$, or $\rm 1.65~kpc$.
\label{fig-VFsMSFR}
}
\vspace{2.5ex}
\end{figure*}

\subsection{Emission and Kinematics Maps}
            \label{Sub-maps}

To extract the flux, velocity, and velocity dispersion maps, we fitted the
spectra of individual pixels.  Prior to the fitting, the data cubes were
lightly smoothed with a median filter width of 3~pixels spectrally and,
for all but five objects, $3 \times 3$ pixels spatially to increase the
S/N without significant loss of resolution (see below).  For five of the
largest, low surface brightness sources (Deep3a-6004, Deep3a-6397, K20-ID6,
K20-ID7, and GMASS-2540), a $5 \times 5$ pixel spatial filter was used.
We first fitted the H$\alpha$ emission line, with the amplitude, central
wavelength, and width of the Gaussian profile as free parameters.  We then
performed fits to the neighbouring [\ion{N}{2}]\,$\lambda 6584$ line, fixing
the central wavelength and width based on the fit to H$\alpha$ at each pixel,
leaving only the amplitude to vary.  Since the [\ion{N}{2}] line is weaker
than H$\alpha$, these constraints helped to extract the flux in the fainter
[\ion{N}{2}] emission regions.  The [\ion{N}{2}] and H$\alpha$ lines are
expected to originate from the same regions within our \sinszc\ galaxies,
and thus to have similar kinematics, justifying this approach.  We verified
this assumption by comparing the results from these constrained fits to those
where amplitude, central wavelength, and width were let free for several of
the galaxies with higher overall S/N on [\ion{N}{2}].  We found no significant
difference within the uncertainties between the respective line flux, velocity,
and dispersion maps.

In the resulting H$\alpha$ kinematic maps, we masked out pixels where the
S/N of H$\alpha$ drops below 5.  We further masked out pixels for which
the line center and width were clearly unreliable based on inspection of
the velocity and velocity dispersion maps.  These outliers were identified
with the following criteria: velocity and dispersion of a given pixel
exceeding by a factor of at least two the typical maximum value over the
maps, a velocity uncertainty of $\rm \ga 100~km\,s^{-1}$, and a relative
dispersion uncertainty of $\ga 50\%$.
For the [\ion{N}{2}] line and \niiha\ ratio maps, relying on constrained
fits, valid pixels were required to satisfy a $\rm S/N > 5$ in both H$\alpha$
and [\ion{N}{2}] flux as well as the same additional criteria as for the
H$\alpha$ kinematic maps.
We note that since [\ion{N}{2}] is always weaker than H$\alpha$ in our
\sinszc\ AO sample galaxies, since the fits to [\ion{N}{2}] rely on the
best-fit kinematic parameters to H$\alpha$, and because of the quality
cuts applied to mask out the emission maps, the resulting valid regions
in the [\ion{N}{2}]/H$\alpha$ maps are biased towards higher ratios, and
the more so for the fainter regions of a galaxy.  This bias is important
to keep in mind when interpreting variations in line ratio maps, as done
in Section~\ref{Sect-metal}.

We also derived continuum maps obtained by summing over the emission-line
free wavelength channels of the SINFONI data cubes, applying a $2\sigma$
clipping to reduce the impact of elevated noise due to night sky lines.
In contrast to the line emission, the continuum in our IFU data is more
sensitive to systematic uncertainties from the background subtraction
and flat-fielding.  The SINFONI-based continuum maps for our targets are
generally of limited usefulness except for the objects with bright or more
centrally concentrated continuum emission.  High-resolution {\em HST\/}
imaging is available in $H$-band for 31 of our \sinszc\ AO targets, and in
at least one additional band for 29 of them, which provides more sensitive
and reliable maps of the stellar continuum emission as well as color- or
SED-based stellar mass maps \citep{FS11a, Lan14, Tacc15a, Tacc15b, Tacc18}.

Figures~\ref{fig-SBHaMSFR} and \ref{fig-VFsMSFR} show the H$\alpha$
emission line maps and velocity fields for all 35 \sinszc\ AO targets.
Additional maps are presented in Appendix~\ref{App-FullDataSets}
(Figure~\ref{fig-bbAO_1}), showing for each galaxy the H$\alpha$ and
[\ion{N}{2}] surface brightness distributions, and the H$\alpha$ velocity
and velocity dispersion maps.  For thirteen objects, the [\ion{N}{2}]
emission is too weak to reliably extract 2D maps.  The near-IR $H$ or $K$
band images (from {\em HST\/} observations where available, otherwise from
SINFONI) are also shown for comparison.  

We quantified the impact of the smoothing applied when extracting the
2D maps as follows.  We re-measured the characteristics of the effective
PSFs associated with individual galaxies after subjecting them to the same
spatial median filtering as the science cubes.
The average increase in the smoothed stellar PSF FWHMs with $3 \times 3$
pixel filter width is only about $10\%$ and the axis ratios vary by $< 2\%$,
significantly smaller than the OB-to-OB variations in PSF properties.
For the five data sets with $5 \times 5$ pixel median filtering, the FWHMs
and axis ratios are on average about $50\%$ and $5\%$ larger, respectively.
Spectrally, the 3-pixel wide median filtering broadens the line profiles
by about $10\%$ in the $K$ band and $5\%$ in the $H$ band, as
determined by a Gaussian fit to the smoothed templates.
In all measurements of galaxy properties based on the 2D maps, we accounted
for the spatial and spectral broadening introduced by the median filtering.

\subsection{Centers, Position-Velocity Diagrams, and Axis Profiles}
            \label{Sub-pv}

We defined the H$\alpha$ morphological center of each galaxy as the
center of ellipses matching the outer isophotes of the emission line maps.
This choice makes the morphological center position less sensitive to bright
asymmetric or small-scale features such as clumps and is more robust in cases
where the line emission is not centrally concentrated (e.g., ZC406690).  For
the objects with the most regular (disk-like) velocity fields and dispersion
maps, the kinematic center, corresponding to the position of steepest velocity
gradient and central peak in dispersion \citep[e.g.,][]{vdK78, Gen14a} is well
constrained and generally coincides within about two pixels ($0\farcs 1$)
with the H$\alpha$ morphological center.  It also generally lies within
the same distance of the rest-optical continuum centroid based on the $H$
or $K$ band images from {\em HST\/} observations or synthetized from the
SINFONI$+$AO cubes, and of the central peak in stellar mass maps.
For the objects with irregular kinematic maps, our adopted center position
relied on the centroid determined from the morphologies and stellar mass maps.

We identified the kinematic major axis as the direction of the largest
observed velocity difference across the source and passing through the
adopted center position.  For most of the galaxies, this axis agrees well
with the direction of the largest extent in H$\alpha$ and rest-optical
continuum emission although there are some notable exceptions (discussed
in Section~\ref{Sub-PAs}).
For some objects with more complex morphologies, or with irregular
kinematics, the velocity extrema are not always symmetrically
located along a given axis through the center (e.g., ZC410123).
For the purpose of extracting position-velocity (p-v) diagrams, we
nevertheless used the kinematic PA as defined here, noting that for
the more irregular cases this does not always probe the full velocity
difference across a source.  We estimated the uncertainty on the kinematic
PA from the difference in angle between lines passing through the center
and the positions of the blue- and redshifted velocity extrema.

We extracted the p-v diagrams from the unsmoothed reduced data cubes
in synthetic slits, integrating the light along the spatial direction
perpendicular to the slit orientation.  We used a slit width of six pixels
($0\farcs 3$, or about two spatial resolution elements) except for four
large disks with low H$\alpha$ surface brightness along the major axis,
where we used a ten-pixel-wide slit ($0\farcs 5$; Deep3a-6004, Deep3a-6397,
GMASS-2540, and K20-ID7).  The p-v diagrams are shown in
Appendix~\ref{App-AOnoAO} (Figure~\ref{fig-AOnoAO_1}).

We computed axis profiles based on spectra integrated in circular apertures
equally spaced along the kinematic major and minor axes.  For the objects
with most irregular velocity fields, we also considered axes passing through
the center and the positions of the blue- and redshifted velocity extrema to
better sample the full velocity difference.  We used apertures with diameter
of six pixels, separated by three pixels.  No median filtering was applied to
the cubes or to the aperture spectra.  The fits to H$\alpha$ and [\ion{N}{2}]
in the aperture spectra were performed as described in Section~\ref{Sub-maps}.
The resulting axis profiles were truncated where the S/N on H$\alpha$
drops below 3, and $3\sigma$ upper limits on [\ion{N}{2}] were calculated 
for the positions where the line is formally undetected ($\rm S/N < 3$).

\subsection{Integrated Spectra and Radial Profiles}
            \label{Sub-spectra}

For each galaxy, we measured the global emission line properties and
radial profiles from spatially-integrated spectra extracted from the
unsmoothed reduced data cubes.  We followed two approaches, differing
in the shape of the apertures employed and in the co-addition of the
spectra of individual pixels within the apertures.  In both cases, the
apertures were positioned at the center of the galaxies as defined in
Section~\ref{Sub-pv}.  Again, no spectral smoothing was applied to the spectra,
and the flux, central wavelength, and width were let free for H$\alpha$
while constrained fits were performed for [\ion{N}{2}]; $3\sigma$ upper
limits were calculated when [\ion{N}{2}] is formally undetected.

In the first approach, measurements were made in circular apertures
of increasing radius, summing the spectra of pixels within the
apertures to obtain the integrated spectrum.  From a curve-of-growth
analysis, we determined the total H$\alpha$ flux and corresponding total
aperture radius.  The curve-of-growth does not always converge within the
formal $1\sigma$ measurement uncertainties at large radii, although it
always exhibits a clear flattening at a radius roughly encompassing most
of the emission seen in the line maps.  In those cases, the choice of
total aperture was guided by the H$\alpha$ line map.
Possible reasons for this divergence could be real signal from the source
at low surface brightness that gets buried in the noisier edges on an
individual pixel basis or is cut off by the effective FOV, or
systematics that affect the accuracy of the line measurements.
This effect is however typically small; the largest difference between
the adopted total H$\alpha$ flux and the flux in the largest aperture
considered (with diameter typically $1.2 \times$ and up to $1.7 \times$
larger, generally limited by the effective FOV) is on average and median
$\sim 10\%$ (at most $25\%$), within $\leq 2\sigma$ of the formal
measurement uncertainties.

In the second approach, we used elliptical apertures with major axis
aligned with the kinematic PA determined in Section~\ref{Sub-pv}.  The
axis ratio was based on the value from the rest-frame optical continuum
maps when available or the H$\alpha$ maps otherwise, accounting for average
beam-smearing effects at the half-light radius of each galaxy.  For most
objects, these aperture parameters match the outer isophotes of the H$\alpha$
line maps well (see Section~\ref{Sect-struct_meas}).  The spectra of
individual pixels were co-added after shifting them (through interpolation)
to a common peak H$\alpha$ wavelength based on the velocity field.
We refer to these spectra as ``velocity-shifted spectra'' to distinguish
them from those extracted in circular apertures.  By better matching the
apertures to the line emitting regions and by concentrating the light in
wavelength space, this approach leads to higher S/N, optimizes measurements
of fainter lines such as [\ion{N}{2}], and increases the contrast between
narrow and broad emission components when present.
The improvement in S/N is most noticeable for the sources with largest
velocity gradients and low axis ratios.  On the other hand, since this
method involves velocity shifting, the resulting spectra probe the line
emission in the regions where the velocity field is reliable and misses
some of the outer parts of galaxies.  A suite of elliptical apertures and
corresponding annuli was employed, with major axis radius $r_{\rm maj}$
increasing in steps of 2 pixels ($0\farcs 1$), up to the annulus at which
$> 40\%$ of the area becomes masked out based on the velocity field.
Radial profiles in [\ion{N}{2}]/H$\alpha$ line ratios were computed from
the velocity-shifted spectra in the elliptical annuli.  Because the outer
annuli exclude some fraction of the pixels, the corresponding ratios are
assumed to be representative of the average ratio along these annuli.

Table~\ref{tab-Hameas} lists the aperture radius $r_{\rm ap}$, the
H$\alpha$ vacuum redshift $z_{\rm H\alpha}$, flux $F({\rm H\alpha})$,
velocity dispersion $\sigma_{\rm tot}({\rm H\alpha})$, and \niiha\ ratio
derived from the integrated spectrum of each source in the ``total''
circular aperture.  The table also reports the H$\alpha$ flux and \niiha\
ratio measurements obtained from the velocity-shifted spectra in the largest
elliptical apertures, along with the parameters of these apertures (major
axis radius $r_{\rm maj,ap}$, minor-to-major axis ratio $q_{\rm ap}$, and
position angle $\rm PA_{\rm ap}$).
For ZC407376, an interacting system, the Table lists the measurements for
each of the clearly separated pair components.  For ZC400569, which exhibits
a chain of bright H$\alpha$ clumps, measurements centered on the northern
component that dominates the stellar light and mass
(Section~\ref{Sub-struct_disc}) are given separately as well.
All uncertainties reported correspond to the formal fitting uncertainties.

On average (and median), the H$\alpha$ flux enclosed in the largest elliptical
aperture is about $80\%$ of that measured in the total circular aperture,
because of the smaller area covered by the former apertures.
The \niiha\ ratios for the 24 sources for which [\ion{N}{2}] is detected in
both circular and elliptical aperture spectra agree to within 10\% on average
(and median).  For four objects, [\ion{N}{2}] is undetected in the circular
aperture spectrum and the $3\sigma$ upper limits on \niiha\ are consistent
within about $30\%$ (mean and median) with the ratio measured in the higher
S/N elliptical aperture spectrum.  For the ten remaining sources, the upper
limits on \niiha\ in the elliptical apertures are typically $35\%$ lower
than those in the circular apertures.

\tabletypesize{\footnotesize}
\begin{deluxetable*}{lcccccccrcc}[p]
\renewcommand\arraystretch{1.00}
\tablecaption{Integrated Line Emission Properties
              \label{tab-Hameas}}
\tablecolumns{11}
\tablewidth{0pt}
\tablehead{
   \multicolumn{11}{c}{} \\[-1.35ex]
   \multicolumn{2}{c}{} &
   \multicolumn{4}{c}{Circular Apertures\,\tablenotemark{a}} &
   \multicolumn{5}{c}{Elliptical Apertures\,\tablenotemark{b}} \\[0.65ex]
    \cline{3-6}  \cline{7-11} \\[-2.5ex]
   \colhead{Source} &
   \colhead{$z_{\rm H\alpha}$\,\tablenotemark{c}} &
   \colhead{$r_{\rm ap}$} &
   \colhead{$F({\rm H\alpha})$} &
   \colhead{$\rm \sigma_{tot}(H\alpha)$} &
   \colhead{[\ion{N}{2}]/$\rm H\alpha$} &
   \colhead{$r_{\rm maj,ap}$} &
   \colhead{$q_{\rm ap}$} &
   \colhead{$\rm PA_{ap}$} &
   \colhead{$F({\rm H\alpha})$} &
   \colhead{[\ion{N}{2}]/$\rm H\alpha$} \\[-1.8ex]
   \colhead{} &
   \colhead{} &
   \colhead{} &
   \colhead{($\rm 10^{-17}\,erg\,s^{-1}\,cm^{-2}$)} &
   \colhead{($\rm km\,s^{-1}$)} &
   \colhead{} &
   \colhead{} &
   \colhead{} &
   \colhead{(deg)} &
   \colhead{($\rm 10^{-17}\,erg\,s^{-1}\,cm^{-2}$)} &
   \colhead{}
}
\startdata
Q1623-BX455         & $2.4078$  & $0\farcs 60$ & $9.5^{+0.5}_{-0.4}$     &
   $145^{+15}_{-16}$     & $<0.22$               &
   $0\farcs 50$ & $0.55$  & $+65$     &
   $6.9^{+0.3}_{-0.2}$     & $0.26^{+0.05}_{-0.04}$\\[0.65ex]
Q1623-BX502         & $2.1557$  & $0\farcs 60$ & $12.7 \pm 0.6$          &
   $66 \pm 5$            & $<0.08$               &
   $0\farcs 40$ & $0.80$  & $+45$     &
   $10.2 \pm 0.3$          & $<0.05$               \\[0.65ex]
Q1623-BX543         & $2.5209$  & $0\farcs 85$ & $16.8^{+0.5}_{-0.6}$    &
   $163^{+8}_{-11}$      & $<0.14$               &
   $0\farcs 60$ & $0.67$  & $0$       &
   $12.1 \pm 0.3$          & $<0.10$               \\[0.65ex]
Q1623-BX599         & $2.3312$  & $0\farcs 90$ & $28.7^{+0.9}_{-0.8}$    &
   $180^{+8}_{-10}$      & $0.19^{+0.02}_{-0.02}$&
   $0\farcs 60$ & $0.85$  & $-55$     &
   $22.7^{+0.5}_{-0.4}$    & $0.19^{+0.02}_{-0.01}$\\[0.65ex]
Q2343-BX389         & $2.1724$  & $0\farcs 95$ & $21.0^{+0.8}_{-0.9}$    &
   $258^{+23}_{-27}$     & $0.21^{+0.03}_{-0.04}$&
   $1\farcs 10$ & $0.35$  & $-50$     &
   $14.1^{+0.4}_{-0.3}$    & $0.19^{+0.03}_{-0.02}$\\[0.65ex]
Q2343-BX513         & $2.1082$  & $0\farcs 60$ & $13.9^{+0.6}_{-0.7}$    &
   $139 \pm 11$          & $0.22^{+0.03}_{-0.03}$&
   $0\farcs 40$ & $0.81$  & $-35$     &
   $9.3^{+0.4}_{-0.3}$     & $0.21^{+0.02}_{-0.02}$\\[0.65ex]
Q2343-BX610         & $2.2107$  & $0\farcs 90$ & $15.4 \pm 0.5$          &
   $164^{+9}_{-10}$      & $0.40^{+0.04}_{-0.03}$&
   $1\farcs 00$ & $0.60$  & $-10$     &
   $14.9^{+0.5}_{-0.3}$    & $0.37^{+0.03}_{-0.02}$\\[0.65ex]
Q2346-BX482         & $2.2571$  & $1\farcs 00$ & $13.5^{+0.6}_{-0.5}$    &
   $123^{+6}_{-7}$       & $0.14^{+0.04}_{-0.02}$&
   $1\farcs 10$ & $0.50$  & $-65$     &
   $14.8^{+0.4}_{-0.2}$    & $0.14^{+0.02}_{-0.01}$\\[0.65ex]
Deep3a-6004         & $2.3871$  & $1\farcs 10$ & $11.7^{+1.1}_{-0.9}$    &
   $142^{+18}_{-19}$     & $0.48^{+0.11}_{-0.08}$&
   $0\farcs 90$ & $0.95$  & $-20$     &
   $10.8^{+0.8}_{-0.6}$    & $0.46^{+0.07}_{-0.07}$\\[0.65ex]
Deep3a-6397         & $1.5133$  & $1\farcs 10$ & $12.7^{+0.8}_{-0.6}$    &
   $142^{+9}_{-14}$      & $0.44^{+0.05}_{-0.04}$&
   $1\farcs 20$ & $0.85$  & $-80$     &
   $13.6^{+0.8}_{-0.6}$    & $0.39^{+0.04}_{-0.04}$\\[0.65ex]
Deep3a-15504        & $2.3830$  & $0\farcs 95$ & $15.7 \pm 0.4$          &
   $181^{+8}_{-9}$       & $0.35^{+0.03}_{-0.02}$&
   $1\farcs 00$ & $0.75$  & $-35$     &
   $14.2^{+0.4}_{-0.3}$    & $0.34^{+0.02}_{-0.02}$\\[0.65ex]
K20-ID6             & $2.2348$  & $1\farcs 00$ & $6.5^{+0.7}_{-0.8}$     &
   $91^{+14}_{-13}$      & $<0.30$               &
   $0\farcs 70$ & $0.90$  & $+60$     &
   $4.5^{+0.5}_{-0.4}$     & $0.25^{+0.06}_{-0.06}$\\[0.65ex]
K20-ID7             & $2.2240$  & $1\farcs 00$ & $8.2 \pm 0.7$           &
   $148^{+14}_{-15}$     & $0.27^{+0.08}_{-0.08}$&
   $1\farcs 40$ & $0.50$  & $+25$     &
   $9.2^{+0.6}_{-0.4}$     & $0.20^{+0.06}_{-0.04}$\\[0.65ex]
GMASS-2303          & $2.4507$  & $0\farcs 60$ & $7.6^{+0.6}_{-0.5}$     &
   $107 \pm 9$           & $<0.24$               &
   $0\farcs 40$ & $0.70$  & $-60$     &
   $4.4^{+0.3}_{-0.2}$     & $<0.18$               \\[0.65ex]
GMASS-2363          & $2.4520$  & $0\farcs 65$ & $4.3^{+0.4}_{-0.3}$     &
   $111 \pm 10$          & $<0.29$               &
   $0\farcs 50$ & $0.67$  & $+55$     &
   $3.4^{+0.1}_{-0.2}$     & $0.16^{+0.04}_{-0.05}$\\[0.65ex]
GMASS-2540          & $1.6149$  & $1\farcs 25$ & $4.3^{+0.7}_{-0.5}$     &
   $76^{+17}_{-18}$      & $<0.56$               &
   $0\farcs 60$ & $0.86$  & $+20$     &
   $1.1 \pm 0.1$           & $<0.54$               \\[0.65ex]
SA12-6339           & $2.2971$  & $0\farcs 75$ & $7.4 \pm 0.2$           &
   $112^{+5}_{-6}$       & $0.18^{+0.04}_{-0.03}$&
   $0\farcs 30$ & $0.80$  & $+40$     &
   $4.7 \pm 0.1$           & $0.15^{+0.02}_{-0.02}$\\[0.65ex]
ZC400528            & $2.3873$  & $0\farcs 85$ & $15.7 \pm 0.8$          &
   $187 \pm 19$          & $0.69^{+0.05}_{-0.06}$&
   $0\farcs 40$ & $0.85$  & $+80$     &
   $9.2 \pm 0.3$           & $0.70^{+0.04}_{-0.04}$\\[0.65ex]
ZC400569            & $2.2405$  & $1\farcs 00$ & $10.0^{+0.7}_{-0.8}$    &
   $199^{+23}_{-29}$     & $0.41^{+0.07}_{-0.06}$&
   $1\farcs 30$ & $0.35$  & $+15$     &
   $9.6^{+0.4}_{-0.3}$     & $0.44^{+0.03}_{-0.03}$\\[0.65ex]
ZC400569N           & $2.2432$  & $1\farcs 00$ & $9.1 \pm 0.9$           &
   $185 \pm 15$          & $0.45 \pm 0.06$       &
   $1\farcs 00$ & $0.95$  & $+70$     &
   $10.5^{+0.4}_{-0.3}$    & $0.42^{+0.03}_{-0.03}$\\[0.65ex]
ZC401925            & $2.1412$  & $0\farcs 75$ & $6.9^{+0.7}_{-0.4}$     &
   $99 \pm 13$           & $<0.25$               &
   $0\farcs 40$ & $0.75$  & $+60$     &
   $3.9^{+0.3}_{-0.2}$     & $<0.16$               \\[0.65ex]
ZC403741            & $1.4457$  & $0\farcs 80$ & $12.3 \pm 0.6$          &
   $93^{+8}_{-7}$        & $0.46^{+0.04}_{-0.04}$&
   $0\farcs 60$ & $0.85$  & $+25$     &
   $10.7^{+0.4}_{-0.3}$    & $0.43^{+0.02}_{-0.03}$\\[0.65ex]
ZC404221            & $2.2199$  & $1\farcs 00$ & $10.5^{+0.8}_{-0.7}$    &
   $86^{+8}_{-9}$        & $0.26^{+0.06}_{-0.07}$&
   $1\farcs 00$ & $0.40$  & $-10$     &
   $8.9 \pm 0.3$           & $0.18^{+0.03}_{-0.03}$\\[0.65ex]
ZC405226            & $2.2870$  & $0\farcs 90$ & $6.2^{+0.4}_{-0.3}$     &
   $97^{+5}_{-6}$        & $0.15^{+0.05}_{-0.04}$&
   $0\farcs 90$ & $0.65$  & $-40$     &
   $6.0^{+0.3}_{-0.2}$     & $0.17^{+0.04}_{-0.03}$\\[0.65ex]
ZC405501            & $2.1539$  & $0\farcs 90$ & $6.9 \pm 0.4$           &
   $85^{+5}_{-6}$        & $<0.14$               &
   $1\farcs 00$ & $0.30$  & $+10$     &
   $5.3 \pm 0.2$           & $<0.09$               \\[0.65ex]
ZC406690            & $2.1950$  & $1\farcs 00$ & $30.1 \pm 0.5$          &
   $140 \pm 3$           & $0.12^{+0.01}_{-0.01}$&
   $1\farcs 10$ & $0.75$  & $-70$     &
   $29.0^{+0.5}_{-0.4}$    & $0.12^{+0.01}_{-0.01}$\\[0.65ex]
ZC407302            & $2.1819$  & $0\farcs 80$ & $16.8^{+0.4}_{-0.3}$    &
   $168 \pm 7$           & $0.25^{+0.02}_{-0.02}$&
   $0\farcs 80$ & $0.60$  & $+55$     &
   $14.7^{+0.3}_{-0.2}$    & $0.23^{+0.01}_{-0.01}$\\[0.65ex]
ZC407376            & $2.1729$  & $0\farcs 95$ & $13.6^{+0.6}_{-0.5}$    &
   $131^{+7}_{-10}$      & $0.25^{+0.05}_{-0.04}$&
   $0\farcs 90$ & $0.35$  & $+20$     &
   $9.2^{+0.3}_{-0.2}$     & $0.21^{+0.03}_{-0.03}$\\[0.65ex]
ZC407376S           & $2.1730$  & $0\farcs 60$ & $9.9^{+0.4}_{-0.3}$     &
   $149^{+9}_{-10}$      & $0.23^{+0.04}_{-0.03}$&
   $0\farcs 50$ & $0.90$  & $-60$     &
   $8.1 \pm 0.2$           & $0.20^{+0.03}_{-0.03}$\\[0.65ex]
ZC407376N           & $2.1728$  & $0\farcs 40$ & $3.2 \pm 0.2$           &
   $95^{+5}_{-8}$        & $0.26^{+0.07}_{-0.05}$&
   $0\farcs 40$ & $0.70$  & $+60$     &
   $2.6^{+0.2}_{-0.1}$     & $0.21^{+0.06}_{-0.05}$\\[0.65ex]
ZC409985            & $2.4569$  & $0\farcs 75$ & $11.2^{+0.8}_{-0.7}$    &
   $70^{+11}_{-7}$       & $0.18^{+0.04}_{-0.05}$&
   $0\farcs 40$ & $0.85$  & $-15$     &
   $8.5 \pm 0.3$           & $0.15^{+0.02}_{-0.02}$\\[0.65ex]
ZC410041            & $2.4541$  & $0\farcs 85$ & $9.1^{+0.8}_{-0.7}$     &
   $86^{+11}_{-8}$       & $<0.25$               &
   $0\farcs 80$ & $0.35$  & $-55$     &
   $6.5 \pm 0.3$           & $<0.14$               \\[0.65ex]
ZC410123            & $2.1986$  & $0\farcs 85$ & $7.0^{+0.8}_{-0.7}$     &
   $118^{+20}_{-18}$     & $<0.36$               &
   $0\farcs 60$ & $0.60$  & $+35$     &
   $4.4^{+0.4}_{-0.3}$     & $<0.22$               \\[0.65ex]
ZC411737            & $2.4442$  & $0\farcs 65$ & $7.5^{+0.7}_{-0.6}$     &
   $105^{+13}_{-11}$     & $<0.24$               &
   $0\farcs 40$ & $0.75$  & $-60$     &
   $5.0 \pm 0.3$           & $<0.14$               \\[0.65ex]
ZC412369            & $2.0281$  & $0\farcs 70$ & $18.0^{+0.6}_{-0.4}$    &
   $152^{+8}_{-7}$       & $0.21^{+0.02}_{-0.02}$&
   $0\farcs 60$ & $0.70$  & $-70$     &
   $13.1 \pm 0.3$          & $0.19^{+0.02}_{-0.02}$\\[0.65ex]
ZC413507            & $2.4800$  & $0\farcs 85$ & $10.1^{+1.1}_{-0.8}$    &
   $114 \pm 17$          & $<0.27$               &
   $0\farcs 50$ & $0.75$  & $-35$     &
   $6.2 \pm 0.4$           & $<0.16$               \\[0.65ex]
ZC413597            & $2.4502$  & $0\farcs 80$ & $8.6^{+0.7}_{-0.6}$     &
   $91^{+11}_{-8}$       & $<0.22$               &
   $0\farcs 50$ & $0.75$  & $+45$     &
   $6.3 \pm 0.3$           & $0.16^{+0.03}_{-0.04}$\\[0.65ex]
ZC415876            & $2.4354$  & $0\farcs 65$ & $14.1^{+0.8}_{-0.7}$    &
   $105^{+8}_{-7}$       & $0.15^{+0.04}_{-0.05}$&
   $0\farcs 50$ & $0.80$  & $-50$     &
   $11.5^{+0.5}_{-0.4}$    & $0.14^{+0.03}_{-0.03}$\\[0.65ex]
\enddata
\parskip=-1.5ex
\tablenotetext{a}
{
Properties from Gaussian line profile fits to the spatially-integrated
spectrum of each galaxy extracted in a circular aperture.  The radius
of the adopted ``total'' aperture, and the total H$\alpha$ flux,
velocity dispersion, and [\ion{N}{2}]/H$\alpha$ ratio are listed.
The velocity dispersion is corrected for the instrumental LSF.
The uncertainties correspond to the formal 68\% confidence intervals
from 100 Monte Carlo simulations; $3\sigma$ upper limits are
given when the [\ion{N}{2}]\,6584\AA emission line is undetected.
}
\tablenotetext{b}
{
Properties from Gaussian line profile fitting to the spatially-integrated
spectrum extracted in an elliptical aperture, corrected for velocity shifts
across the aperture.  The major axis radius, axis ratio, and PA of the
elliptical aperture, and the H$\alpha$ flux, and [\ion{N}{2}]/H$\alpha$
ratio are listed.
Uncertainties and upper limits were computed as for the properties in
the circular apertures.
}
\tablenotetext{c}
{
Redshift (vacuum) from the H$\alpha$ line fits to the spectrum of each
galaxy integrated in the circular aperture.
}
\parskip=0ex
\end{deluxetable*}

\section{MEASUREMENTS OF THE H$\alpha$ SIZES AND GLOBAL SURFACE
         BRIGHTNESS PROFILES}
         \label{Sect-struct_meas}

\looseness=-2
In this Section, we describe the measurements of the H$\alpha$ sizes
and global surface brightness distributions of the \sinszc\ AO sample.
It is obvious from, e.g., Figure~\ref{fig-SBHaMSFR}, that the observed
H$\alpha$ morphologies of the galaxies are more complex than smooth
single-component models.
Examination of the azimuthally-averaged radial profiles in elliptical
annuli indicates that a majority of the objects (26/35) exhibit a clear
off-center bump or upturn, reflecting ring-like structures or bright
clumps within the galaxies, and asymmetric extensions or possible faint
companions around the outer isophotes.
For sixteen sources, these features are superposed on an otherwise
centrally peaked profile whereas for the other ten objects, the central
surface brightness drops towards the center\,\footnote{
 For ZC407376, an interacting system, the profiles of the
 compact individual components are centrally peaked, and we also derived
 the sizes for each of them separately.  For ZC400569, we measured the
 sizes for the entire H$\alpha$ emission as well as the mass-dominating
 northern component.}
\citep[see also][]{Gen14a,Tacc15a}.

In view of these morphologies, we followed two different approaches to
derive the H$\alpha$ sizes of the objects.  The first method is based on
curve-of-growth analysis, the second one on parametric fits to the H$\alpha$
surface brightness distribution.  Although the choice of method and the
interpretation of the results should be tailored to the particular aim
of an analysis, we preliminarily note that the two approaches give very
consistent results for our \sinszc\ AO galaxies.
To assess the impact of the PSF, we derived the sizes and global profile
properties using the parameters of the PSF associated with each individual
galaxy and of the higher S/N average PSF.

\subsection{Size Estimates from Curve-of-Growth Analysis}
            \label{Sub-sizes_cog}

In the simplest approach, we derived the intrinsic half-light radius 
from the curve-of-growth in circular apertures, hereafter denoted
$r_{1/2}^{\rm circ}$.  We corrected the observed half-light radius for
beam smearing by subtracting the PSF half-light radius in quadrature,
i.e., $\rm 0.5\,\times\,FWHM$ of the \psfgal\ based on circular Gaussian
fits (Table~\ref{tab-obs}), or as determined from the curve-of-growth for
the \psfave\ (Appendix~\ref{App-AOperf}).
The uncertainties take into account those stemming from the curve-of-growth
behaviour at large radii and those of the effective PSF parameters.
The maximum deviation from convergence of the curve-of-growth of individual
galaxies beyond the radius adopted for the total H$\alpha$ flux measurement
(Section~\ref{Sub-spectra}) implies an average and median difference of
about $5\%$ in beam-smeared half-light radius (maximum difference of
$22\%$).  The uncertainties of the PSF are more important, and we adopt
a conservative $30\%$ to account for the fact that the PSF stars are not
observed simultaneously with, and along the same optical path as the
galaxies, and for the deviations of the PSF shape from a pure circular
Gaussian profile (Section~\ref{Sub-PSF})

Overall, the $r_{1/2}^{\rm circ}$ estimates obtained with the different
choices of PSF agree within the formal $1\sigma$ uncertainties.
As expected, however, the differences are systematic, with the values
derived with the \psfave\ lower than those obtained with the \psfgal\
by $\sim 10\%$ on average and in the median.
These differences depend on galaxy size and are most significant for
objects with $r_{1/2}^{\rm circ} \la 2.5~{\rm kpc}$ (corresponding to
the half-light radius of the \psfave\ at $z \sim 2$), for which the mean
and median differences are about $15\%$.  The largest difference is
35\% for BX502; SA12-6339, the smallest source, is formally unresolved
when using the \psfave.

\subsection{Parametric Fits to the H$\alpha$ Line Maps}
            \label{Sub-sizes_galfit}

In the alternative approach, we derived the sizes by fitting intrinsic 2D
\citet{Ser68} profiles convolved with the PSF to the observed H$\alpha$
surface brightness distributions.  We followed a similar procedure as
described by \citet{FS11a}.  We used the code GALFIT \citep{Pen02} and,
for simplicity, considered single-component models.  The free parameters
in the fits were the major axis effective radius $R_{e}$, the S\'ersic index
$n$, the projected minor-to-major axis ratio $q$, the PA of the major axis,
and the total flux.  The S\'ersic index was allowed to vary in the range
$0.1 < n < 4.0$.  The input error maps accounted for the background rms
noise as well as the Poisson noise from the sources.  The input PSFs were
noiseless models created with the parameters derived for the \psfgal\ and
\psfave\ cases, based on elliptical Gaussian fits
(see Appendix~\ref{App-AOperf}).  Since asymmetric and/or clumpy distributions
can heavily bias the results and the profile parameters tend to be degenerate
with the background level \citep[e.g.,][]{Pen02, Dok08, FS11a}, we fixed the
center position and the ``sky'' level in the fits.

For realistic estimates of the uncertainties on the derived parameters, we
ran 200 fits for every galaxy, varying in each iteration the center position,
the sky level, and the PSF.  The center coordinates were drawn randomly from
a uniform distribution in a square box of typically $4 \times 4$ pixels around
the adopted position (Section~\ref{Sub-pv}).  An accurate determination of the
background level in our H$\alpha$ maps is complicated by the limited FOV and
by possible residual systematics (as discussed in Section~\ref{Sub-spectra}).
The sky values were conservatively drawn from a Gaussian distribution
based on the fluxes of pixels sufficiently well outside of the regions
where H$\alpha$ is formally detected at $\rm S/N > 3$.
The input PSFs were generated by varying the $\rm FWHM_{maj}$ and axis
ratio according to Gaussian distributions centered at, and with dispersions
of $30\%$ and $10\%$ the nominal values, and the PA's were drawn from a
Gaussian distribution with dispersion of $50^{\circ}$ around the nominal
PA (the parameters of the narrow and broad \psfave\ components
were varied independently).

The best-fit parameters for a given galaxy were taken as the median
of the results for all 200 iterations, and the $1\sigma$ uncertainties
from the central $68\%$ of the distribution.
For the northern component of ZC407376,
fits were unsuccessful because of the faintness of the source.
For GMASS-2540, the fits are unreliable because of its low average
surface brightness, strongly asymmetric clumpy ring-like H$\alpha$
morphology, and the fact that its large extent is not fully covered by
the effective SINFONI FOV.  The parametric fit results for both objects
were thus excluded in our subsequent quantitative analysis.
As for the size estimates from the H$\alpha$ curve-of-growth, the
$R_{\rm e}$ values from the S\'ersic model fits with the \psfave\ are
systematically lower than those with the \psfgal.  The differences are
approximately $10\%$ on average and median, and increase towards smaller
sources (mean and median of around $15\%$, and up to $45\%$)
but remain within the formal $\approx 1\sigma$ uncertainties.
Differences in derived S\'ersic indices, axis ratios, and PA's are small
(on average $6\%$, $14\%$, and $2^{\circ}$, respectively) and well within
the $1\sigma$ uncertainties.

\tabletypesize{\footnotesize}
\begin{deluxetable*}{lcccccccccc}[p]
\renewcommand\arraystretch{1.00}
\tablecaption{H$\alpha$ Sizes and Global Structural Properties
              \label{tab-struct}}
\tablecolumns{11}
\tablewidth{0pt}
\tablehead{
   \multicolumn{11}{c}{} \\[-1.35ex]
   \multicolumn{1}{c}{} &
   \multicolumn{5}{c}{Single-Gaussian PSF\,\tablenotemark{a}} &
   \multicolumn{5}{c}{Double-Gaussian PSF\,\tablenotemark{b}} \\[0.65ex]
    \cline{2-6}  \cline{7-11} \\[-2.5ex]
   \colhead{Source} &
   \colhead{$r_{1/2}^{\rm circ}$} &
   \colhead{$R_{\rm e}$} &
   \colhead{$n$} &
   \colhead{$q$} &
   \colhead{$\rm PA$} &
   \colhead{$r_{1/2}^{\rm circ}$} &
   \colhead{$R_{\rm e}$} &
   \colhead{$n$} &
   \colhead{$q$} &
   \colhead{$\rm PA$} \\[-1.8ex]
   \colhead{} &
   \colhead{(kpc)} &
   \colhead{(kpc)} &
   \colhead{} &
   \colhead{} &
   \colhead{(deg)} &
   \colhead{(kpc)} &
   \colhead{(kpc)} &
   \colhead{} &
   \colhead{} &
   \colhead{(deg)}
}
\startdata
Q1623-BX455  & $1.9^{+0.2}_{-0.3}$   & $2.6^{+1.2}_{-0.5}$   & $0.80^{+0.69}_{-0.28}$   & $0.49^{+0.08}_{-0.04}$   & $+77^{+10}_{-5}$  &
               $1.4 \pm 0.6$         & $2.0^{+0.8}_{-0.4}$   & $0.90^{+2.37}_{-0.55}$   & $0.27^{+0.16}_{-0.20}$   & $+75^{+13}_{-5}$  \\[0.65ex]
Q1623-BX502  & $1.7 \pm 0.3$         & $2.1^{+0.8}_{-0.4}$   & $1.02^{+0.66}_{-0.27}$   & $0.82^{+0.07}_{-0.08}$   & $-23^{+20}_{-22}$ &
               $1.1^{+0.6}_{-0.4}$   & $1.7^{+0.7}_{-0.5}$   & $0.96^{+1.47}_{-0.67}$   & $0.70^{+0.17}_{-0.13}$   & $-20^{+31}_{-19}$ \\[0.65ex]
Q1623-BX543  & $2.5 \pm 0.3$         & $3.7^{+1.5}_{-0.9}$   & $1.64^{+1.40}_{-0.64}$   & $0.48^{+0.10}_{-0.14}$   & $-4^{+6}_{-4}$    &
               $2.4^{+0.3}_{-0.4}$   & $3.3^{+1.2}_{-0.8}$   & $1.37^{+1.02}_{-0.43}$   & $0.48^{+0.08}_{-0.11}$   & $-4^{+6}_{-4}$    \\[0.65ex]
Q1623-BX599  & $2.5 \pm 0.2$         & $2.6 \pm 0.4$         & $0.67^{+0.28}_{-0.15}$   & $0.87^{+0.08}_{-0.12}$   & $-33^{+16}_{-13}$ &
               $2.3^{+0.3}_{-0.4}$   & $2.2^{+0.4}_{-0.5}$   & $0.64^{+0.22}_{-0.29}$   & $0.87^{+0.08}_{-0.10}$   & $-31^{+29}_{-16}$ \\[0.65ex]
Q2343-BX389  & $4.0 \pm 0.2$         & $5.9^{+0.4}_{-0.2}$   & $0.29^{+0.07}_{-0.11}$   & $0.41^{+0.04}_{-0.03}$   & $-47 \pm 2$       &
               $3.9^{+0.3}_{-0.4}$   & $5.8^{+0.2}_{-0.1}$   & $0.16^{+0.14}_{-0.06}$   & $0.39^{+0.05}_{-0.08}$   & $-47 \pm 2$       \\[0.65ex]
Q2343-BX513  & $2.1 \pm 0.3$         & $3.2^{+0.9}_{-0.6}$   & $1.11^{+0.48}_{-0.26}$   & $0.63^{+0.04}_{-0.05}$   & $+5^{+4}_{-10}$   &
               $1.7^{+0.6}_{-1.1}$   & $2.6^{+1.5}_{-0.8}$   & $1.71^{+2.19}_{-0.73}$   & $0.48^{+0.12}_{-0.15}$   & $+5^{+5}_{-6}$    \\[0.65ex]
Q2343-BX610  & $3.6^{+0.1}_{-0.2}$   & $4.7^{+0.1}_{-0.3}$   & $0.13^{+0.09}_{-0.03}$   & $0.56 \pm 0.04$          & $+14^{+1}_{-2}$   &
               $3.5^{+0.2}_{-0.3}$   & $4.6^{+0.1}_{-0.2}$   & $0.10^{+0.05}_{-0.01}$   & $0.50^{+0.06}_{-0.08}$   & $+14 \pm 2$       \\[0.65ex]
Q2346-BX482  & $4.1 \pm 0.3$         & $5.3^{+0.2}_{-0.3}$   & $0.10^{+0.04}_{-0.01}$   & $0.61 \pm 0.03$          & $-57^{+2}_{-3}$   &
               $3.9 \pm 0.4$         & $5.4 \pm 0.3$         & $0.10 \pm 0.01$          & $0.55^{+0.04}_{-0.06}$   & $-56^{+3}_{-4}$   \\[0.65ex]
Deep3a-6004  & $4.7 \pm 0.5$         & $5.1^{+0.4}_{-1.6}$   & $0.16^{+0.55}_{-0.06}$   & $0.59^{+0.20}_{-0.08}$   & $+58^{+5}_{-8}$   &
               $4.5 \pm 0.6$         & $5.1^{+0.3}_{-1.5}$   & $0.10^{+0.62}_{-0.01}$   & $0.49^{+0.11}_{-0.21}$   & $+61^{+4}_{-6}$   \\[0.65ex]
Deep3a-6397  & $4.9 \pm 0.1$         & $6.1^{+0.5}_{-0.6}$   & $0.76^{+0.32}_{-0.46}$   & $0.56 \pm 0.05$          & $-74 \pm 3$       &
               $4.8 \pm 0.2$         & $5.9^{+0.6}_{-0.9}$   & $0.52^{+0.57}_{-0.38}$   & $0.54^{+0.08}_{-0.16}$   & $-73^{+4}_{-3}$   \\[0.65ex]
Deep3a-15504 & $3.7 \pm 0.2$         & $6.6^{+1.5}_{-1.0}$   & $1.01^{+0.28}_{-0.23}$   & $0.46^{+0.02}_{-0.03}$   & $-13^{+3}_{-2}$   &
               $3.5^{+0.3}_{-0.4}$   & $6.4^{+2.4}_{-1.3}$   & $1.25^{+0.69}_{-0.42}$   & $0.40^{+0.06}_{-0.09}$   & $-12^{+3}_{-2}$   \\[0.65ex]
K20-ID6      & $4.0 \pm 0.1$         & $4.3^{+1.6}_{-1.1}$   & $0.51^{+0.43}_{-0.39}$   & $0.53^{+0.26}_{-0.17}$   & $+50^{+7}_{-6}$   &
               $3.8 \pm 0.2$         & $4.2^{+1.2}_{-1.3}$   & $0.37^{+0.94}_{-0.26}$   & $0.43^{+0.21}_{-0.20}$   & $+50 \pm 6$       \\[0.65ex]
K20-ID7      & $5.1 \pm 0.3$         & $5.5^{+0.3}_{-1.2}$   & $0.10^{+0.36}_{-0.01}$   & $0.75^{+0.12}_{-0.04}$   & $-2^{+37}_{-6}$   &
               $5.0 \pm 0.4$         & $5.5^{+0.4}_{-1.2}$   & $0.10^{+0.92}_{-0.01}$   & $0.72^{+0.16}_{-0.07}$   & $0^{+27}_{-5}$    \\[0.65ex]
GMASS-2303   & $2.2 \pm 0.6$         & $2.4^{+1.3}_{-0.6}$   & $1.00^{+0.63}_{-0.42}$   & $0.58^{+0.12}_{-0.08}$   & $-12^{+6}_{-14}$  &
               $1.9^{+0.8}_{-1.6}$   & $2.2^{+1.0}_{-0.9}$   & $0.91^{+1.67}_{-0.71}$   & $0.47^{+0.13}_{-0.12}$   & $-10^{+9}_{-13}$  \\[0.65ex]
GMASS-2363   & $2.0^{+0.3}_{-0.4}$   & $2.1^{+1.2}_{-0.2}$   & $0.50^{+0.87}_{-0.14}$   & $0.70^{+0.21}_{-0.07}$   & $+37^{+13}_{-32}$ &
               $1.5^{+0.6}_{-1.3}$   & $1.9^{+2.0}_{-0.4}$   & $0.50^{+2.75}_{-0.40}$   & $0.53^{+0.17}_{-0.15}$   & $+34^{+14}_{-12}$ \\[0.65ex]
GMASS-2540   & $7.9 \pm 0.2$         & \ldots                & \ldots                   & \ldots                   & \ldots            &
               $7.8^{+0.2}_{-0.3}$   & \ldots                & \ldots                   & \ldots                   & \ldots            \\[0.65ex]
SA12-6339    & $1.5 \pm 0.3$         & $2.3^{+1.0}_{-0.6}$   & $2.18^{+2.20}_{-0.92}$   & $0.81^{+0.12}_{-0.11}$   & $-3^{+29}_{-33}$  &
               $<1.4$                & $1.2^{+1.4}_{-0.6}$   & $3.75^{+1.25}_{-2.23}$   & $0.65^{+0.20}_{-0.26}$   & $-11^{+32}_{-31}$ \\[0.65ex]
ZC400528     & $2.4 \pm 0.2$         & $3.2^{+1.6}_{-0.5}$   & $1.03^{+0.67}_{-0.36}$   & $0.54^{+0.08}_{-0.06}$   & $-25^{+7}_{-5}$   &
               $2.1^{+0.4}_{-0.5}$   & $2.7^{+1.5}_{-0.8}$   & $1.50^{+1.35}_{-0.52}$   & $0.40^{+0.12}_{-0.17}$   & $-24^{+7}_{-5}$   \\[0.65ex]
ZC400569     & $4.4 \pm 0.5$         & $7.6^{+0.8}_{-1.8}$   & $0.40^{+0.26}_{-0.09}$   & $0.33^{+0.05}_{-0.03}$   & $+4^{+4}_{-3}$    &
               $4.2^{+0.6}_{-0.7}$   & $7.4^{+1.3}_{-1.5}$   & $0.44^{+0.46}_{-0.15}$   & $0.28^{+0.06}_{-0.05}$   & $+5 \pm 3$        \\[0.65ex]
ZC400569N    & $3.9^{+1.1}_{-1.2}$   & $3.2^{+0.7}_{-0.4}$   & $0.76^{+0.21}_{-0.13}$   & $0.98^{+0.02}_{-0.09}$   & $+70^{+7}_{-6}$   &
               $3.7^{+1.3}_{-1.5}$   & $2.8^{+0.9}_{-0.6}$   & $0.95^{+0.37}_{-0.24}$   & $0.95^{+0.06}_{-0.11}$   & $+70^{+7}_{-6}$   \\[0.65ex]
ZC401925     & $2.2^{+0.4}_{-0.5}$   & $2.6^{+1.3}_{-0.5}$   & $0.88^{+0.74}_{-0.45}$   & $0.65 \pm 0.09$          & $-9 \pm 7$        &
               $1.9^{+0.5}_{-0.9}$   & $2.5^{+1.2}_{-0.5}$   & $0.65^{+0.98}_{-0.55}$   & $0.62^{+0.09}_{-0.10}$   & $-10^{+8}_{-11}$  \\[0.65ex]
ZC403741     & $2.5^{+0.1}_{-0.2}$   & $2.5^{+0.3}_{-0.2}$   & $0.42^{+0.23}_{-0.07}$   & $0.89 \pm 0.04$          & $+52^{+23}_{-21}$ &
               $2.2^{+0.3}_{-0.5}$   & $2.2^{+0.4}_{-0.3}$   & $0.29^{+0.39}_{-0.19}$   & $0.84^{+0.07}_{-0.10}$   & $+45 \pm 16$      \\[0.65ex]
ZC404221     & $1.7^{+0.1}_{-0.2}$   & $2.2^{+1.6}_{-0.5}$   & $1.89^{+2.81}_{-0.68}$   & $0.70^{+0.14}_{-0.15}$   & $-9^{+6}_{-7}$    &
               $1.3^{+0.4}_{-0.8}$   & $1.5^{+1.1}_{-0.5}$   & $1.71^{+2.79}_{-1.06}$   & $0.68 \pm 0.20$          & $-9^{+13}_{-12}$  \\[0.65ex]
ZC405226     & $3.6 \pm 0.2$         & $5.2^{+0.7}_{-0.6}$   & $0.56^{+0.17}_{-0.14}$   & $0.67^{+0.05}_{-0.02}$   & $-42^{+4}_{-2}$   &
               $3.4 \pm 0.3$         & $5.1^{+0.7}_{-0.5}$   & $0.53^{+0.26}_{-0.14}$   & $0.65 \pm 0.05$          & $-43 \pm 2$       \\[0.65ex]
ZC405501     & $4.1 \pm 0.2$         & $6.2^{+0.7}_{-0.9}$   & $0.10^{+0.69}_{-0.01}$   & $0.29^{+0.14}_{-0.05}$   & $+13 \pm 2$       &
               $3.9^{+0.3}_{-0.4}$   & $6.3^{+0.3}_{-0.6}$   & $0.10^{+0.81}_{-0.01}$   & $0.23 \pm 0.09$          & $+12 \pm 3$       \\[0.65ex]
ZC406690     & $4.8 \pm 0.1$         & $5.4 \pm 0.1$         & $0.10 \pm 0.01$          & $0.76 \pm 0.02$          & $-87 \pm 2$       &
               $4.6^{+0.1}_{-0.2}$   & $5.3 \pm 0.1$         & $0.10 \pm 0.01$          & $0.73^{+0.03}_{-0.06}$   & $-86^{+2}_{-3}$   \\[0.65ex]
ZC407302     & $3.1 \pm 0.2$         & $4.1^{+0.4}_{-0.3}$   & $0.47^{+0.14}_{-0.11}$   & $0.53^{+0.04}_{-0.03}$   & $+28^{+3}_{-5}$   &
               $2.9^{+0.3}_{-0.4}$   & $3.9 \pm 0.3$         & $0.35^{+0.22}_{-0.25}$   & $0.46^{+0.07}_{-0.09}$   & $+29 \pm 4$       \\[0.65ex]
ZC407376     & $4.0 \pm 0.1$         & $5.5^{+0.5}_{-0.3}$   & $0.10^{+0.34}_{-0.01}$   & $0.38^{+0.10}_{-0.05}$   & $+15^{+3}_{-2}$   &
               $3.9 \pm 0.2$         & $5.5^{+0.7}_{-0.3}$   & $0.10^{+0.34}_{-0.01}$   & $0.32^{+0.08}_{-0.10}$   & $+15 \pm 2$       \\[0.65ex]
ZC407376S    & $2.1^{+0.3}_{-0.4}$   & $3.1^{+2.6}_{-0.7}$   & $1.16^{+1.33}_{-0.58}$   & $0.58 \pm 0.12$          & $-23^{+19}_{-13}$ &
               $1.8^{+0.5}_{-0.8}$   & $2.6^{+3.5}_{-0.9}$   & $1.17^{+1.89}_{-0.56}$   & $0.55^{+0.14}_{-0.15}$   & $-19^{+21}_{-13}$ \\[0.65ex]
ZC407376N    & $1.6^{+0.4}_{-0.6}$   & \ldots                & \ldots                   & \ldots                   & \ldots            &
               $1.2^{+0.7}_{-1.0}$   & \ldots                & \ldots                   & \ldots                   & \ldots            \\[0.65ex]
ZC409985     & $1.9 \pm 0.1$         & $2.5^{+0.9}_{-0.4}$   & $0.86^{+0.56}_{-0.29}$   & $0.60^{+0.16}_{-0.04}$   & $-10^{+9}_{-5}$   &
               $1.4^{+0.4}_{-0.8}$   & $2.1^{+0.6}_{-0.4}$   & $0.81^{+1.17}_{-0.61}$   & $0.46^{+0.18}_{-0.14}$   & $-11^{+8}_{-5}$   \\[0.65ex]
ZC410041     & $3.8 \pm 0.3$         & $5.2^{+0.6}_{-0.4}$   & $0.22^{+0.46}_{-0.12}$   & $0.31^{+0.12}_{-0.07}$   & $-55 \pm 3$       &
               $3.6^{+0.4}_{-0.5}$   & $5.3^{+0.5}_{-0.6}$   & $0.10^{+0.63}_{-0.01}$   & $0.25^{+0.11}_{-0.14}$   & $-56^{+3}_{-2}$   \\[0.65ex]
ZC410123     & $3.3^{+0.1}_{-0.2}$   & $4.0^{+0.9}_{-0.5}$   & $0.38^{+0.50}_{-0.20}$   & $0.38^{+0.12}_{-0.06}$   & $+36^{+3}_{-4}$   &
               $3.1 \pm 0.3$         & $3.8^{+0.6}_{-1.3}$   & $0.33^{+1.08}_{-0.23}$   & $0.31^{+0.28}_{-0.12}$   & $+36^{+4}_{-6}$   \\[0.65ex]
ZC411737     & $2.0^{+0.2}_{-0.3}$   & $2.8^{+1.2}_{-0.4}$   & $0.57^{+0.74}_{-0.40}$   & $0.56^{+0.13}_{-0.07}$   & $-6^{+8}_{-6}$    &
               $1.7^{+0.5}_{-0.8}$   & $2.4^{+0.9}_{-0.3}$   & $0.18^{+1.46}_{-0.08}$   & $0.45^{+0.14}_{-0.18}$   & $-6^{+7}_{-9}$    \\[0.65ex]
ZC412369     & $2.5 \pm 0.3$         & $3.7^{+0.9}_{-0.6}$   & $0.93^{+0.35}_{-0.25}$   & $0.54^{+0.09}_{-0.06}$   & $-35^{+4}_{-3}$   &
               $2.1^{+0.5}_{-0.7}$   & $3.1^{+1.0}_{-0.9}$   & $1.22^{+1.00}_{-0.64}$   & $0.41^{+0.13}_{-0.14}$   & $-34^{+6}_{-4}$   \\[0.65ex]
ZC413507     & $2.8 \pm 0.2$         & $3.3^{+1.6}_{-0.6}$   & $0.64^{+0.38}_{-0.32}$   & $0.79^{+0.14}_{-0.10}$   & $-35^{+78}_{-10}$ &
               $2.5^{+0.3}_{-0.4}$   & $2.9^{+1.1}_{-0.5}$   & $0.65^{+0.65}_{-0.55}$   & $0.72 \pm 0.17$          & $-31^{+50}_{-13}$ \\[0.65ex]
ZC413597     & $2.1 \pm 0.1$         & $2.4^{+1.4}_{-0.5}$   & $1.09^{+0.67}_{-0.44}$   & $0.66^{+0.18}_{-0.12}$   & $+54^{+17}_{-5}$  &
               $1.8^{+0.3}_{-0.5}$   & $2.1^{+0.9}_{-0.6}$   & $0.98^{+1.51}_{-0.66}$   & $0.52^{+0.23}_{-0.18}$   & $+51^{+15}_{-6}$  \\[0.65ex]
ZC415876     & $2.0 \pm 0.2$         & $2.2^{+0.5}_{-0.3}$   & $0.70^{+0.44}_{-0.21}$   & $0.77^{+0.11}_{-0.08}$   & $-74 \pm 8$       &
               $1.5^{+0.4}_{-0.8}$   & $1.8^{+0.6}_{-0.4}$   & $0.52^{+1.04}_{-0.42}$   & $0.70^{+0.15}_{-0.17}$   & $-74^{+19}_{-24}$ \\[0.65ex]
\enddata
\parskip=-1.5ex
\tablecomments
{
The measurements reported are the intrinsic half-light radius
computed from the H$\alpha$ curve-of-growth in circular apertures
$r_{1/2}^{\rm circ}$, and the intrinsic major axis effective radius
$R_{\rm e}$, S\'ersic index $n$, projected minor-to-major axis ratio
$q$, and PA of the major axis (in degrees east of north) from 2D
S\'ersic model fits to the H$\alpha$ surface brightness distributions.
}
\tablenotetext{a}
{
Measurements derived with the best-fit single-Gaussian fits
to the PSFs associated with each individual galaxy.
}
\tablenotetext{b}
{
Measurements derived with the best-fit double-Gaussian fit
to the average, high $\rm S/N$ PSF.
}
\parskip=0ex
\end{deluxetable*}

\subsection{Consistency of the H$\alpha$ Size Estimates}
            \label{Sub-sizes_disc}

Table~\ref{tab-struct} gives the $r_{1/2}^{\rm circ}$ estimates from the
curve-of-growth analysis and the $R_{\rm e}$, $n$, $q$, and PA obtained
from the S\'ersic profile fits to the H$\alpha$ line maps.  The results
are reported for both choices of PSF for comparison.
The $r_{1/2}^{\rm circ}$ values derived with the \psfave\ are in the
range $\rm 1.1 - 5.0~kpc$, with a mean of 2.8~kpc and median of 2.5~kpc.
The major axis effective radii $R_{\rm e}$ are between 1.2 and 7.4~kpc,
with mean and median of 3.7 and 3.0~kpc, respectively.
Since the curves-of-growth were measured in circular apertures, the
$r_{1/2}^{\rm circ}$ values should be compared to the circularized
effective radii $R_{\rm e,circ} \equiv R_{\rm e}\,\sqrt{q}$ for better
consistency.  The $r_{1/2}^{\rm circ}$ are on average (and median)
$10\%$ larger than the $R_{\rm e,circ}$, with a scatter of $15\%$.
When using the individual \psfgal, the $r_{1/2}^{\rm circ}$ and
$R_{\rm e,circ}$ agree within $4\%$ on average (and median),
also with a scatter of $15\%$.

The size estimates obtained from the curve-of-growth and from the parametric
fits are thus in very good agreement, in particular considering some of the
inherent differences between the two approaches.
The curves-of-growth were computed directly from the data cubes and so rely
on H$\alpha$ line fits from higher S/N spectra than those at the individual
pixel level when making the line maps, although they are more prone to
limitations due to the small effective FOV of our SINFONI$+$AO data sets
for the largest sources (see Appendix~\ref{App-AOnoAO}).  The parametric fits
can mitigate both the S/N and FOV limitations at large radii if the global
profiles are sufficiently well represented by a simple (single-component)
model.  Both methods are sensitive to uncertainties from the background
subtraction.  The tight correlation between the $r_{1/2}^{\rm circ}$ and
$R_{\rm e,circ}$ estimates suggest that these sources of uncertainty do
not affect importantly our measurements.

The curve-of-growth method has the main advantage of accounting properly
for all the light irrespective of the details of (potentially complex)
surface brightness distributions.  On the other hand, in this approach
we treated the beam smearing simplistically (subtracting in quadrature the
PSF half-light radius) and the use of circular apertures would overestimate
the sizes because projection effects from galaxy inclination are neglected.
The parametric approach accounts more accurately for projected axis ratios,
and for the impact of beam smearing for non-Gaussian intrinsic galaxy profiles
and non-Gaussian PSFs, although single-component S\'ersic models as adopted
here obviously do not account for the detailed substructure of the galaxies.
Based on a large suite of 2D S\'ersic models created with varying
$R_{\rm e}$, $n$, $q$, the PSF, and spatial sampling in the ranges spanned
by our AO sample, we found that differences between $r_{1/2}^{\rm circ}$
and $R_{\rm e,circ}$ are mostly affected by the projected axis ratio, with
little dependence on galaxy profile and little impact of the simplistic
beam smearing correction in the curve-of-growth approach.
The trend in $r_{1/2}^{\rm circ}/R_{\rm e,circ}$ among the galaxies is
consistent with the model expectations, with an average ratio of 1.04
for the sources with $q > 0.5$ and 1.15 for those with $q < 0.5$ in the
\psfave\ case (1.01 and 1.11, respectively, in the \psfgal\ case).
We conclude that the (small) differences in H$\alpha$ sizes obtained
from the two methods for our sample can be attributed partly to galaxy
inclination effects, along with other factors such as complex morphologies
that are more difficult to quantify.

\subsection{Global H$\alpha$ Surface Brightness Distributions
            and Comparison to $H$-band Continuum Results}
            \label{Sub-struct_disc}

The simple S\'ersic models adopted in Section~\ref{Sub-sizes_galfit} lead in
many cases to significant fit residuals on small scales left as imprint of
bright clumps, ring-like features, or other prominent irregular substructure
in H$\alpha$ light, and obviously provide a poor representation of systems
with spatially resolved interacting units.
In particular, rings and bright off-center clumps drive the best-fit S\'ersic
indices to low values.  The mean and median indices are $n \sim 0.7$ and
$0.5$, respectively, using the \psfave\ (and essentially identical mean
and median of $n \sim 0.7$ in the \psfgal\ case).
Best-fit indices of around 0.1, the lower limit allowed in our parametric
fits, are reached by eight objects with the most prominent rings or off-center
clumps (Q2343-BX610, Q2346-BX482, Deep3a-6004, K20-ID7, ZC405501, ZC406690,
ZC410041, and the merger ZC407376 when modeled as a single system).
Only SA12-6339 has $n > 2$ (though with large uncertainties because of the
compactness of this source).  Taken at face value, these results imply that
the \sinszc\ AO galaxies have global observed H$\alpha$ surface brightness
distributions consistent with morphologically late-type, disk-dominated
systems, usually defined as having $n < 2 - 2.5$ \citep[e.g.,][]{Bel04,Tru06}.
Even for SA12-6339, the derived $n$ for either PSF choice is within
$1\sigma$ of $n=2$.
The sample exhibits a trend of decreasing $n$ with larger $R_{\rm e}$
(Spearman rank correlation coefficient about $\rho -0.6$, significant
at the $3.5\sigma$ level), reflecting the qualitative trend
of shallower or more ring-like profiles towards larger galaxies that is
apparent from the H$\alpha$ maps and profiles.

A similar parametric analysis was performed on the $H$-band maps for the
subset of 29 galaxies with high-resolution {\em HST\/} near-IR imaging
\citep{FS11a, Tacc15b}.
Qualitatively, the $H$-band morphologies often exhibit similar noticeable
substructure as seen in H$\alpha$ but they tend to be overall smoother and
more centrally peaked (see Appendix~\ref{App-FullDataSets}).
For ZC400569 and ZC407376, the $H$-band parameters refer to the northern
and southern components, respectively.
Quantitatively, and excluding GMASS-2540 because of the unreliable fits
to the H$\alpha$ line maps, the rest-optical and H$\alpha$ sizes are
essentially identical within about $5\%$ on average, both in terms of major
axis and circularized effective radius.  The axis ratios and PA's agree on
average within about $10\%$ and 18 degrees, respectively.  For a majority of
the sources, the rest-optical light also globally follows a disk-like profile,
with $80\%$ having a best-fit $n < 2.5$.  However, systematically higher
S\'ersic indices are derived, with a mean $n = 1.6$ (median $n = 1.2$)
compared to $n = 0.7$ (median $n = 0.5$) from the H$\alpha$ maps of the
same 28 objects.  This differentiation is more pronounced at higher
stellar masses; galaxies with $\log(M_{\star}/{\rm M_{\odot}}) > 10.7$
have an average $n$ of 2.5 (median $n \approx 2.3$) in rest-optical
light, and an average and median $n$ of about 0.8 in H$\alpha$ light.

The trend of increasingly more centrally peaked morphologies in
$H$-band versus H$\alpha$ towards higher masses is consistent with
the presence of a significant bulge along with suppressed star formation
activity in the central regions of these massive galaxies, as discussed
by \citet{Gen14a} and \citet{Tacc15a,Tacc18}.  This trend also echoes
findings from the average H$\alpha$ and rest-optical continuum profiles
from much larger samples of SFGs, albeit at lower $z \sim 1$, from the
3D-HST {\em HST\/}/WFC3 grism survey \citep{Nel12, Nel13, Nel16b, Wuy13}.
Such differences in global light profiles could be indicative of quenching
in the central regions of high-mass galaxies, caused by a massive bulge
stabilizing the gas against the formation of massive star-forming clumps
\citep[e.g.,][]{Mar09, Gen14a}, efficient removal of gas via powerful
nuclear outflows
\citep[][see also 
\citealt{Can12,Cre15,Bru15,Bru16,Per15,Car16}]{FS14, Gen14b},
or central gas consumption \citep[e.g.,][]{Tacc16, Tad17}.

An important uncertainty in interpreting the sizes and global profiles
from H$\alpha$ (and continuum) emission is the possible effects of
spatially-variable dust extinction \citep[e.g.,][]{Wuy13, Nel16a}.
Higher extinction in the inner regions would imply more centrally peaked
intrinsic profiles than observed, such that the effective radii inferred
here would overestimate the true intrinsic sizes.  This effect could be
exacerbated if the H$\alpha$ emission from the \ion{H}{2} regions were more
attenuated relative to the stellar continuum light along the line-of-sight
(e.g., \citealt{Cal00}; \citeauthor*{FS09}; \citeauthor*{Man11}; 
\citealt{Ly12,Wuy13,Kas13,Pri14,Koy15,Pug16}).
Spatially-resolved maps of the extinction towards the stars and the
nebular line emission are very challenging to obtain for distant galaxies.
Exploiting new {\em HST\/} imaging probing the near- and far-UV emission
of ten of the \sinszc\ AO galaxies together with the H$\alpha$ maps,
\citet{Tacc18} showed that the dust attenuation peaks at the center, with an
$A_{V}$ on average $\rm \sim 1~mag$ higher than in the outer disk regions.
Most importantly, while the central dust attenuation is higher in the
four galaxies with $M_{\star} > 10^{11}~{\rm M_{\odot}}$ (Q2343-BX610,
Deep3a-15504, ZC400569, and ZC400528), the dust-corrected specific SFR
profiles still drop in the inner regions and thus support quenching of
star formation at the center of these galaxies.

\section{MEASUREMENTS OF THE H$\alpha$ KINEMATIC PROPERTIES}
         \label{Sect-kin_meas}

In this Section, we present the measurements of the intrinsic rotation
velocity and local velocity dispersion from the H$\alpha$ data of the
\sinszc\ AO sample, after summarizing the choice of galaxy parameters
necessary for the beam smearing and inclination corrections.
These measurements assume a rotating disk framework, which is
justified in more detail in the next Section
(see also \citealt{Sha08,Cre09}; \citeauthor*{FS09}; \citealt{Gen14a}).

\subsection{Beam Smearing and Inclination Corrections}
            \label{Sub-BS_sini}

We applied the beam smearing corrections for velocity and velocity
dispersion (denoted $C_{\rm PSF,v}$ and $C_{\rm PSF,\sigma}$)
presented by \citet[][Appendix A]{Bur16}.
These corrections were derived from rotating disk models ($n = 1$) with a
range of sizes, inclinations, and masses appropriate for our sample and
convolved with the \psfave.  As discussed in Section~\ref{Sect-struct_meas}
and further in Section~\ref{Sect-class}, the disk assumption is valid for
the vast majority of our galaxies but the corrections are necessarily more
uncertain for the few non-disk systems.
The beam smearing depends on the ratio of galaxy to PSF sizes as well as the
radius at which the observed velocity and $\sigma_{0}$ values are measured.
In velocity, there is very little dependence on galaxy inclination and mass
at fixed size.  In contrast, the effects on $\sigma_{0}$ also depend fairly
sensitively on galaxy inclination and mass.
It is also important to account for the extended wings of the AO PSF.
For instance, the beam smearing corrections in velocity for our objects
are $10 - 40\%$ larger than those for a single-Gaussian PSF with
FWHM equal to that of the core component of the \psfave.

For the purpose of beam smearing corrections (and kinematic analysis), we
adopted the $R_{\rm e}$ values from the single-component S\'ersic model fits
to the {\em HST\/} $H$-band imaging when available \citep[from][]{Tacc15b},
and from the H$\alpha$ maps for the other sources (from
Section~\ref{Sub-sizes_galfit}).
The $H$-band light being dominated by the continuum from stars making up
the bulk of the stellar mass, it is less likely to be affected by sites of
on-going intense star formation, and the parametric fits to $H$-band maps
are arguably more robust because of the higher Strehl ratio, simultaneous
PSF, and much wider FOV of the {\em HST\/} imaging compared to the SINFONI
H$\alpha$ data.  As noted in the previous Section, the difference in sizes
between H$\alpha$ and $H$-band emission is small such that the choice made
here has little consequence on the results.  For reference, the $R_{\rm e}$
adopted here for each galaxy is given in Table~\ref{tab-kin}.

Similarly, we adopted the inclinations inferred from the axis ratios $q$
from the fits to the $H$-band maps whenever possible and from the H$\alpha$
maps otherwise, with the exceptions detailed below (and indicated in
Table~\ref{tab-kin}).  We assumed a disk geometry with finite intrinsic
thickness $q_{0}$ and computed the inclination $i$ via
$\sin^{2}(i) = (1 - q^{2}) / (1 - q_{0}^{2})$.
We used $q_{0} = 0.20$, motivated by the typical local random motions
relative to rotational motions in previous work and structural studies
from deep high resolution rest-UV/optical imaging of $z \sim 2$ disks
\citep[e.g.,][]{Elm05,Elm17,Gen08,Cre09,Law12b,Law12a,New13,vdW14a,Wis15}.
The differences with inclination corrections in the thin disk approximation
are however $< 5\%$, so the exact choice has little impact on our analysis.
For eight of the most compact sources, the $q$ estimates are quite uncertain.
We assigned to these objects the average for a distribution of randomly
inclined disks corresponding to $\langle \sin(i) \rangle = \pi / 4$
\citep[e.g.,][]{Law09}.
For four large disks with bright off-center clumps and non-axisymmetric
features in both H$\alpha$ and $H$-band emission, the inclinations implied
by the morphological axis ratios lead to models that do not reproduce the
observed kinematics well and poorly match the mass priors from the stellar
and gaseous components \citep[][]{Gen14a, Gen17}.
For those cases, which include Deep3a-6004 with nearly orthogonal kinematic
and morphological major axes, we adopted the inclination from the best-fit
disk modeling.

\subsection{Intrinsic Rotation Velocity and Velocity Dispersion}
            \label{Sub-vobs_sig0}

We determined the maximum observed velocity difference across the galaxies,
$\Delta v_{\rm obs}$, based on the profiles extracted along the kinematic
major axis as defined in Section~\ref{Sub-pv}.  For objects that do not
exhibit regular disk-like kinematics, we evaluated the $\Delta v_{\rm obs}$
from the velocity maps and from the identification of the bluest and reddest
spectral channels at which H$\alpha$ emission is still detected over at least
one spatial resolution element.
We then calculated the intrinsic rotation velocity through
$V_{\rm rot} \times \sin(i) = C_{\rm PSF,v} \times \Delta\,v_{\rm obs}/2$.
The beam smearing correction factors $C_{\rm PSF,v}$ for our galaxies
are 1.3 on average (and median), and at most 2 for the smallest objects
(Q1623-BX502, SA12-6339, ZC404221).

In the framework of rotating disks, the observed line width at a given
position reflects the intrinsic local velocity dispersion $\sigma_{0}$
(related to the disk thickness) as well as the beam-smeared line-of-sight
velocity distribution.  For the galaxies with most regular kinematics, 
we determined $\sigma_{\rm 0,obs}$ from the velocity dispersion profiles
along the kinematic major axis at the largest radii possible, away from
the central peak caused by the steep inner disk velocity gradient.
For the other sources, we estimated $\sigma_{\rm 0,obs}$
from the dispersion maps and from the line widths in the outer parts
of the galaxies from inspection of the data cubes.  We then computed the
intrinsic $\sigma_{0} = C_{\rm PSF,\sigma} \times \sigma_{\rm 0,obs}$,
where the correction factors also take into account galaxy inclination,
size, and mass relevant to the spatial beam smearing in dispersion.
The corrections for our sample have an average and median of $\approx 0.85$,
and range between 0.3 and 1.0 (three objects have $C_{\rm PSF,\sigma} < 0.6$,
driven by their compactness: ZC400528, ZC404221, and SA12-6339).

Two underlying assumptions in our $\sigma_{0}$ estimates (and beam smearing
corrections) are that the intrinsic velocity dispersion is isotropic and
that it is constant throughout the regions probed by our SINFONI$+$AO data.
The latter assumption is motivated by the detailed analysis of the residuals
in velocity dispersion between the data and disk models for several \sinszc\
galaxies with highest quality AO and seeing-limited observations
\citep[see][]{Gen06, Gen08, Gen17, Cre09}.
Although the impact of beam smearing is reduced at the higher resolution
of AO-assisted data compared to seeing-limited observations, the derived
$\sigma_{0}$ can still include contributions from non-circular motions on
$\rm \la 1 - 2~kpc$ scales.  Naturally, for non-disk systems, $\sigma_{0}$
may also be more related to kinematic perturbations than to support from
random motions in a disk-like geometry.

Table~\ref{tab-kin} reports the derived $\Delta v_{\rm obs}$, $\sin(i)$,
$V_{\rm rot}$, and $\sigma_{0}$ values along with the $V_{\rm rot}/\sigma_{0}$
ratios as a measure of the relative amount of dynamical support from ordered
rotation (or orbital motions) versus random (and non-circular) motions.
The uncertainties are taken as the 68\% confidence intervals derived from a
Monte Carlo approach (with 1000 realizations), propagating the errors of the
measurements, galaxy inclinations, sizes, and masses, and \psfave\ parameters
to better account for the non-Gaussian nature of the uncertainties of input
properties and the non-linear functions involved.
For our sample, $V_{\rm rot}$ ranges from 38 to $\rm 364~km\,s^{-1}$,
with a mean and median of 181 and $\rm 141~km\,s^{-1}$, respectively.
The $\sigma_{0}$ estimates are between 20 and $\rm 77~km\,s^{-1}$, with
a mean of 49 and nearly identical median of $\rm 51~km\,s^{-1}$.
The $V_{\rm rot}/\sigma_{0}$ ratios span the range from 0.97 to 13,
with an average of 4.1 and median of 3.2.
These values compare well with the ensemble properties of $1.4 < z < 2.6$
disks that are sufficiently resolved in the seeing-limited data of the
parent \sinszc\ sample and the $\rm KMOS^{3D}$ survey with the VLT/KMOS
multi-IFU (\citeauthor*{FS09, Man11}; \citealt{Wis15, Bur16, Wuy16, Ueb17}).
Thus, our AO sample does not stand out in its global kinematic properties
from larger and more complete SFG samples at $z \sim 2$.

\tabletypesize{\footnotesize}
\begin{deluxetable*}{lccccccccccl}[p]
\renewcommand\arraystretch{0.95}
\tablecaption{Kinematic Properties and Dynamical Masses
              \label{tab-kin}}
\tablecolumns{12}
\tablewidth{0pt}
\setlength{\tabcolsep}{4pt}
\tablehead{
   \colhead{Source} &
   \colhead{$R_{\rm e}$\,\tablenotemark{a}} &
   \colhead{$\sin(i)$\,\tablenotemark{a}} &
   \colhead{$\rm PA_{kin}$} &
   \colhead{$\rm \Delta PA$\,\tablenotemark{b}} &
   \colhead{$\Delta v_{\rm obs}/2$} &
   \colhead{$V_{\rm rot}$} &
   \colhead{$\sigma_{0}$} &
   \colhead{$V_{\rm rot}/\sigma_{0}$} &
   \colhead{$V_{\rm c}$\,\tablenotemark{c}} &
   \colhead{$M_{\rm dyn}$\,\tablenotemark{c}} &
   \colhead{Disk criteria\,\tablenotemark{d}} \\[-1.8ex]
   \colhead{} &
   \colhead{(kpc)} &
   \colhead{} &
   \colhead{(deg)} &
   \colhead{(deg)} &
   \colhead{($\rm km\,s^{-1}$)} &
   \colhead{($\rm km\,s^{-1}$)} &
   \colhead{($\rm km\,s^{-1}$)} &
   \colhead{} &
   \colhead{($\rm km\,s^{-1}$)} &
   \colhead{($\rm 10^{10}~M_{\odot}$)} &
   \colhead{}
}
\startdata
Q1623-BX455    & $2.1^{+0.8}_{-0.4}$   & $0.98^{+0.01}_{-0.10}$ & $65 \pm 15$            & $10^{+20}_{-16}$       &
     $125 \pm 15$       & $175^{+54}_{-17}$   & $56^{+15}_{-24}$   & $3.1^{+2.4}_{-0.5}$   &
     $203^{+51}_{-15}$     & $3.9^{+2.1}_{-0.7}$   &  1,2,3,4,5        \\[0.65ex]
Q1623-BX502    & $1.1 \pm 0.7$         & $0.77^{+0.12}_{-0.18}$ & $45 \pm 15$            & $44 \pm 16$            &
     $33 \pm 10$        & $77^{+50}_{-32}$    & $40^{+16}_{-14}$   & $2.0^{+1.5}_{-0.8}$   &
     $106^{+45}_{-21}$     & $0.58^{+0.77}_{-0.21}$&  1,2,3,5          \\[0.65ex]
Q1623-BX543    & $3.3^{+1.2}_{-0.8}$   & $0.90 \pm 0.05$        & $0 \pm 25$             & $4^{+26}_{-25}$        &
     $93 \pm 15$        & $128^{+29}_{-21}$   & $70^{+15}_{-25}$   & $1.8^{+0.9}_{-0.3}$   &
     $181^{+28}_{-27}$     & $5.0^{+2.4}_{-2.0}$   &  Irr              \\[0.65ex]
Q1623-BX599    & $2.4 \pm 0.6$         & $0.74^{+0.12}_{-0.15}$ & $-55 \pm 25$           & $2 \pm 28$             &
     $80 \pm 20$        & $139^{+62}_{-36}$   & $71^{+18}_{-27}$   & $2.0^{+1.4}_{-0.4}$   &
     $191^{+55}_{-35}$     & $4.1^{+3.3}_{-1.7}$   &  Irr              \\[0.65ex]
Q2343-BX389    & $6.2 \pm 1.2$         & $0.98^{+0.01}_{-0.06}$ & $-50 \pm 5$            & $2 \pm 5$              &
     $260 \pm 23$       & $299^{+40}_{-21}$   & $56^{+13}_{-15}$   & $5.3^{+1.7}_{-0.7}$   &
     $316^{+39}_{-20}$     & $29^{+9}_{-6}$        &  1,2              \\[0.65ex]
Q2343-BX513    & $2.6^{+1.5}_{-0.8}$   & $0.78^{+0.14}_{-0.19}$ & $-35 \pm 10$           & $40^{+11}_{-12}$       &
     $60 \pm 15$        & $102^{+64}_{-26}$   & $55^{+24}_{-28}$   & $1.8^{+2.1}_{-0.5}$   &
     $144^{+64}_{-30}$     & $2.5^{+3.2}_{-1.2}$   &  Irr              \\[0.65ex]
Q2343-BX610    & $4.5 \pm 1.1$         & $0.86^{+0.08}_{-0.12}$ & $-10 \pm 10$           & $27 \pm 10$            &
     $180 \pm 25$       & $241^{+62}_{-38}$   & $64^{+17}_{-24}$   & $3.8^{+2.0}_{-0.6}$   &
     $268^{+60}_{-36}$     & $15^{+8}_{-5}$        &  1,2,3,4,5        \\[0.65ex]
Q2346-BX482    & $6.0 \pm 0.8$         & $0.88^{+0.08}_{-0.14}$ & $-65 \pm 5$            & $3 \pm 10$             &
     $225 \pm 20$       & $287^{+63}_{-30}$   & $58^{+14}_{-15}$   & $4.9^{+1.6}_{-0.7}$   &
     $306^{+65}_{-29}$     & $26^{+12}_{-5}$       &  1,2,3,4          \\[0.65ex]
Deep3a-6004    & $5.1 \pm 0.5$         & $0.44^{+0.23}_{-0.09}$ & $-20 \pm 10$           & $75 \pm 12$            &
     $135 \pm 10$       & $362^{+109}_{-126}$ & $55^{+11}_{-17}$   & $6.5^{+2.4}_{-1.8}$   &
     $376^{+106}_{-123}$   & $34^{+23}_{-19}$      &  1,2,3,5          \\[0.65ex]
Deep3a-6397    & $5.9^{+0.6}_{-0.9}$   & $0.50^{+0.21}_{-0.13}$ & $-80 \pm 5$            & $7^{+7}_{-6}$          &
     $150 \pm 25$       & $351^{+138}_{-107}$ & $59^{+13}_{-17}$   & $6.0^{+2.6}_{-1.6}$   &
     $367^{+134}_{-101}$   & $37^{+33}_{-18}$      &  1,2,3,4,5        \\[0.65ex]
Deep3a-15504   & $6.0 \pm 0.4$         & $0.57^{+0.18}_{-0.16}$ & $-35 \pm 5$            & $7 \pm 5$              &
     $150 \pm 20$       & $305^{+138}_{-80}$  & $63^{+13}_{-15}$   & $4.8^{+1.8}_{-1.1}$   &
     $327^{+130}_{-74}$    & $30^{+28}_{-12}$      &  1,2,3,4,5        \\[0.65ex]
K20-ID6        & $3.9 \pm 0.3$         & $0.52^{+0.22}_{-0.14}$ & $60 \pm 10$            & $23 \pm 11$            &
     $100 \pm 30$       & $236^{+128}_{-95}$  & $29 \pm 13$        & $8.1^{+6.0}_{-3.1}$   &
     $242^{+124}_{-90}$    & $11^{+14}_{-6}$       &  1,2,3,4          \\[0.65ex]
K20-ID7        & $8.4 \pm 1.1$         & $0.91^{+0.06}_{-0.16}$ & $25 \pm 5$             & $7 \pm 5$              &
     $210 \pm 25$       & $254^{+69}_{-28}$   & $49 \pm 15$        & $5.2^{+2.3}_{-0.8}$   &
     $270^{+68}_{-27}$     & $28^{+17}_{-6}$       &  1,2              \\[0.65ex]
GMASS-2303     & $1.6 \pm 0.5$         & $0.78^{+0.11}_{-0.17}$ & $-60 \pm 20$           & $79 \pm 24$            &
     $63 \pm 10$        & $127^{+64}_{-26}$   & $40 \pm 14$        & $3.2^{+2.2}_{-0.7}$   &
     $146^{+59}_{-22}$     & $1.6^{+1.4}_{-0.4}$   &  1,2              \\[0.65ex]
GMASS-2363     & $2.3 \pm 0.6$         & $0.87^{+0.07}_{-0.10}$ & $55 \pm 10$            & $6 \pm 10$             &
     $105 \pm 10$       & $168^{+45}_{-21}$   & $32^{+19}_{-17}$   & $5.3^{+5.0}_{-1.7}$   &
     $178^{+45}_{-13}$     & $3.4^{+1.9}_{-0.6}$   &  1,2,3,4,5        \\[0.65ex]
GMASS-2540     & $8.5 \pm 1.0$         & $0.52^{+0.21}_{-0.14}$ & $20 \pm 30$            & $66 \pm 35$            &
     $125 \pm 40$       & $266^{+122}_{-107}$ & $20^{+11}_{-9}$    & $13^{+9}_{-6}$        &
     $269^{+123}_{-105}$   & $29^{+33}_{-18}$      &  \ldots            \\[0.65ex]
SA12-6339      & $1.2^{+1.4}_{-0.6}$   & $0.78^{+0.14}_{-0.23}$ & $40 \pm 20$            & $51^{+38}_{-37}$       &
     $25 \pm 8$         & $58^{+44}_{-24}$    & $25^{+42}_{-8}$    & $2.3 \pm 1.5$         &
     $74^{+81}_{-6}$       & $0.31^{+1.84}_{-0.08}$&  Irr              \\[0.65ex]
ZC400528       & $2.4 \pm 0.7$         & $0.61^{+0.17}_{-0.18}$ & $80 \pm 20$            & $35 \pm 22$            &
     $150 \pm 25$       & $341^{+184}_{-89}$  & $28^{+23}_{-15}$   & $12^{+14}_{-5}$       &
     $344^{+186}_{-84}$    & $13^{+15}_{-6}$       &  1,2,3,5          \\[0.65ex]
ZC400569       & $7.4^{+1.3}_{-1.5}$   & $0.98^{+0.01}_{-0.02}$ & $15 \pm 15$            & $11 \pm 15$            &
     $263 \pm 10$       & $312^{+19}_{-13}$   & $41^{+23}_{-21}$   & $7.6^{+5.0}_{-2.1}$   &
     $321^{+25}_{-12}$     & $36^{+8}_{-5}$        &  Irr              \\[0.65ex]
ZC400569N      & $7.1 \pm 1.4$         & $0.71^{+0.13}_{-0.17}$ & $70 \pm 15$            & $32 \pm 18$            &
     $220 \pm 25$       & $364^{+138}_{-64}$  & $43^{+16}_{-21}$   & $8.5^{+7.2}_{-1.8}$   &
     $372^{+136}_{-63}$    & $46^{+38}_{-16}$      &  1,2,3,5          \\[0.65ex]
ZC401925       & $2.6 \pm 0.6$         & $0.78^{+0.13}_{-0.21}$ & $60 \pm 10$            & $60 \pm 12$            &
     $55 \pm 20$        & $96^{+62}_{-34}$    & $59^{+18}_{-19}$   & $1.6^{+1.2}_{-0.5}$   &
     $145^{+57}_{-30}$     & $2.5^{+2.6}_{-1.0}$   &  1                \\[0.65ex]
ZC403741       & $2.2^{+0.4}_{-0.3}$   & $0.55^{+0.11}_{-0.13}$ & $25 \pm 10$            & $20 \pm 19$            &
     $70 \pm 8$         & $189^{+73}_{-36}$   & $36^{+13}_{-12}$   & $5.2^{+2.9}_{-1.1}$   &
     $201^{+74}_{-34}$     & $4.1^{+3.4}_{-1.3}$   &  1,2,3,4,5        \\[0.65ex]
ZC404221       & $0.8 \pm 0.6$         & $0.78^{+0.14}_{-0.18}$ & $90 \pm 20$            & $77 \pm 20$            &
     $38 \pm 13$        & $100^{+44}_{-46}$   & $29^{+40}_{-8}$    & $3.4^{+1.1}_{-2.2}$   &
     $114^{+70}_{-20}$     & $0.48^{+1.43}_{-0.15}$&  Irr              \\[0.65ex]
ZC405226       & $5.4 \pm 0.8$         & $0.82^{+0.10}_{-0.12}$ & $-40 \pm 10$           & $26 \pm 14$            &
     $100 \pm 23$       & $143^{+45}_{-36}$   & $58^{+19}_{-18}$   & $2.5^{+1.1}_{-0.6}$   &
     $178^{+45}_{-32}$     & $8.0^{+5.0}_{-2.6}$   &  1,2,3,4,5        \\[0.65ex]
ZC405501       & $5.8 \pm 0.9$         & $0.97^{+0.02}_{-0.07}$ & $10 \pm 5$             & $1 \pm 5$              &
     $85 \pm 15$        & $101^{+23}_{-16}$   & $52^{+16}_{-13}$   & $1.9^{+0.6}_{-0.4}$   &
     $139^{+28}_{-17}$     & $5.2^{+2.3}_{-1.4}$   &  1,2,3,4,5        \\[0.65ex]
ZC406690       & $7.0 \pm 1.2$         & $0.44^{+0.23}_{-0.09}$ & $-70 \pm 10$           & $7 \pm 12$             &
     $120 \pm 10$       & $313^{+88}_{-107}$  & $60^{+16}_{-15}$   & $5.3^{+1.6}_{-1.7}$   &
     $332^{+88}_{-97}$     & $36^{+21}_{-18}$      &  1,2,3,4          \\[0.65ex]
ZC407302       & $3.6 \pm 1.2$         & $0.91^{+0.05}_{-0.09}$ & $55 \pm 5$             & $8 \pm 6$              &
     $165 \pm 35$       & $217^{+71}_{-40}$   & $56^{+11}_{-25}$   & $3.9^{+2.8}_{-0.6}$   &
     $240^{+62}_{-39}$     & $9.6^{+6.0}_{-3.8}$   &  1,2,3,4,5        \\[0.65ex]
ZC407376       & $5.5^{+0.7}_{-0.3}$   & $0.97^{+0.02}_{-0.04}$ & $20 \pm 25$            & $5 \pm 25$             &
     $73 \pm 18$        & $86^{+24}_{-20}$    & $56^{+28}_{-27}$   & $1.6^{+1.0}_{-0.5}$   &
     $134^{+46}_{-32}$     & $4.6^{+3.8}_{-2.0}$   &  Irr              \\[0.65ex]
ZC407376S      & $1.6 \pm 0.5$         & $0.60 \pm 0.18$        & $-60 \pm 10$           & $8 \pm 14$             &
     $35 \pm 18$        & $89^{+65}_{-45}$    & $77^{+37}_{-41}$   & $1.2^{+1.4}_{-0.5}$   &
     $167^{+86}_{-57}$     & $2.1^{+3.1}_{-1.3}$   &  Irr              \\[0.65ex]
ZC407376N      & $1.4 \pm 0.3$         & $0.83^{+0.09}_{-0.13}$ & $60 \pm 10$            & $6 \pm 14$             &
     $58 \pm 15$        & $113^{+49}_{-30}$   & $35 \pm 14$        & $3.2^{+2.3}_{-0.9}$   &
     $130^{+48}_{-24}$     & $1.1^{+0.9}_{-0.3}$   &  1,2              \\[0.65ex]
ZC409985       & $1.9 \pm 0.2$         & $0.76^{+0.11}_{-0.16}$ & $-15 \pm 20$           & $1 \pm 20$             &
     $20 \pm 8$         & $38^{+20}_{-14}$    & $39^{+15}_{-13}$   & $0.97^{+0.57}_{-0.30}$&
     $80^{+30}_{-20}$      & $0.57^{+0.52}_{-0.26}$&  Irr              \\[0.65ex]
ZC410041       & $4.7 \pm 1.2$         & $1.00^{+0.00}_{-0.03}$ & $-55 \pm 10$           & $9 \pm 10$             &
     $88 \pm 10$        & $101^{+15}_{-9}$    & $48^{+14}_{-16}$   & $2.1^{+0.9}_{-0.4}$   &
     $134^{+22}_{-15}$     & $3.9^{+1.7}_{-1.3}$   &  1,2,3,4,5        \\[0.65ex]
ZC410123       & $3.2 \pm 0.7$         & $0.94^{+0.03}_{-0.07}$ & $35 \pm 25$            & $19 \pm 26$            &
     $63 \pm 15$        & $81^{+24}_{-17}$    & $60^{+25}_{-26}$   & $1.4^{+0.9}_{-0.4}$   &
     $137^{+44}_{-32}$     & $2.8^{+2.4}_{-1.3}$   &  1                \\[0.65ex]
ZC411737       & $1.8 \pm 0.2$         & $0.78^{+0.14}_{-0.18}$ & $-60 \pm 10$           & $41 \pm 12$            &
     $73 \pm 18$        & $139^{+56}_{-36}$   & $38^{+20}_{-18}$   & $3.7^{+2.8}_{-1.1}$   &
     $156^{+61}_{-32}$     & $2.0^{+2.0}_{-0.8}$   &  1,2              \\[0.65ex]
ZC412369       & $3.1 \pm 1.1$         & $0.90^{+0.05}_{-0.08}$ & $-70 \pm 20$           & $16 \pm 22$            &
     $80 \pm 20$        & $120^{+49}_{-27}$   & $75^{+13}_{-27}$   & $1.6^{+1.0}_{-0.3}$   &
     $182^{+37}_{-27}$     & $4.8^{+2.8}_{-1.9}$   &  1                \\[0.65ex]
ZC413507       & $2.6 \pm 0.5$         & $0.77^{+0.13}_{-0.20}$ & $-35 \pm 10$           & $13 \pm 12$            &
     $70 \pm 20$        & $132^{+69}_{-40}$   & $42^{+21}_{-19}$   & $3.1^{+2.5}_{-0.9}$   &
     $153^{+70}_{-35}$     & $2.8^{+3.1}_{-1.2}$   &  1,2              \\[0.65ex]
ZC413597       & $1.6 \pm 0.5$         & $0.78^{+0.13}_{-0.18}$ & $45 \pm 40$            & $30 \pm 41$            &
     $48 \pm 10$        & $92^{+52}_{-24}$    & $38^{+21}_{-18}$   & $2.4^{+2.6}_{-0.8}$   &
     $115^{+54}_{-19}$     & $0.98^{+1.24}_{-0.34}$&  Irr              \\[0.65ex]
ZC415876       & $2.4 \pm 1.0$         & $0.64^{+0.16}_{-0.19}$ & $-50 \pm 15$           & $30 \pm 41$            &
     $73 \pm 15$        & $153^{+105}_{-41}$  & $47^{+13}_{-19}$   & $3.2^{+3.0}_{-0.7}$   &
     $176^{+98}_{-34}$     & $3.5^{+4.4}_{-1.3}$   &  1,2,3,4,5        \\[0.65ex]
\enddata
\parskip=-2.0ex
\tablenotetext{a}
{
Adopted half-light radius and galaxy inclination, from the best-fit Sersic
model to {\em HST\/} $H$-band imaging when available \citep{Tacc15b} or to
the H$\alpha$ maps otherwise (Table~\ref{tab-struct}, using the \psfave\ case).
Exceptions are five compact sources (Q2343-BX513, ZC401925, ZC404221,
ZC411737, and ZC413597) for which we adopted the average
$\langle \sin(i) \rangle = \pi / 4$ for randomly inclined disks, and
four large disks (Deep3a-6004, Deep3a-6397, Deep3a-15504, and ZC406690)
for which we adopted the values inferred from detailed kinematic modeling
\citep{Gen14a, Gen17}, as discussed in Section~\ref{Sub-BS_sini}.
}
\tablenotetext{b}
{
Kinematic misalignment, based on the position angle derived from the
structural fits to {\em HST\/} $H$-band imaging when
available or to the H$\alpha$ maps otherwise.
}
\tablenotetext{c}
{
Circular velocity and total dynamical mass in the rotating disk framework.
Arguably, the $V_{\rm rot}$ and $\sigma_{0}$ estimates become more uncertain
for the smallest sources due to large beam smearing, or for objects with
observed irregular kinematics.
Assuming that rotational motions still intrinsically dominate the gravitational
support
for the 10 objects with irregular kinematics (indicated in the rightmost
column), an alternative estimate based on the integrated H$\alpha$ line
width with
$V_{\rm c} = \sigma_{\rm tot}({\rm H\alpha}) / 0.8\,\sin(i)$
would yield the following values of
$V_{\rm c}$ and $M_{\rm dyn}/10^{10}~{\rm M_{\odot}}$:
Q1623-BX543: $227^{+22}_{-46}$, $7.9^{+0.4}_{-4.1}$;  
Q1623-BX599: $305^{+37}_{-90}$, $10^{+2}_{-6}$;
Q2343-BX513: $224^{+17}_{-66}$, $6.1^{+0.2}_{-3.7}$;
SA12-6339  : $180^{+11}_{-51}$, $1.8^{+0.7}_{-1.1}$;
ZC400569 (full system): $254^{+3}_{-57}$, $22^{+1}_{-10}$;
ZC404221   : $138^{+7}_{-42}$, $0.71^{+0.24}_{-0.40}$;
ZC407376 (full system): $170^{+10}_{-29}$, $7.4^{+0.7}_{-2.4}$;
ZC407376S  : $311^{+52}_{-126}$, $7.2^{+1.7}_{-5.0}$;
ZC409985   : $116^{+9}_{-38}$, $1.2^{+0.1}_{-0.7}$;
ZC413597   : $147^{+8}_{-44}$, $1.6^{+0.1}_{-0.9}$.
}
\tablenotetext{d}
{
List of disk criteria fulfilled by each galaxy, as described in
Section~\ref{Sub-kinclass}.  Sources for which the observed velocity field
is irregular and without clear monotonic gradient (our disk criterion 1)
are indicated with ``Irr.''
GMASS-2540 is not classified because of the FOV and S/N
limitations of the SINFONI AO data.
}
\parskip=0ex
\end{deluxetable*}

\subsection{Kinematic and Morphological Alignment}
            \label{Sub-PAs}

From simple geometrical arguments, the line of nodes of an axisymmetric
oblate rotating disk with a smooth light distribution is expected to be
aligned with its projected morphological major axis.  The coincidence
in position angles from the kinematics and morphology is thus generally
a criterion in determining the nature of a galaxy.
In practice, complications arise for a variety of physical reasons
(e.g., presence of non-axisymmetric structures, nearby companions
with overlapping outer isophotes, spatially non-uniform extinction),
compounded by surface brightness sensitivity and spatial resolution
limitations for high redshift galaxies.

\begin{figure}[!ht]
\begin{center}
\includegraphics[scale=0.73,clip=1,angle=0]{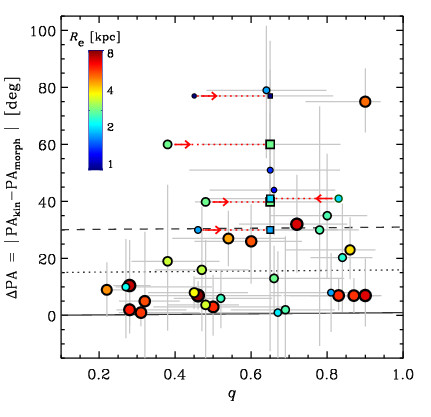}
\end{center}
\vspace{-0.7cm}
\renewcommand\baselinestretch{0.5}
\caption{
\small
Kinematic misalignment $\rm \Delta PA$ as a function of morphological
axis ratio $q$ for the \sinszc\ AO sample.
The symbols are color-coded and sized according to the effective radius
$R_{\rm e}$ of the galaxies.
The five compact objects with ill-defined axis ratios and to which we
assigned $q = 0.64$, corresponding to the $\langle \sin(i) \rangle = \pi / 4$
for randomly oriented disks (Section~\ref{Sub-BS_sini}), are plotted as
squares connected with red dotted lines to their nominal best-fit $q$
shown as circles with a red arrow pointing towards the adopted value.
Most of the well resolved objects show very good agreement between their
kinematic and morphological position angles, especially for the more
edge-on systems.  A larger proportion of the compact sources have
$\rm \Delta PA > 30^{\circ}$, reflecting ill-defined PA's due to
irregular or fairly featureless morphologies and/or velocity fields.
\label{fig-dPA}
}
\end{figure}

We examined the agreement between the derived kinematic position angle
$\rm PA_{kin}$ (Section~\ref{Sub-pv}) and the morphological position angle
$\rm PA_{morph}$, quantified via the ``kinematic misalignment''
$\rm \Delta PA = \vert PA_{kin} - PA_{morph} \vert$.
GMASS-2540 is excluded here because it is not fully covered by the SINFONI AO
data and because of its low H$\alpha$ surface brightness emission hampering a
reliable $\rm PA_{kin}$ determination.
Following the same motivation as in Section~\ref{Sub-BS_sini}, we adopted as
$\rm PA_{morph}$ the position angle from the $H$-band morphologies when
available, and from the H$\alpha$ maps otherwise.
As noted in Section~\ref{Sub-struct_disc}, the $\rm PA_{H\alpha}$ and
${\rm PA}_{H}$ are similar within $1.5\sigma$, and we verified that our
conclusions would not be affected by adopting instead $\rm PA_{H\alpha}$
for all objects.

For our sample, the mean and median $\rm \Delta PA$ are $23^{\circ}$ and
$13^{\circ}$, respectively.  For 73\% of the galaxies, the alignment
is better than $30^{\circ}$.  Figure~\ref{fig-dPA} shows the position
angle offsets as a function of the adopted axis ratio $q$ and effective
radius $R_{\rm e}$.  Not unexpectedly, and as seen in other high-redshift
samples \citep[e.g.,][]{Wis15, Har17}, our galaxies broadly follow a
trend of increasing $\rm \Delta PA$ for more face-on and smaller sources.
Of the ten objects with $\rm \Delta PA > 30^{\circ}$, eight are compact
sources with distortions/extensions in their outer isophotes, asymmetric
light distributions in their inner brighter regions, or little or irregular
velocity structure (Q1623-BX502, Q1623-BX513, GMASS-2303, SA12-6339, ZC400528,
ZC401925, ZC404221, ZC411737).  The other two sources are large, massive and
fairly face-on disks for which the misalignment can be attributed to
morphological features (a non-circular ring with a clump on one side
for Deep3a-6004, and clumps in a tail-like structure for ZC400569N;
see Section~\ref{Sub-notes} for more details).

From the case-by-case inspection of our galaxies, it is clear that while
it provides a useful indication, the kinematic misalignment should be used
with caution when classifying high-redshift objects as disks versus non-disks.
Asymmetric morphological features can drive up the $\rm \Delta PA$ in
otherwise kinematically regular disks.  The finite angular resolution and
S/N will limit the ability to determine reliable PA's for compact objects.
For similar reasons, the inclinations based on morphological axis ratios,
in particular for cases with a large misalignment or a more face-on
orientation, may be very uncertain or even wrong (see also, e.g., the
discussions by \citealt{Wuy16} and \citealt{Rod17}).
Ideally, inclinations should be validated with kinematic modeling when
possible, as done in Section~\ref{Sub-BS_sini}.

\subsection{Circular Velocities and Dynamical Masses}
            \label{Sub-masses}

With the intrinsic structural and kinematic properties derived in the
previous Sections, we calculated the circular velocities and dynamical
masses in the rotating disk framework as:
\begin{equation}
 V_{\rm c} = \left(V_{\rm rot}^2 + 3.36\,\sigma_{0}^2\right)^{0.5}
\label{Eq-Vc}
\end{equation}
and
\begin{equation}
 M_{\rm dyn} = 2 \times R_{\rm e}\,V_{\rm c}^2 / G
\label{Eq-Mdyn}
\end{equation} 
\citep{Bin08, Bur10, Bur16},
where $G$ is the gravitational constant.
In equation~(\ref{Eq-Vc}), the dispersion term accounts for pressure support
in an exponential distribution, which is significant for our galaxies
given their typically low $V_{\rm rot}/\sigma_{0} \sim 3 - 4$ ratios.
Equation~(\ref{Eq-Mdyn}) corresponds to the total disk mass in the spherical
approximation (for an infinitely thin Freeman disk, the $M_{\rm dyn}$ values
would scale down by a factor of 0.8; \citealt{Bin08}).
The circular velocities and dynamical masses are reported in
Table~\ref{tab-kin}, with uncertainties derived from a Monte Carlo
approach as for the $V_{\rm rot}$ and $\sigma_{0}$ estimates.

\looseness=-2
Equations~(\ref{Eq-Vc}) and (\ref{Eq-Mdyn}) rely on two simplifying but
important assumptions.
Firstly, it is assumed that all galaxies are rotating disks.
As discussed in more detail in Section~\ref{Sect-class}, the
kinematic properties of the majority of the galaxies are disk-like.
If rotational motions still intrinsically dominate the gravitational
support (either disk rotation or orbital motions) for the ten objects
with observed irregular (or featureless) kinematics, estimates can
be obtained for instance from the integrated H$\alpha$ line width
with $V_{\rm c} = \sigma_{\rm tot}({\rm H\alpha}) / 0.8\,\sin(i)$
(following, e.g., \citealt{Dok15, Wis18} but neglecting the contribution
from outflowing gas and the explicit split of pressure support; see also,
e.g., \citealt{Rix97, Wei06}).  The $V_{\rm c}$ values in that case are
typically a factor of 1.4 higher and the $M_{\rm dyn}$ are about twice higher.

Secondly, it is assumed that the mass distribution is exponential.
Again, this is roughly the case for most of the galaxies in our sample
(see Section~\ref{Sect-struct_meas}).  To gauge the impact of deviations
from $n = 1$, we re-computed $V_{\rm c}$ and $M_{\rm dyn}$ applying the
corrections for Sersic index and dispersion truncation described by
\citet{Rom12} and \citet{Bur16}.  The $V_{\rm c}$ and $M_{\rm dyn}$
values scale by factors in the range $0.6 - 1.9$ and $0.4 - 3.6$,
respectively, but the overall changes are small with an average increase
by $8\%$ in $V_{\rm c}$ and $22\%$ in $M_{\rm dyn}$, and a median increase
by only $\sim 1\%$ in both quantities (reflecting the mean and median
$n$ close to 1 of the sample).

The $M_{\rm dyn}$ estimates derived here agree well with the
results obtained from the more detailed kinematic modeling of 19 of the
galaxies presented by \citet{Gen14a,Gen17}.  Excluding GMASS-2540, the
average (median) differences amount to 0.11 (0.17)~dex, smaller than the
typical uncertainties of our estimates (0.34~dex) and comparable to those
of the modeling results (0.15~dex).
Whereas a modeling approach is more accurate by better taking into account
details of the mass distribution (such as possible bulge and disk components),
the agreement between the simpler approach and the modeling-based results is
reassuring, and indicates that the beam smearing corrections applied and the
galaxy parameters adopted (Sections~\ref{Sub-BS_sini} and \ref{Sub-vobs_sig0})
are overall satisfactory.

\section{NATURE OF THE GALAXIES}
            \label{Sect-class}

In this Section, we revisit the classification of the \sinszc\ AO targets
and highlight features of interest in several of the targets.
This analysis updates previous results with the now complete sample and
SINFONI AO data sets, and summarizes findings from more detailed case
studies presented elsewhere
(\citeauthor*{FS09}; \citealt{Gen06,Gen08,Gen11,Gen14a,Gen17};
\citealt{Sha08,Cre09}; \citealt{New13}; \citealt{Tacc15a,Tacc15b,Tacc18}).

\subsection{Kinematic Classification}
           \label{Sub-kinclass}

The higher resolution of the AO observations allows us to better
assess the nature of compact objects and reveals more detail in
the H$\alpha$ morphologies and kinematics of the larger sources.
To characterize these details quantitatively and accurately relies on very
high S/N over many resolution elements such that all components of the mass
model (e.g., bulge and disk) can be well determined and perturbations can
be well constrained \citep[e.g.,][]{Kra06}.
This is only possible for a small subset of our sample
\citep[e.g.,][]{Sha08,Gen14a,Gen17}.
We focus here on a simpler approach that can be applied to every galaxy,
with the aim of distinguishing between disks and non-disks.
The full system ZC400569 and its massive northern component ZC400569N are
considered individually, and so are the full ZC407376 and its two clearly
separated components.  The very large and low surface brightness GMASS-2540
is excluded but we note that despite the FOV and S/N limitations, the
SINFONI AO data suggest it is a large low-inclination clumpy disk.

We classified an object as disk-like (hereafter simply ``disk'') if at least
one of the following criteria is satisfied (analogous to the set presented
by \citealt{Wis15}):
\\[1.2ex]
\parbox[t]{0.3cm}{1.}
\parbox[t]{8.2cm}{
 The velocity field exhibits a monotonic gradient with well-defined kinematic
 position angle; in the larger systems with high S/N data, this corresponds
 to the detection of a ``spider'' diagram \citep{vdK78}.
}
\\[1.2ex]
\parbox[t]{0.3cm}{2.}
\parbox[t]{8.2cm}{
 The ratio $V_{\rm rot}/\sigma_{0} > \sqrt{3.36}$,
 corresponding to the value at which rotation starts to dominate over
 velocity dispersion in the dynamical support within $R_{\rm e}$ for
 $n = 1$ turbulent disks (Section~\ref{Sub-masses}).
}
\\[1.2ex]
\parbox[t]{0.3cm}{3.}
\parbox[t]{8.2cm}{
 The position of steepest velocity gradient, at the mid-point between
 the velocity extrema along the kinematic axis, coincides with the
 observed peak of velocity dispersion within the uncertainties
 ($\rm \approx 2~pixels$); this position defines the kinematic center
 (Section~\ref{Sub-pv}).
}
\\[1.2ex]
\parbox[t]{0.3cm}{4.}
\parbox[t]{8.2cm}{
 The morphological and kinematic position angles agree, with
 $\rm \Delta PA \leq 30^{\circ}$.
}
\\[1.2ex]
\parbox[t]{0.3cm}{5.}
\parbox[t]{8.2cm}{
 The kinematic center coincides within 2 pixels ($0\farcs 1$) with the
 light-weighted center of the continuum emission (or mass-weighted center
 of the stellar mass distribution when available), as a proxy for the
 gravitational potential; in most cases, the continuum and stellar mass
 maps are fairly smooth and regular such that the weighted center is located
 at/near the central peak, and usually corresponds to a bulge-like component
 in the more massive galaxies.
}
\\[1.2ex]
Table~\ref{tab-kin} lists the disk criteria fulfilled by each galaxy;
in case a source does not meet the necessary first criterion, it is
marked as irregular (``Irr'').

Not counting the full systems ZC400569 and ZC407376, 27 of the other 35
classified objects satisfy the first criterion of a disk-like velocity
gradient.
In a majority of them (16), a flattening or even a turnover is observed in
the velocity curve along the kinematic axis; the more demanding detection
of a spider diagram is fulfilled typically -- but not exclusively -- by
the larger objects with higher S/N.  
When adding the second criterion, 24 of the objects are rotation-dominated
disks.  Various thresholds in $V_{\rm rot}/\sigma_{0}$ have been used in the
literature; all objects with disk-like velocity gradients in our sample have
a ratio above 1, and a large majority (20) still has a ratio above 3.
Among the 24 rotation-dominated disks, 15 exhibit a clear peak in their
velocity dispersion at/near the position of steepest velocity gradient, and
thus meet the third criterion.  In three additional cases, the dispersion
profile is fairly flat, which can be attributed primarily to the low mass
and low to modest central mass concentration of these objects.

Considering the last two criteria involving morphologies, 14 of the 18
rotation-dominated disks with centrally peaked or flat dispersion profile
satisfy $\rm \Delta PA \leq 30^{\circ}$, and 11 of them also have their
kinematic and continuum/mass centers coincident.  The four galaxies that
drop because of a large position angle misalignment show however nearly
coincident centers.  The misalignment is driven by extended low-level
emission affecting the outer isophotes (Q1623-BX502, ZC400528, ZC400569N)
or bright off-center clumps and possible deviations from disk circularity
(Deep3a-6004).  Two of the three galaxies that have aligned PA's but offset
centers exhibit a prominent clumpy ring-like structure even in $H$-band light
and stellar mass maps (Q2346-BX482, ZC406690); the light-weighted center
is just marginally off the kinematic center (by about 2.5~pixels, or
$0\farcs 13$) and obviously affected by the asymmetric clump distribution
along the ring.  When considering the geometric center (i.e., unweighted) or
the center of the outer isophotes, these two galaxies would then satisfy the
fifth criterion as well.  The third source that meets all but this criterion
has a significant projected neighboring source in continuum light and overall
lower H$\alpha$ S/N ratio such that the center positions are more uncertain
(K20-ID6).

Ten objects do not exhibit a disk-like velocity gradient, including the full
systems ZC400569 and ZC407376 (although one component of each is classified
as a disk).   The other eight objects tend to lack any kinematic structure,
with some of them having nearby companions and others showing also little
morphological structure.  The lack of kinematic structure could reflect an
intrinsic dynamical property or result from projection effects (e.g., in
nearly face-on disks) and beam-smearing.  All of these eight sources
are compact ($R_{\rm e} = 0.8 - 3.3~{\rm kpc}$),
with low $\Delta v_{\rm obs}/2 \sim 20 - 90~{\rm km\,s^{-1}}$ and
inferred $V_{\rm rot}/\sigma_{0}$ ratios below or within $1\sigma$ of
1.8.

In summary, $\sim 70\%$ of the objects in our AO sample satisfy criteria
1$-$3 and are thus kinematically-classified rotation-dominated disks.
The other $\sim 30\%$ have observed properties that are not compatible
with signatures of rotation in an inclined disk, i.e. exhibit clearly
irregular or fairly featureless 2D kinematics in the data.

\subsection{Features of Note in Individual Galaxies}
           \label{Sub-notes}

Obviously, the set of disk criteria above is somewhat simplistic.
It is designed assuming ideal smooth rotating disks and neglects the
signatures of internally- or externally-driven perturbations caused by
non-axisymmetric substructures such as massive disk clumps, bars, and
spiral arms, or induced by minor mergers and interactions
\citep[e.g.,][]{Bour07, Genel12, Cev12}, which may also be difficult to
disentangle from each other.
Low-inclination disks may appear as fairly featureless in their velocity
field and dispersion maps, and the kinematic misalignment may be more
affected by asymmetric features, star-forming clumps, or
possible deviations from circularity \citep[e.g.,][]{Wuy16}.
Some merger configurations can mimic the smooth monotonic disk-like velocity
gradients of ordered disk rotation, or out-of-equilibrium disks can form in
late merger stages \citep[e.g.,][]{Law06,Sha08,Rob06,Rob08}.

\begin{figure}[!ht]
\begin{center}
\includegraphics[scale=0.79,clip=1,angle=0]{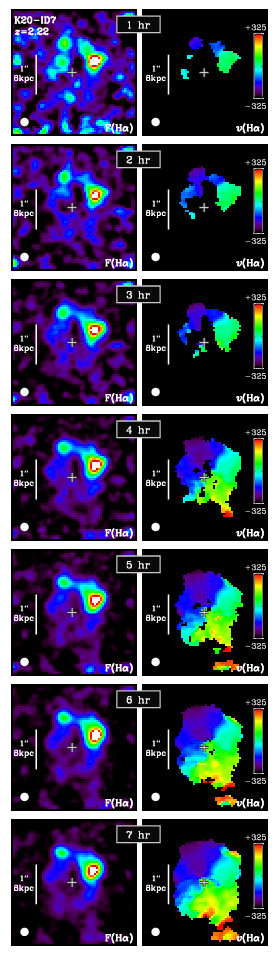}
\end{center}
\vspace{-0.6cm}
\renewcommand\baselinestretch{0.5}
\caption{
\small
SINFONI AO-assisted maps of K20-ID7 as a function of integration time.
Each row shows the H$\alpha$ line map ({\em left panel\/}) and the
velocity field ({\em right panel\/}) extracted from the data cube after
each hour of integration, up to the total time of 7\,hr obtained for
this source.
The (linear) color-coding for the line maps and for the velocity fields
is the same for all rows.
The angular scale and the PSF FWHM are indicated in each panel by the
vertical bar and the white-filled circle, respectively.
With just a few hours of integration, only the brightest regions are
detected and one would infer that this object is morphologically and
kinematically irregular.  The full extent of the source and its nature
become apparent only after $\rm \sim 5\,hr$.
\label{fig-TintSeq}
}
\vspace{-1ex}
\end{figure}

Equally importantly, however, the simple classification does not capture
the richness of information about individual galaxies provided by the
high quality AO data sets.
The comparison presented in Appendix~\ref{App-AOnoAO} illustrates the gain
in detail for each galaxy between the seeing-limited and AO data, from both
the higher resolution and our observing strategy emphasizing S/N.
Figure~\ref{fig-TintSeq} further demonstrates the importance of sensitivity
in recovering the nature and properties of a source, especially when surface
brightness variations are large and asymmetric.  The Figure shows the
H$\alpha$ line map and velocity field of K20-ID7 extracted from data cubes
in a sequence of increasing on-source integration time.  With a few hours
integration, the source would be classified as irregular and rather small;
only after 5 hours or more does the large extent and overall regular velocity
field of the main source become apparent along with the connection to the
small physically associated low-mass companion to the south.

Given the amount of detail seen in many of the AO data sets, attempting a
more refined classification as done in Section~\ref{Sub-kinclass} becomes
arguably subjective without detailed quanditative analysis and modeling,
which is beyond the scope of the present paper.  Nonetheless, it is
instructive to highlight selected objects that exhibit most prominently
several of the features mentioned above, and whose nature is best revealed
by the high-resolution AO data.

{\em $\bullet$ Deep3a-15504 ---}
This galaxy was the first object we observed with SINFONI$+$AO and
has the deepest data of our survey (23\,h).  It is a large
high-mass rotating disk with a massive bulge seen in rest-optical
continuum light and stellar mass map causing a steep inner velocity
gradient, several modestly bright H$\alpha$ emitting star-forming
clumps are distributed across the disk, and an AGN revealed by diagnostic
rest-UV and optical line emission drives a nuclear outflow that is
spatially extended and anisotropic in broad $\rm H\alpha +$[\ion{N}{2}]
emission \citep{Gen06, Gen08, Gen14a, FS14, Tacc15a}.  It is one of the
objects for which the detected emission extends sufficiently far out to
probe the decline in the outer disk rotation curve recently discovered in
$z \sim 1 - 2.5$ SFGs \citep{Gen17,Lan17}.  The kinematics show a strong
velocity gradient along the minor axis towards the center, characteristic
of gas inflow in the disk and/or outflow.
A small, faint, low-mass source is detected in continuum and H$\alpha$ at
the north edge of the disk, redshifted by $\rm 140~km\,s^{-1}$
relative to the disk rotation at that location, consistent with a
$\sim 1 : 30$ minor merging event
\citep{Gen17}.

{\em $\bullet$ ZC\,400569 ---}
This target is one with most complex structure in H$\alpha$ among our
sample, and also one of the deepest SINFONI$+$AO data sets (22.5\,h).
The velocity difference across the full extent of the source reaches
$\rm 570~km\,s^{-1}$ with a change of orientation of the isovelocity
contours from North to South and a chain of H$\alpha$ knots southward
of the brightest northern peak.  The northern component is a massive
rotating disk hosting a significant bulge, while the main southern knots
$1\farcs 0$ and $1\farcs 5$ to the south are associated with lower
mass satellites with $\sim 5\%$ and $2.3\%$, respectively,
of the stellar mass of the main disk \citep{Tacc15a, Gen17}.
It is another of the individual galaxies with best evidence for a dropping
outer rotation curve discussed in detail by \citet{Gen17}, and is also one
of the objects where broad line emission signatures of an AGN-driven wind
are detected and spatially-resolved in the AO data \citep{FS14}.

{\em $\bullet$ ZC\,406690 ---}
This disk has the most prominent clumpy ring among our targets,
with very little H$\alpha$ and continuum emission detected inside
the ring.  Intense star formation in the clumps drives powerful outflows,
whose broad and blueshifted emission in the southwest part of the ring
\citep{New12b} causes the local perturbations apparent in our velocity
and dispersion maps (extracted with single-Gaussian line fits).
The kinematics in the inner regions of the galaxy suggest the presence
of a significant central mass concentration, although its non-detection
at rest-optical wavelengths would imply it is very highly dust-obscured
or mostly in cold gas form \citep{Gen17, Tacc15b}.
ZC406690 exhibits the steepest fall-off detected in the outer disk
rotation curve among the sample discussed by \citet{Gen17}.
The properties of the faint continuum and H$\alpha$ source $1\farcs 6$
west of the center imply it is a small satellite with $\sim 15\%$ of the
stellar mass of the main disk.

{\em $\bullet$ Q2343-BX610 ---}
Another of the best rotating disk examples among our sample, Q2343-BX610
further has no evidence for a physically associated neighbour in the
available data.  It exhibits signatures of gas outflows driven by a
weak or obscured AGN in the center and by star formation at the location
of the bright southern clump \citep{FS14}.  Its rest-optical morphology
shows bar- and spiral-like features \citep{FS11a}, which could cause the
deviations from pure rotation along the minor axis observed in the kinematics.

{\em $\bullet$ Deep3a-6004 ---}
This massive galaxy exhibits most compelling characteristics of a disk
but with nearly orthogonal kinematic and morphological major axes
($\rm \Delta PA = 75^{\circ} \pm 12^{\circ}$).
The inclination inferred from kinematic disk modeling \citep{Gen08, Gen14a}
is significantly different from that implied by the H$\alpha$ axis ratio and
indicates a quite face-on orientation ($i \sim 25^{\circ}$) such that the
bright clumpy ring-like structure and possible deviations from circularity
affect the kinematic misalignment more importantly.
Broad $\rm H\alpha + $ [\ion{N}{2}] with elevated \niiha\ ratio associated
with an AGN-driven outflow strongly dominates the line emission inside the
star-forming ring, causing the apparent twist in isovelocity contours and
the high velocity dispersions at the center, where a massive bulge component
is in place \citep{FS14, Tacc15a}.

{\em $\bullet$ Q2346-BX482 ---}
This galaxy is another example of a disk with prominent star-forming ring
strongly dominated by one bright large clump although stellar light and
mass at faint levels is detected around the kinematic center based on the
near-IR {\em HST\/} data \citep{Gen08, FS11a, Gen14a, Tacc15b}.
It is associated with a source $\rm 27~kpc$ in projection to the
south-east at a relative velocity of $\rm +630~km\,s^{-1}$, which is
detected in H$\alpha$ in the wider FOV of the seeing-limited SINFONI
data and in CO~$3-2$ line emission in IRAM Plateau de Bure interferometric
observations \citep{Tac13}.
With a stellar mass of $\sim 25\%$ that of Q2346-BX482, this companion
may have induced the small kinematic perturbations on the north-western
edge of the main galaxy discussed by \citet{Gen08}.

{\em $\bullet$ Q2343-BX389 ---}
This large, nearly edge-on disk has a small southern companion at a projected
distance of 5~kpc and at the same redshift within $\rm 20~km\,s^{-1}$ with
estimated mass $\sim 10$ times lower than that of the main northern component
\citep{FS11a, Tacc15b}.
It thus represents a minor merger, consistent with the regular velocity
field of the primary component although the interaction could explain the
irregularities in the dispersion map.

{\em $\bullet$ ZC\,407302 ---}
A bright compact source lies about 4~kpc from the
center, on the northeast edge of the main body of the galaxy.  Because its
line-of-sight velocity is fully consistent with the extension of the velocity
field of the main disk part out to this radius, it could plausibly be a massive
clump that formed in-situ although we cannot rule out that it may be a small
accreted satellite.  The clump contains $\sim 1/20$ of the total stellar mass,
which could have induced the perturbations seen in velocity dispersion.

{\em $\bullet$ K20-ID7 ---}
Based on seeing-limited SINFONI data, this source had been classified
as major merger mainly because of the non-axisymmetry of the velocity
dispersion map \citep{Sha08}.  The AO-assisted data resolve this source
into a prominent ring structure with similarly regular velocity field and
little variations in velocity dispersion across the system.  The H$\alpha$
and rest-optical continuum emission trace extended material to the south
beyond the ring, connected to a smaller source at a projected separation
of $1\farcs 6$ and a relative velocity of about $\rm +350~km\,s^{-1}$,
roughly in line with the extension of the velocity gradient, and with
an inferred stellar mass of $\sim 15\%$ that of the main source.
Overall, the properties of K20-ID7 at 1.5~kpc resolution
appear more consistent with a large low-inclination disk undergoing a
$\sim 1:7$ minor interaction.
The stellar mass map for K20-ID7 is less centrally peaked compared to
the massive disks with bright star-forming ring-like structures, which
may reflect the decrease in bulge-to-total mass ratios at lower galaxy
masses \citep[e.g.,][]{Lan14}.

{\em $\bullet$ ZC407376 ---}
The two components of this interacting pair have a projected
separation of $1^{\prime\prime}$ (8~kpc), the same redshift within
$\rm 20~km\,s^{-1}$, nearly equal stellar masses, and derived
dynamical masses within a factor of two of each other, making this
system a major merger.  Both components appeared as dispersion-dominated
objects at seeing-limited resolution; the AO data resolves ZC407376N
into a small disk with projected velocity difference of
$\rm 120~km\,s^{-1}$ and $V_{\rm rot}/\sigma_{0} \sim 3$
while ZC407376S still has irregular kinematics.

{\em $\bullet$ Compact sources ---}
Many of the small, lower-mass sources in our sample exhibit characteristics
of disk rotation in the high-resolution AO data (most notably GMASS-2363,
Q1623-BX455, and ZC403741), as discussed in detail by \citet{New13}.
Among the 18 objects with $R_{\rm e} < 3~{\rm kpc}$, $55\%$ show a
monotonic velocity gradient and have $V_{\rm rot}/\sigma_{0} > 1.8$.
In contrast, and similarly to ZC407376S, SA12-6339 and ZC409985 show the
least amount of or no clear 2D velocity structure in the AO data.  This
may suggest that they are genuinely dispersion-dominated systems as a
result of strongly dissipative mass assembly, although it is not possible
to rule out that beam smearing and/or a nearly face-on orientation could
cause the lack of observed structure.

\section{SPATIAL VARIATIONS IN \niiha\ RATIOS}
         \label{Sect-metal}

\begin{figure*}[!ht]
\begin{center}
\includegraphics[scale=0.80,clip=1,angle=0]{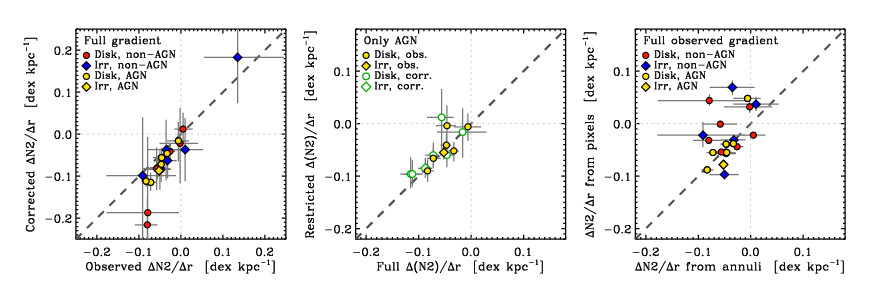}
\end{center}
\vspace{-0.5cm}
\renewcommand\baselinestretch{0.5}
\caption{
\small
Comparison of measurements of \niigrad\ for the \sinszc\ AO sample.
In all panels, the dashed line shows the 1:1 relation, and dotted lines
indicate a flat slope.
{\em Left:\/} Observed versus beam smearing-corrected
gradients over the full radial range.  Different symbols correspond to
galaxies with disk-like or irregular kinematics, and with or without
evidence for an AGN, as labeled in the plot.  The gradients are modestly
negative or flat for all but one galaxy (the compact ZC413597), and beam
smearing effects are overall small (typically $\rm -0.024~dex~kpc^{-1}$,
with the largest differences reaching about $\rm -0.1~dex~kpc^{-1}$ for
two of the smaller galaxies).
{\em Middle:\/} Full radial gradients versus gradients fitted over the
restricted $r > 0\farcs 3$ range for the galaxies with AGN.
Disk-like and irregular systems are distinguished by different symbol
shapes, and observed and beam smearing-corrected gradients are plotted
as filled and open symbols following the legend in the panel.  The full
and restricted gradients are essentially identical except for one object
(Deep3a-6004, strongly dominated by broad $\rm H\alpha + $[\ion{N}{2}]
emission from an AGN-driven outflow in the central regions).
{\em Right:\/} Full radial gradients derived from the spectra in elliptical
annuli versus gradients fitted to the distribution of pixels
with H$\alpha$ flux above the threshold avoiding the bias towards higher
\niiha\ ratios (see Section~\ref{Sub-metal_maps}).
No beam smearing correction is applied here (and ZC413597 is omitted
because too few pixels can be included).
The kinematic nature and presence or not of an AGN are indicated with
different symbols and colors as in the left panel.  The largest differences
between annuli- and pixel-based gradients are driven by the particularly
limited spatial coverage of the unbiased pixels in some of the objects.
\label{fig-metal_meas}
}
\vspace{3ex}
\end{figure*}

In this Section, we exploit our AO-assisted data sets to characterize the
spatial variations in \niiha\ ratio.  This ratio has been used in various
studies of $z \sim 0.7 - 2.7$ SFGs to investigate radial variations in
gas-phase oxygen abundances (in short, ``metallicity'').  The radial
gradients have been found to be overall fairly flat,
with some trends reported of shallower or positive gradients among
interacting/merging systems and kinematically disturbed disks, and
with higher redshift, lower mass, or higher specific SFR
\citep{Yua11,Yua12,Swi12a,Que12,Jon10a,Jon13,Sto14,Lee16,EWuy16}.
Results at $z > 3$ from bluer rest-optical strong line diagnostics still
accessible from the ground suggest more prevalent positive gradients and
anticorrelation with the SFR distribution \citep{Cre10, Tro14}.
By analogy with results in nearby SFGs
\citep[e.g.][]{Chi07, Kew06, Kew10, Rup10, Ric12},
and in line with predictions from theoretical models and simulations
\citep[e.g.,][]{Rah11, Kob11, Few12, Pil12, Tor12, Gib13, Mot13},
these findings have been interpreted as resulting from externally-driven
(interactions/mergers, enhanced halo gas accretion) or internally-driven
(feedback in the form of outflows, increased gas turbulence) mixing
efficiently redistributing metals within galaxies and diluting the gas
metallicity.

The picture remains unclear, however, because of the complicating
effects of AGN and shock excitation, ISM conditions, and possible
N/O abundance variations on the \niiha\ ratio, as amply discussed
in the literature\,\footnote{
 \citet{Lee16} found however consistent metallicity gradients obtained
 from the $\rm N2 = \log([N\,II]\,\lambda 6584 / H\alpha)$ and
 $\rm O3N2 = \log\left\{([O\,III]\,\lambda 5007 / H\beta) /
                 ([N\,II]\,\lambda 6584 / H\alpha)\right\}$
 indices when employing the same set of calibrators (such as proposed
 e.g., by \citealt{Pet04}) for eight of their $z \sim 2$ lensed targets
 (without evidence for AGN or strong outflows) for which AO-assisted maps
 of all lines were obtained, suggesting no major impact from N/O and ISM
 conditions variations within galaxies.
}
\citep[e.g.,][]
{Kew02,Kew13,Per09,Yua12,And13,Ste14,Sha15,San15,San17,Kas17,Str17,Cre17}.
In addition, high redshift samples with gradient measurements at 
high spatial resolution are still small especially at $z > 2$ and
larger seeing-limited samples are more affected by beam smearing.
Our \sinszc\ AO sample enables us to significantly expand on existing
measurements at $2 \la z \la 2.6$ by more than doubling the number of
galaxies with \niiha\ observations resolved on $\rm \sim 1~kpc$ scales.
Despite the above limitations, the \niiha\ ratio has the significant
advantage of being insensitive to extinction and both lines are observed
simultaneously in the same band under the same conditions.

\tabletypesize{\footnotesize}
\begin{deluxetable*}{llll}[!ht]
\renewcommand\arraystretch{0.95}
\tablecaption{Radial \niiha\ Gradients
              \label{tab-metal}}
\tablecolumns{4}
\tablewidth{0pt}
\setlength{\tabcolsep}{12pt}
\tablehead{
   \colhead{Source} &
   \colhead{Observed \niigrad\,\tablenotemark{a}} &
   \colhead{Restricted observed \niigrad\,\tablenotemark{b}} &
   \colhead{Corrected \niigrad\,\tablenotemark{c}} \\[-1.8ex]
   \colhead{} &
   \colhead{($\rm dex~kpc^{-1}$)} &
   \colhead{($\rm dex~kpc^{-1}$)} &
   \colhead{($\rm dex~kpc^{-1}$)}
}
\startdata
Q2343-BX389    & $-0.047^{+0.019}_{-0.018}$  & \ldots                       & $-0.084^{+0.032}_{-0.030}$   \\[0.95ex]
Q1623-BX455    & $-0.079^{+0.074}_{-0.098}$  & \ldots                       & $-0.187^{+0.180}_{-0.238}$   \\[0.95ex]
Q2346-BX482    & $+0.005^{+0.023}_{-0.020}$  & \ldots                       & $+0.012^{+0.039}_{-0.034}$   \\[0.95ex]
Q2343-BX513    & $+0.010 \pm 0.043$          & \ldots                       & $-0.037 \pm 0.075$           \\[0.95ex]
Q1623-BX599    & $-0.032^{+0.022}_{-0.025}$  & \ldots                       & $-0.063^{+0.037}_{-0.042}$   \\[0.95ex]
Q2343-BX610    & $-0.072^{+0.014}_{-0.012}$  & $-0.066^{+0.019}_{-0.016}$   & $-0.115^{+0.023}_{-0.020}$   \\[0.95ex]
Deep3a-15504   & $-0.033 \pm 0.010$          & $-0.052^{+0.017}_{-0.020}$   & $-0.046 \pm 0.013$           \\[0.95ex]
Deep3a-6004    & $-0.046^{+0.017}_{-0.021}$  & $-0.004^{+0.039}_{-0.044}$   & $-0.056^{+0.023}_{-0.028}$   \\[0.95ex]
Deep3a-6397    & $-0.047^{+0.008}_{-0.011}$  & $-0.041^{+0.019}_{-0.020}$   & $-0.072^{+0.011}_{-0.015}$   \\[0.95ex]
ZC400528       & $-0.006^{+0.025}_{-0.027}$  & \ldots                       & $-0.016^{+0.045}_{-0.049}$   \\[0.95ex]
ZC400569N      & $-0.083^{+0.012}_{-0.011}$  & $-0.090^{+0.022}_{-0.021}$   & $-0.112^{+0.016}_{-0.014}$   \\[0.95ex]
ZC400569       & $-0.052^{+0.008}_{-0.009}$  & $-0.055^{+0.014}_{-0.013}$   & $-0.087^{+0.013}_{-0.014}$   \\[0.95ex]
ZC403741       & $-0.080^{+0.024}_{-0.030}$  & \ldots                       & $-0.216^{+0.065}_{-0.081}$   \\[0.95ex]
ZC404221       & $-0.002^{+0.043}_{-0.049}$  & \ldots                       & $-0.022^{+0.084}_{-0.095}$   \\[0.95ex]
ZC406690       & $-0.056^{+0.019}_{-0.018}$  & \ldots                       & $-0.080^{+0.028}_{-0.027}$   \\[0.95ex]
ZC407302       & $-0.026 \pm 0.012$          & \ldots                       & $-0.041 \pm 0.020$           \\[0.95ex]
ZC407376S      & $-0.035^{+0.042}_{-0.043}$  & \ldots                       & $-0.037^{+0.071}_{-0.073}$   \\[0.95ex]
ZC407376       & $-0.050^{+0.027}_{-0.029}$  & \ldots                       & $-0.084^{+0.044}_{-0.048}$   \\[0.95ex]
ZC409985       & $-0.091^{+0.079}_{-0.087}$  & \ldots                       & $-0.099^{+0.139}_{-0.153}$   \\[0.95ex]
ZC412369       & $-0.058^{+0.032}_{-0.033}$  & \ldots                       & $-0.082^{+0.050}_{-0.051}$   \\[0.95ex]
ZC413597       & $+0.135^{+0.110}_{-0.080}$  & \ldots                       & $+0.183^{+0.150}_{-0.109}$   \\[0.95ex]
\enddata
\parskip=-2.0ex
\tablecomments
{
The gradients are expressed in terms of the \niiha\ ratio, the measured
quantity.  Gradients in terms of inferred metallicity are prone to
uncertainties in calibrations, which are known to be very uncertain
\citep[e.g.,][]{Kew08, Mai08}.
For comparison with results at $z \sim 1 - 3$ in the literature,
which often use the linear calibration proposed by \citet{Pet04},
with $\rm 12 + \log(O/H) = 8.90 + 0.57\,\log([N\,II]/H\alpha)$,
the metallicity gradients would correspond to
$\rm \Delta(O/H)/\Delta r = 0.57 \times \Delta(O/H)/\Delta r$.
}
\tablenotetext{a}
{
Full range radial \niiha\ gradient based on integrated spectra in
elliptical apertures.
}
\tablenotetext{b}
{
Radial \niiha\ gradient fitted over the restricted range $r > 0\farcs 3$
for sources with evidence of an AGN.
}
\tablenotetext{c}
{
Full range radial gradients corrected for beam smearing effects.
}
\parskip=0ex
\vspace{-5ex}
\end{deluxetable*}

\begin{figure*}[!ht]
\begin{center}
\includegraphics[scale=0.63,clip=1,angle=0]{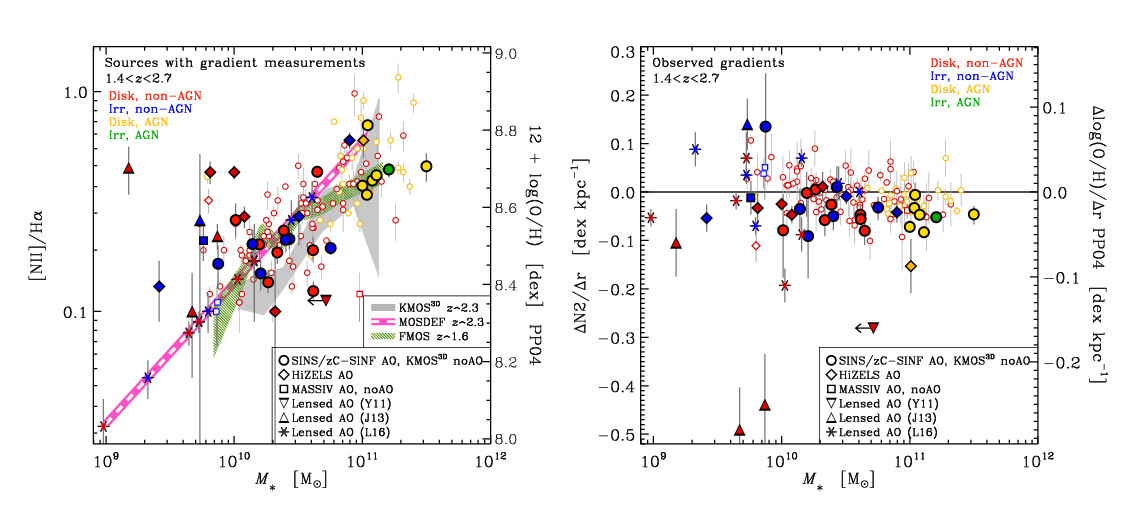}
\end{center}
\vspace{-0.8cm}
\renewcommand\baselinestretch{0.5}
\caption{
\small
\niiha\ properties versus stellar mass for $1.4 < z < 2.6$ galaxies
with a \niigrad\ measurement from IFU observations.
Large symbols represent measurements from AO-assisted data and
small open symbols show those from seeing-limited data, as labeled at
the bottom of each panel.
The \sinszc\ AO results are compared to those of the $\rm KMOS^{3D}$
\citep{EWuy16}, HiZELS \citep{Swi12a, Mol17}, and MASSIV \citep{Que12}
surveys, and from the lensed samples presented by \citet{Yua11},
\citet{Jon13}, and \citet{Lee16}.
Different colors distinguish between systems with disk-like or irregular
kinematics, and those hosting or not an AGN, as shown by the legend at
the top of the panels.
{\em Left:\/} Global \niiha\ ratios, and corresponding O/H abundance
inferred from the linear N2 calibration of \citet{Pet04}.  In addition,
we plot the \niiha\ vs $M_{\star}$ relationships derived from stacked
spectra in mass bins from 124 $1.9 < z < 2.7$ galaxies from the $\rm KMOS^{3D}$
multi-IFU survey \citep[][including AGNs; grey-shaded area]{EWuy16} and
642 $1.4 < z < 1.7$ galaxies from the FMOS multi-slit spectroscopic
survey \citep[][green-hatched area]{Kas17}, along with the linear fit
to the stacking results from 195 $z < z < 2.5$ galaxies from the MOSDEF
multi-slit spectroscopic survey \citep[][pink dashed line]{San18}.  The
latter fit was used to assign a stellar mass to the lensed objects from
\citet[][plotted as skeletal stars; see text]{Lee16}.
{\em Left:\/} Observed \niigrad\ measurements.  Beam smearing corrections
cannot be applied to all samples because for some of them, sizes and other
relevant galaxy properties are not available.  Overall, the radial gradients
are tightly clustered around a flat slope of zero, with large overlap between
the various samples.  The \sinszc\ sample roughly doubles the number of
published \niigrad\ measurements from AO-assisted IFU data at $2 < z < 2.6$,
and covers more than an order of magnitude in stellar mass.
\label{fig-metal_comp}
}
\vspace{2ex}
\end{figure*}

\subsection{Radial Gradients in [\ion{N}{2}]/H$\mathit{\alpha}$}
            \label{Sub-metal_prof}

We constructed radial profiles in \niiha\ ratio from the velocity-shifted
spectra extracted in elliptical annuli (Section~\ref{Sub-spectra}).  The
annuli have a width of $0\farcs 1$ and a major axis radius $r$ increasing
by $0 \farcs 1$ up to the largest aperture considered for the integrated
measurements reported in Table~\ref{tab-Hameas}.
For reliable gradient determinations, we considered only galaxies for which
at least three annuli have \niiha\ measured with $\rm S/N > 3$.  Nineteen
of the targets satisfy this reliability criterion; this number increases
to 21 when counting the two sub-components ZC400569N and ZC407376S where
we can also extract a sufficiently sampled profile.
The galaxies span nearly the entire range in stellar mass of our full AO
sample ($M_{\star} = 7.5 \times 10^{9} - 3.2 \times 10^{11}~{\rm M_{\odot}}$),
and the \niiha\ profiles extend over physical deprojected radii ranging
from 3.2 to 10.7~kpc, and 6.5~kpc on average (and median).
The radial profiles are presented in Appendix~\ref{App-metal}
(Figure~\ref{fig-metal_1}) together with the H$\alpha$ and \niiha\ maps,
and the velocity-shifted integrated spectra of these sources.

We quantified the gradients \niigrad\ from linear
regression with censored data to account for upper limits, where
$\rm N2 = \log([N{\small II}]/H\alpha)$ and $\rm r$ is in physical units.
We derived the uncertainties from 500 Monte-Carlo iterations.  For the
galaxies with evidence of an AGN, we also fitted the gradients restricted
to the outer $r > 0\farcs 3$ parts only, motivated by the extent of broad
$\rm H\alpha\,+$\,[\ion{N}{2}] line emission associated with nuclear outflows
in the most massive galaxies \citep{FS14}; this radius corresponds to 2.5~kpc
and encloses $\sim 80\%$ of the total light for a point-like source with the
average \psfave.

Beam smearing has a small but non-negligible effect in our AO data
because of the broad halo of the PSF (Appendix~\ref{App-AOperf}).
To estimate its impact, we created for each galaxy a suite of simulated
$\rm H\alpha$+[\ion{N}{2}] data cubes consisting of model rotating disks
with intrinsic \niigrad\ between $-0.2$ and $\rm +0.2~dex\,kpc^{-1}$.
The models were generated for the size, Sersic index, inclination, and
dynamical mass derived from the data, with the line emission normalized
to the source-integrated H$\alpha$ flux, \niiha\ ratio, and assuming
[\ion{N}{2}]\,$\lambda\,6548/$[\ion{N}{2}]\,$\lambda\,6584=0.34$ \citep{Sto00}.
The intrinsic models were convolved with the LSF and \psfave, noise was
added based on the reduced cubes, and the \niigrad\ were measured
from the mock data in the same way as for the observations.  Corrected
gradients were then derived from the relation between mock-observed and
intrinsic values of each galaxy.  Although these simulations are simplistic,
they highlight cases for which corrections may become quite important.

The best-fit slopes and 68\% confidence intervals for the full, restricted,
and corrected gradients are given in Table~\ref{tab-metal}.
Figure~\ref{fig-metal_meas} (left and middle panels) compares the full
observed and beam smearing-corrected \niigrad\ for all sources,
and those over the restricted radial range for the AGN sources.
The full observed gradients range from $-0.091$ to $\rm +0.135~dex~kpc^{-1}$,
with a mean of $-0.035$ and median of $\rm -0.047~dex~kpc^{-1}$.  
All sources have a best-fit slope that is negative or consistent with
zero within the $1\sigma$ uncertainties, with the exception of the
compact and kinematically irregular ZC413597.
The inferred intrinsic gradients are on average (and median) only modestly
steeper by about $\rm 0.025~dex~kpc^{-1}$, which is comparable to the
typical measurement uncertainties ($\rm 0.030~dex~kpc^{-1}$).
As expected, the largest corrections are for the smaller sources, reaching
differences of $-0.11$ and $\rm -0.14~dex~kpc^{-1}$ for Q1623-BX455
and ZC403741, respectively, implying that our observations recover for these
objects $\approx 40\%$ of the gradient assuming it is intrinsically smooth
and monotonic.
Among the seven sources with an AGN, the restricted gradients remain
overall the same (mean difference of $\rm +0.004~dex~kpc^{-1}$); the
largest difference is for Deep3a-6004, where the inner regions are
strongly dominated by the broad line emission from its AGN-driven
outflow \citep{FS14}.

For our sample, we find no significant trend in \niigrad\ with
galaxy stellar, star formation, and structural properties, or with
global \niiha\ ratio\,\footnote{
 In observed \niigrad, there may be weak trends of shallower \niigrad\
 at lower $M_{\star}$, \uvrest, sSFR and size, with Spearman rank correlation
 coefficient $\rho \approx 0.2 - 0.3$ and low $1\sigma$ significance.
 However, the trends are driven by one object with positive \niigrad,
 by the AGN-hosting galaxies, and by beam smearing effects since for
 our sample mass, color, and specific sSFR vary with effective radius.
 In corrected \niigrad, the trend with $R_{\rm e}$ disappears while the
 strength and significance of those with $M_{\star}$, \uvrest, and sSFR
 marginally increase ($\rho \approx 0.25 - 0.4$ at the $1 - 2\,\sigma$
 level), but they all weaken again when using the restricted gradients
 for the AGN or removing them altogether from the sample, and excluding
 ZC413597.}.
Considering the kinematic nature and nuclear activity of the sources, while
the mean and median \niigrad\ (observed or corrected) may suggest shallower
or more positive gradients among kinematically irregular systems versus
disks, and among non-AGN versus AGN, the differences have $< 1\sigma$
significance (and Kolmogorov-Smirnov [K-S] tests do not support that the
\niigrad\ distributions between these subsets significantly differ).

\subsection{Comparison with Other $\mathit{z \sim 2}$ AO Samples}
            \label{Sub-metal_comp}

Figure~\ref{fig-metal_comp} plots the source-integrated \niiha\ ratio and
\niigrad\ measurements as a function of galaxy stellar mass for our sample
along with those of other published AO samples.  The comparison is restricted
to the redshift range $1.4 < z < 2.6$ spanned by the \sinszc\ AO galaxies to
mitigate evolutionary effects
\citep[e.g.][]{Erb06a,Jon10a,Jon13,EWuy14a,EWuy16,Tro14,Zah14,San15,Lee16},
and to results obtained from the same N2 indicator as used here for
consistency.  The relevant published AO samples include 16 lensed objects
\citep{Yua11, Jon13, Lee16}, eight galaxies drawn from the HiZELS survey
\citep{Swi12a, Mol17}, and one observed as part of the MASSIV survey
\citep{Que12}.  For the lensed galaxy of \citet{Yua11}, the quoted
dynamical mass is used as upper limit on the stellar mass.  For the
\citet{Lee16} sample, no mass estimate is listed or can be inferred
based on the published data; the range of 
$\log(M_{\star}/{\rm M_{\odot}}) \sim 9 - 9.6$ is mentioned for the
subset of galaxies with available near-IR photometry.  For the purpose
of Figure~\ref{fig-metal_comp}, we assigned $M_{\star}$ values for the
11 objects of \citeauthor{Lee16} calculated from their \niiha\ ratios
using the linear N2$-$$M_{\star}$ relation for $z \sim 2.3$ SFGs given
by \citet{San18}, yielding a range of
$\log(M_{\star}/{\rm M_{\odot}}) \sim 9 - 10.6$.

In Figure~\ref{fig-metal_comp}, we show the data in terms of \niiha\ ratio
and gradient, the measured quantities.  As an indication, the right-hand
axes show the corresponding values for oxygen abundance adopting the linear
calibration of \citet[][``PP04'']{Pet04} that is commonly used in high
redshift studies, $\rm 12 + \log(O/H) = 8.90 + 0.57 \times N2$.
Since galaxy sizes are not available for all the literature samples
considered here but would be essential for beam smearing corrections,
the comparison is made for the observed (uncorrected) \niigrad.  
As a further comparison, we include results derived based on seeing-limited
observations from the $\rm KMOS^{3D}$ survey, the largest sample with
\niigrad\ measurements \citep{EWuy16}, and the four $z \sim 1.5$
galaxies from the MASSIV survey that fall in the redshift range
considered here \citep{Que12}.
In the background of the galaxy-integrated \niiha $-$ $M_{\star}$ panel,
we further plot the relationships derived from stacked spectra in mass
bins of much larger samples from the $\rm KMOS^{3D}$
\citep[including AGNs;][]{EWuy16} and MOSDEF 
\citep[excluding AGNs;][]{San18} surveys,
and at $z \sim 1.6$ from the FMOS survey
\citep[excluding AGNs;][]{Kas17}.

The objects with \niigrad\ measurements broadly follow the \niiha\ versus
$M_{\star}$ relationships delineated by $\rm KMOS^{3D}$, MOSDEF, and FMOS
SFGs down to $\log(M_{\star}/{\rm M_{\odot}}) \sim 10$.  Objects around
and below this mass almost all lie above these relationships, a trend that
partly reflects the observational surface brightness limitations associated
with the \niiha\ employed for the gradient measurements.

In \niigrad, the distributions between the various AO samples at
$1.4 < z < 2.6$ largely overlap.
Our \sinszc\ AO sample represents an important extension and strengthens
the previous findings of typically weak radial variations in \niiha.
It dominates at the higher masses and extends down to the regime probed so
far largely by lensed and/or lower $z$ galaxies; 19 of our sources are at
$2 < z < 2.6$ compared to 15 of the published AO targets (the other ten
discussed here being at $1.4 < z < 1.8$).
From the combined AO samples (46 objects), the average and median
\niigrad\ are $-0.053$ and $\rm -0.044~dex~kpc^{-1}$, with a scatter of
$\rm 0.115~dex~kpc^{-1}$; excluding the five galaxies with steepest negative
and positive gradients ($> 1.5\sigma$ away from the mean) increases the
average to $-0.036$ and reduces the scatter to $\rm 0.054~dex~kpc^{-1}$.
The most prominent feature in the \niigrad\ distributions is the wider
range spanned by the lensed objects, which \citet{Lee16} argued could
reflect large variations in gas and metal mixing due to feedback between
different galaxies, which are better probed with the aid of gravitational
magnification.

Even with the increased statistics, no strong trend with galaxy properties
emerges when adding the published AO targets to ours.  There is a possible
distinction between disks and non-disks, with median \niigrad\ of $-0.047$
and $\rm -0.009~dex~kpc^{-1}$, respectively, but the difference is at the
$1.8\sigma$ level and a K-S test does not indicate that the distributions
are significantly distinct.
The nine galaxies with evidence for an AGN differ even less from the non-AGN
ones (median of $-0.047$ and $\rm -0.033~dex~kpc^{-1}$, respectively, a
$1.2\sigma$ difference).  No significant correlation is seen in \niigrad\
as a function of $M_{\star}$ (excluding the \citealt{Lee16} sample, for which
no stellar mass is given as explained above).
Although the AO and seeing-limited measurements overlap, the latter have
fairly uniformly flatter \niigrad\ at fixed stellar mass over the better
sampled $\log(M_{\star}/{\rm M_{\odot}}) \ga 10$ range (the mean, median,
and scatter are $-0.001$, $0.000$, and $\rm 0.037~dex~kpc^{-1}$, respectively),
consistent with expectations from beam smearing
(see also Appendix~\ref{App-AOnoAO}).

Comparisons between different samples as presented in
Figure~\ref{fig-metal_comp} still have important caveats.
In particular, the physical resolution between the data sets varies
widely, from $\rm \sim 100~pc$ for the most strongly magnified objects
to $\rm \sim 1 - 2~kpc$ for the non-lensed AO samples, and up to
$\rm \sim 5~kpc$ for the seeing-limited data
(see also the discussions by \citealt{Yua13} and \citealt{Car17}).
Even for strongly lensed galaxies, the stretch is generally non-uniform
and can be negligible along the least magnified direction, such that beam
smearing may still play a role for these sources.

\subsection{Pixel Distributions in [\ion{N}{2}]/H$\mathit{\alpha}$ Ratio}
            \label{Sub-metal_maps}

The \niiha\ maps presented in Appendix~\ref{App-metal} suggest possibly
more complex spatial variations than simple smooth gradients.  We examined
the distributions of \niiha\ ratio of individual pixels as a function of
radius to assess the degree of azimuthal scatter and the consistency between
azimuthally-averaged and pixel-based gradients.

Before interpreting the data, it is important to account for the surface
brightness effects that can bias the \niiha\ ratio towards higher values
in regions of fainter H$\alpha$ line emission.  We empirically derived the
$3\sigma$ limit in \niiha\ as a function of H$\alpha$ flux based on the
S/N(H$\alpha$) vs $F({\rm H\alpha})$ relationship for each galaxy.  We then
determined the threshold in H$\alpha$ flux, $F({\rm H\alpha})_{\rm pix,thresh}$,
below which the pixel distribution in \niiha\ becomes biased towards higher
values.  This process is illustrated in the rightmost panels of
Figure~\ref{fig-metal_1}.
The pixels with a ratio measurement $> 3\sigma$ but an H$\alpha$ flux
$< F({\rm H\alpha})_{\rm pix,thresh}$ are marked on the ratio maps, and are
distinguished from those in the unbiased regime in the radial distributions
of pixel ratios shown in the fifth column of the Figures.  It is apparent
from these Figures that accounting for the biased regime is essential in
discussing trends in 2D and with line flux based on \niiha.  For instance,
an anticorrelation between \niiha\ and observed $F({\rm H\alpha})$ is seen
in many objects but is largely driven by this bias.  

Keeping in mind the bias just described, the data nonetheless reveal
the presence of an important scatter in \niiha\ at fixed radius in
several of the galaxies.  This scatter may arise from the contribution
of shock excitation in ionized gas outflows, reflect localized enrichment
or dilution of the ISM in metals on short timescales, or be caused by
the possible presence of recently accreted, metal-poor low-mass companions.
One particularly striking example is ZC406690, where the
$\rm H\alpha + $[\ion{N}{2}] emission from star formation-driven outflows
exhibits different properties (line widths and ratios) between different
clumps along the bright ring-like structure, at least in part due to shocks
\citep{New12b}.  Another interesting example is ZC400528, where the broad
$\rm H\alpha + $[\ion{N}{2}] component associated with an AGN-driven outflow
is anisotropic and extends primarily along one side of the kinematic minor
axis \citep{FS14}, opposite an off-center clump or close-by low-mass satellite
with very weak [\ion{N}{2}] emission.

We compared the \niigrad\ derived from the spectra of elliptical annuli
with those from the distributions of unbiased pixels following a similar
fitting procedure (denoted \niigradpix).
The right panel of Figure~\ref{fig-metal_meas} shows the comparison, and
the best-fit \niigradpix\ are overplotted in the fifth column of
Figure~\ref{fig-metal_1}.
In half of the cases, the annuli- and pixel-based gradients agree within
about $1\sigma$.  The largest differences (at the $1.5 - 3\,\sigma$ level)
occur for six objects and can be attributed to:
(i) the smaller radial coverage of the pixels missing the decrease in
\niiha\ compared to the measurements in annuli extending further out in
compact objects with significantly negative gradients (ZC407376S, ZC412369,
and especially ZC403741 where the more abrupt drop in the outermost annulus
is not probed by the pixels),
(ii) the brighter asymmetric regions with outflow emission dominating
the off-center pixel distribution in ZC400528, and
(iii) the limited spatial coverage of the pixels being particularly
affected by the bright asymmetric features in the full systems ZC407376
and ZC400569.

We conclude from the comparison presented here that while the more detailed
information from 2D maps allows in principle a more direct association of
\niiha\ variations with specific sub-galactic regions, surface brightness
limitations inherent to this ratio can still make the interpretation quite
challenging even for our deep AO data sets, especially when seeking to
constrain radial variations.
Although the summation of the spectra of individual pixels applied in
extracting the spectra in annuli (Section~\ref{Sub-spectra}) is inherently
light-weighted, it better probes the regions of fainter line emission.
The \niiha\ profiles derived from these spectra thus enable a more reliable
estimate of \niigrad, albeit in an azimuthally-averaged sense and with
the potential caveats of annuli binning \citep[e.g.,][]{Yua13}.

\section{SUMMARY}   \label{Sect-conclu}

We presented sensitive, high resolution SINFONI$+$AO observations of 35
star-forming, high-redshift galaxies, the result of a 12-year series of
observing campaigns at the ESO VLT carried out between April 2005 and
August 2016.  Most galaxies (32 out of 35) are at $2 < z < 2.6$ and
constitute the largest sample of galaxies with AO-assisted, near-IR
integral field spectroscopy targeting this specific redshift range.
As a legacy to the community, we make the reduced data cubes
publicly available.

Some of the targets were selected from the literature for having both a
suitable redshift and a nearby star usable for the AO wavefront correction.
Others, those coming from the zCOSMOS-Deep project (Lilly et al. 2007),
were specifically targeted as fulfilling the above AO requirements and 
provide targets for this SINS/zC-SINF survey.  As such, the whole sample
was not assembled following a single selection criterion.  However, we
demonstrated that this sample is largely representative in terms of SFR,
size, and rest-optical colors, of main sequence SFGs in the same redshift
and mass ranges (as illustrated in Section 2.3, but see also discussions
by \citeauthor*{FS09} and \citeauthor*{Man11}).
The angular resolution of the AO data span a fairly narrow range, with
PSF FWHM from $0\farcs 13$ to $0\farcs 33$ and a median of $0\farcs 18$,
corresponding to 1.5~kpc at the redshift of these galaxies.
The prime target of the observations was the line emission from H$\alpha$
and [\ion{N}{2}], allowing us to construct high-resolution maps of the SFR
surface density, velocity field, and velocity dispersion for all galaxies,
and \niiha\ maps for a subset of 19 of them.

About $70\%$ of the galaxies show ordered disk rotation patterns in their
kinematics, including several of the compact sources, as well as individual
components in two larger interacting systems.
The other sources have irregular or nearly featureless kinematics on
resolved scales of $\rm 1 - 2~kpc$.
In the larger disks, the AO data reveal second-order features that are
reminiscent of perturbations induced by massive bulges and star-forming
disk clumps, possible bar-/spiral-like structure, or low-mass nearby
satellite galaxies, or that are related to the presence of strong
ionized gas outflows driven by star formation or AGN.
The sensitive AO data from our survey highlights the richness of detail and
diversity of processes that become apparent at higher angular resolution.
Simple classification schemes will not capture this richness but it is
nonetheless clear that the majority of the galaxies are observed in a disk
configuration (further supported in most cases by the global stellar light
and H$\alpha$ structure).  This result implies a fairly stable dynamical
state, consistent with the notion that the mere existence of the ``main
sequence'' requires SFGs to be in a quasi-steady state equilibrium between
gas accretion, ejection and star formation \citep[e.g.,][]{Lil13}.

Our \sinszc\ AO sample more than doubled the number of galaxies with high
resolution \niiha\ gradient measurements at $2 < z < 2.7$.  The radial
gradients are fairly shallow (similar to findings from other studies),
and almost exclusively negative.  The shallow slopes may reflect
efficient gas and metal mixing from various internal and external
processes, but can also be directly affected by the contribution from
shock and AGN excitation, as seen in some of the galaxies.
The maps and pixel distributions show that azimuthal variations are present
in several cases, associated with ionized gas outflows or more metal-poor
regions in the disks or nearby companions.
Future progress will rely importantly on mapping multiple diagnostic
line ratios fully in 2D to disentangle metallicity, excitation, and
physical conditions.

It will be important to take the next steps not only by increasing
sample size and coverage of galaxy parameter space, but also by pushing
quantitative analyses to the characterization of substructure in kinematics,
morphologies, and line ratio properties and to establish connections with
variations in stellar population and reddening within galaxies.
For instance, a robust identification of distinct residuals in velocity
and dispersion maps together with the breaking of age-reddening degeneracies
in the central regions of high redshift galaxies will be crucial in mapping
the emergence of galactic bulges.
Also, future efforts to map outflows from disks, clumps, and nuclear
regions on $\rm \sim 1~kpc$ scales are crucially needed for many more
{\em typical\/} high redshift SFGs in order to constrain more accurately
physical parameters such as the mass loading factor and establish the
time-averaged impact of feedback.
Our work and other IFU studies show that it is feasible given sufficient
integration time to reach the necessary sensitivity.  Such observations
will be critical in providing much-needed quantitative constraints for
theory and numerical simulations of galaxy evolution.
Another key goal includes the systematic exploration of the properties
of the progenitor population of $z \sim 2$ galaxies, using the same
diagnostics for consistency.

Such studies will be facilitated by upcoming instruments such as ERIS
at the VLT and NIRSpec on board the {\em James Webb Space Telescope\/},
affording near-IR IFU capabilities with improved sensitivity, AO performance,
and wavelength coverage.  Combining the resolved distribution and kinematics
of the warm ($\rm \sim 10^{4}~K$) ionized gas as probed by H$\alpha$ with
those of the dominant (and unobscured) cold molecular and atomic gas with
NOEMA and ALMA will be essential to assess the baryonic mass budget and
the extent to which outflows are able to entrain the neutral ISM phase.

Given the twelve year period it took to complete the SINFONI AO-assisted
observations of our 35 targets, it appears unlikely that another comparable
sample with similar deep and high-quality kinematic and structural
information will be produced anytime soon.  Accordingly, this sample
offers the best candidate targets at $z \sim 2$ for further investigations
at other wavelengths and at higher spatial resolution.

\acknowledgments

We wish to thank the ESO staff for excellent support during the
many observing campaigns over which the SINFONI data were obtained.
Special thanks to J. Navarrete for always enthusiastic help and
discussions.
We also thank the SINFONI and PARSEC teams for their hard work on the
instrument and laser, which allowed our programs to be carried out
successfully.
This paper and the SINS/zC-SINF survey have benefitted from many
stimulating discussions with many colleagues, especially
M. Franx, P. van Dokkum, M. Bureau, S. Courteau, L. Kewley, 
A. Dekel, and E. Wisnioski.
N. M. F. S. acknowledges support by the Minerva Program of the
Max-Planck-Gesellschaft and by the Schwerpunkt Programme SPP1177
of the Deutsche Forschungsgemeinschaft.
A. R. and C. M. acknowledge support from INAF-PRIN 2008 and 2012;
A. R. also acknowledges support from INAF-PRIN 2014 and 2017, and
is grateful for the kind hospitality of the Excellence Cluster Universe
(Garching) and the Max-Planck-Institut f\"ur Extraterrestrische Physik
while working on the early and final versions of this paper.
Y. P. acknowledges support from the National Key Program for Science and
Technology Research and Development under grant number 2016YFA0400702, and
the NSFC grant no. 11773001.
A. C. acknowledges the grants ASI n.I/023/12/0 ``Attivit\`a relative
alla fase B2/C per la missione Euclid'' and PRIN MIUR 2015 ``Cosmology
and Fundamental Physics: illuminating the Dark Universe with Euclid.''

\appendix

\section{AO PERFORMANCE AND POINT SPREAD FUNCTION}
         \label{App-AOperf}

\subsection{Impact of Reference Star Brightness and Observing Conditions
            on the AO Performance}
            \label{Appsub-AOperf}

We inspected variations of the effective angular resolution of our data
sets with the properties of the AO reference star, and with the seeing
at visible wavelengths, atmospheric coherence time $\tau_{0}$, and
airmass during the observations.  The seeing, $\tau_{0}$, and airmass
are taken from the Paranal observatory's data recorded for all individual
exposures.  For this trending analysis, we use the conditions during the
PSF star observations, noting that variations between the PSF star and
science target observations are very similar to those within individual
OBs and between different OBs.
We also estimated the Strehl ratio from the peak-to-total flux ratio of the
effective PSF image, accounting for the pixel sampling, the seeing at
the observed wavelength of H$\alpha$ and average airmass of each object,
and the outer scale correction for the VLT pupil of 8\,m (following
\citealt{Sar00}).

\begin{figure}[!ht]
\begin{center}
\includegraphics[scale=0.55,clip=1,angle=0]{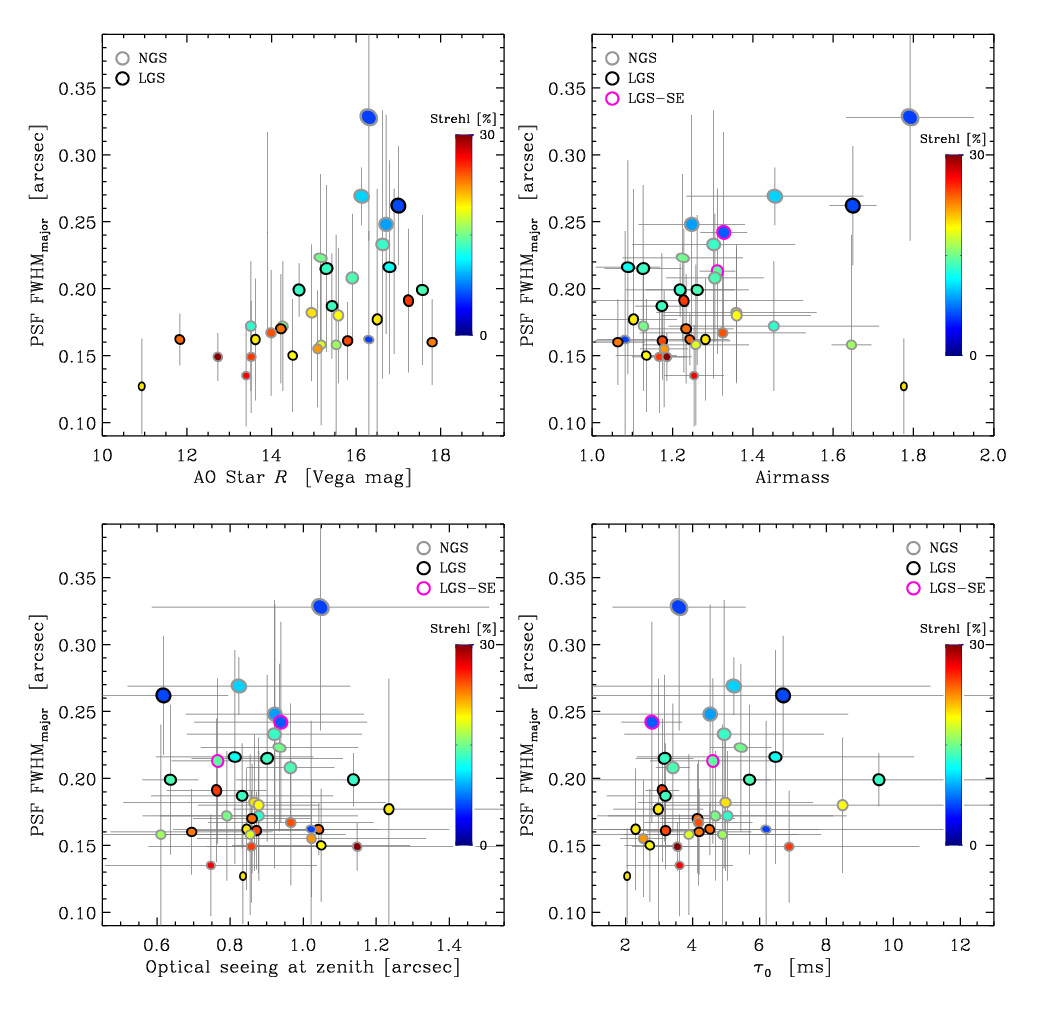}
\end{center}
\vspace{-0.8cm}
\renewcommand\baselinestretch{0.5}
\caption{
\small
Impact of tip-tilt star brightness and observing conditions on the
AO performance for our data sets.  Individual elliptical symbols have
sizes proportional to the major and minor axis FWHMs of the PSF, and
orientation corresponding to its PA  The color coding scales linearly
with the estimated Strehl ratio as shown by the color bar in each panel.
{\em Top left:\/} PSF properties as a function of the AO star $R$-band
magnitude.
{\em Top right:\/} PSF properties as a function of the average airmass
of the individual PSF star observations.
{\em Bottom left:\/} PSF properties as a function of the average optical
($\rm \lambda = 0.5\,\mu m$) seeing measured at zenith.
{\em Bottom right:\/} PSF properties as a function of the average
atmospheric coherence time $\tau_{0}$.
The vertical error bars correspond to the OB-to-OB variations in PSF FWHMs,
and the horizontal error bars indicate the standard deviation of the observing
conditions recorded for the individual PSF star exposures.
Data taken in NGS, LGS, and LGS-SE modes are indicated with grey,
black, and pink symbol outlines, respectively.
The FWHM and Strehl ratio are more closely coupled with the brightness of
the AO star, less with the airmass and $\tau_{0}$, and little if at all with
the optical seeing at zenith.
\label{fig-obscond}
}
\end{figure}

Table~\ref{tab-AOperf} lists the reference star's $R$-band magnitude and
separation from the science target, and the average and standard deviation
of the observing conditions (seeing, $\tau_{0}$, airmass) along with the
parameters derived from 2D elliptical Gaussian fits to the effective PSF
image and estimated Strehl ratio.  Figure~\ref{fig-obscond} shows the
distributions of effective PSF parameters and Strehl ratio as a function
of the reference AO star brightness, average airmass, optical seeing, and
coherence time.

As expected, the effective PSF major axis FWHM ($\rm FWHM_{major}$) and the
Strehl ratio are tightly coupled.  The estimated Strehl ratios range from
$5.8\%$ to $32\%$, with an average of $17\%$ and median of $16\%$.
As for the angular resolution (see Section~\ref{Sub-PSF}), there is
no significant distinction in Strehl ratio between the data taken in NGS
and LGS mode (the mean and median differ by $<5\%$) except for the
two LGS-SE data sets (Strehl ratio of $7\%$ and $14\%$).
The $\rm FWHM_{major}$ correlates most strongly with the magnitude of the
star (the Spearman rank correlation is $\rho = 0.55$ with significance of
$3.2\sigma$), and so does the Strehl ratio ($\rho = 0.42$ and $2.5\sigma$).
Both quantities also vary with $\tau_{0}$ and airmass but less
significantly ($\rho = 0.2 - 0.3$ at the $1.2 - 1.9\,\sigma$ level).
There is no trend with the seeing for our data, either in the optical
at zenith or corrected to the near-IR wavelength and airmass of the
observations.

\subsection{Averaged AO Point Spread Function}
            \label{Appsub-AOPSF}

Because of the non-ideal AO performance, reflected in the Strehl ratios
from the AO reference star images, broad wings in the PSF shape are
expected from the uncorrected seeing.  While the wings can be discerned
in most of the PSF images, the data of individual stars are too noisy for
an accurate profile characterization.  To obtain a high S/N PSF profile,
we thus spatially registered, normalized by the peak flux, and co-averaged
(with $3.5\sigma$ clipping) the PSF images associated with the different
galaxies.  We excluded the PSF associated with $\rm Deep3a-15504$ because
the star is in a double system and, while it is resolved in our data, the
secondary component is relatively bright and may affect the average.
We fitted a single and a double-component 2D elliptical Gaussian model
to the resulting averaged AO PSF.  Averaging instead the PSFs with a FWHM
within the central 68\% of the distribution, or the NGS or LGS subsets,
makes no significant difference ($\leq 0\farcs 01$ and $\leq 0\farcs 07$
in FWHM for the narrow core and broad halo, respectively, and $< 5\%$ in
their total flux contributions).

\begin{figure}[!ht]
\begin{center}
\includegraphics[scale=0.88,clip=1,angle=0]{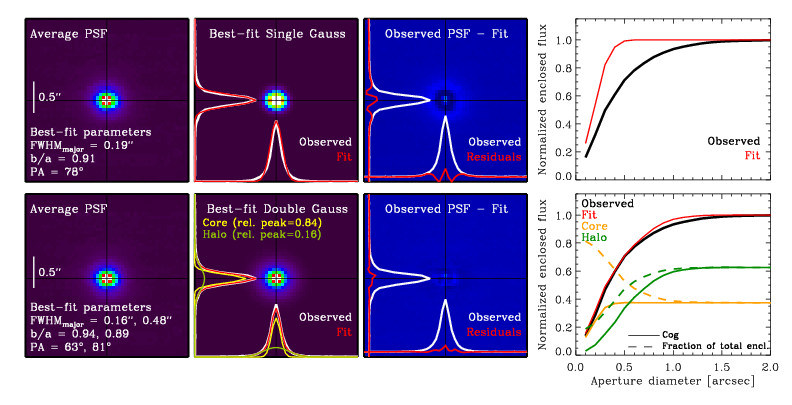}
\end{center}
\vspace{-0.8cm}
\renewcommand\baselinestretch{0.5}
\caption{
\small
Detailed profile characteristics derived from the high S/N averaged
image of all PSFs associated with the combined data cubes of the sample.
From left to right, the panels in each row show the average PSF image,
the best-fitting model, the fit residuals, and the curves-of-growth of
the data and model.
{\em Top row:\/} Results for a single-component 2D elliptical Gaussian fit.  
In the middle-left panel, cuts through the PSF image (white solid line)
and best-fit model (red solid line) are plotted on the bottom and left
axes.  In the middle-right panel, cuts on the residual image are plotted
(red solid line) on the same scale as for the PSF image profiles (white
solid line).  In the right panel, the normalized curves-of-growth in
circular apertures of the average PSF and the single-component model
are plotted in black and red, respectively.
{\em Bottom row:\/} Similar series of panels showing the results for a
double-component 2D elliptical Gaussian fit.  In the middle-left panel,
the axis profiles for the observed average PSF, total model fit, and
individual narrow core and broad halo components are plotted (white,
red, yellow, and green solid lines, respectively).  In the right panel,
the individual curves-of-growth of the core and halo are shown (yellow and
green solid lines) along with those of the observed and total model fit
(black and red solid lines).  In addition, the variations of the fractional
contributions to the total flux of the core and halo are plotted (yellow
and green dashed lines).
The double-component model provides a much better representation of the
PSF profile and curve-of-growth, and the low-amplitude but wide halo
dominates the total flux for aperture diameters $> 0\farcs 6$.
\label{fig_PSFstack}
}
\vspace{-6ex}
\end{figure}

The images and axis profiles of the adopted average PSF, best-fit
models, and residuals along with the curves-of-growth are shown in
Figure~\ref{fig_PSFstack}.  The best-fit parameters are given in the
Figure panels.  Even in the high S/N average PSF, a single Gauss fit
provides a reasonable description of the inner PSF profile, and the
core component in the double-Gauss fit has an amplitude about 5 times
higher than the broad component.  For aperture diameters larger than
$0\farcs 6$, the three times wider broad component starts to dominate
the enclosed flux and reaches a total of $63\%$.
Because of the significant power in the PSF halo, it is important to
assess its impact in the derivation of intrinsic galaxy properties
from the data (such as sizes, rotation velocities, and intrinsic
velocity dispersions), albeit in an average sense since it does not
capture galaxy-to-galaxy PSF variations.

\tabletypesize{\footnotesize}
\begin{deluxetable}{lllrcccccrr}[!ht]
\renewcommand\arraystretch{0.8}
\tablecaption{Observational Parameters and AO Performance 
              \label{tab-AOperf}}
\tablecolumns{11}
\tablewidth{0pt}
\tablehead{
   \multicolumn{11}{c}{} \\[-0.50ex]
   \multicolumn{2}{c}{Science target} &
   \multicolumn{2}{c}{Reference Star\,\tablenotemark{a}} &
   \multicolumn{4}{c}{Observing Conditions\,\tablenotemark{b}} &
   \multicolumn{3}{c}{Effective PSF\,\tablenotemark{c}} \\[0.65ex]
\cline{1-2} \cline{3-4}  \cline{5-7}  \cline{8-11} \\[-2.5ex]
   \colhead{Source} &
   \colhead{AO mode} &
   \colhead{$R_{\rm Vega}$} &
   \colhead{Distance} &
   \colhead{Optical seeing} &
   \colhead{$\tau_{0}$} &
   \colhead{Airmass} &
   \colhead{$\rm FWHM_{major}$} &
   \colhead{Axis ratio} &
   \colhead{PA} &
   \colhead{Strehl ratio} \\[-0.8ex]
   \multicolumn{1}{c}{} &
   \multicolumn{1}{c}{} &
   \multicolumn{1}{c}{(mag)} &
   \multicolumn{1}{c}{} &
   \multicolumn{1}{c}{} &
   \multicolumn{1}{c}{(ms)} &
   \multicolumn{1}{c}{} &
   \multicolumn{1}{c}{} &
   \multicolumn{1}{c}{} &
   \multicolumn{1}{c}{(deg)} &
   \multicolumn{1}{c}{($\%$)}
}
\startdata
$\rm Q1623-BX455$   & LGS       & 10.9    & $48^{\prime\prime}$    &
  $0\farcs 84$ ($0\farcs 00$) & 2.0 (0.0)     & 1.78 (0.00)   & $0\farcs 13$  & 0.71  & $-3$  & 20            \\
$\rm Q1623-BX502$   & NGS       & 15.5    & $8^{\prime\prime}$     &
  $0\farcs 61$ ($0\farcs 27$) & 4.9 (2.9)     & 1.65 (0.05)   & $0\farcs 16$  & 0.91  & $ +71$& 16            \\
$\rm Q1623-BX543$   & NGS       & 16.3    & $25^{\prime\prime}$    &
  $1\farcs 05$ ($0\farcs 46$) & 3.6 (2.0)     & 1.79 (0.16)   & $0\farcs 33$  & 0.84  & $ +51$& 5.8           \\
$\rm Q1623-BX599$   & LGS       & 17.0    & $50^{\prime\prime}$    &
  $0\farcs 76$ ($0\farcs 17$) & 9.6 (11.9)    & 1.64 (0.02)   & $0\farcs 18$  & 0.97  & $ +75$& 13            \\
$\rm Q2343-BX389$   & LGS-SE    & \ldots  & \ldots                 &
  $0\farcs 77$ ($0\farcs 13$) & 4.6 (1.9)     & 1.31 (0.05)   & $0\farcs 21$  & 0.92  & $ +19$& 14            \\
$\rm Q2343-BX513$   & LGS       & 13.6    & $51^{\prime\prime}$    &
  $0\farcs 85$ ($0\farcs 00$) & 2.3 (0.0)     & 1.28 (0.00)   & $0\farcs 16$  & 0.88  & $ +1$ & 20            \\
$\rm Q2343-BX610$   & LGS-SE    & \ldots  & \ldots                 &
  $0\farcs 94$ ($0\farcs 24$) & 2.8 (0.9)     & 1.33 (0.06)   & $0\farcs 24$  & 0.95  & $ +47$& 6.7           \\
$\rm Q2346-BX482$   & LGS       & 15.4    & $50^{\prime\prime}$    &
  $0\farcs 83$ ($0\farcs 25$) & 3.2 (1.8)     & 1.17 (0.07)   & $0\farcs 19$  & 0.91  & $-85$ & 13            \\
$\rm Deep3a-6004$   & LGS       & 15.8    & $49^{\prime\prime}$    &
  $0\farcs 87$ ($0\farcs 16$) & 3.2 (1.0)     & 1.17 (0.17)   & $0\farcs 16$  & 0.98  & $ +87$& 24            \\
$\rm Deep3a-6397$   & LGS       & 16.8    & $26^{\prime\prime}$    &
  $0\farcs 81$ ($0\farcs 22$) & 6.5 (4.1)     & 1.09 (0.08)   & $0\farcs 22$  & 0.77  & $ +83$& 12            \\
$\rm Deep3a-15504$  & NGS       & 16.3    & $17^{\prime\prime}$    &
  $1\farcs 02$ ($0\farcs 38$) & 6.2 (5.5)     & 1.08 (0.07)   & $0\farcs 16$  & 0.74  & $ +75$& 5.8           \\
$\rm K20-ID6$       & LGS       & 15.3    & $55^{\prime\prime}$    &
  $0\farcs 90$ ($0\farcs 21$) & 3.2 (0.9)     & 1.13 (0.10)   & $0\farcs 22$  & 0.88  & $-71$ & 13            \\
$\rm K20-ID7$       & LGS       & 17.8    & $39^{\prime\prime}$    &
  $0\farcs 70$ ($0\farcs 22$) & 4.2 (1.6)     & 1.06 (0.07)   & $0\farcs 16$  & 0.88  & $ +90$& 23            \\
$\rm GMASS-2303$    & LGS       & 17.2    & $25^{\prime\prime}$    &
  $0\farcs 76$ ($0\farcs 14$) & 3.1 (0.5)     & 1.23 (0.30)   & $0\farcs 19$  & 0.81  & $ +17$& 25            \\
$\rm GMASS-2363$    & NGS       & 13.5    & $16^{\prime\prime}$    &
  $0\farcs 88$ ($0\farcs 27$) & 5.0 (3.6)     & 1.45 (0.26)   & $0\farcs 17$  & 0.94  & $ +84$& 13            \\
$\rm GMASS-2540$    & LGS       & 16.5    & $42^{\prime\prime}$    &
  $1\farcs 24$ ($0\farcs 43$) & 3.0 (1.3)     & 1.10 (0.11)   & $0\farcs 18$  & 0.88  & $-8$  & 20            \\
$\rm SA12-6339$     & LGS       & 14.5    & $57^{\prime\prime}$    &
  $1\farcs 05$ ($0\farcs 24$) & 2.7 (0.8)     & 1.14 (0.08)   & $0\farcs 15$  & 0.89  & $ +58$& 19            \\
$\rm ZC400528$      & NGS       & 15.1    & $7^{\prime\prime}$     &
  $1\farcs 02$ ($0\farcs 31$) & 2.5 (0.7)     & 1.18 (0.06)   & $0\farcs 16$  & 0.94  & $ +30$& 22            \\
$\rm ZC400569$      & NGS       & 15.2    & $20^{\prime\prime}$    &
  $0\farcs 86$ ($0\farcs 26$) & 3.9 (3.3)     & 1.26 (0.13)   & $0\farcs 16$  & 0.89  & $-73$ & 18            \\
$\rm ZC401925$      & NGS       & 16.1    & $24^{\prime\prime}$    &
  $0\farcs 82$ ($0\farcs 31$) & 5.2 (5.9)     & 1.46 (0.22)   & $0\farcs 27$  & 0.85  & $ +71$& 10            \\
$\rm ZC403741$      & NGS       & 14.3    & $15^{\prime\prime}$    &
  $0\farcs 79$ ($0\farcs 23$) & 4.7 (3.5)     & 1.13 (0.00)   & $0\farcs 17$  & 0.89  & $ +51$& 15            \\
$\rm ZC404221$      & NGS       & 15.9    & $8^{\prime\prime}$     &
  $0\farcs 97$ ($0\farcs 12$) & 3.4 (0.5)     & 1.31 (0.12)   & $0\farcs 21$  & 0.96  & $ +89$& 14            \\
$\rm ZC405226$      & NGS       & 16.7    & $5^{\prime\prime}$     &
  $0\farcs 92$ ($0\farcs 24$) & 4.5 (4.1)     & 1.25 (0.13)   & $0\farcs 25$  & 0.96  & $ +83$& 8.7           \\
$\rm ZC405501$      & NGS       & 15.2    & $21^{\prime\prime}$    &
  $0\farcs 94$ ($0\farcs 22$) & 5.4 (0.9)     & 1.23 (0.15)   & $0\farcs 22$  & 0.60  & $ +76$& 15            \\
$\rm ZC406690$      & NGS       & 14.9    & $18^{\prime\prime}$    &
  $0\farcs 87$ ($0\farcs 36$) & 5.0 (2.6)     & 1.36 (0.20)   & $0\farcs 18$  & 0.89  & $ +69$& 20            \\
$\rm ZC407302$      & LGS       & 14.2    & $39^{\prime\prime}$    &
  $0\farcs 86$ ($0\farcs 16$) & 4.1 (1.1)     & 1.23 (0.12)   & $0\farcs 17$  & 0.89  & $ +75$& 23            \\
$\rm ZC407376$      & NGS       & 16.6    & $22^{\prime\prime}$    &
  $0\farcs 92$ ($0\farcs 24$) & 4.9 (3.0)     & 1.30 (0.20)   & $0\farcs 23$  & 0.88  & $ +82$& 13            \\
$\rm ZC409985$      & NGS       & 13.4    & $17^{\prime\prime}$    &
  $0\farcs 75$ ($0\farcs 29$) & 3.6 (1.6)     & 1.25 (0.07)   & $0\farcs 14$  & 0.90  & $ +78$& 27            \\
$\rm ZC410041$      & NGS       & 14.0    & $21^{\prime\prime}$    &
  $0\farcs 97$ ($0\farcs 23$) & 4.2 (1.6)     & 1.33 (0.21)   & $0\farcs 17$  & 0.93  & $ +60$& 24            \\
$\rm ZC410123$      & LGS       & 17.6    & $25^{\prime\prime}$    &
  $0\farcs 64$ ($0\farcs 08$) & 5.7 (0.6)     & 1.26 (0.13)   & $0\farcs 20$  & 0.88  & $ +76$& 14            \\
$\rm ZC411737$      & LGS       & 14.6    & $29^{\prime\prime}$    &
  $1\farcs 14$ ($0\farcs 32$) & 9.6 (9.0)     & 1.22 (0.14)   & $0\farcs 20$  & 0.95  & $ +40$& 13            \\
$\rm ZC412369$      & LGS       & 11.8    & $29^{\prime\prime}$    &
  $1\farcs 04$ ($0\farcs 16$) & 4.5 (3.3)     & 1.24 (0.10)   & $0\farcs 16$  & 0.85  & $ +48$& 23            \\
$\rm ZC413507$      & NGS       & 12.7    & $29^{\prime\prime}$    &
  $1\farcs 15$ ($0\farcs 41$) & 3.5 (1.4)     & 1.19 (0.06)   & $0\farcs 15$  & 0.93  & $ +40$& 32            \\
$\rm ZC413597$      & NGS       & 15.6    & $28^{\prime\prime}$    &
  $0\farcs 88$ ($0\farcs 17$) & 8.5 (5.5)     & 1.36 (0.19)   & $0\farcs 18$  & 0.93  & $ +12$& 19            \\
$\rm ZC415876$      & NGS       & 13.5    & $27^{\prime\prime}$    &
  $0\farcs 86$ ($0\farcs 28$) & 6.9 (3.9)     & 1.17 (0.07)   & $0\farcs 15$  & 0.91  & $-83$ & 25            \\
\enddata
\parskip=-2.0ex
\tablenotetext{a}
{
$R$-band magnitude of the reference star used for the AO
correction in NGS mode, or for tip-tilt correction in LGS mode.
In LGS-SE mode (used for $\rm Q2343-BX389$ and $\rm Q2343-BX610$),
no tip-tilt correction was applied.
}
\tablenotetext{b}
{
The average optical seeing measured at zenith, coherence time
$\tau_{0}$, and airmass over all individual exposures of the
PSF calibration star, with the standard deviation given in
parenthesis (standard deviation of 0 indicates only a single
measurement/exposure is available).
}
\tablenotetext{c}
{
Characteristics of the PSF associated with the final reduced data
of each source (see Section~\ref{Sub-PSF}): the major axis FWHM,
axis ratio, and PA (in degrees east of north) of the best-fit 2D
elliptical Gaussian, and the Strehl ratio measured based on the PSF
star's image.
}
\parskip=0ex
\vspace{-5ex}
\end{deluxetable}

\begin{figure}[!t]
\begin{center}
\includegraphics[scale=0.85,clip=1,angle=0]{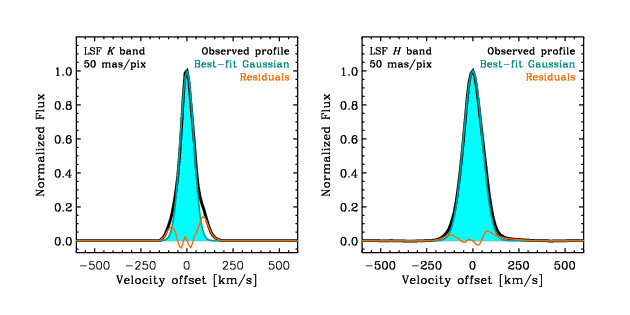}
\end{center}
\vspace{-1.0cm}
\renewcommand\baselinestretch{0.5}
\caption{
\small
Line spread functions (LSFs) of our reduced SINFONI data at
$\rm 50~mas\,pixel^{-1}$ from the average profile of several telluric
OH lines.
{\em Left:\/} LSF in $K$ band.
{\em Right:\/} LSF in $H$ band.
In each panel, the average observed profile from the data (black line),
the best-fit single Gaussian profile (cyan-filled curve), and the
residuals (orange line) are plotted.
The best-fit Gaussian has a velocity $\rm FWHM = 85~km\,^{-1}$ in $K$
and $\rm 120~km\,^{-1}$ in $H$.  Broader wings are apparent in the
observed LSFs, which contain 12\% and 5\% of the total flux in $K$
and $H$, respectively.
\label{fig_LSF}
}
\end{figure}

\section{LINE SPREAD FUNCTIONS}
         \label{App-LSF}

To characterize the effective spectral resolution of our SINFONI
$\rm 50~mas\,pixel^{-1}$ $K$ and $H$ band data, we extracted spectra of
the night sky emission in circular apertures of 1\arcsec\ radius in ``sky''
cubes, obtained by reducing the data but without background subtraction
(see \citeauthor*{FS09}).  We co-averaged the profiles in velocity space of
several of the brighter emission features in each of $K$ and $H$ that have
no neighbouring line within $\rm \pm 300~km\,s^{-1}$ with peak amplitude
greater than 1\% of the lines considered.  Most of these features consist
of blended pairs but their separations are $\rm < 5~km\,s^{-1}$ in $K$ and
$\rm < 18~km\,s^{-1}$ in $H$ such that they approximate a single line at
SINFONI's resolution.
The average LSFs are plotted in Figure~\ref{fig_LSF}.  Their profile is close
to Gaussian with $\rm FWHM = 85~km\,s^{-1}$ and $\rm 120~km\,s^{-1}$ in $K$
and $H$, respectively.  Slight deviations from a pure Gaussian are nonetheless
apparent but the power in these wings compared to the total in the empirical
profile is small: about $12\%$ in $K$, and $5\%$ in $H$.
Variations in the best-fit FWHM of the effective LSFs are $< 15\%$ across
the $K$ band and $< 10\%$ across the $H$ band.
We note that the low-level wings have a negligible impact on single-Gauss
spectral fits as used throughout this paper but their effects should be
considered when investigating detailed line profile properties.

\section{IMPACT OF BROAD UNDERLYING EMISSION ON THE
         EXTRACTED EMISSION LINE PROPERTIES}
         \label{App-broad}

Our line fits assumed a single Gaussian component for the emission lines
of interest.  This approach provides a satisfactory representation of the
observed line profiles for most sources on an individual basis.  The typical
line widths, profiles, and \niiha\ ratios indicate the emission is generally
dominated by star-forming regions across the galaxies.
However, a broader underlying component attributed to ionized gas outflows
discovered in our \sinszc\ galaxies and driven by star formation and/or an
AGN \citep{Sha09,Gen11,Gen14b,New12b,New12a,FS14} is thus not taken into
account.  Because of its large width and low amplitude, the broad component
is generally unnoticed in the spectrum of individual pixels or small regions
within the galaxies.  It becomes more clearly apparent in the
spatially-integrated spectrum of some galaxies or brightest clumps
individually and in co-added spectra of galaxies.  From the latter higher S/N
data, the derived flux and width relative to the narrow component together
with the modest velocity offsets by up to several tens of $\rm km\,s^{-1}$
indicate that such a broad component could make up a significant fraction of
the fluxes from single-component fits and affect the inferred line widths.

To estimate the possible impact of a broad component to our H$\alpha$
measurements, we generated sets of $5000 - 6000$ simulated spectra as
follows.  We created mock line profiles consisting of a narrow and broad
component with ranges of intrinsic line widths $\rm \sigma_{narrow}$ and
$\rm \sigma_{broad}$, velocity offsets $dv_{\rm broad}$ of the broad
component relative to the narrow component, and broad-to-narrow flux
ratios $F_{\rm broad} / F_{\rm narrow}$.  The values were drawn randomly
from uniform distributions based on our analyses of the best individual
cases and stacked spectra with high S/N exhibiting broad emission
\citep{New12a, FS14, Gen14b}: 
$\rm \sigma_{narrow} = 50 - 120~km\,s^{-1}$,
$\rm \sigma_{broad} = 130 - 1300~km\,s^{-1}$,
$dv_{\rm broad} = -100$ to $\rm +100~km\,s^{-1}$,
and $F_{\rm broad} / F_{\rm narrow} = 0 - 3$.
In order to simulate realistic noise properties, we inserted the mock line
profiles convolved with the instrumental LSF at random wavelength positions
in actual spectra taken from our data sets, avoiding the intervals with
emission lines from the real galaxies.
The total line fluxes were scaled randomly so as to cover uniformly the
typical ranges of fluxes and S/N ratios of our H$\alpha$ maps and aperture
spectra ($\rm S/N \sim 3 - 50$).  We then fitted the resulting mock
spectra assuming a single Gaussian line profile following the method
applied for the real data (Section~\ref{Sub-meth}).

\begin{figure}[!t]
\begin{center}
\includegraphics[scale=0.60,clip=1,angle=0]{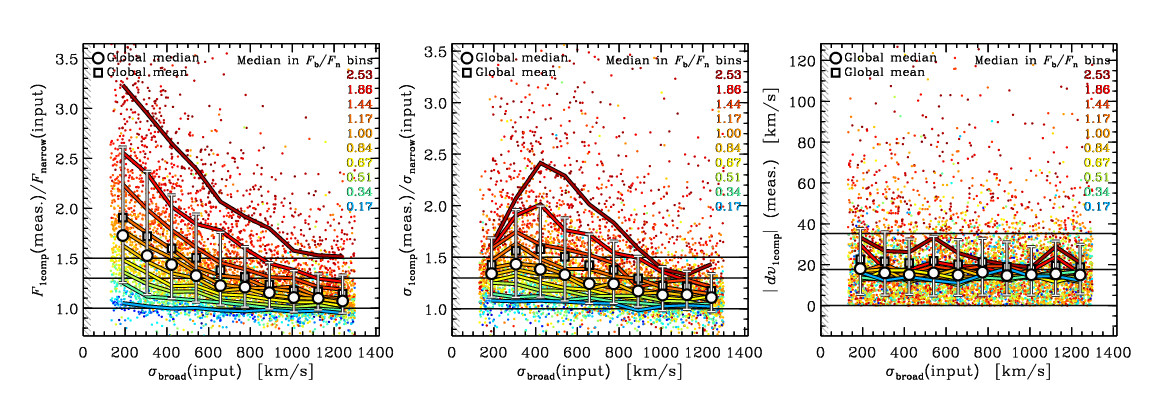}
\end{center}
\vspace{-0.85cm}
\renewcommand\baselinestretch{0.5}
\caption{
\small
Impact of the presence of a broad emission component underneath narrow
line emission on measurements based on single-Gaussian profile fits.
The results plotted are for a subset of the simulations carried
out, for the case of the spectral resolution of the SINFONI $K$-band
data taken in the $\rm 50\,mas\,pixel^{-1}$ scale, with simulated
double Gaussian profiles inserted at random wavelengths in the real
data of Q2343-BX610, and other assumed line parameters as described
in the text of Appendix~\ref{App-broad}.
{\em Left:\/} Ratio of the measured line flux from the single Gaussian
fit to the input flux in the narrow component
($F_{\rm 1comp}/F_{\rm narrow}$) versus input velocity width
of the broad component ($\sigma_{\rm broad}$).
{\em Middle:\/} Same as the left panel but for the ratio of the measured
velocity width (corrected for LSF smearing) to the input velocity width
of the narrow component ($\rm \sigma_{1comp}/\sigma_{narrow}$).
{\em Right:\/} Same as the other panels but for the absolute value of the
velocity offset in measured line centroid relative to the input narrow
component ($dv_{\rm 1comp}$).
In all panels, individual points show the results of 6000 simulated
spectra, color-coded according to the input broad-to-narrow
component flux ratio as labeled in the top right.
Large white-filled circles with error bars show the median and central
$68\%$ of the full distribution in bins of input $\sigma_{\rm broad}$;
large white-filled squares show the average in the same bins.
Thick colored lines indicate the median trends split in bins of
broad-to-narrow component flux ratios ($F_{\rm b}/F_{\rm n}$) centered
on the values labeled in the plots.
The grey-hatched band shows the dispersion corresponding to the LSF.
In the left and middle panel, black solid horizontal lines indicate
ratios of 1, 1.3, and 1.5; in the right panel, they show velocity
offsets of 0, 0.5, and 1.0 times the spectral resolution element.
The presence of a broad component leads to overestimates in line flux and
width exceeding $30\%$ only when $F_{\rm broad}/F_{\rm narrow}$ approaches
or exceeds unity, and $\rm \sigma_{broad} \sim 200-500~km\,s^{-1}$.
\label{fig-sims_broad}
}
\end{figure}

The results are shown in Figure~\ref{fig-sims_broad} for one set of
6000 simulations inserted into the real data of Q2343-BX610 (results
using the data of other galaxies observed in $K$ or $H$ band are very
similar).  The plots show the effects on the flux, LSF-corrected width,
and velocity centroid from the single-Gaussian fits to the mock spectra
($F_{\rm 1comp}$, $\rm \sigma_{1comp}$, $dv_{\rm 1comp}$) relative to
the input narrow component parameters as a function of the broad component
width and broad-to-narrow flux ratio.
The line fluxes and widths are most affected by the underlying broad
component when it is not explicitly accounted for in the fitting.
As expected, the impact becomes more important towards higher
broad-to-narrow flux ratios and narrower widths of the broad component.
For broad emission with $F_{\rm broad} / F_{\rm narrow} \sim 0.7$
and $\rm \sigma_{broad} \sim 200~km\,s^{-1}$, characteristic of star
formation-driven outflows as derived from co-added spectra of clumps,
$\log(M_{\star}/{\rm M_{\odot}}) \la 10.6$ galaxies, and outer disk
regions of $\log(M_{\star}/{\rm M_{\odot}}) \ga 10.6$ galaxies
\citep{New12b, New12a, FS14, Gen14b},
the $F_{\rm 1comp}$ typically overestimates $F_{\rm narrow}$ by
$50\%$ and the $\rm \sigma_{1comp}$ typically is $30\%$ larger
than $\rm \sigma_{narrow}$.
In the presence of a broader AGN-driven nuclear outflow,
with representative $F_{\rm broad} / F_{\rm narrow} \sim 0.7$ and
$\rm \sigma_{broad} \sim 700~km\,s^{-1}$ based on co-added nuclear
spectra of $\log(M_{\star}/{\rm M_{\odot}}) \ga 10.6$ galaxies
\citep{FS14, Gen14b},
the $F_{\rm 1comp}$ and $\rm \sigma_{1comp}$ are typically $20\%$
larger than $F_{\rm narrow}$ and $\rm \sigma_{narrow}$.
The largest effects are seen for $F_{\rm broad} / F_{\rm narrow} > 1$
and $\rm \sigma_{broad} \sim 200 - 500~km\,s^{-1}$, where typically
$F_{\rm 1comp}$ exceeds $F_{\rm narrow}$ by factors of at least 1.5
and $\rm \sigma_{1comp}$ exceeds $\rm \sigma_{narrow}$ by factors of
$1.3$ or more.
In contrast, the impact on $dv_{\rm 1comp}$ is small and comparable
to half the resolution element in velocity; a larger (but still typically
modest) velocity offset requires $F_{\rm broad} / F_{\rm narrow} > 2$.
There is no strong trend with input $\rm \sigma_{narrow}$, or with the
flux level and S/N although the scatter in $F_{\rm 1comp}/F_{\rm narrow}$,
$\rm \sigma_{1comp}/\sigma_{narrow}$, and $v_{\rm 1comp}-v_{\rm narrow}$
increases by a factor of $\sim 2 - 4$ from the highest to lowest S/N ratios.

Despite the sensitivity of our AO data sets, the S/N is still insufficient
to reliably fit double Gaussians to the line emission at the pixel level
and for small apertures.
The analysis above suggests that the impact of a broad component on our
single-Gaussian fits could be non-negligible but overall modest and it
is not expected to strongly bias the sample results.  Indeed, from stacking
analysis, the broad emission from star formation-driven outflows was found
to depend most strongly on the SFR surface density and to become important
($F_{\rm broad}/F_{\rm narrow}>0.5$) at
$\rm \Sigma(SFR) \ga 1~M_{\odot}\,yr^{-1}\,kpc^{-2}$ \citep{New12a}.
For our \sinszc\ AO sample, SFR surface densities are below this threshold
over $\ga 70\%$ of the area.
Nuclear AGN-driven outflows are only detected in the inner $r \la 0\farcs 3$
regions of the six most massive galaxies of our AO sample \citep{FS14}.

Two examples from our sample where the broad emission affects noticeably
the emission and kinematic maps as extracted in this paper are ZC\,406690
and $\rm Deep3a-6004$.  In the former galaxy, the brightest among our AO
sample, particularly strong broad blueshifted emission with
$\rm FWHM \sim 500 - 600~km\,s^{-1}$ tracing outflowing gas in/around
clumps on the southwest side of the H$\alpha$ and rest-UV/optical ring-like
structure visibly distorts the velocity and dispersion maps obtained from
single-Gauss fits (see Figure~\ref{fig-bbAO_1}), leading to significant local
residuals relative to the best-fit kinematic disk model \citep{Gen11, Gen17}.
For $\rm Deep3a-6004$, our most massive galaxy, broad
$\rm FWHM \sim 900~km\,s^{-1}$ and high excitation (\niiha\ $\sim 1$)
emission tracing an AGN-driven outflow dominates around the kinematic
center, causing the twist of the isovelocity lines and the slight asymmetry
in the flux and dispersion maps presented here (see Figure~\ref{fig-bbAO_1}).
This object has by far the largest nuclear $F_{\rm broad}/F_{\rm narrow}$
ratio ($\sim 3$), such that narrower line emission that would trace star
formation around the center is comparatively very weak \citep{FS14, Gen14a}.
In other galaxies of our sample where broad emission is detected, its effects
are less important.

\section{PRESENTATION OF THE FULL AO DATA SETS}
         \label{App-FullDataSets}

Figure~\ref{fig-bbAO_1} shows the main line emission and kinematic maps
extracted from the SINFONI data cubes for our \sinszc\ AO sample, along
with high-resolution near-IR broad-band maps.
Each row presents the maps of a given galaxy.
The leftmost panels show near-IR broad-band maps, probing the
rest-optical continuum at $\sim 0\farcs 2$ resolution.
For 31 objects, the imaging was obtained with {\em HST\/}
in the $H$-band using either the WFC3/IR or the NICMOS/NIC2 camera
\citep[][see also \citealt{FS11a, Law12b}]{Tacc15b}.
For the other four objects, the maps were synthesized from the SINFONI$+$AO
data cubes (see Section~\ref{Sub-maps}) in the $H$ band for $z < 2$ sources
(D3a-6397 and ZC403741) and in the $K$ band for those at $z > 2$ (Q2343-BX513
and SA12-6339).
The next two panels show the velocity-integrated flux for H$\alpha$ and
[\ion{N}{2}]\,$\lambda 6584$ at each pixel position.  The two rightmost
panels show the velocity field (relative to the systemic velocity) and
velocity dispersion maps of H$\alpha$.  The velocity dispersion map is
corrected for the instrumental spectral resolution.  

The galaxy, its H$\alpha$ redshift, the AO observing mode, and the total
on-source integration time with SINFONI and with {\em HST\/} are given at
the top of the two first panels on the left.
The colour coding for the emission maps scales linearly with flux from dark
blue to red/white for the minimum to maximum levels displayed (varying for
each galaxy).  The scaling between the H$\alpha$ and [NII] maps is tied such
that the minimum is the same and the maximum level of the [NII] map is half
that of the H$\alpha$ map.
The color coding for the velocity and dispersion maps scales linearly
from blue to red following the color bar in the respective panels.
Black contours in all three maps correspond to H$\alpha$ fractional
flux levels relative to the maximum of 0.2, 0.4, 0.6, 0.8, and 0.95.
The FWHM spatial resolution is represented by the filled circle at the
bottom left of the {\em HST\/} maps and by the filled ellipse for the
SINFONI$+$AO maps; the latter reflects the slightly asymmetric AO PSFs
and represents the {\em effective\/} resolution including the effects of
the median filtering applied spatially in extracting the maps from the data
cubes.  The angular scale is indicated by the vertical bars in the leftmost
panels.  The black cross indicates the center of the galaxy.

\section{COMPARISON OF DATA AND MEASUREMENTS FROM
         AO-ASSISTED AND SEEING-LIMITED OBSERVATIONS}
         \label{App-AOnoAO}

Since every galaxy of our \sinszc\ AO sample was first observed under
natural seeing conditions, we examined the consistency of the basic data
and measurements obtained at a different resolution.  This comparison
entails different integration times (hence sensitivities) and often long
time periods covered by the observations of an object (in between which
instrumental interventions were carried out).
Therefore, it is not strictly an assessment of the effects of resolution
but it provides also an instructive view on how various observational
factors may impact data obtained with different strategies or different
IFU instruments.

\subsection{Visual Inspection}
            \label{AppSub-AOvsNoAO_visual}

Figure~\ref{fig-AOnoAO_1} compares the SINFONI H$\alpha$ line, velocity,
and velocity dispersion maps, position-velocity (p-v) diagrams along the
major axis, and source-integrated spectra extracted in circular apertures
obtained with AO-assisted observations and in seeing-limited mode.
Successive rows are grouped pairwise for each galaxy, with the AO and
seeing-limited data shown in the top and bottom row of a pair, respectively.
Q1623-BX502 is excluded as it was only observed in AO mode.  The area covered
by the AO maps is outlined with the dotted yellow square on the seeing-limited
maps.  The labels, color coding, contour levels, and orientation of the maps
follow the same scheme as used in Figure~\ref{fig-bbAO_1}.
In addition, the dashed rectangle and solid circle overlaid on the H$\alpha$
line maps show the synthetic slit and the aperture used to extract the p-v
diagrams and the integrated spectra, respectively.
In the p-v diagrams, the horizontal axis corresponds to the velocity relative
to the systemic velocity, and the vertical axis corresponds to the spatial
position along the synthetic slit, with bottom to top running from the south
to the north end of the slit.  The angular scale is indicated by the vertical
bar on the left, and the color scale is identical to that of the H$\alpha$
line map of the object.  The wavelength range displayed for the integrated
spectra corresponds to the velocity range shown for the p-v diagrams
($\rm \pm 2500~km\,s^{-1}$ around H$\alpha$).  The error bars show the
$1\sigma$ uncertainties derived from the noise properties of each
data set and include the scaling with aperture size described in
Section~\ref{Sub-datared}, which accounts for the fact that the effective
noise is not purely Gaussian.  Vertical green hatched bars show the location
of bright night sky lines, with width corresponding to the FWHM of the
effective spectral resolution of the data.

In general, and considering the differences in angular resolution and
integration times, the AO and seeing-limited results are very consistent
with each other.  In all cases, the integrated line profile shapes agree
well, and the more so for the objects with highest S/N spectra.  Clearly,
the level to which the maps and p-v diagrams agree depends primarily on
the S/N achieved, resulting from the combination of integration time and
surface brightness distribution of the galaxies.
The typically three times higher resolution and roughly 3.5 times higher
sensitivity of the AO data sets obviously reveal in more detail the
morphology and kinematics of the H$\alpha$ line emission.
On the other hand, the wider field of view of the no-AO data probes the
extended emission (or lack thereof) out to $1.5 - 2$ times larger radii.
As long as a source is sufficiently well resolved in the no-AO data
($r_{1/2}^{\rm circ} \ga 3~{\rm kpc}$, about half of the galaxies),
the broad features of the velocity field can be recognized (presence and
direction of velocity gradients), and the p-v diagrams show qualitatively
similar spatial variations in velocities and line widths along the major
axis.  Lower S/N sometimes also leads to more important differences in
the H$\alpha$ morphologies (e.g., ZC410041, ZC400569).

The largest diversity of changes in the appearance of the maps and
p-v diagrams occur (unsurprisingly) for the smaller sources, with
$r_{1/2}^{\rm circ} \la 3~{\rm kpc}$.  These cases can be split as follows.
For six objects, significant velocity gradients were detected in the no-AO
data and are confirmed in the AO data, which further resolve clumps and/or
more diffuse extensions in the H$\alpha$ maps
(Q1623-BX543, GMASS-2363, ZC400528, ZC403741, ZC412369, ZC415876).
For eight other objects, structure is resolved in the H$\alpha$ morphology
and/or velocity field and p-v diagrams from the AO data that was not, or
only marginally, apparent in the no-AO observations
(Q1623-BX455, Q1623-BX599, Q2343-BX513, GMASS-2303, ZC401925, ZC409985,
ZC411737, ZC413507).
For the remaining three objects, both AO and no-AO data show compact
and fairly featureless H$\alpha$ morphologies and kinematics
(SA12-6339, ZC404221, ZC413597).

\vspace{-1.0ex}

\subsection{Quantitative Comparison of Extracted Properties}
            \label{Sub-AOvsNoAO_quant}

\begin{figure*}[!ht]
\begin{center}
\includegraphics[scale=0.65,clip=1,angle=0]{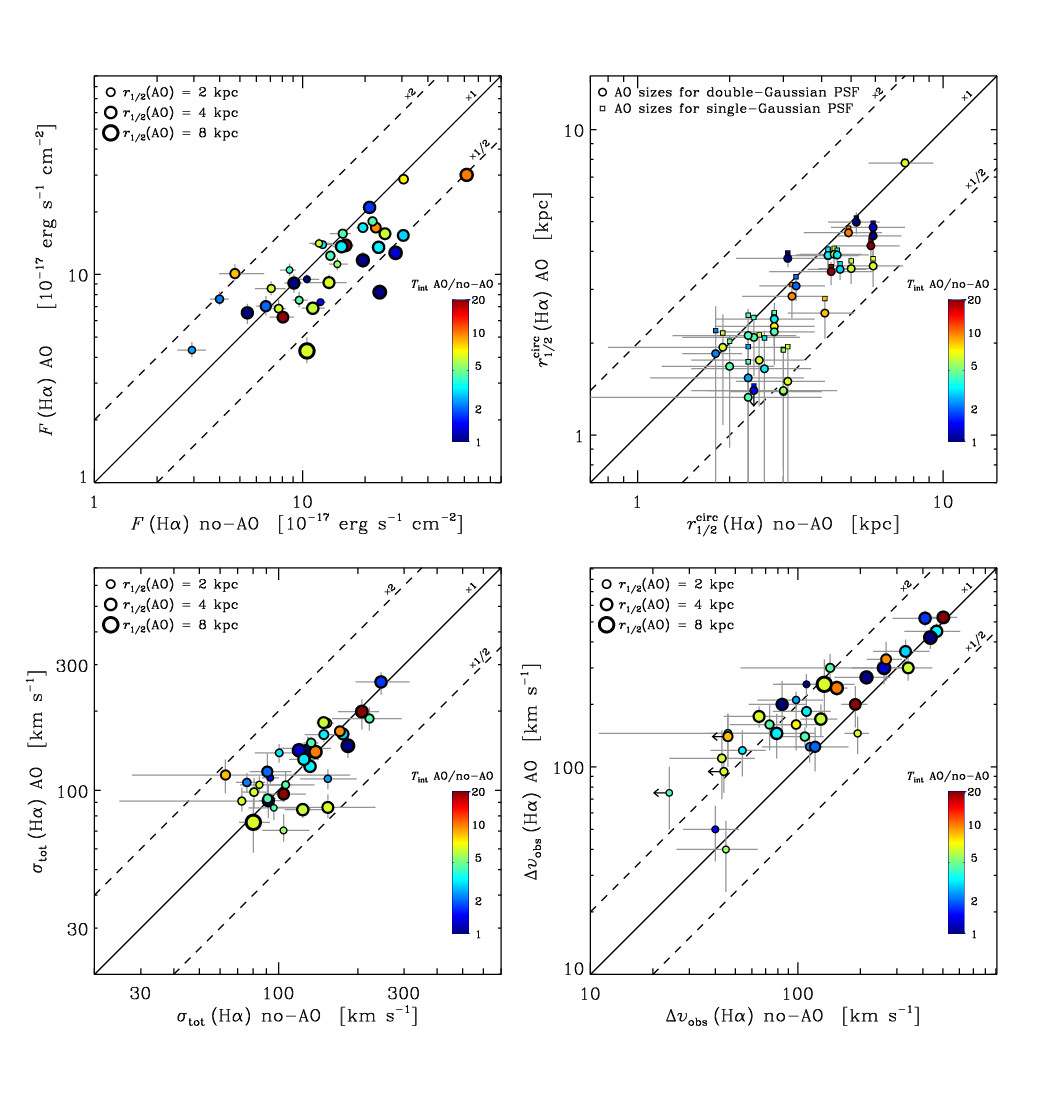}
\end{center}
\vspace{-1.5cm}
\renewcommand\baselinestretch{0.5}
\caption{
\small
Comparison of global galaxy properties measured from the AO-assisted
versus seeing-limited SINFONI data sets.
{\em Top left:\/} H$\alpha$ fluxes from the integrated spectra in the
total circular apertures.
{\em Top right:\/} H$\alpha$ half-light radii from the curve-of-growth
analysis in circular apertures.
{\em Bottom left:\/} Velocity dispersions from the line widths measured
in the integrated spectra in the total circular apertures.
{\em Bottom right:\/} Observed maximum velocity difference from the
H$\alpha$ kinematics.
In all panels, the solid line corresponds to equality between the
quantities compared, and the dashed lines indicate the range by a
factor of two about the one-to-one relationship.
The symbols are color-coded according to the logarithm of the ratio
of integration times in AO and no-AO mode as shown by the color bars.
For the comparisons of total fluxes, velocity dispersions, and maximum
observed velocity differences, the size of the symbols is proportional
to the logarithm of the H$\alpha$ half-light radii measured from the AO
data using the \psfave; reference symbol sizes are plotted
at the top left of the panels for three values of $r_{1/2}^{\rm circ}$.
For the comparison of half-light radii, the large circles correspond to the
AO-based sizes derived using the \psfave\ parameters for every galaxy, and
the small squares represent those computed with the parameters of the \psfgal\
associated with each individual galaxy.
\label{fig-AOvsNoAO_Ha}
}
\vspace{2ex}
\end{figure*}

Figure~\ref{fig-AOvsNoAO_Ha} compares the results derived from the AO and
seeing-limited data for selected basic measurements: the total H$\alpha$
flux and line width from the integrated spectrum, the half-light radius from
curve-of-growth analysis, and the maximum observed velocity difference, all
extracted following the procedures described in Sections~\ref{Sect-extract},
\ref{Sect-struct_meas}, and \ref{Sect-kin_meas}.

The best agreement is seen for the integrated H$\alpha$ line widths.
On average (and median) the same values are obtained within 5\% from the
AO and seeing-limited data.  The objects with largest differences (up to
a factor of $\approx 2$) typically have short (1\,hr) integrations and
higher noise in their no-AO data.
The AO-based total H$\alpha$ fluxes tend to be lower (by $10\%$ on average
and $14\%$ on median), and the half-light radii $r_{1/2}^{\rm circ}$ smaller
(by $20\%$ on average and median) than those measured in the no-AO data.
These differences are comparable to the measurement uncertainties, typically
dominated by those of the flux calibration and continuum subtraction
(Section~\ref{Sub-meth}) and PSF determination (Section~\ref{Sub-PSF}).
Part of the bulk offsets is driven by the smaller effective FOV of the
AO-assisted observations limiting measurements of the most extended line
emission in larger targets, notably for K20-ID7, GMASS-2540, and ZC406690
where the fluxes from the AO data miss up to roughly half the total flux.
Another factor is the impact of the broad wings of the \psfave\ used in
computing the intrinsic radii, possibly leading to underestimates in the
intrinsic size of the smallest sources.  Using the \psfgal\ associated with
individual galaxies for the AO data instead brings the $r_{1/2}^{\rm circ}$
for the compact objects in better agreement with those from the no-AO data,
resulting in a smaller average (and median) size difference of $10\%$ for
the full sample.
The choice of circular apertures to derive the total fluxes and half-light
radii may also introduce some differences.  Although they were chosen to
enclose as best as possible the total H$\alpha$ emission based mainly on
the curve-of-growth behaviour (see Section~\ref{Sub-spectra}), they may miss
more of the extended emission for sources with highest isophotal ellipticity,
and were different between the AO and seeing-limited data in most cases.
To gauge the impact of these factors on the measurements, we estimated
aperture corrections based on the 2D S\'ersic models described in
Section~\ref{Sub-sizes_disc}, for the H$\alpha$ structural parameters
derived in Section~\ref{Sub-sizes_galfit} and convolved with the appropriate
PSFs.  With these corrections, the AO-based fluxes are then on average $4\%$
(on median $8\%$) lower than the no-AO-based ones while the size differences
remain essentially the same.  The largest discrepancies are only modestly
reduced because such simple aperture corrections do not capture the impact
of S/N differences and of clumpy irregular morphologies.

The comparison in observed velocity difference shows the expected effects of
beam smearing.  The agreement is best among the largest sources, with AO-based
$\Delta v_{\rm obs}$ on average $\approx 1.4$ times higher (on median,
$\approx 1.2$) for the targets with $r_{1/2}^{\rm circ} \ga 3~{\rm kpc}$.
The difference among the smaller sources increases to an average and median
factor of $\approx 1.7$.  Two of the most compact objects (SA12-6339
and ZC409985) deviate from this beam smearing-driven trend: the same
low $\Delta v_{\rm obs}$ is measured within the $1\sigma$ uncertainties
between the AO and no-AO data, suggesting that these sources are genuinely
dispersion-dominated or have very low inclination.
With the beam smearing corrections described in Section~\ref{Sub-BS_sini},
the systematic offset is reduced to a mean and median factor of $\approx 1.2$
and $\approx 1.6$ for the larger and smaller sources, respectively.  The AO
versus no-AO differences also reflect in part the shallower seeing-limited
data not probing fully the velocity gradients in fainter or lower-surface
brightness objects (explaining for instance the nearly twice higher AO-based
$\Delta v_{\rm obs}$ for GMASS-2540 despite its large extent).

\begin{figure*}[!ht]
\begin{center}
\includegraphics[scale=0.85,clip=1,angle=0]{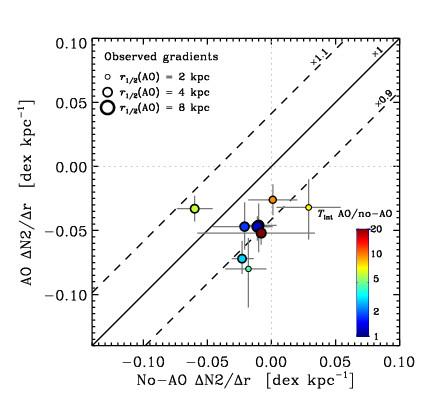}
\end{center}
\vspace{-0.7cm}
\renewcommand\baselinestretch{0.5}
\caption{
\small
Comparison of radial \niiha\ gradients measured from the
AO-assisted versus seeing-limited SINFONI data sets.
The gradients are derived from the \niiha\ ratio measured in the
co-averaged spectra of individual pixels within elliptical annuli.
The data are shown for the nine targets for which the ratio can be
measured over at least three annuli in both the AO and no-AO data.
The solid line corresponds to the one-to-one relationship, and the
dashed lines indicate offsets by $\rm \pm 0.04~dex~kpc^{-1}$, or
factors of 0.9 and 1.1, about this relationship.
The symbols are color-coded according to the logarithm of the ratio of
AO to no-AO integration times as shown by the color bar, and their size
is proportional to the logarithm of the curve-of-growth-based H$\alpha$
half-light radii measured from the AO data using the \psfave;
reference symbol sizes are plotted at the top left of the panel.
\label{fig-AOvsNoAO_metal}
}
\vspace{5ex}
\end{figure*}

As a last comparison, Figure~\ref{fig-AOvsNoAO_metal} shows the
radial \niiha\ gradients obtained from the AO and no-AO data sets.
The measurements from the seeing-limited data were made in elliptical annuli
with same PA and similar axis ratio as used for the AO data, and 2-pixel
width and separation giving a 2.5 times coarser sampling (i.e., $0\farcs 25$).
The comparison is restricted to the nine objects for which a \niiha\ ratio
can be measured in at least three annuli for both observing modes
(Q2343-BX389, Q2343-BX599, Q2343-BX610, Deep3a-6004, Deep3a-6397,
Deep3a-15504, ZC400569, ZC403741, and ZC407302).
The \niigrad\ from the higher resolution observations are on average (and
median) more negative by $\rm \approx 0.035~dex~kpc^{-1}$.  The one exception
is Deep3a-15504, where the gradient derived from the AO data is shallower
($-0.033 \pm 0.010$ compared to $\rm -0.060 \pm 0.014~dex~kpc^{-1}$).
The strong, spatially-resolved and asymmetric broad H$\alpha +$ [\ion{N}{2}]
emission associated with the AGN-driven outflow in this galaxy \citep{FS14}
could cause this difference, given the simple single-Gaussian fits to the
lines applied here.
Aperture positioning may be an issue; for instance, small offsets could
flatten a steep inner gradient, but we verified that this has little
impact for this galaxy and the restricted gradient (at $r > 0\farcs 3$;
Section~\ref{Sub-metal_prof}) is also shallower in the AO data.

In summary, the comparisons between AO and seeing-limited data presented
in this Appendix show an overall good agreement, with differences consistent
with expectations from beam smearing, and from varying S/N and effective FOV
between the observations obtained in each mode.  Although the agreement
generally improves when applying aperture and beam smearing corrections,
there still remains noticeable differences.  These corrections obviously
do not, or poorly, account for emission components that may be undetected
in the data.  In addition, they are derived from simple models that do not
capture the complexity of the galaxies that appears most prominently in the
higher resolution, and typically higher S/N, AO-assisted data sets.

\clearpage


\begin{figure}[p]
\begin{center}
\includegraphics[scale=0.75,clip=1,angle=0]{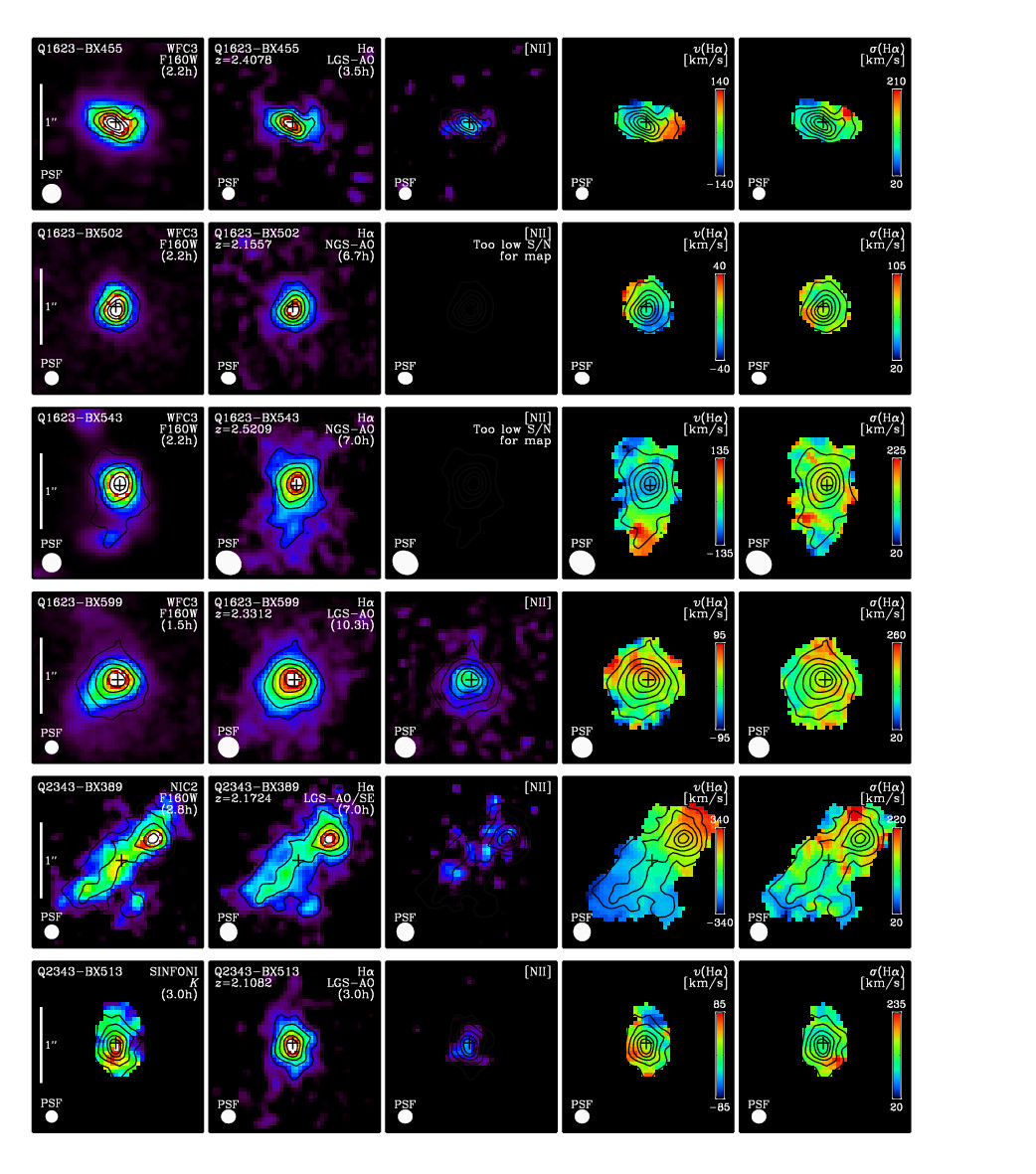}
\end{center}
\vspace{-0.6cm}
\renewcommand\baselinestretch{0.5}
\caption{
\small
High-resolution near-IR broad-band images (in $H$ or $K$ band), H$\alpha$
and [\ion{N}{2}] emission line maps, and H$\alpha$ velocity and velocity
dispersion maps of the \sinszc\ AO sample, as described in
Appendix~\ref{App-FullDataSets}.
\label{fig-bbAO_1}
}
\end{figure}

\clearpage

\begin{figure}[p]
\addtocounter{figure}{-1}
\begin{center}
\includegraphics[scale=0.75,clip=1,angle=0]{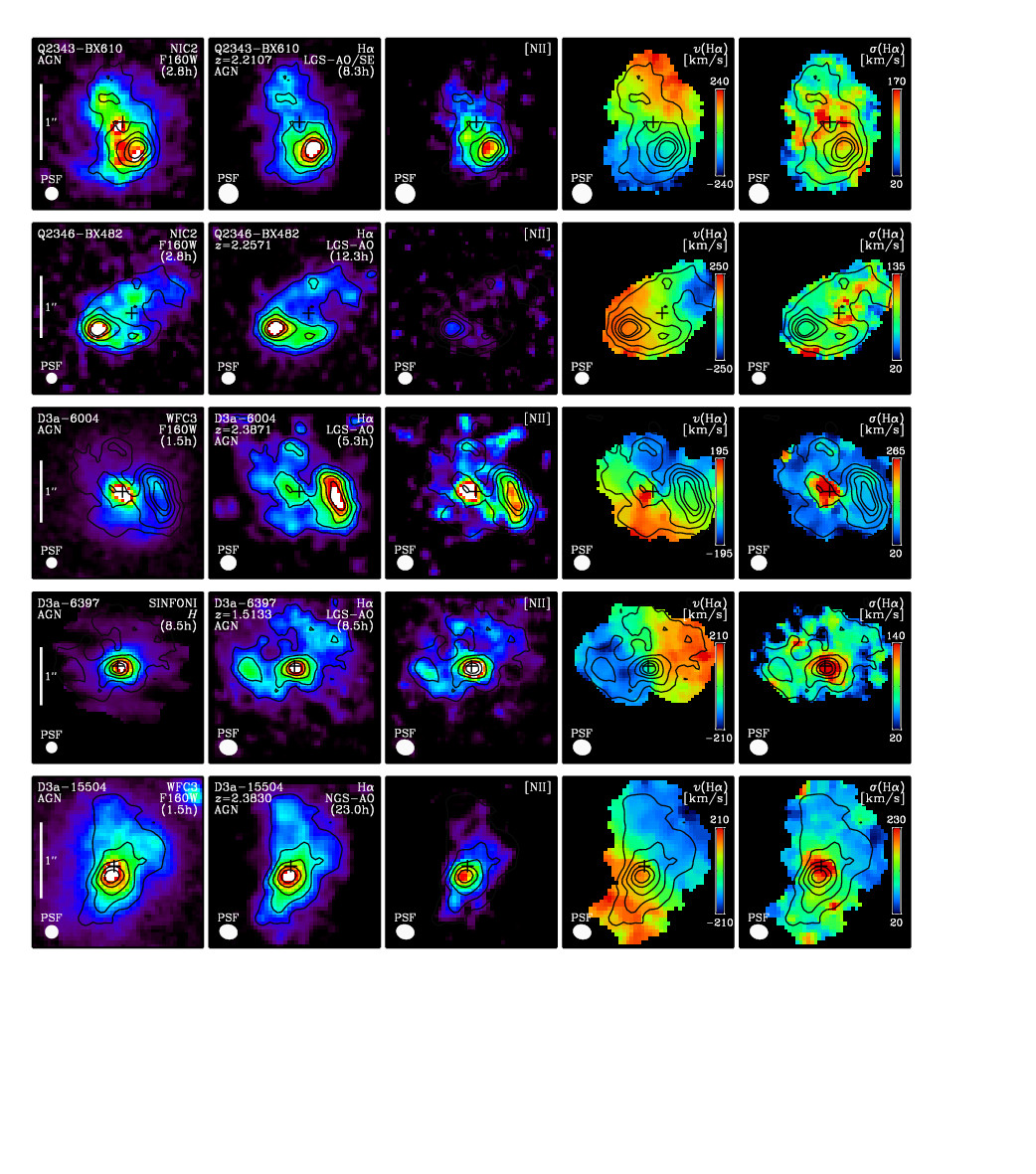}
\end{center}
\vspace{-4.1cm}
\renewcommand\baselinestretch{0.5}
\caption{
\small
(Continued.)
}
\end{figure}

\clearpage

\begin{figure}[p]
\addtocounter{figure}{-1}
\begin{center}
\includegraphics[scale=0.75,clip=1,angle=0]{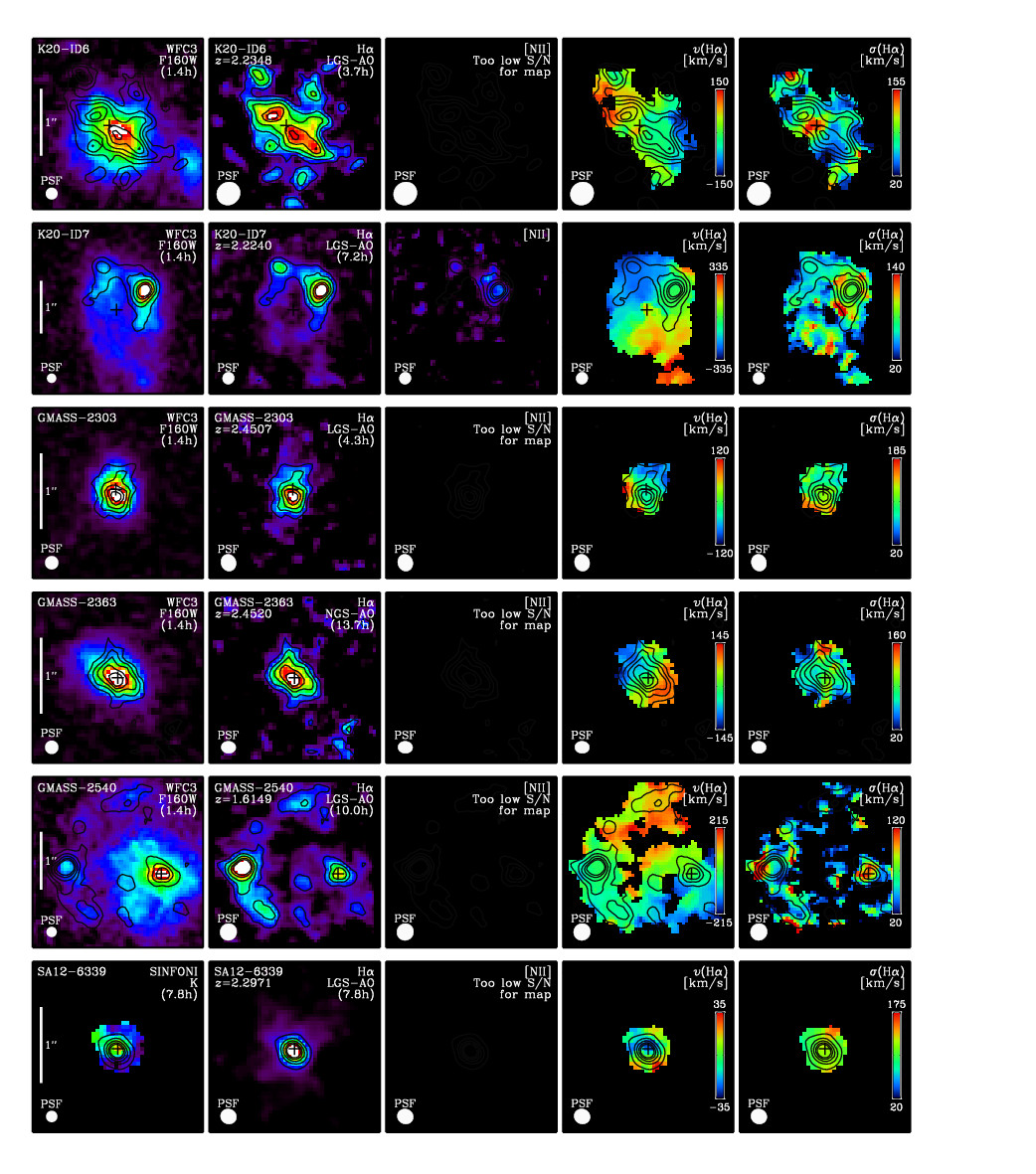}
\end{center}
\vspace{-0.6cm}
\renewcommand\baselinestretch{0.5}
\caption{
\small
(Continued.)
}
\end{figure}

\clearpage

\begin{figure}[p]
\addtocounter{figure}{-1}
\begin{center}
\includegraphics[scale=0.75,clip=1,angle=0]{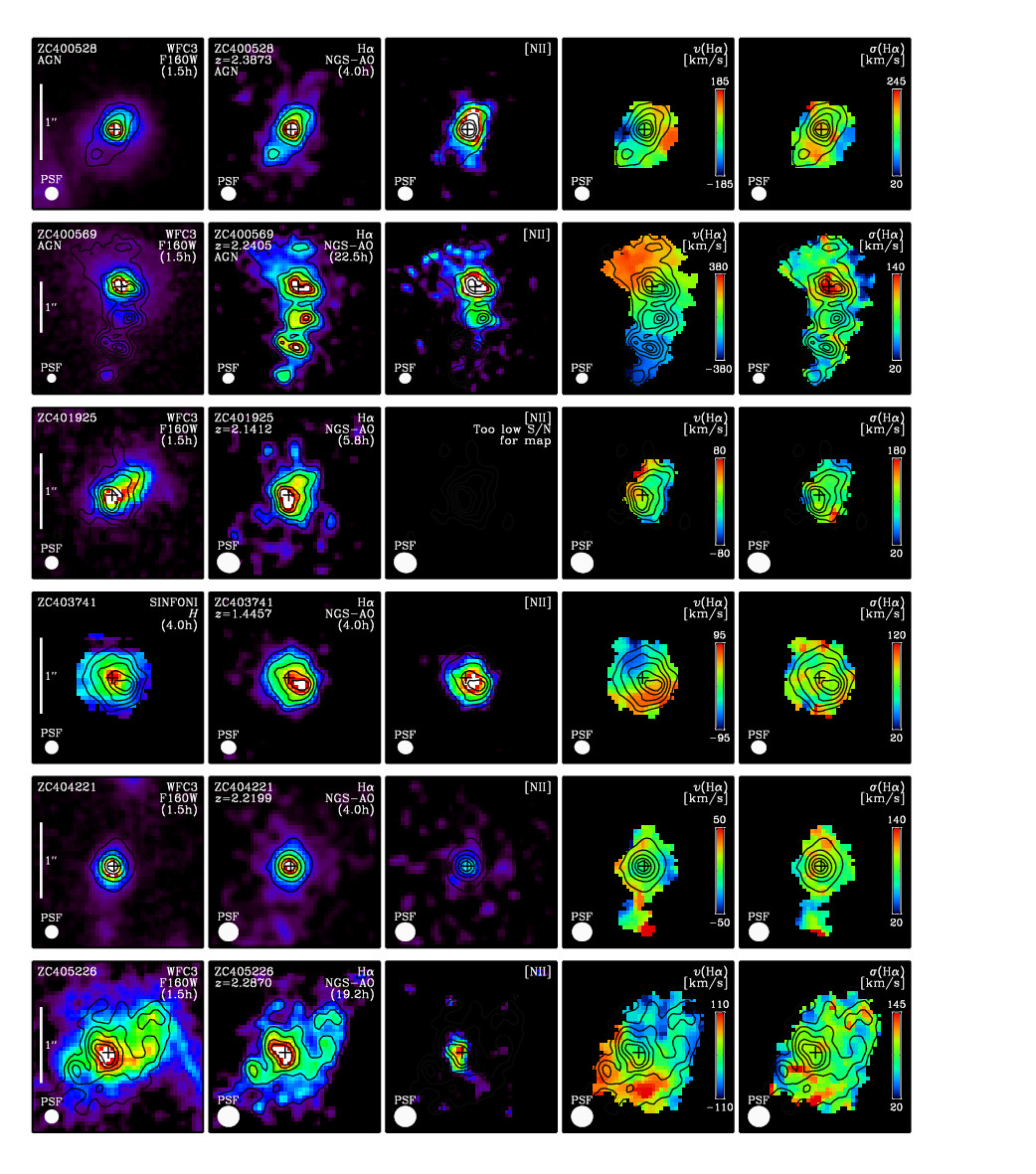}
\end{center}
\vspace{-0.6cm}
\renewcommand\baselinestretch{0.5}
\caption{
\small
(Continued.)
}
\end{figure}

\clearpage

\begin{figure}[p]
\addtocounter{figure}{-1}
\begin{center}
\includegraphics[scale=0.75,clip=1,angle=0]{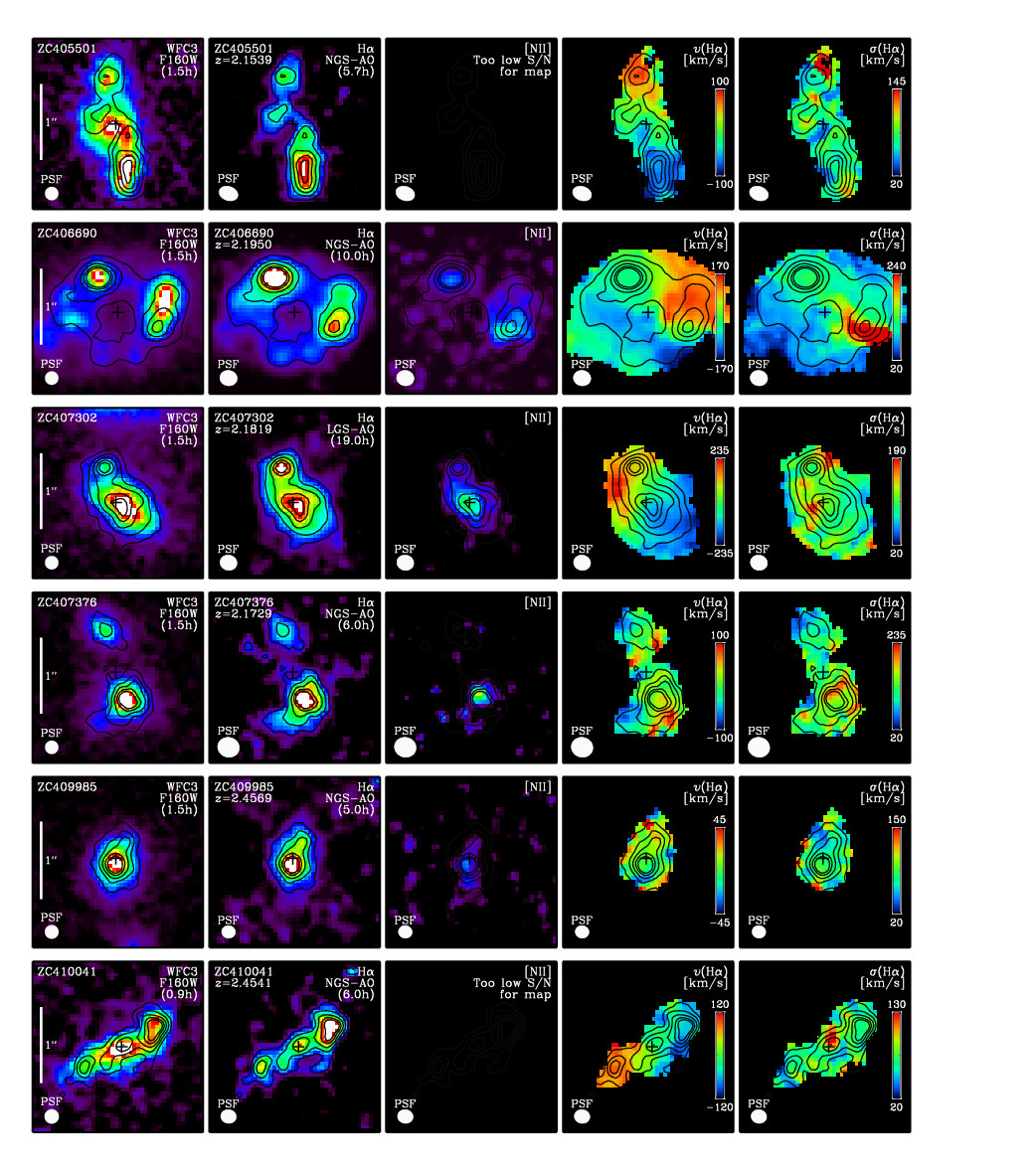}
\end{center}
\vspace{-0.6cm}
\renewcommand\baselinestretch{0.5}
\caption{
\small
(Continued.)
}
\end{figure}

\clearpage

\begin{figure}[p]
\addtocounter{figure}{-1}
\begin{center}
\includegraphics[scale=0.75,clip=1,angle=0]{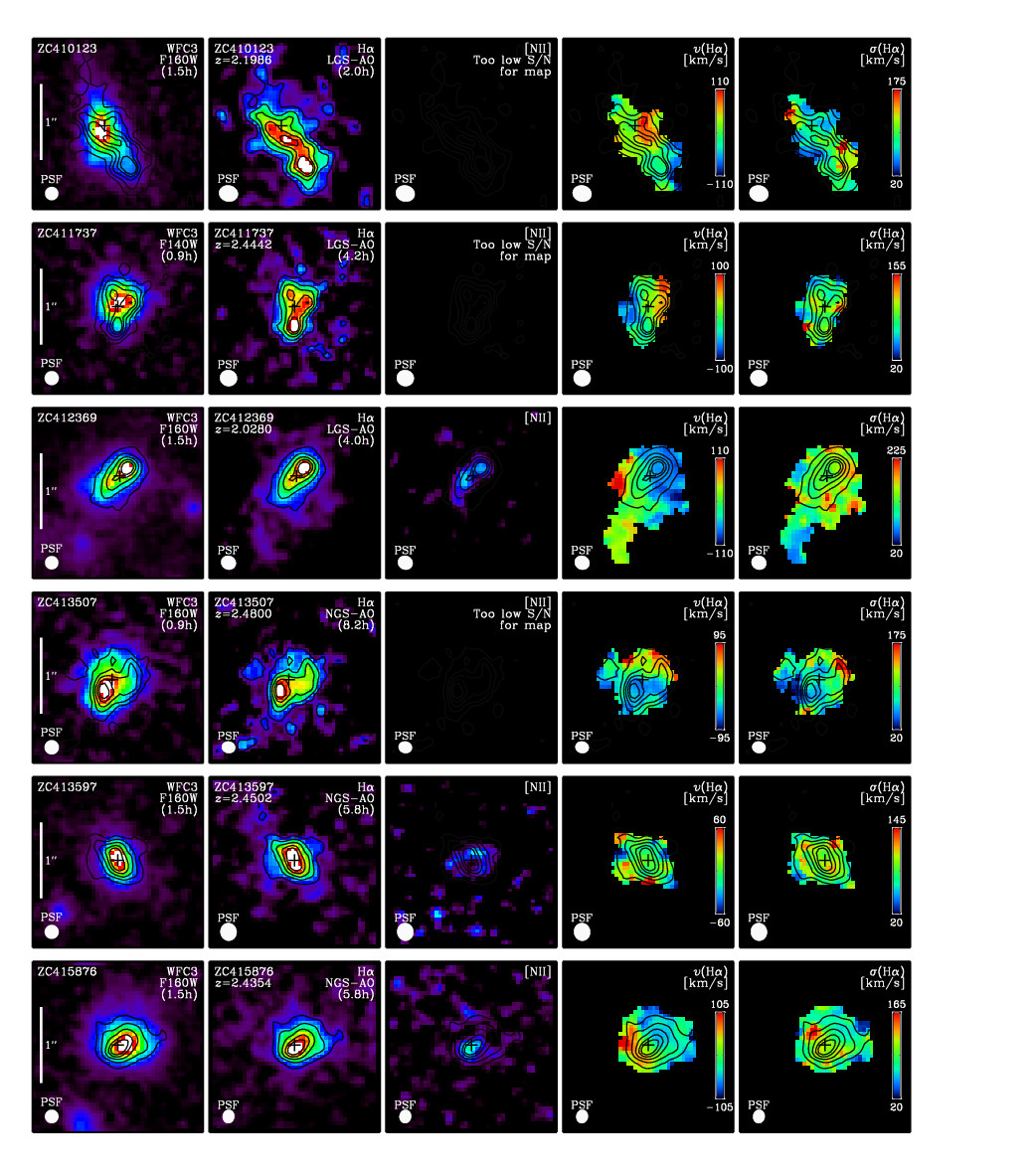}
\end{center}
\vspace{-0.6cm}
\renewcommand\baselinestretch{0.5}
\caption{
\small
(Continued.)
}
\end{figure}

\clearpage


\begin{figure}[p]
\begin{center}
\includegraphics[scale=0.56,clip=1,angle=0]{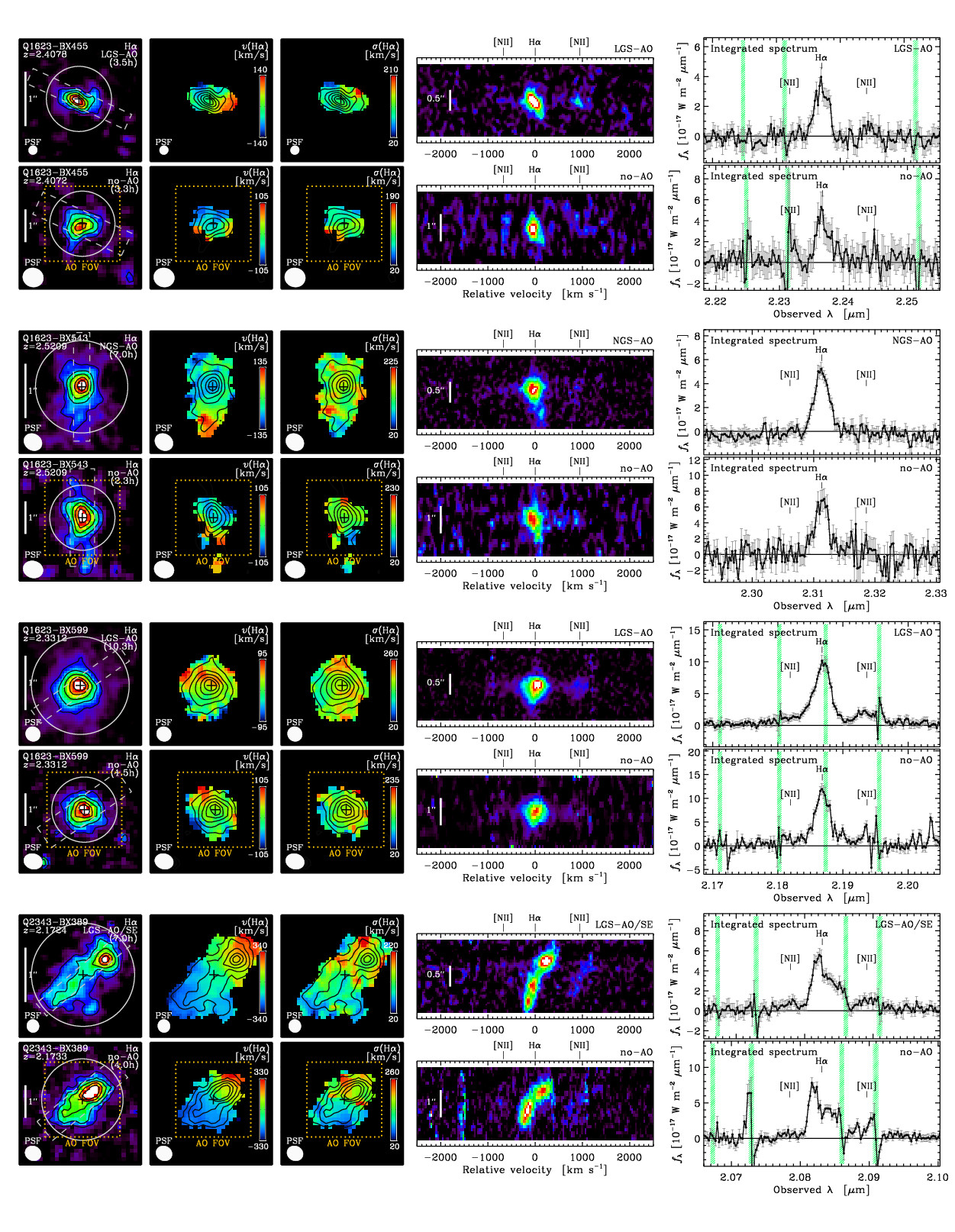}
\end{center}
\vspace{-1.1cm}
\renewcommand\baselinestretch{0.5}
\caption{
\small
H$\alpha$ line flux, velocity, and velocity dispersion maps,
p-v diagrams, and integrated spectra of the \sinszc\ AO sample.
Rows are grouped pairwise for each galaxy, showing the AO and
seeing-limited data (top and bottom row, respectively) as
described in Appendix~\ref{App-AOnoAO}.
\label{fig-AOnoAO_1}
}
\end{figure}

\clearpage

\begin{figure}[p]
\addtocounter{figure}{-1}
\begin{center}
\includegraphics[scale=0.56,clip=1,angle=0]{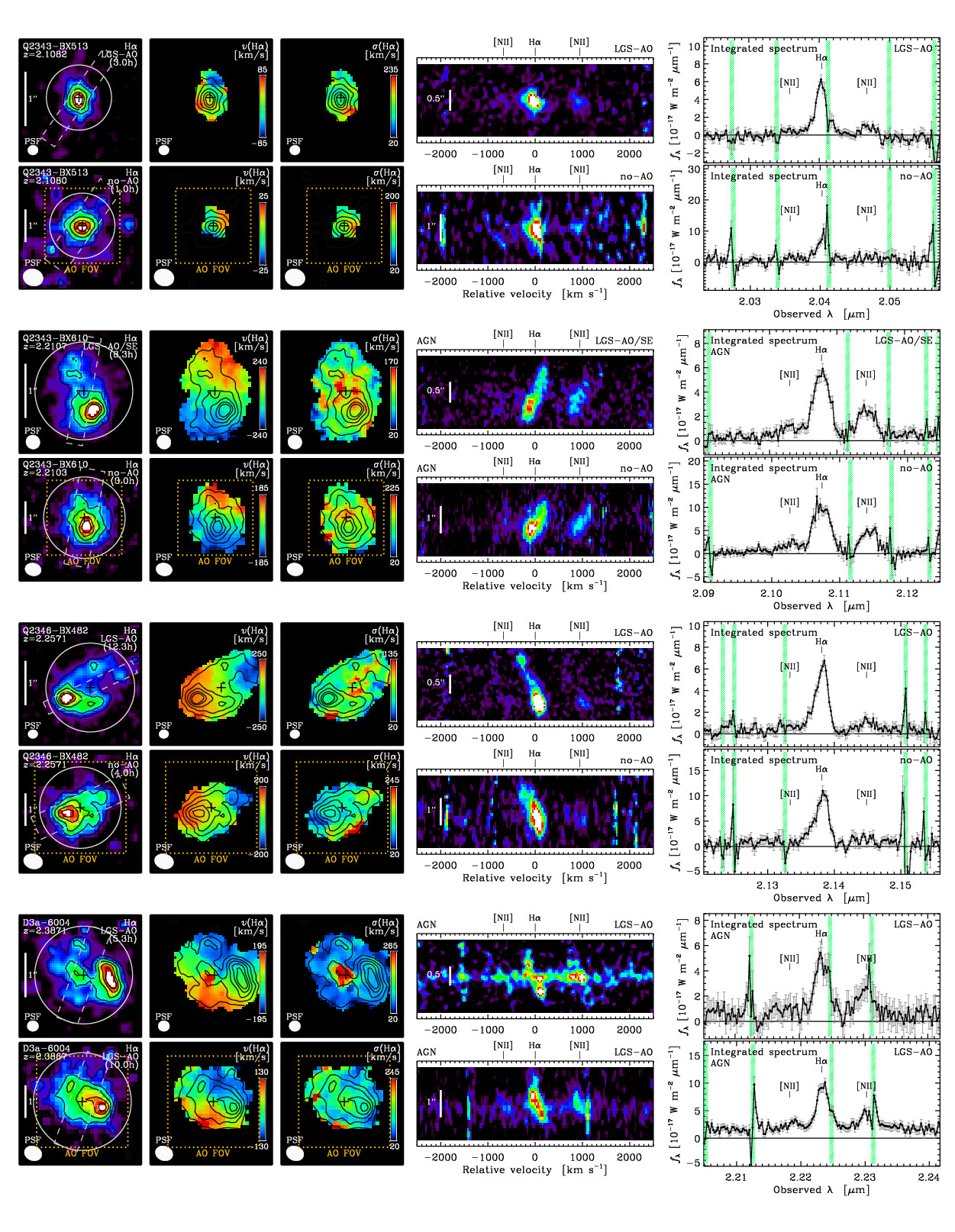}
\end{center}
\vspace{-1.1cm}
\renewcommand\baselinestretch{0.5}
\caption{
\small
(Continued.)
}
\end{figure}

\clearpage

\begin{figure}[p]
\addtocounter{figure}{-1}
\begin{center}
\includegraphics[scale=0.56,clip=1,angle=0]{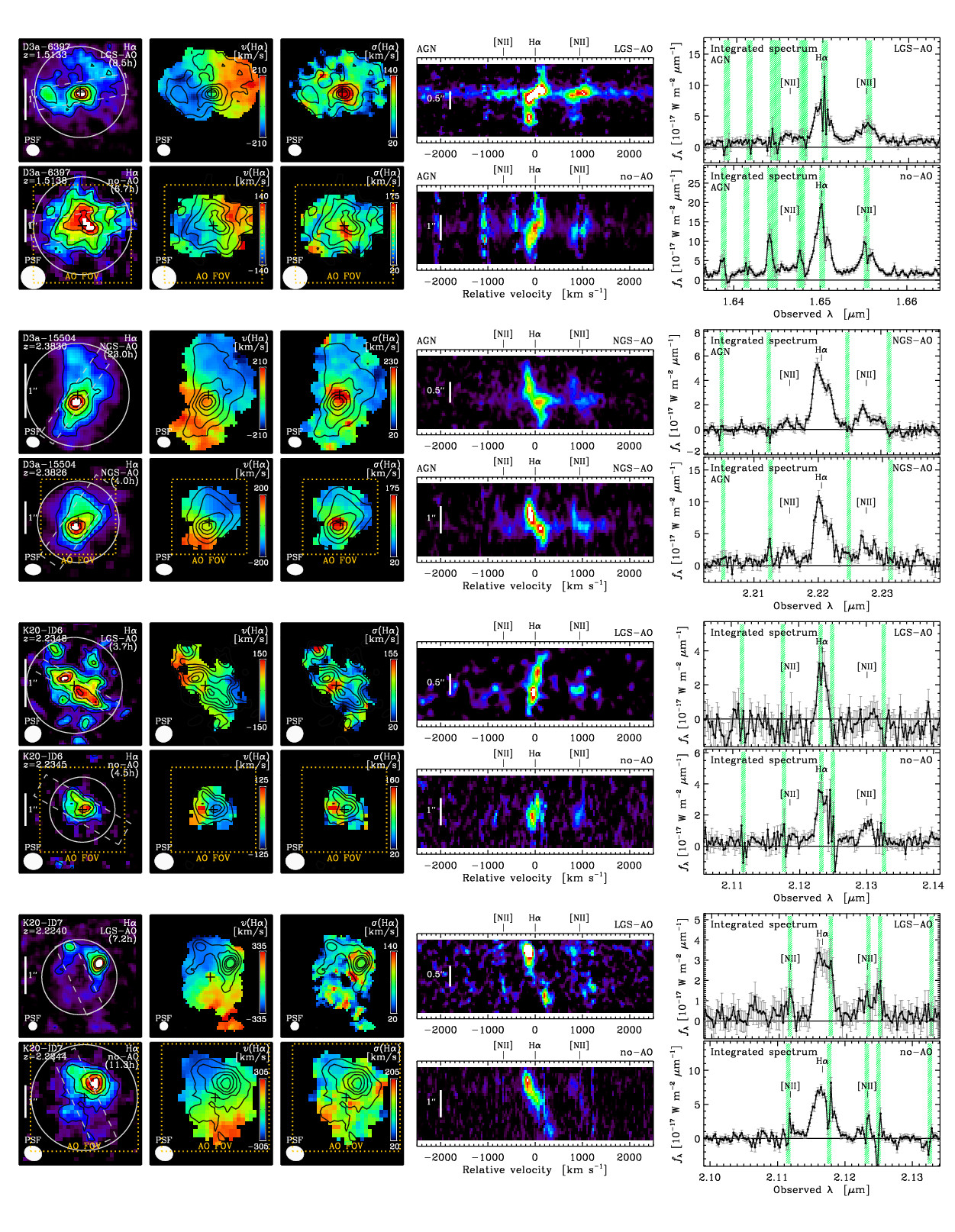}
\end{center}
\vspace{-1.1cm}
\renewcommand\baselinestretch{0.5}
\caption{
\small
(Continued.)
}
\end{figure}

\clearpage

\begin{figure}[p]
\addtocounter{figure}{-1}
\begin{center}
\includegraphics[scale=0.56,clip=1,angle=0]{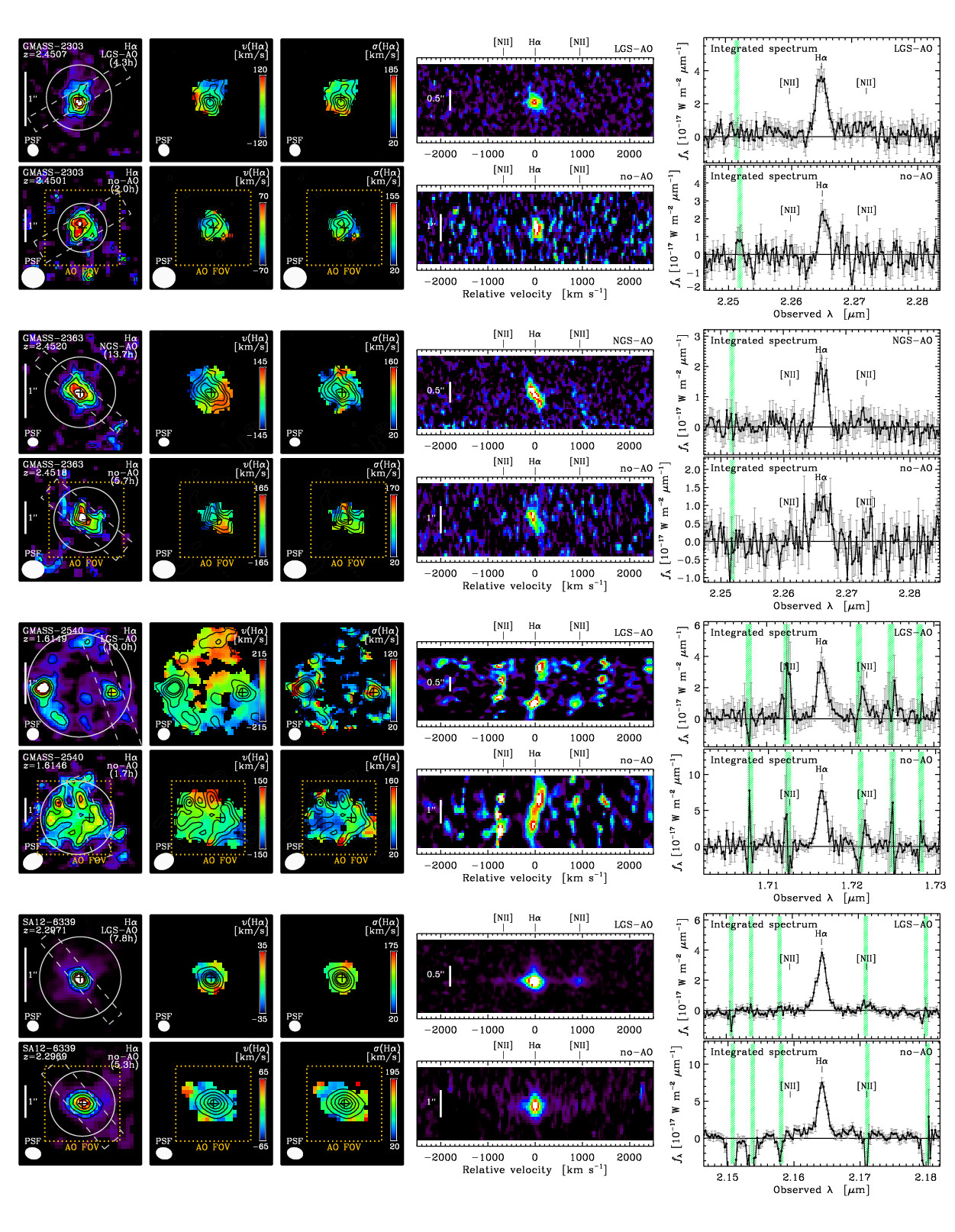}
\end{center}
\vspace{-1.1cm}
\renewcommand\baselinestretch{0.5}
\caption{
\small
(Continued.)
}
\end{figure}

\clearpage

\begin{figure}[p]
\addtocounter{figure}{-1}
\begin{center}
\includegraphics[scale=0.56,clip=1,angle=0]{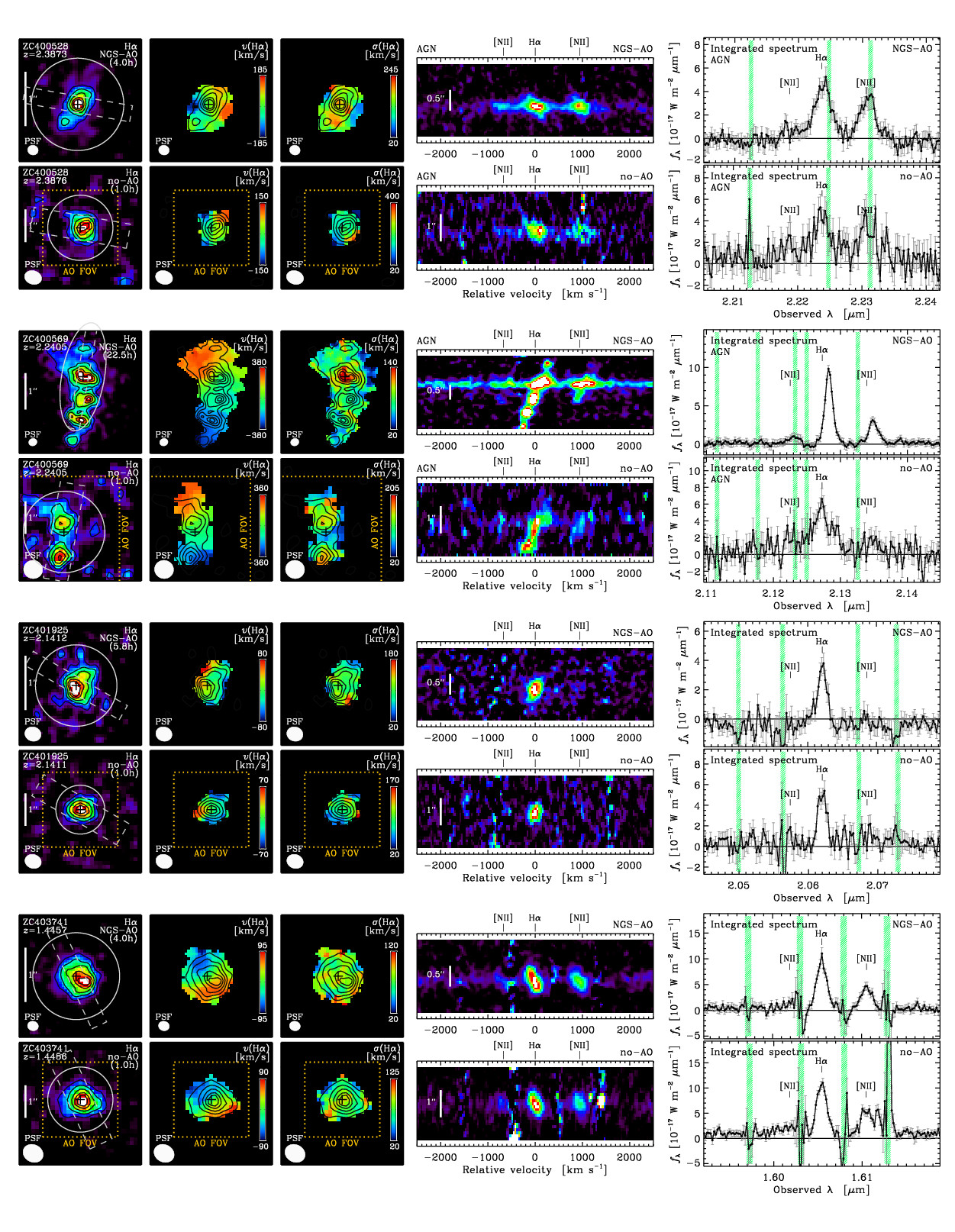}
\end{center}
\vspace{-1.1cm}
\renewcommand\baselinestretch{0.5}
\caption{
\small
(Continued.)
}
\end{figure}

\clearpage

\begin{figure}[p]
\addtocounter{figure}{-1}
\begin{center}
\includegraphics[scale=0.56,clip=1,angle=0]{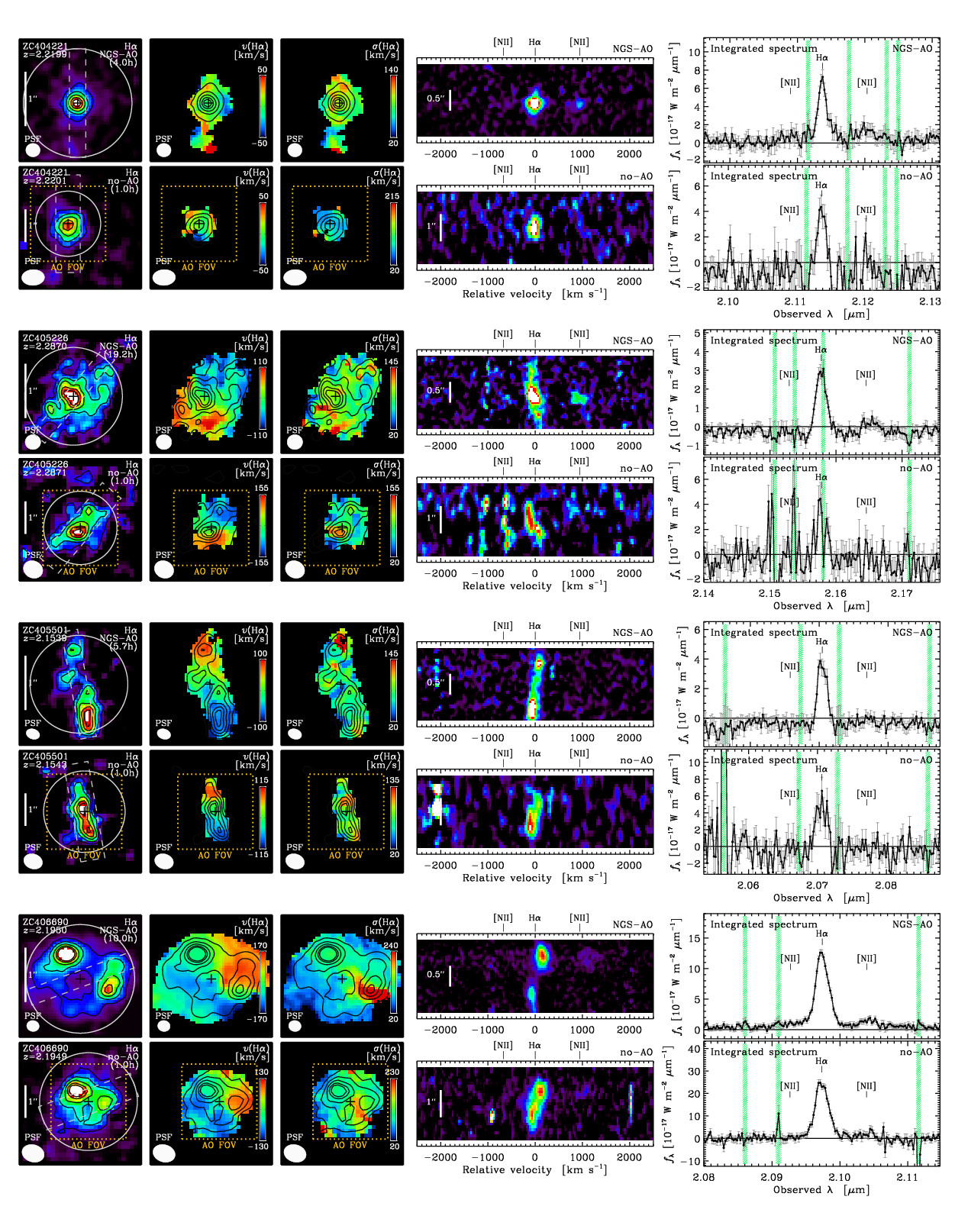}
\end{center}
\vspace{-1.1cm}
\renewcommand\baselinestretch{0.5}
\caption{
\small
(Continued.)
}
\end{figure}

\clearpage

\begin{figure}[p]
\addtocounter{figure}{-1}
\begin{center}
\includegraphics[scale=0.56,clip=1,angle=0]{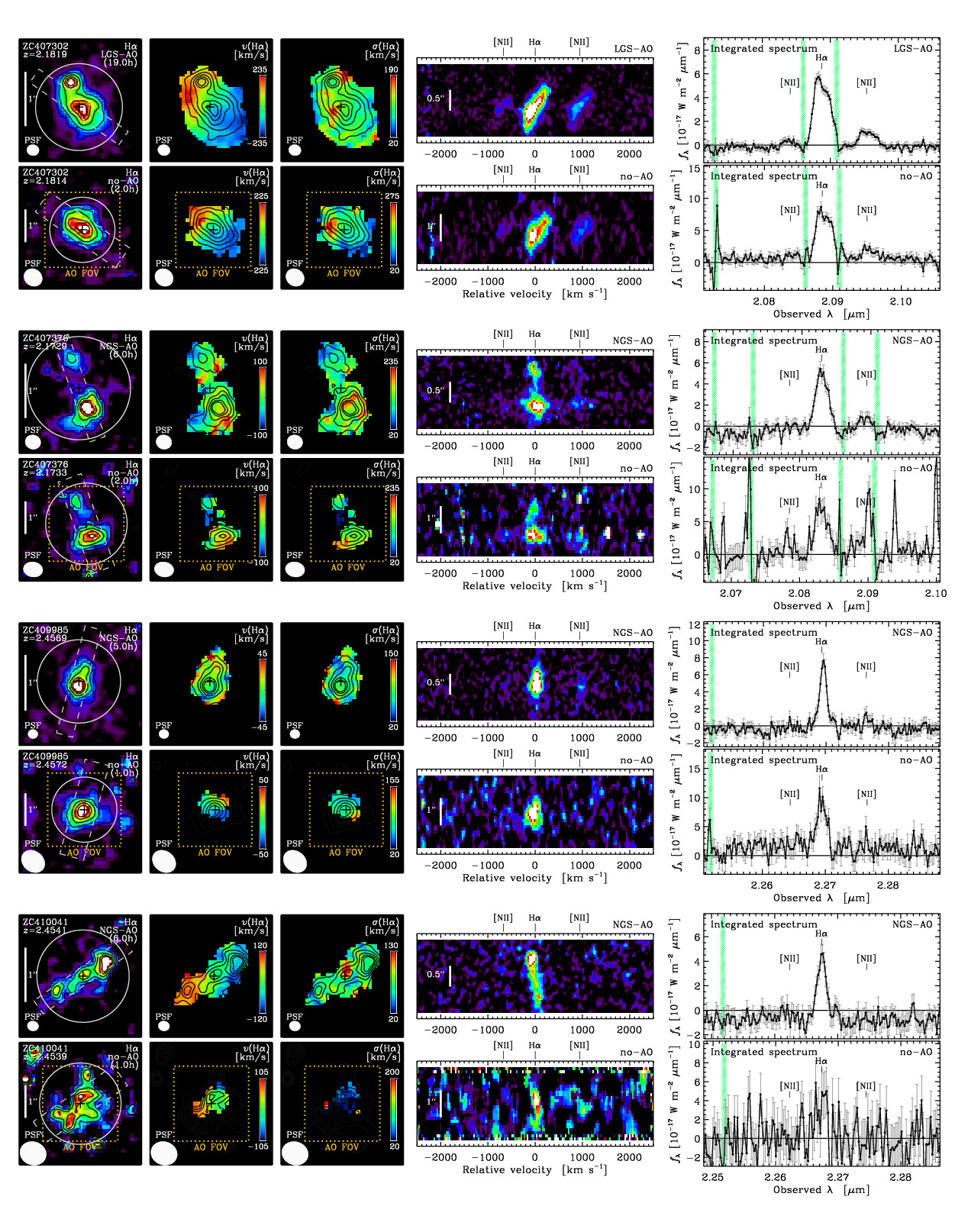}
\end{center}
\vspace{-1.1cm}
\renewcommand\baselinestretch{0.5}
\caption{
\small
(Continued.)
}
\end{figure}

\clearpage

\begin{figure}[p]
\addtocounter{figure}{-1}
\begin{center}
\includegraphics[scale=0.56,clip=1,angle=0]{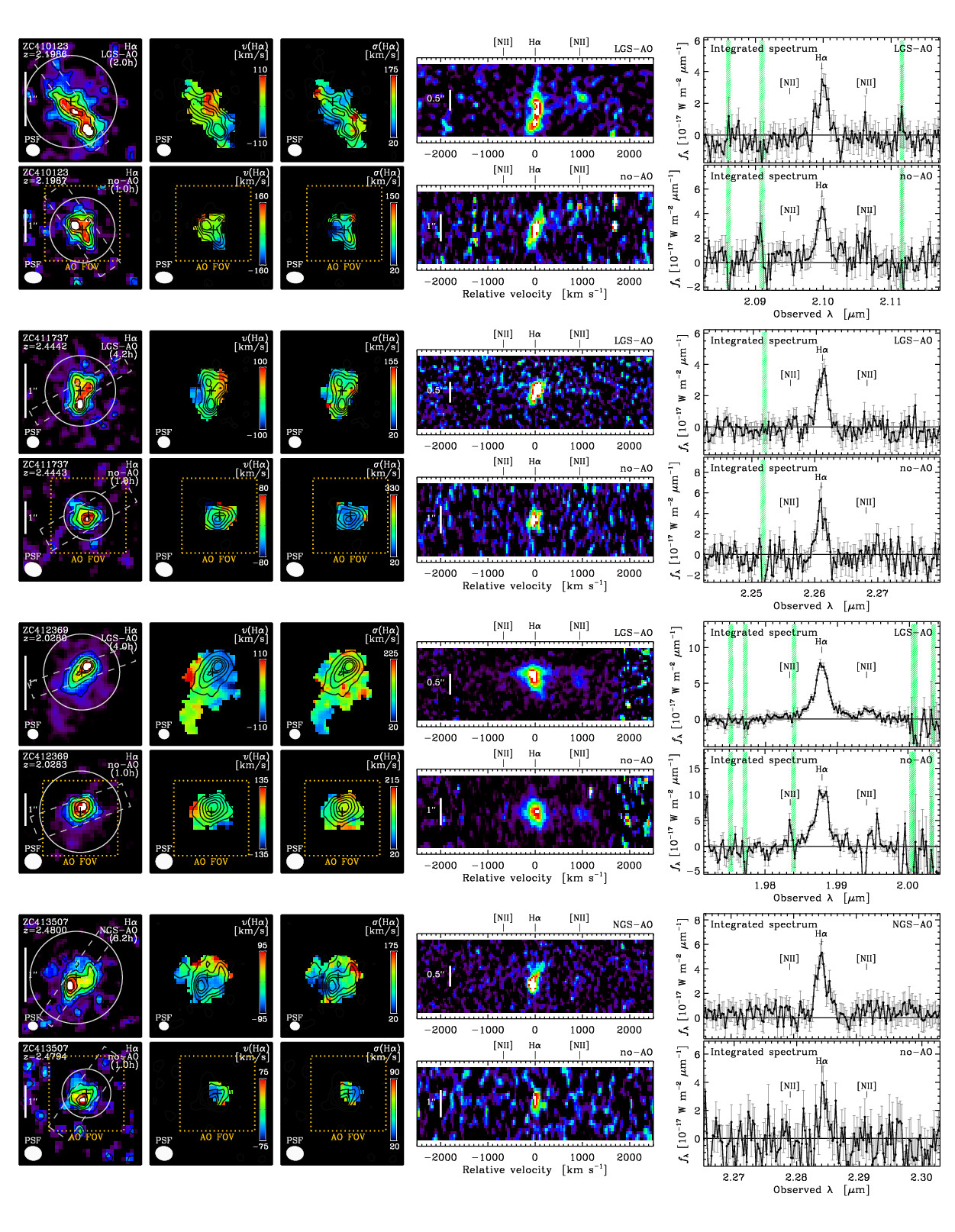}
\end{center}
\vspace{-1.1cm}
\renewcommand\baselinestretch{0.5}
\caption{
\small
(Continued.)
}
\end{figure}

\clearpage

\begin{figure}[!t]
\addtocounter{figure}{-1}
\begin{center}
\includegraphics[scale=0.56,clip=1,angle=0]{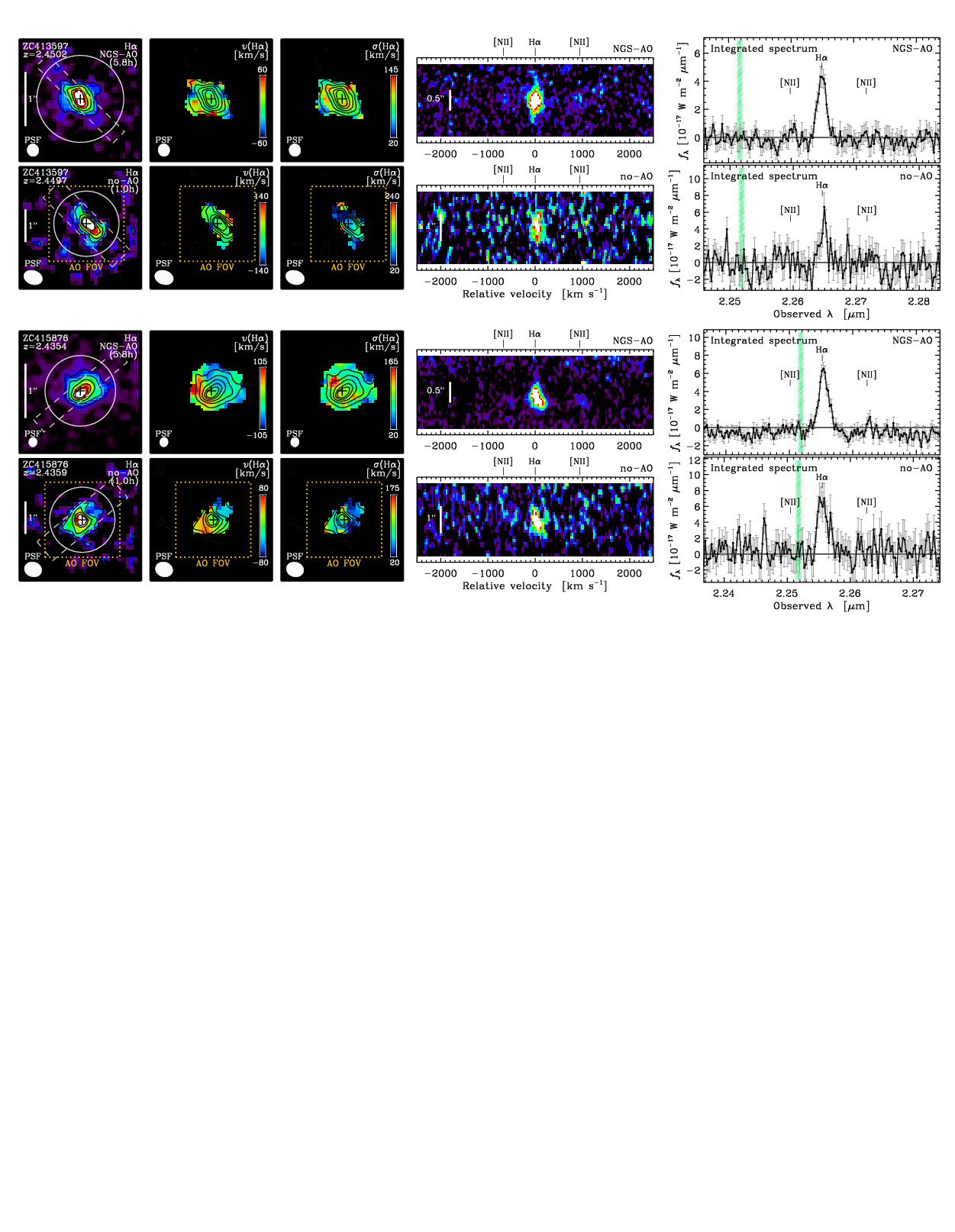}
\end{center}
\vspace{-12.0cm}
\renewcommand\baselinestretch{0.5}
\caption{
\small
(Continued.)
}
\end{figure}


\hphantom{hhh}
\bigskip
\bigskip
\section{MAPS AND RADIAL PROFILES OF THE [\ion{N}{2}]/H$\alpha$ RATIO}
         \label{App-metal}

Figure~\ref{fig-metal_1} shows for the individual
galaxies the spatially-resolved and integrated information from the
[\ion{N}{2}]/H$\alpha$ ratios extracted from the SINFONI AO data.

The first two panels from left to right show the velocity-integrated
H$\alpha$ flux and \niiha\ ratio maps.  Pixels for which the ratio
has a $\rm S/N < 3$ are masked out.
The galaxy, its H$\alpha$ redshift, the AO observing mode, and the total
on-source integration time are given at the top of the H$\alpha$ line map.
Galaxies hosting an AGN have the corresponding label at the top left of
the [\ion{N}{2}]/H$\alpha$ ratio map.
The colour coding for the H$\alpha$ maps scales linearly with flux from dark
blue to red/white for the minimum to maximum levels displayed (varying for
each galaxy).  The color coding for the \niiha\ ratio maps scales linearly
from blue to red following the color bar in the respective panels.  Black
contours in all three maps correspond to H$\alpha$ fractional flux levels
relative to the maximum of 0.2, 0.4, 0.6, 0.8, and 0.95.
The spatial resolution is represented by the filled ellipse at the bottom
left; the diameters along each axis correspond to the FWHMs and characterize
the {\em effective} PSF, including the median filtering applied spatially in
extracting the maps from the data cubes.  The angular scale is indicated by
the vertical bars on the left of the H$\alpha$ line map.  The solid white
ellipse overlaid on the maps shows the aperture used to extract the spectrum
in the third panel.  The black cross indicates the center of the galaxy.
In all maps, north is up and east is to the left.

The third panel from left shows the velocity-shifted integrated spectrum
extracted in the elliptical aperture marked on the maps.  Before co-adding,
the spectra of individual pixels, they are shifted according to the velocity
field (see Figure~\ref{fig-bbAO_1}) to bring the H$\alpha$ line centroid at
zero offset relative to the systemic redshift.
The elliptical apertures trace roughly the outer H$\alpha$ isophotes,
further enhancing the resulting S/N compared to circular apertures.
The error bars correspond to the $1\sigma$ uncertainties derived from the
noise properties of each data set, and include the scaling with aperture
size following the model described in Section~\ref{Sub-datared}, which
accounts for the fact that the effective noise is not purely Gaussian.
Vertical green hatched bars show the locations of bright night sky lines,
with width corresponding to the FWHM of the effective spectral resolution
of the data.

The fourth panel from the left shows the radial variations of the \niiha\
ratio, measured from velocity-shifted spectra extracted in elliptical annuli
of the same axis ratio and position angle as for the integrated spectrum
plotted in the previous panel.  The width of the annuli is two pixels,
or $0\farcs 1$.  The half-width at half-maximum of the PSF core is plotted
in units of the physical radius of the horizontal axis.  The black dots and
error bars give the \niiha\ ratio measurements and $1\sigma$ uncertainties;
downward arrows correspond to $3\sigma$ upper limits.
In the fifth panels, the \niiha\ ratio or upper limits thereof are plotted
for individual pixels as a function of the deprojected radius from the center,
where the parameters of the ellipse shown on the maps define the projection
assuming a disk geometry.  Ratios are plotted at the $3\sigma$ level if
the formal $\rm S/N > 3$.
The color-coding corresponds to the same flux levels as used for the
H$\alpha$ maps.
In both these panels, the horizontal thin orange and thick yellow line
indicates the \niiha\ ratio and $1\sigma$ uncertainties derived from
the velocity-shifted integrated spectrum.
The black line and grey-shaded area corresponds to the best-fit linear
gradient in $N2 \equiv \log({\rm [N{\small II}]/H\alpha})$ versus radius
$r$ and range of slopes from the $68\%$ confidence intervals, obtained
from a linear regression with censored data.
The integrated \niiha\ ratio and the best-fit linear gradient are labeled
in each plot.
For objects with evidence for an AGN, fits were also performed excluding
the central $r < 0.3\prime\prime$ regions (shown as hatched vertical bar);
the best-fit line and its uncertainties are overplotted, and the slope is
labeled in magenta.

In the rightmost panel, the \niiha\ ratio of the pixels are plotted
versus their H$\alpha$ flux, with $1\sigma$ uncertainties.
The black-and-white curve shows the $3\sigma$ limiting \niiha\ ratio
as a function of H$\alpha$ flux derived for each object.
The flux at which the pixel distribution starts to scatter below this
limit then corresponds to the level at/below which the measurements
become biased towards the higher \niiha\ ratios; the hatched region
marks this flux regime.  In this and the fifth panel, the large filled
symbols show the pixels in the unbiased flux regime, and the small filled
and open symbols show those in the biased regime with a $> 3\sigma$ and
$< 3\sigma$ \niiha\ measurement, respectively.  The average ratio for
the unbiased pixels is labeled in the fifth panel.
Pixels with a formal $> 3\sigma$ \niiha\ measurement but a
$F({\rm H\alpha})$ flux in the biased regime are also indicated with a
superposed cross in the ratio maps of the second panel.

\clearpage


\begin{figure}[p]
\begin{center}
\includegraphics[scale=0.47,clip=1,angle=90]{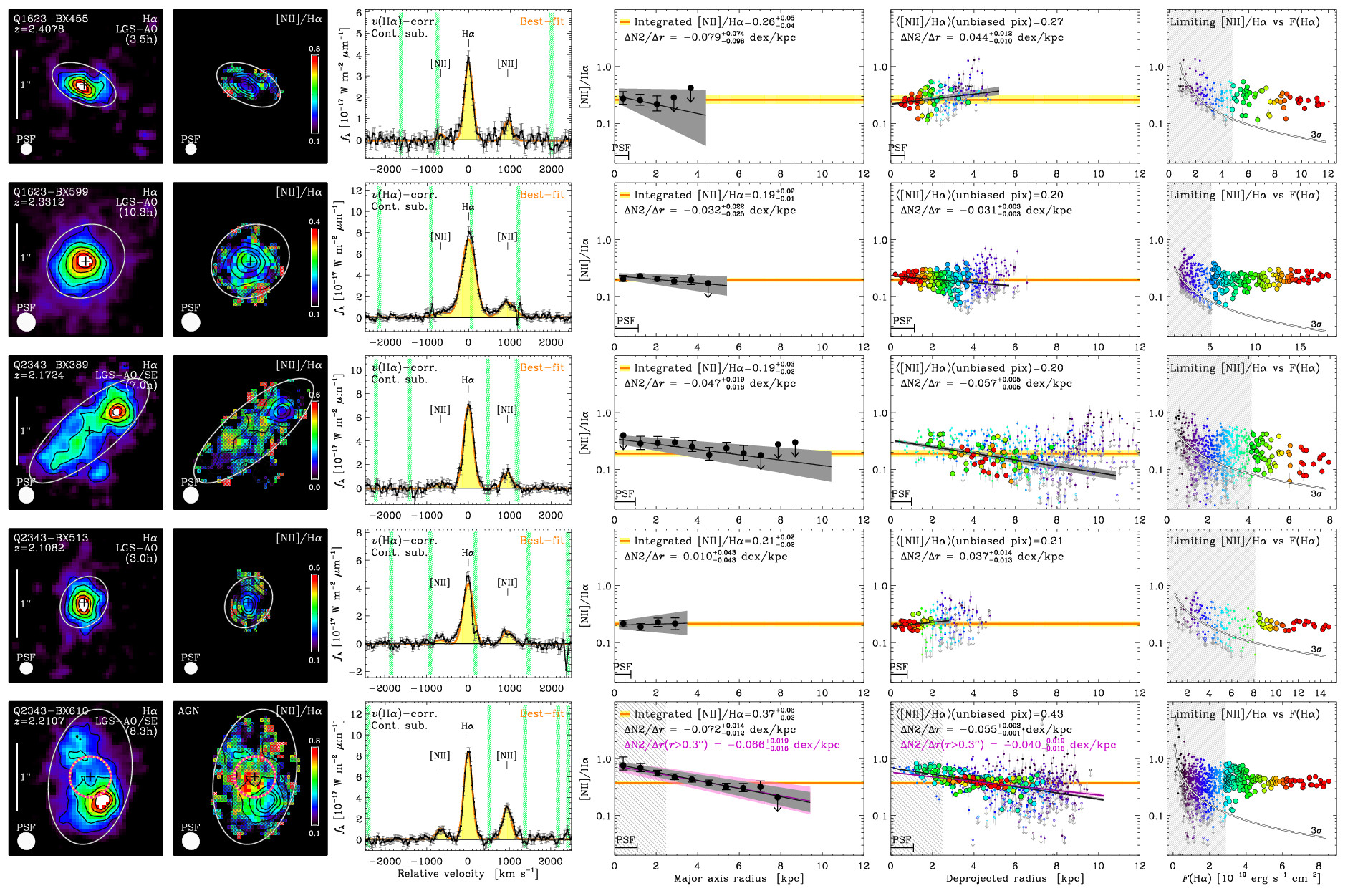}
\end{center}
\vspace{-0.4cm}
\renewcommand\baselinestretch{0.5}
\caption{
\small
H$\alpha$ line maps, [\ion{N}{2}]/H$\alpha$ ratio maps, velocity-shifted
integrated spectra, [\ion{N}{2}]/H$\alpha$ radial profiles in elliptical
apertures, and distribution of [\ion{N}{2}]/H$\alpha$ ratios in individual
pixels versus (deprojected) radius and measured H$\alpha$ line flux,
as described in Appendix~\ref{App-metal}.
\label{fig-metal_1}
}
\end{figure}


\begin{figure}[p]
\addtocounter{figure}{-1}
\begin{center}
\includegraphics[scale=0.47,clip=1,angle=90]{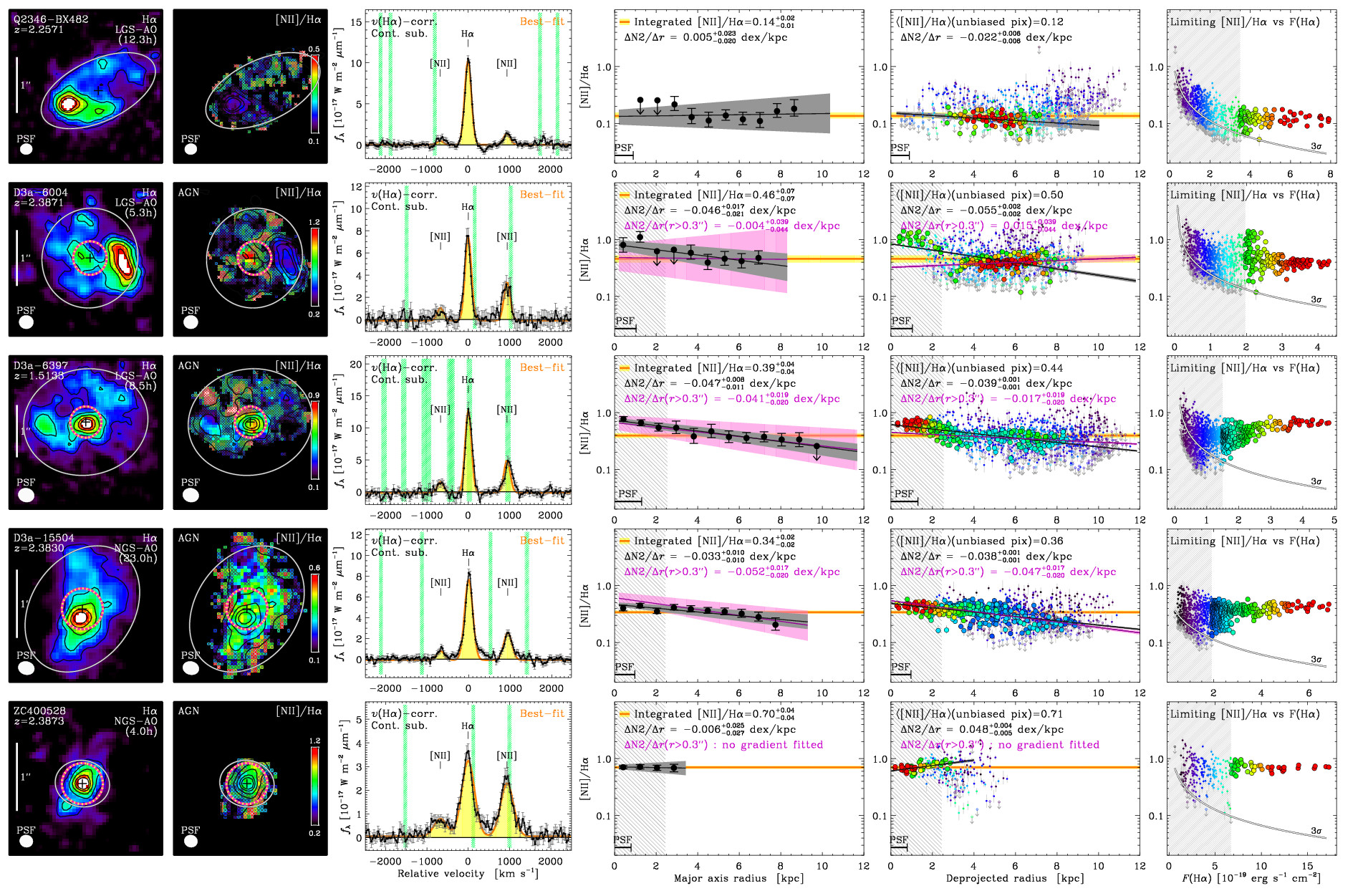}
\end{center}
\vspace{-0.4cm}
\renewcommand\baselinestretch{0.5}
\caption{
\small
(Continued.)
}
\end{figure}


\begin{figure}[p]
\addtocounter{figure}{-1}
\begin{center}
\includegraphics[scale=0.47,clip=1,angle=90]{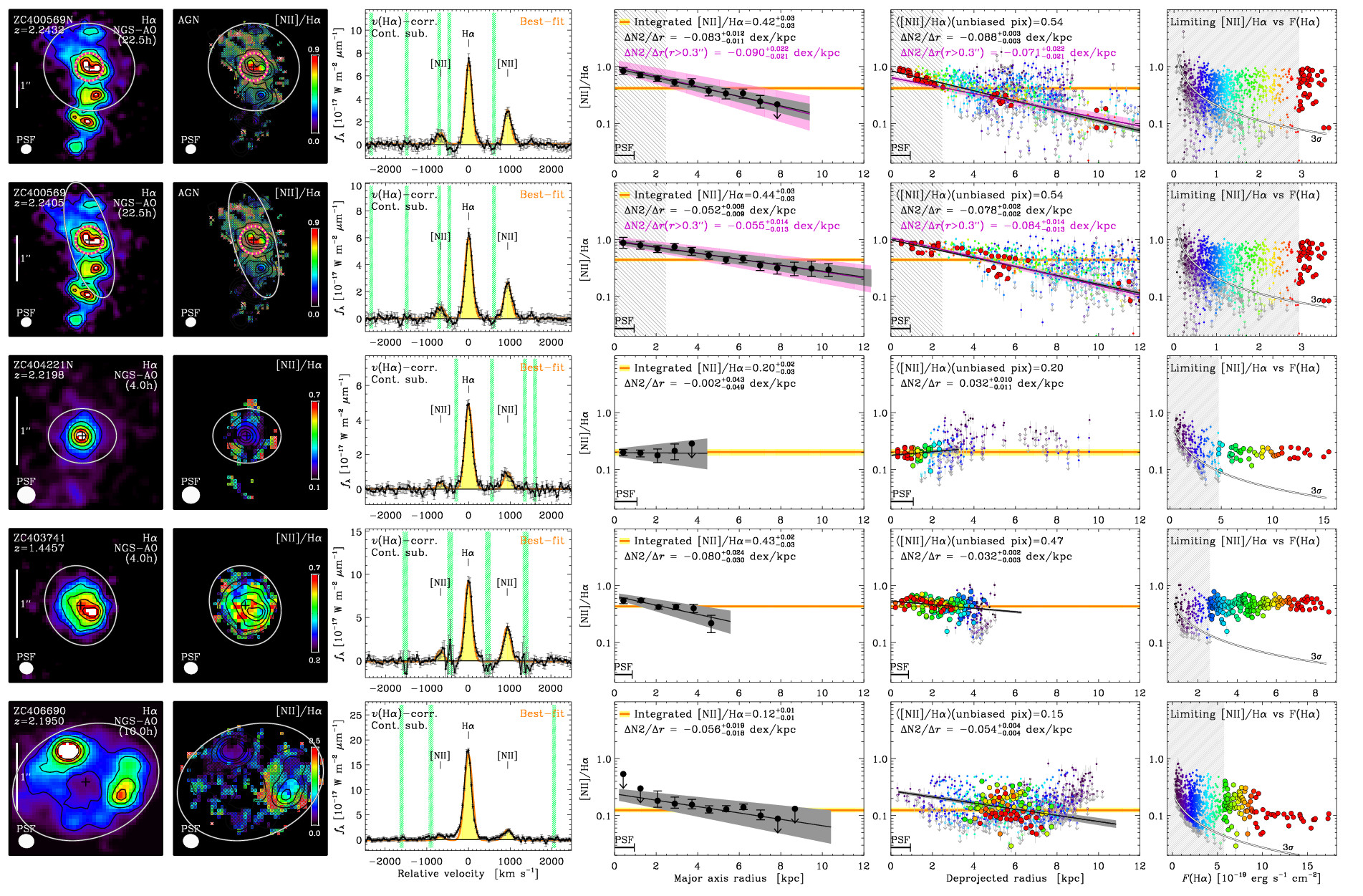}
\end{center}
\vspace{-0.4cm}
\renewcommand\baselinestretch{0.5}
\caption{
\small
(Continued.)
}
\end{figure}


\begin{figure}[p]
\addtocounter{figure}{-1}
\begin{center}
\includegraphics[scale=0.47,clip=1,angle=90]{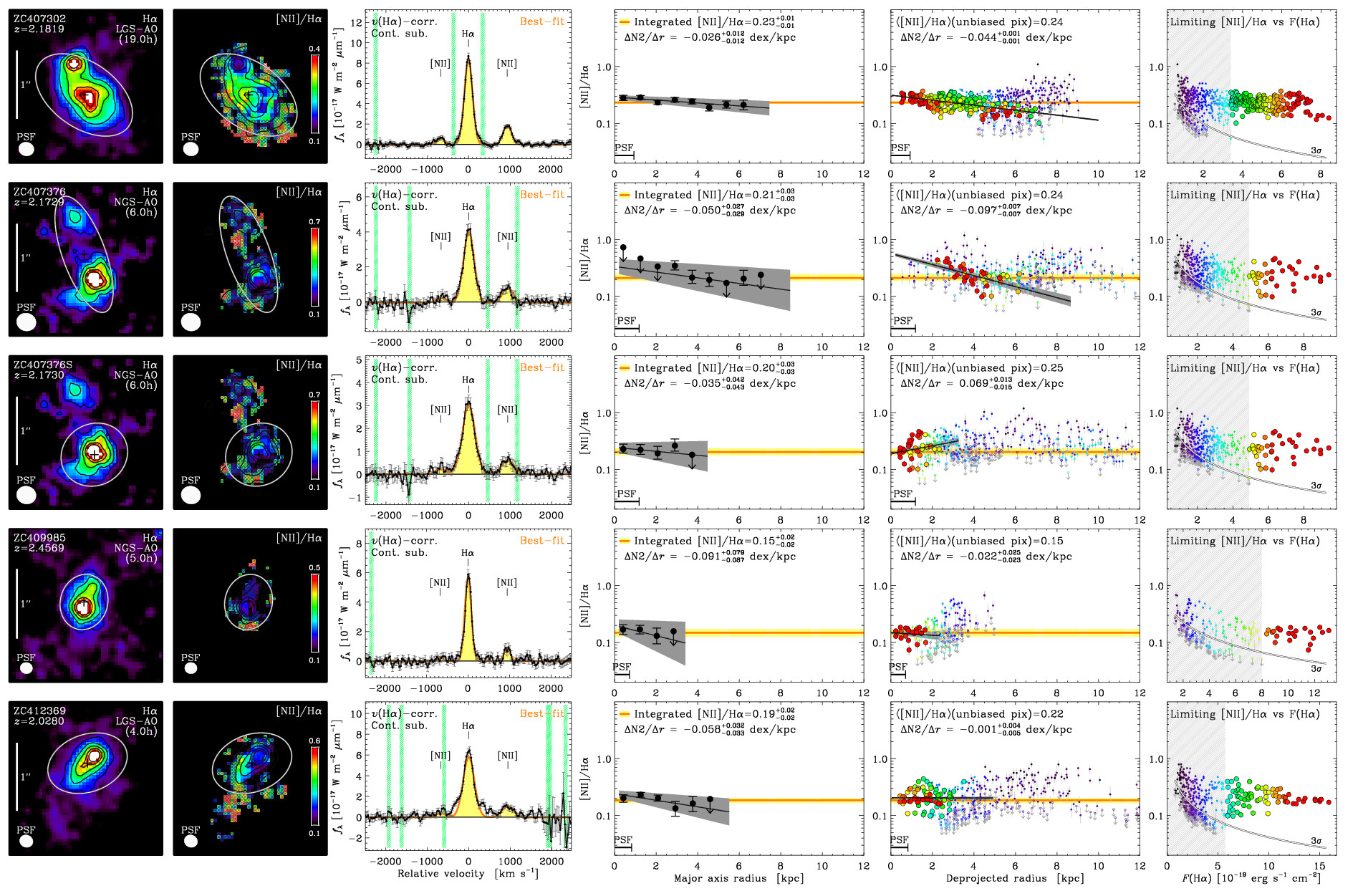}
\end{center}
\vspace{-0.4cm}
\renewcommand\baselinestretch{0.5}
\caption{
\small
(Continued.)
}
\end{figure}

\clearpage

\textwidth=7.45in
\columnwidth=3.45in

\end{document}